\documentclass[degree=doctor,language=english,fontset=fandol]{thuthesis}

\thusetup{
  output = electronic,
  language = english,
  title*  = {Industrial Load Modeling and Optimization for Market-Based Interaction with Power Systems},
  degree-category*  = {Doctor of Philosophy},
  department = {Department of Electrical Engineering},
  discipline*  = {Electrical Engineering},
  author*  = {Ruike Lyu},
  supervisor*  = {Professor Chongqing Kang},
  date = {2026-06-01},
  include-spine = true,
}

\usepackage{amsthm}

\thusetup{
  math-font  = xits,  
}

\usepackage{threeparttable}

\usepackage{multirow}

\usepackage{longtable}

\usepackage{algorithm}
\usepackage{algorithmic}

\usepackage{siunitx}

\usepackage[sort]{natbib}
\graphicspath{{figures/}}

\makeatletter
\newcommand\dif{
  \mathop{}\!%
  \ifthu@math@style@TeX
    d%
  \else
    \mathrm{d}%
  \fi
}
\makeatother

\usepackage{hyperref}

\usepackage[sort]{natbib}

\begin{document}

\makeatletter
\renewcommand\maketitle{%
  \cleardoublepage
  \pagenumbering{Alph}%
  \thu@pdfbookmark{-1}{\thu@title@en}%
  \thu@titlepage@en
  \clearpage
}
\makeatother
\maketitle


\begin{committee}[name={List of public reviewers and defense committee members}]

  \newcolumntype{C}[1]{@{}>{\centering\arraybackslash}p{#1}}

  \section*{Public reviewers}

  \begin{center}
    \begin{tabular}{C{3cm}C{3cm}C{9cm}@{}}
      Chen Shen   & Professor & Department of Electrical Engineering, Tsinghua University \\
      Zechun Hu   & Professor & Department of Electrical Engineering, Tsinghua University \\
      Zhaohao Ding & Professor & North China Electric Power University \\
    \end{tabular}
  \end{center}

  \section*{Defense committee}

  \begin{center}
    \begin{tabular}{C{2.75cm}C{2.98cm}C{4.63cm}C{4.63cm}@{}}
      Chair     & Zongxiang Lu  & Professor            & Tsinghua University \\
      Member    & Chongqing Kang & Professor           & Tsinghua University \\
                & Ning Zhang     & Associate Professor & Tsinghua University \\
                & Yi Wang        & Assistant Professor & The University of Hong Kong \\
                & Zhaohao Ding   & Professor           & North China Electric Power University \\
                & Qingchun Hou   & Research Professor  & Zhejiang University \\
      Secretary & Yanghao Yu     & Assistant Researcher & Tsinghua University \\
    \end{tabular}
  \end{center}

\end{committee}



\frontmatter
\begin{abstract*}
  The transition toward a renewable-dominated power system increases the flexibility needed to balance net-load variations. Industrial loads account for over 40\% of the world's electricity consumption and can shift large amounts of demand through automated production scheduling. However, making industrial processes a dependable flexibility resource for power systems remains challenging, because industrial process constraints are non-convex, equipment parameters are often private, and device-level models of thousands of facilities are too detailed for aggregator and power-system optimization. This dissertation builds computationally efficient methods for modeling and optimizing industrial flexibility for power-system interaction. 
  
  First, this dissertation reformulates two widely used industrial process models---the State Task Network (STN) and Resource Task Network (RTN)---and proposes the Linearized State Task Network (LSTN) and Continuous Resource Task Network (cRTN), which largely reduce the number of integer variables while preserving the accuracy of the original constraints. The reformulation shortens the solution time of a standard grid-interaction case from 24 hours to 30 minutes and thus increases the supported number of coordinated industrial users from 20 to 2,000.

  To estimate private model parameters, this dissertation develops a Production Scheduling Identification (PSI) framework that incorporates industrial process physics and cost-minimizing behavior to narrow the parameter search space. Without internal information, PSI can use hourly smart meter data to establish ready-to-use industrial process models with errors of 5.2\% and 8.5\% using 21 days of training data from cement and steel-powder loads under historical PJM prices; the compared machine-learning models yield errors of 13.4\%--19.2\%.

  To further reduce the model dimensionality, this dissertation proposes a Data-Driven Dimension Reduction (D3R) framework that learns low-dimensional linear constraints from operating samples generated by the original industrial process models. For steelmaking, this method can replace the original model containing 10,208 integer variables with a linear model containing 24--48 variables. Across cement, steelmaking, and steel-powder cases, D3R yields errors of 3.6\%--10.3\%, lower than the comparison methods and within or close to the 10\%--20\% tolerances reported in current demand-side participation rules. 

  For market-based interaction, this dissertation develops a ``dimension-reduced bidding and precise disaggregation'' framework for load aggregators. The bidding model incorporates reduced constraints, multi-period coupling, and stochastic regulation-signal scenarios for global optimization. Shadow prices calculated during bidding are then reused to allocate real-time commands among tens of thousands of resources through millisecond-level arithmetic calculations. In the representative comparison, the framework reduces interaction costs by 40\% relative to a simplified strategy while retaining the solution quality of the joint optimization model.

  The dissertation thus links industrial process physics to the models and decisions that can be integrated into power-system operations.

  \thusetup{
    keywords* = {industrial load, demand response, load modeling, parameter identification, market-based interaction},
  }
\end{abstract*}


\begin{acknowledgements}
  I would like to express my deepest gratitude to my advisor, Professor Chongqing Kang, for his invaluable guidance. Professor Kang taught me the true meaning of academic rigor. His exceptionally high standards, insightful questioning, and continuous push to explore the boundaries of scientific inquiry prevented me from becoming complacent and were instrumental to the completion of this research.

I am also deeply indebted to my co-advisor, Dr.\ Hongye Guo, for his unwavering mentorship and support over the years. Dr.\ Guo has shown me the profound impact a dedicated mentor can have on a student's development. It has been an immense privilege to be advised by Professor Kang and Dr.\ Guo, and I will carry their teachings with me throughout my career.

I would like to thank Professor Qing Xia for his constructive feedback and persistent encouragement, which challenged me to continuously deepen the analytical insights of my research---a habit that will benefit me for a lifetime. My sincere thanks also go to the faculty and members of EILab, particularly Professor Ning Zhang, Professor Haiwang Zhong, and Dr.\ Ershun Du. I am grateful to Professor Zhang and Dr.\ Du for their invaluable assistance during my visiting research abroad, and to Professor Zhong for his continuous support of my role as a student counselor and my overall personal development.

I also wish to thank Professor Daniel Kirschen, Professor Goran Strbac, Dr.\ Jianxiao Wang, Dr.\ Kedi Zheng, Dr.\ Yuxuan Gu, Dr.\ Qinghu Tang, Dr.\ Tengmu Li, and my labmates Chuyi Li, Xiangbo Su, and Yan Shen for their valuable discussions and direct contributions to my doctoral research. I am deeply grateful to all my colleagues in EILab, who not only fostered an exceptional research environment but also shared countless late nights in the lab with me.

During my time as a Visiting Student Research Collaborator at Princeton University, I was fortunate to receive tremendous guidance and support from my host advisor, Professor Jesse Jenkins, and Dr.\ Eric Larson, for which I am truly grateful. I extend my thanks to all the researchers in the ZERO Lab and ESAG, with special thanks to Dr.\ Hongxi Luo and Anna Li for their direct assistance with my work.

I would like to thank the members of my dissertation defense committee for their time and insightful comments on this work. On a personal note, I am endlessly grateful to my parents and family for their unconditional love and support, which allowed me to focus entirely on my research. I also thank my friends for bringing joy and balance to my life. While there are too many people to list individually, I will always remember that it is your collective support that has made this journey possible. Thank you all!

This dissertation was supported by the National Natural Science Foundation of China (Key Program No.\ 52130702 and Young Scientists Fund No.\ 52107102) and the National Key Research and Development Program of China (Projects No.\ 2023YFB2407300 and No.\ 2021YFB2401201).

\end{acknowledgements}

\tableofcontents





\newcommand{\denogroup}[1]{%
  \item[]\hspace*{-\leftmargin}\textbf{#1}
}

\begin{denotation}[4.2cm]

  \denogroup{Abbreviations}
  \item[VPP] Virtual Power Plant
  \item[LA] Load Aggregator
  \item[DER] Distributed Energy Resource
  \item[STN] State Task Network
  \item[RTN] Resource Task Network
  \item[LSTN] Linearized State Task Network
  \item[cRTN] Continuous Resource Task Network
  \item[mSTN] Modified State Task Network
  \item[PSI] Production Scheduling Identification
  \item[D3R] Data-Driven Dimension Reduction
  \item[SCUC] Security Constrained Unit Commitment
  \item[MILP] Mixed Integer Linear Programming
  \item[KKT] Karush--Kuhn--Tucker conditions
  \item[ZO-SGD] Zeroth-Order Stochastic Gradient Descent
  \item[MLP] Multi-Layer Perceptron
  \item[LSTM] Long Short-Term Memory
  \item[SVR] Support Vector Regression
  \item[nRMSE] Normalized Root Mean Square Error

  \denogroup{Notation Conventions}
  \item[Underline $\underline{\cdot}$] Lower bound
  \item[Overline $\overline{\cdot}$] Upper bound
  \item[Superscript $*$] Optimal value
  \item[Superscripts ${\rm ch}$/${\rm dis}$] Charging/discharging, respectively
  \item[Superscript ${\rm init}$] Initial value
  \item[Superscripts ${\rm o}$/${\rm r}$/${\rm a}$] Original constraints, reduced constraints, and approximate model, respectively
  \item[Superscripts ${\rm e}$/${\rm r}$/${\rm cap}$/${\rm mil}$] Energy, regulation, capacity, and mileage, respectively
  \item[Parenthesized superscripts $(n)$/$(k)$] Data sample $n$ or operating scenario $k$, respectively

  \denogroup{Sets and Indices}
  \item[$t$] Time-period index
  \item[$T$] Total number of time periods
  \item[$\mathcal{T}$] Set of time periods
  \item[$i$] Task, device, or resource index
  \item[$I$] Total number of tasks, devices, or resources
  \item[$f$] Factory index
  \item[$\mathcal{F}$] Set of factories
  \item[$k$] State, scenario, or iteration index
  \item[$K$] Total number of states, scenarios, or iterations
  \item[$s$] Regulation-signal scenario index
  \item[$S$] Total number of regulation-signal scenarios
  \item[$\mathcal{S}$] Set of regulation-signal scenarios
  \item[$n$] Data-sample or date index
  \item[$N$] Total number of data samples
  \item[$\Delta t$] Length of a time period
  \item[$j$] Resource index paired with $i$ to represent coupling

  \denogroup{Chapter 2: LSTN Model}
  \item[$i^{\rm end}$] Number of production tasks
  \item[$\mathcal{I}^{\rm P}$] Set of production tasks
  \item[$\mathcal{I}^{\rm S}$] Set of states, including raw materials and products
  \item[$t^{\rm end}$] Final time-period index
  \item[$\mathcal{K}_i$] Set of operating points of task $i$
  \item[$Cost$] Total energy cost
  \item[$Pr_t$] Electricity price in period $t$
  \item[$E_t$] Factory energy consumption in period $t$
  \item[$P_{i,k}$] Power of task $i$ at operating point $k$
  \item[$\Delta t_{t,i,k}$] Continuous time for which task $i$ operates at point $k$ during period $t$
  \item[$S_{t,i}$] Quantity of material $i$ at the end of period $t$
  \item[$S_i^0$] Initial quantity of material $i$
  \item[$S_i^{\rm max}$] Storage limit of material $i$
  \item[$S_i^{\rm tar}$] Target quantity of material $i$
  \item[$G_{i,k}$] Production rate of material by task $i$ at operating point $k$
  \item[$C_{i,k}$] Consumption rate of material by task $i$ at operating point $k$

  \denogroup{Chapter 2: Conventional RTN Model}
  \item[$E_i^{\rm req}$] Energy required to complete task $i$
  \item[$i+$] Index of the task following task $i$
  \item[$N_{i,t}$] Integer variable indicating the start of task $i$
  \item[$P_i^{\rm L}$, $P_i^{\rm H}$] Minimum and maximum processing power of task $i$
  \item[$P_{i,t}$] Power of task $i$ in period $t$
  \item[$E_t^{\rm EL}$] Energy consumption of the production process at time $t$
  \item[$r$] Resource index
  \item[$r^{\rm d}$] Product index at the transfer destination
  \item[$r^{\rm f}$] Product index at the final stage
  \item[$r^{\rm s}$] Product index at the transfer origin
  \item[$R_{r,t}$] Quantity of resource $r$ at time $t$
  \item[$S_{i,t}$] Processing state of task $i$ in period $t$
  \item[$\tau_i^{\rm L}$, $\tau_i^{\rm H}$] Minimum and maximum processing times of task $i$
  \item[$W_{r^{\rm d}}$] Maximum waiting time of resource $r^{\rm d}$
  \item[$w_{r^{\rm d}}$] Transfer time of resource $r^{\rm d}$
  \item[$\gamma_{{\rm EL},i,\theta}$] Energy consumed $\theta$ periods after task $i$ starts
  \item[$\gamma_{r,i,\theta}$] Resource $r$ consumed or produced $\theta$ periods after task $i$ starts

  \denogroup{Chapter 2: Continuous RTN Model}
  \item[$D_{i,k,t}$] Time for which task $i$ remains in state $k$ during period $t$
  \item[$\mathbf{G}$] Task--resource incidence matrix
  \item[$g_{r,i,k}$] Resource $r$ produced per unit time by task $i$ in state $k$; negative values denote consumption
  \item[$R_{i^{\rm end}}^{\rm tg}$] Target amount of the final-stage product
  \item[$R_{r^{i+},t}$] Progress of the task that produces resource $r^{i+}$
  \item[$R_{r^{i-},t}$] Resource consumed by task $i$
  \item[$r^{i+}$] Resource produced by task $i$
  \item[$r^{i-}$] Resource consumed by task $i$
  \item[$u_{i,t}$] Processing state of task $i$ in period $t$; $u_{i,t}=1$ ($0$) denotes an ongoing (completed) process
  \item[$w_i$, $W_i$] Transfer time and maximum waiting time of task $i$

  \denogroup{Chapter 3: Parameters and Variables}
  \item[$Pr_t^{(d)}$] Electricity price in period $t$ on day $d$
  \item[$E_t^{(d)}$] Metered electricity consumption in period $t$ on day $d$
  \item[$E_{dt}$] Modeled electricity consumption in period $t$ on day $d$
  \item[$Cost_d$] Energy cost on day $d$
  \item[$g_i$] Production output achieved by task $i$ per unit of energy consumed
  \item[$P_{dti}$] Operating power of task $i$ in period $t$ on day $d$
  \item[$P_i^{\rm max}$] Rated power of task $i$
  \item[$S_{dti}$] State or quantity of material $i$ at the end of period $t$ on day $d$
  \item[$S_i^{\rm 0/max/tar}$] Initial, maximum, and target state of material $i$
  \item[$\theta^{(d)}$] Model parameters updated for day $d$
  \item[$\mu_{di}^{\rm Star}$] Dual variable of the production-target constraint
  \item[$\mu_{dti}^{\rm Smin/Smax}$] Dual variables of the lower and upper state constraints
  \item[$\mu_{dti}^{\rm Pmin/Pmax}$] Dual variables of the lower and upper power constraints
  \item[$\lambda_{ti}^{S}$] Dual variable of the state-transition constraint
  \item[$\alpha$] Adaptive decay parameter in model identification

  \denogroup{Chapter 4: Industrial-Load Constraint Reduction}
  \item[$x_t$] Net electricity consumption in period $t$ (kWh)
  \item[$\mathbf{x}^{(n)}$] Vector of hourly electricity consumption on day $n$
  \item[$\mathbf{Pr}^{(n)}$] Vector of hourly electricity prices on day $n$
  \item[$h(\cdot)$] Original high-dimensional constraint function
  \item[$\mathbf{y}^{\rm o}$] Internal variables of the original constraints
  \item[$\mathbf{y}^{\rm r}$] Internal variables of the reduced constraints
  \item[$\mathbf{A}$, $\mathbf{b}$] Parameter matrix and right-hand side of $\mathbf{A}\mathbf{x}\leq\mathbf{b}$
  \item[$J(\cdot)$] Loss function measuring approximation error
  \item[$HD$] Historical dataset $\{(\mathbf{Pr}^{(n)},\mathbf{x}^{(n)})\}$
  \item[$p_{t,i}$] Average power consumption of adjustable load $i$ in period $t$ (kW)
  \item[$\underline{P}_i$, $\overline{P}_i$] Lower and upper power bounds of adjustable load $i$ (kW)
  \item[$\underline{E}_i$, $\overline{E}_i$] Lower and upper daily-energy bounds of adjustable load $i$ (kWh)
  \item[$\underline{\mu}^{\rm P}_{t,i}$, $\overline{\mu}^{\rm P}_{t,i}$] Lagrange multipliers of the power-limit constraints
  \item[$\underline{\mu}^{\rm E}_{i}$, $\overline{\mu}^{\rm E}_{i}$] Lagrange multipliers of the energy-limit constraints
  \item[$z_{t,i}$] Binary variable used to reformulate complementary slackness
  \item[$M$] Large positive constant in the Fortuny--Amat reformulation
  \item[$B$] Batch size in the iterative algorithm
  \item[$\alpha$] Learning rate in the iterative algorithm

  \denogroup{Chapter 4: Distributed-Resource Aggregation}
  \item[$p_t$] Aggregator energy output or baseline power in period $t$ (kW)
  \item[$r_t$] Aggregator regulation capacity in period $t$ (kW)
  \item[$p^{\rm dis}_{t,i}$, $p^{\rm ch}_{t,i}$] Discharging and charging power of resource $i$ in period $t$ (kW)
  \item[$\overline{p}^{\rm dis(ch)}_{t,i}$] Charging or discharging power limit of resource $i$ in period $t$ (kW)
  \item[$e_{t,i}$] Energy state of resource $i$ at the end of period $t$ (kWh)
  \item[$e_{0,i}$] Initial energy of resource $i$ (kWh)
  \item[$\underline{e}_{t,i}$, $\overline{e}_{t,i}$] Lower and upper energy bounds of resource $i$ in period $t$ (kWh)
  \item[$\eta^{\rm ch}$, $\eta^{\rm dis}$] Charging and discharging efficiencies
  \item[$\Delta t^{\rm req}$] Required duration of maximum regulation output (h)
  \item[$\mathcal{P}$] Original feasible region of the load aggregator
  \item[$\mathcal{P}^{\rm a}$] Approximate feasible region of the load aggregator
  \item[$h^{\rm a}(\cdot)$] Constraint function of the approximate model
  \item[$\theta^{\rm a}$] Parameter set of the approximate model
  \item[$I^{\rm a}$] Number of resources in the approximate model
  \item[$Pr_t^{\rm r}$] Equivalent regulation-capacity price in period $t$
  \item[$Pr_t^{\rm cap}$, $Pr_t^{\rm mil}$] Regulation-capacity and mileage prices in period $t$
  \item[$s^{\rm perf}$] Aggregator performance score
  \item[$a_t^{\rm mil}$] Expected regulation mileage in period $t$
  \item[$Pr^{\rm deg}(\cdot)$] Resource-degradation cost function
  \item[$\lambda^{\rm e}_{t,i}$] Lagrange multiplier of the energy-transition constraint
  \item[$\lambda^{\rm e0}_{i}$] Lagrange multiplier of the initial-energy constraint
  \item[$\underline{\mu}^{\rm pd(c)}_{t,i}$, $\overline{\mu}^{\rm pd(c)}_{t,i}$] Lagrange multipliers of the charging/discharging power limits
  \item[$\underline{\mu}^{\rm e}_{t,i}$, $\overline{\mu}^{\rm e}_{t,i}$] Lagrange multipliers of the energy limits
  \item[$\underline{\mu}^{r}_{t,i}$] Lagrange multiplier of the nonnegative regulation-capacity constraint
  \item[$\mu^{rpd}_{t,i}$, $\mu^{rpc}_{t,i}$] Lagrange multipliers of the regulation power-capacity limits
  \item[$\mu^{rec}_{t,i}$, $\mu^{red}_{t,i}$] Lagrange multipliers of the regulation-duration constraints
  \item[$D$] Price--dispatch dataset
  \item[$g(\mu,\lambda)$] Lagrangian dual function

  \denogroup{Chapter 5: Parameters and Variables}
  \item[$\hat{t}$] Current time
  \item[$\Delta\hat{t}$] Effective duration of the regulation signal
  \item[$p_{t,s,i}$] Net output of resource $i$ in period $t$ under scenario $s$
  \item[$p^{\rm dis}_{t,s,i}$] Discharging power of resource $i$ in period $t$ under scenario $s$
  \item[$p^{\rm ch}_{t,s,i}$] Charging power of resource $i$ in period $t$ under scenario $s$
  \item[$\underline{p}^{\rm dis(ch)}_{t,i}$, $\overline{p}^{\rm dis(ch)}_{t,i}$] Lower and upper charging/discharging power bounds
  \item[$p_k$] Output of power segment $k$
  \item[$\underline{p}_k$, $\overline{p}_k$] Lower and upper bounds of power segment $k$
  \item[$\tilde{p}_t$] Aggregator energy bid or baseline power in period $t$
  \item[$p^{\rm req}$] Required net output of the load aggregator
  \item[$e_{t,i}$] State of resource $i$ at the beginning of period $t$
  \item[$e_i^{\rm init}$] Initial state of resource $i$
  \item[$\underline{e}_{t,i}$, $\overline{e}_{t,i}$] Lower and upper state bounds of resource $i$ in period $t$
  \item[$\rho_i$] State-dissipation rate of resource $i$
  \item[$w_{t,i}$] Effect of the external environment on the state of resource $i$
  \item[$\delta$] Regulation signal in $[-1,1]$
  \item[$\delta_s$] Representative regulation-signal value of scenario $s$
  \item[$\pi_s$] Probability of scenario $s$
  \item[$\boldsymbol{\Pi}_t$] Vector of scenario probabilities in period $t$
  \item[$\eta^{\rm ch}_{ij}$] Effect of charging resource $j$ on the state of resource $i$
  \item[$\eta^{\rm dis}_{ij}$] Effect of discharging resource $j$ on the state of resource $i$
  \item[$\mathbf{H}^{\rm ch}$, $\mathbf{H}^{\rm dis}$] Charging/discharging efficiency incidence matrices
  \item[$\mathbf{F}$] State-transition matrix
  \item[$Pr_i^{\rm dis}$, $Pr_i^{\rm ch}$] Discharging and charging cost coefficients of resource $i$
  \item[$Pr_t^{\rm e}$] Energy-market price in period $t$
  \item[$c_k$] Equivalent marginal-cost coefficient of power segment $k$
  \item[$Cost_t$] Operating cost in period $t$
  \item[$Cost_i^{\rm imd}$] Immediate operating cost of resource $i$
  \item[$Cost_i^{\rm prf}$] Opportunity cost of the state change of resource $i$
  \item[$Income_t^{\rm e}$] Energy-market revenue in period $t$
  \item[$Income_t^{\rm r}$] Regulation-market revenue in period $t$
  \item[$Profit$] Aggregator profit over the study horizon
  \item[$\lambda^e_{t,i}$] Lagrange multiplier or shadow price of the state constraint of resource $i$
  \item[$\lambda^{\rm bal}_{t,s}$] Lagrange multiplier of the power-balance constraint in period $t$ under scenario $s$
  \item[$\underline{\mu}^{\rm dis(ch)}_{t,s,i}$, $\overline{\mu}^{\rm dis(ch)}_{t,s,i}$] Lagrange multipliers of the power limits
  \item[$\mathbf{1}_{\rm I}$] All-ones vector

\end{denotation}

\mainmatter

\chapter{Introduction}

\section{Background and Motivation}

\begin{figure}[htbp]
    \centering
    \includegraphics[width=\textwidth]{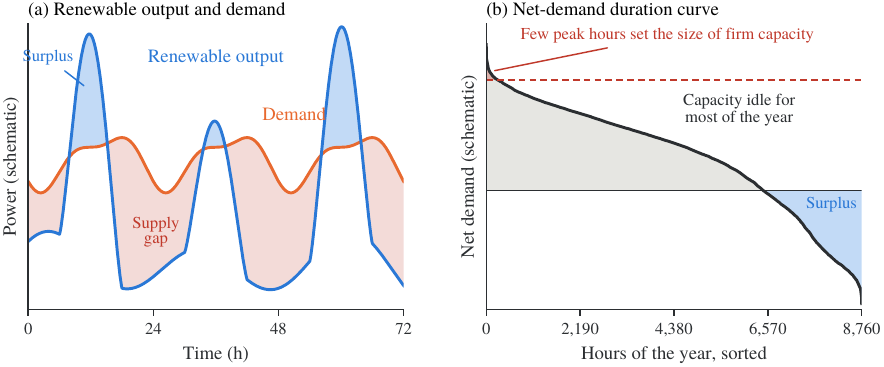}
    \caption[Net load with high renewable generation]{Schematic of net load with high renewable generation: (a) renewable output and electricity demand over three days; (b) annual net load sorted in descending order (duration curve). The net load is defined as the difference between electricity demand and renewable generation.}
    \label{fig:net_load_curve}
\end{figure}

The global power grid is accelerating the transition toward a renewable-dominated, new-type power system. However, the continuously increasing share of renewable energy introduces significant operational challenges~\cite{ref1,ref2}. If electricity demand remains fixed, the fluctuating output of renewable generation will cause steep peak-to-valley differences in the system's net load curve. This necessitates substantial flexible resources to balance electricity supply and demand (as illustrated schematically in Fig.~\ref{fig:net_load_curve}; see also~\cite{ref3}). Since the duration of peak loads or extreme ramping events is relatively short, building redundant flexible resources solely to meet these demands would significantly inflate power system investment costs. Furthermore, these resources would suffer from low utilization rates during off-peak periods, resulting in resource waste. As an alternative, developing demand-side flexibility to ``smooth out'' the net load curve has emerged as a cost-effective strategy for reducing the cost of ensuring reliable power supply.

In recent years, a series of national-level policies have been issued to promote the coordinated interaction among sources, grids, loads, and energy storage within China's power system~\cite{ref5,ref6,ref7}. In 2024, the National Energy Administration issued the \textit{Action Plan for Accelerating the Construction of a New-Type Power System (2024--2027)} and the \textit{Guiding Opinions on Supporting the Innovative Development of New Business Entities in the Electricity Sector}. Both documents emphasize strengthening grid-load interaction and mandate the participation of new entities such as VPPs in the electricity spot market~\cite{ref5,ref6}. Furthermore, the \textit{Implementation Plan for the Special Action on Optimizing Power System Regulation Capability (2025--2027)}, released in late 2024, proposed that by deeply tapping into the flexibility potential of load-side resources, the system should ensure the effective utilization of at least 200 GW of newly installed renewable energy capacity annually~\cite{ref7}. These policies make the market-based use of demand-side flexibility a stated direction of China's power-system transition.

\begin{figure}[htbp]
    \centering
    \includegraphics[width=\textwidth]{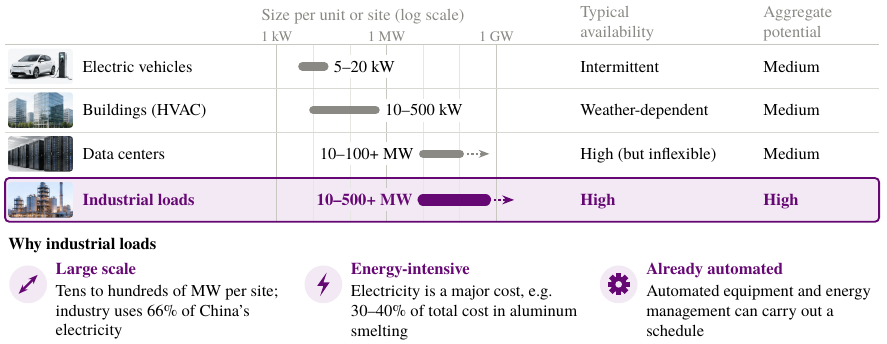}
    \caption{Industrial loads as a source of power-system flexibility, with typical unit-capacity ranges}
    \label{fig:industrial_load_motivation}
\end{figure}

This dissertation focuses on industrial users because their large loads and price sensitivity give them both the motivation and the capability to participate in grid interaction. First, the industrial sector accounts for 66\% of China's total electricity consumption~\cite{ref8} and exhibits a concentrated consumption pattern (for instance, electric arc furnaces account for over 80\% of the total electricity used in the secondary steelmaking process~\cite{ref10}). Despite this, the development of industrial demand-side flexibility in China remains in its early stages, leaving interaction potential untapped~\cite{ref9}. Second, electricity expenses constitute a substantial proportion of the product cost structure for industrial users---accounting for 30\%--40\% of total costs in the aluminum smelting industry, for example~\cite{ref11}. This financial aspect motivates them to engage in demand response programs to reduce operational costs. Furthermore, industrial users generally maintain a high degree of equipment automation. This provides a technological foundation for power system interaction, enabling them to adjust their consumption patterns in response to grid requirements. For these reasons, industrial users are the main subject of this dissertation (Fig.~\ref{fig:industrial_load_motivation}).

Although the numerical studies use secondary steelmaking, steel-powder manufacturing, and cement production as representative cases, the scope of this dissertation is industrial loads in a broader sense. The LSTN/cRTN framework in Chapter 2 reformulates the general STN/RTN process-model families at the time scale of grid interaction and is not tied to one steel grade or industrial park. Loads with similar material--task network structures can reuse the model form by replacing process and equipment parameters. Processes with strong temperature--efficiency coupling, such as aluminum smelting, or highly customized production routes require dedicated physical models before entering the subsequent identification and dimension-reduction workflow. Table~\ref{tab_industrial_load_scope} in Chapter 2 summarizes the industrial-load types covered by, and currently outside, the proposed modeling scope.

However, industrial loads are consumers rather than generating units; it is technically difficult to embed complex industrial technical constraints directly into economic dispatch or market clearing models for joint optimization. Also, grid operators cannot directly control the production operations of industrial facilities. A more feasible approach is market-based grid-load interaction. Under this framework, intermediaries such as VPPs optimize the industrial users' energy consumption and electricity purchasing within the electricity market.
When such a VPP further coordinates heterogeneous resources---including distributed generation, energy storage, and electric vehicles alongside industrial loads---it is capable of bidding into both energy and ancillary service markets.

Adopting the perspective of a VPP, this dissertation investigates the challenges of integrating industrial users into grid interactions within a market environment. These challenges, spanning facility-level modeling and system-level interaction (Fig.~\ref{fig:intro_challenges}), are summarized in the following four aspects:

\begin{figure}[htbp]
    \centering
    \includegraphics[width=\textwidth]{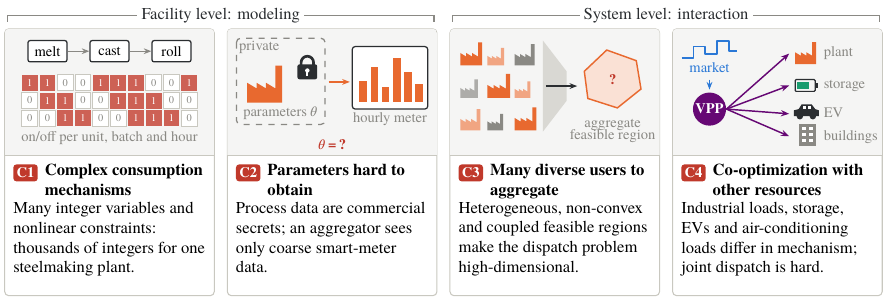}
    \caption{Schematic of the main challenges in using industrial loads for power-system flexibility}
    \label{fig:intro_challenges}
\end{figure}

(1) \textbf{Challenge 1: Characterizing complex electricity consumption mechanisms}. The production processes of industrial users are governed by energy and material conversion mechanisms. Accurately describing their operating dynamics requires numerous integer variables and nonlinear constraints, often leading to combinatorial explosions. These complexities make it difficult to embed industrial models directly into upper-level market clearing and power system dispatch problems. For example, modeling a single user's typical secondary steelmaking process could require thousands of integer variables, leading to the ``curse of dimensionality'' when scaled for grid interaction. Even for a VPP, integrating these detailed industrial load models into operational optimization remains computationally challenging.

(2) \textbf{Challenge 2: Acquiring accurate equipment parameters}. Production parameters often involve commercial secrets, making industrial users reluctant to share process data with VPPs. Furthermore, the diversity of industrial users results in heterogeneous data structures, hindering the creation of unified datasets required for market participation. Without extensive sub-metering, the data a VPP can typically access is coarse-grained smart meter data of individual facilities. This aggregate data is generally insufficient to reveal internal process characteristics or equipment states, limiting modeling accuracy. Consequently, the operational flexibility of industrial users is difficult to assess reliably.

(3) \textbf{Challenge 3: Describing massive demand-side resource constraints}. Aggregating the flexibility of industrial users is complicated by their heterogeneity. The operational feasible regions and adjustment costs of individual industrial loads exhibit non-convex characteristics and are coupled through storage limits, temperature ranges, and assembly line coordination. This complexity renders precise, adaptive aggregation difficult. In grid-load interaction, this causes the dimensionality of the co-optimization problem to increase, making it challenging to solve within acceptable timeframes.

(4) \textbf{Challenge 4: Coordinating diverse demand-side resources}. The control mechanisms governing different industrial loads and other demand-side resources vary significantly, complicating coordinated optimization. Different industrial loads operate under distinct mechanisms that are difficult to standardize. Moreover, the differences between industrial loads and other flexible resources (e.g., energy storage, electric vehicles, and thermostatically controlled loads) limit the ability to fully exploit the overall flexibility of the demand side. This heterogeneity makes it challenging to dispatch disparate resources uniformly in market-based VPP operations.

Unlike generating units, for which mature and standardized operating models are available, industrial loads have process-specific constraints and limited parameter transparency. The central problem addressed in this dissertation is therefore \textbf{how to map industrial energy-use mechanisms into a concise dispatch space using coarse-grained smart meter data as the external measurement interface.} Such a representation must be simple enough for aggregator and power-system optimization but detailed enough to produce executable industrial schedules.

This dissertation connects industrial processes to market operation in four steps: mechanism modeling, parameter identification, constraint dimension reduction, and optimal decision-making. It uses 15-minute or hourly smart meter data as the user-to-grid measurement interface and retains precise process models for final execution. Industrial flexibility can therefore be estimated and aggregated without placing every device-level constraint in the market model.

Academically, this work develops a connected route for making high-dimensional industrial mechanism models identifiable from coarse external measurements and embeddable in electricity-market optimization. Practically, it targets applications that load aggregators and other new market entities can implement under existing metering conditions, with the potential to lower data and modeling barriers and improve the efficiency and predictability of organizing industrial flexibility.

\section{Literature Review}
As the deployment of the new-type power system accelerates, research into industrial user participation in grid interaction has increased globally. This section reviews the literature across two dimensions: facility-level industrial user modeling/identification, and system-level resource aggregation/optimization. First, the theoretical landscape of industrial user modeling is analyzed, encompassing flexibility characterization, cost modeling, parameter identification, and potential assessment. Second, progress in aggregation and optimization decision-making methods for distributed resources is summarized.

\subsection{Industrial Load Modeling for Electricity Market Participation}
To harness industrial flexibility without violating technical constraints, mathematical models must capture the energy-material conversion processes at each production stage. Unlike residential users, industrial loads are characterized by complex mechanisms that introduce nonlinearity into the models~\cite{ref17,ref171}. This is particularly true for energy-intensive industries (e.g., steel, cement, and aluminum) that possess incentives for demand response. These industries feature high stage-specific energy consumption, complex workflows, and modeling challenges arising from the discontinuity of material conversions~\cite{ref18,ref29}. Consequently, the resulting mathematical formulations are often complex, rendering optimization computationally demanding~\cite{ref19,ref30}.

To balance accuracy and computational tractability, existing research frequently employs generic industrial process models to establish unified mathematical formulations. Two widely adopted frameworks are the State Task Network (STN) and the Resource Task Network (RTN). \inlinecite{ref20} proposed the STN model, which represents materials as states and production processes as tasks. As a standardized pipeline modeling approach, STN has been applied in the cement and chemical industries~\cite{ref21,ref22,ref31}. Building upon this, \inlinecite{ref23} introduced the RTN model, which uniformly treats equipment and intermediate products as resources. RTN characterizes practical constraints in discrete processes, such as startup/shutdown sequences and waiting times, improving modeling accuracy for processes like secondary steelmaking. Recently, these models have been leveraged for research in electricity-carbon co-optimization and multi-energy coupling~\cite{ref24,ref25,ref26,ref32}.

However, conventional STN and RTN models face limitations for large-scale coordinated optimization. They rely heavily on integer variables to describe discrete operating states and nonlinear process switching, limiting their applicability for grid interaction. As industrial demand response scales, the mathematical formulations of STN and RTN require refinement, particularly for scenarios involving multi-user coordination or integration into upper-level dispatch. Although some studies, such as \inlinecite{ref26}, have explored adding cutting planes to RTN-based scheduling to improve tractability, the core modeling technology remains largely unchanged. Similarly, \inlinecite{ref27} proposed continuous-time modeling, but focused on load-following accuracy rather than model scalability.

Overall, the scalability of STN and RTN models requires more attention. Few studies appear to have addressed this computational complexity through advanced modeling techniques. \textbf{An ongoing challenge is improving the computational efficiency of these generic models—especially RTN for discrete processes—without sacrificing accuracy.}

\subsection{Parameter Identification for Industrial Load Modeling}
In grid-interactive settings, industrial load model parameters (e.g., rated power, yield rates, production targets) are highly specific to individual facilities and often protected as commercial secrets. For VPP operators, the lack of parameter transparency makes it difficult to aggregate flexibility or optimize interactions effectively~\cite{ref33}. Therefore, identifying production parameters and assessing operational flexibility under incomplete information is a major challenge.

Regarding the types of parameters to be identified, the inherent diversity of industrial processes means that models for different users possess distinct structures. In large-scale scenarios, model structures must be standardized before parameter identification can be efficiently executed in batch. While \inlinecite{ref34} successfully constructed standardized models for residential and commercial resources participating in frequency regulation, the continuous, coupled nature of industrial production makes such similarity-based standardization challenging to apply.

In terms of available information, the objective is to model industrial users using accessible external data, bypassing privacy barriers. The deployment of smart meters offers a potential approach~\cite{ref36,ref37}. VPPs typically monitor hourly electricity consumption at the facility level. While this coarse-grained, aggregated data protects privacy, it harbors valuable information for user modeling~\cite{ref39,ref41}. Consequently, inferring internal parameters from smart meter data has become an active research direction.

Methodologically speaking, Non-Intrusive Load Monitoring (NILM) is related to this objective, aiming to disaggregate facility-level meter data into device-level load curves~\cite{ref38}. However, industrial NILM cannot extract key modeling parameters like material storage limits and typically requires high-frequency data (e.g., 60 Hz)~\cite{ref40}, which standard smart meters cannot provide. Due to these data resolution limits, current smart-meter-based research primarily relies on techniques like classification and clustering~\cite{ref42,ref43,ref44,ref45}. For instance, \inlinecite{ref46} utilized decision trees to estimate demand response potential, while \inlinecite{ref48} proposed a multi-time-scale assessment framework based on load steps. While effective for behavioral classification, these methods cannot infer specific equipment constraints. \textbf{Balancing privacy and implementability while extracting key parameters (e.g., device power, efficiency, inventory limits) solely from hourly aggregate meter data remains a challenge.}

\subsection{Constraint Dimension Reduction for Demand-Side Resource Models}
To manage load populations, the flexibility of various demand-side resources must be quantified. Because the number of demand-side resources exceeds that of generators, modeling each entity individually results in an unmanageable state space. Consequently, demand-side resources must be simplified into low-dimensional constraints for optimization~\cite{ref70}. ``Dimension reduction'' encompasses all levels of constraint simplification, while ``aggregation'' specifically refers to combining multiple devices into virtual entities.

The standard approach to dimension reduction involves establishing standardized models and then aggregating them. \inlinecite{ref51} pioneered the concept of polyhedral feasible regions, abstracting the flexibility of distributed generators, storage, and thermostatically controlled loads into polyhedra—a method widely adopted in subsequent research~\cite{ref52,ref53,ref55,ref56}. However, the mathematical abstraction of polyhedra often lacks explicit physical meaning, complicating its integration into economic dispatch frameworks. Alternatively, resources can be mapped to generic virtual devices (e.g., virtual generators or batteries) with unified parameters~\cite{ref54}. \inlinecite{ref57} expanded on this by standardizing multi-time-scale regulation characteristics. Yet, fitting complex industrial loads into predefined virtual device templates often yields external characteristic errors and poor adaptability.

Regarding aggregation algorithms, combining flexibility resources without considering network topology is mathematically equivalent to computing the Minkowski sum of polyhedra~\cite{ref60}, a classic NP-hard problem~\cite{ref601}. Consequently, research relies on approximation algorithms, broadly categorized as outer approximation~\cite{xu_hierarchical_2016,ref58} and inner approximation~\cite{ref59,ref72}. When network constraints are introduced~\cite{chen_aggregate_2020}, the problem evolves into projecting high-dimensional feasible regions onto lower-dimensional subspaces~\cite{ref61,ref68}. Researchers have attempted to simplify this via heuristic assumptions~\cite{ref65}, Fourier-Motzkin elimination~\cite{ref67}, clustering~\cite{ref70}, and time-period decomposition~\cite{ref68}. 

Fundamentally, these analytical methods face limitations in industrial scenarios. First, predefined geometric templates lack the adaptability required to handle the diverse constraint structures of different industries~\cite{chen_leveraging_2021}. Second, analytical aggregation requires convex constraints. Because industrial models rely on integer variables to capture discrete states, their feasible regions are inherently non-convex, rendering traditional analytical methods ineffective.

Data-driven methods offer an alternative. A common approach involves generating operating points from the precise model, labeling their feasibility, and training a convex quadratic classifier to approximate the aggregated region~\cite{ref621}. However, due to temporal coupling~\cite{ref62}, the required training data scales with the number of time periods. Furthermore, the resulting classifiers often require additional transformations to be compatible with market bidding models, introducing further errors. \textbf{While data-driven methods hold promise for overcoming the adaptability issues of analytical aggregation, handling the non-convexities and unique geometric shapes characteristic of industrial loads requires specialized research.}

\subsection{Grid-Load Coordinated Optimization Considering Demand-Side Flexibility}
In direct load control scenarios, characterized user behavior models can be embedded directly into grid optimization models~\cite{ref75}. For instance, \inlinecite{ref76} integrated industrial production constraints into a park-level energy system dispatch, while \inlinecite{ref77} embedded prospect theory-based EV user behavior into building cluster optimizations. However, for large numbers of small-capacity controllable loads, embedding individual models increases computational complexity, making hierarchical control architectures more appropriate~\cite{ref84}.

Price incentives represent one hierarchical control approach, wherein operators use pricing strategies to guide user behavior toward peak shaving~\cite{ref86}, voltage regulation~\cite{ref85}, and reliability enhancement~\cite{ref87}. Success relies on accurately modeling user price responsiveness. Due to computational limits, system optimizations often rely on simplified statistical models, such as price elasticity matrices~\cite{ref87}, cross-substitution elasticity~\cite{ref88}, or load transfer probabilities~\cite{ref85}. Recognizing diverse user sensitivities, studies have refined these models by categorizing users~\cite{ref89} or applying differentiated elasticity parameters~\cite{ref90}. Others formulate bilevel Stackelberg games between operators and users~\cite{ref91}. However, this body of literature predominantly targets residential and commercial buildings, largely overlooking the specific physical mechanisms of industrial production.

The second hierarchical approach utilizes an aggregation-disaggregation framework via VPPs. VPPs must first determine their market bids by assessing regulation capacity across time periods~\cite{he_optimal_2016,sortomme_optimal_2011,sortomme_optimal_2012, jin_optimizing_2013, vagropoulos_optimal_2013, ansari_coordinated_2015, sarker_optimal_2016, vatandoust_risk-averse_2019, duan_bidding_2021}. Upon receiving grid dispatch signals, they must disaggregate the required net power adjustment among internal resources~\cite{escudero-garzas_fair_2012, sun_real-time_2014, peng_optimal_2017, vagropoulos_real-time_2016}. To optimize economic performance, an ideal disaggregation strategy should prioritize the deployment of lower-cost resources~\cite{he_cooperation_2017}.

Achieving optimal power disaggregation is difficult due to temporal coupling. Adjusting the output of storage-like resources (e.g., EVs) in one period alters their future states, impacting subsequent revenues. For industrial users, this coupling is magnified: stages are linked sequentially, meaning an adjustment in one stage alters intermediate inventories, thereby constraining the facility's future flexibility. Theoretically, optimal response requires continuous, full-horizon stochastic optimization. Given that dispatch signals arrive frequently, solving such online optimizations is computationally prohibitive.

To address this, proportional allocation is widely used, dispatching resources according to their preset capacity shares~\cite{sortomme_optimal_2012,vatandoust_risk-averse_2019,sadeghi_optimal_2021, yi_robust_2022}. While mathematically simple, this strategy fails to exploit resource heterogeneity, suboptimizing economic performance. It also requires simultaneous adjustment of all resources, increasing communication overhead. Consequently, proportional disaggregation is less suited for comprehensive demand response.

Other studies use heuristic priority rules targeting capacity maximization~\cite{peng_optimal_2017}, profit fairness~\cite{escudero-garzas_fair_2012}, or minimal operational disruption~\cite{vagropoulos_real-time_2016}. However, lacking full-horizon foresight, these heuristics do not guarantee long-term optimality. Optimal disaggregation requires comprehensive modeling of resource differences, temporal coupling, bidding revenues, and signal uncertainty, while maintaining low computational times~\cite{vagropoulos_real-time_2016}. \textbf{For VPPs managing industrial users, achieving rapid, optimal power disaggregation while honoring complex production constraints remains an area for ongoing research.}

\section{Research Approach and Main Work of This Dissertation}

\subsection{Research Approach}
\begin{figure}[htbp]
    \centering
    \includegraphics[width=\textwidth]{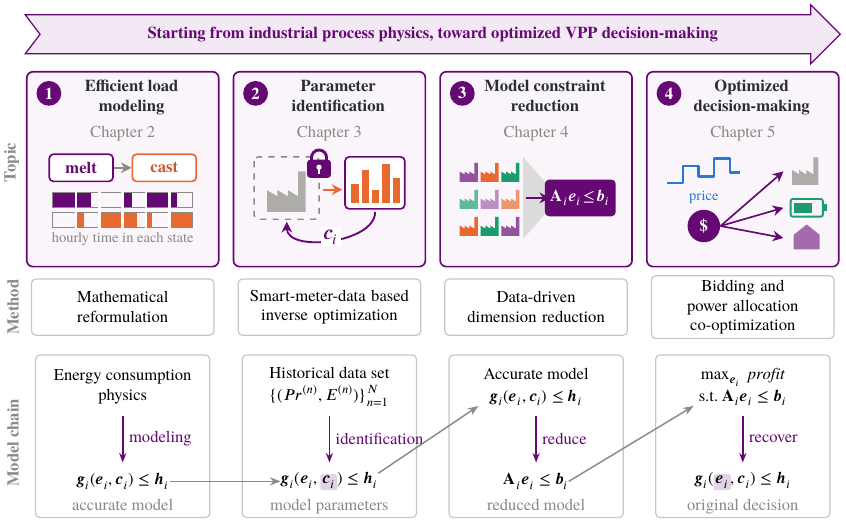}
    \caption{Research approach of this dissertation}
    \label{fig:route}
\end{figure}

Industrial users offer substantial demand-response capacity, but differences across production processes and limited access to private parameters prevent direct use of that capacity in market optimization. This dissertation adopts a four-step research approach (Fig.~\ref{fig:route}). Facility-level modeling and parameter identification provide the basis for system-level dimension reduction and coordinated optimization:

(1) \textbf{Computationally Efficient Load Modeling:} This dissertation captures the physical mechanisms of industrial processes and reformulates their mathematical structures to improve solver compatibility without discarding the accuracy of process constraints.

(2) \textbf{Load Parameter Identification:} The proposed identification method uses hourly smart meter data to estimate the parameters needed to describe the load flexibility.

(3) \textbf{Model Constraint Reduction (Aggregation):} Given the 10\%--20\% error margins for typical grid-load interaction, this dissertation reduces high-dimensional constraints to linear approximations, with an acceptable degree of accuracy but substantially lower computational complexity.

(4) \textbf{Optimized Decision-Making (Operation):} The dimension-reduced linear models are embedded in the VPP's operational framework for coordinated optimization with diverse resources. During power disaggregation, mechanism-based models recover executable decisions using a tailored fast algorithm that can be executed in real time.

\subsection{Main Work and Organization of This Dissertation}

The four steps form a connected approach to market-based interaction between industrial users and power systems. Facility-level mechanism models describe industrial flexibility, while the parameter identification method estimates the required parameters from coarse-grained smart meter data. The dimension-reduction method then converts complex industrial constraints into forms suitable for system-level optimization, and the ``dimension reduction--recovery'' framework uses these models in aggregator decision-making.

Apart from Chapter 1 (Introduction) and Chapter 6 (Conclusion), the core chapters are organized as follows:

\textbf{Chapter 2: General-Purpose Modeling of Industrial Loads for Electricity Market Participation.}
Traditional general-purpose industrial load models require large numbers of integer variables and are therefore poorly suited to grid interaction at scale. This chapter proposes LSTN for continuous industrial processes and cRTN for discrete batch processes. LSTN solves a coordinated case with 2,000 industrial users within minutes, while cRTN reduces the mean solution time of a secondary-steelmaking case from 156.7 to 17.1 minutes. The chapter also identifies a common mathematical structure between STN and RTN.

\textbf{Chapter 3: Industrial Load Parameter Identification Adapted to Coarse-Granularity Measurements.}
Practical use of the models requires facility-specific parameters, which industrial users may be unwilling to disclose. This chapter develops PSI to estimate a load model from hourly price and meter data by combining process constraints with cost-minimizing scheduling behavior. In the reported cement and steel-powder tests, the identified models reproduce external load responses with errors of 5.2\% and 8.5\%, less than half the errors of the compared machine-learning models.

\textbf{Chapter 4: Data-Driven Dimension Reduction for Industrial Load Modeling.}
Coordinated optimization of large user populations requires simpler grid-facing constraints, whereas analytical dimension-reduction methods cannot directly handle the integer variables in industrial process models. This chapter develops D3R, which learns least-error linear constraints from operating samples generated by the precise model. Across three industrial cases, it represents the original response with errors of 3.6\%--10.3\%; for steelmaking, it replaces 10,208 integer variables with 24--48 continuous variables. A national aluminum-smelting application further demonstrates how this advanced computational method can help evaluate the system value of industrial flexibility.

\textbf{Chapter 5: Optimization Decision-Making for Aggregators Coordinating Massive Resources.}
Building on the preceding chapters, this chapter jointly optimizes market bidding and real-time power disaggregation for a load aggregator. Reduced constraints keep the bidding problem tractable, while precise models and shadow prices translate market decisions into millisecond-level resource commands. In the PJM case study, the proposed strategy increases market income by 53.3\% and reduces operating costs by 11.4\% relative to proportional disaggregation, while supporting real-time operations with tens of thousands of resources.


\chapter{General-Purpose Modeling of Industrial Loads for Electricity Market Participation}

\section{Overview}

\subsection{Background}

Exploiting demand-side flexibility potential to mitigate load and net load fluctuations can effectively reduce the generation and transmission capacity required by power systems to ensure electricity supply, thereby improving system investment and operational efficiency.

Demand-side flexibility manifests at two levels: first, hourly load shifting through production schedule adjustments to optimize electricity consumption across time periods; second, sub-hourly real-time response capability, which can even follow renewable energy output fluctuations at the minute level.

This chapter focuses on hourly load shifting because it provides the scheduling foundation for industrial participation in energy markets. Chapter 5 later considers how industrial loads and other demand-side resources can also provide sub-hourly regulation through an aggregator.

To incorporate industrial loads into power system dispatch, the idealistic approach is to embed industrial process technical constraints within the existing Security Constrained Unit Commitment (SCUC) and Economic Dispatch (ED) framework.
This is equivalent to extending the traditional unit commitment problem into a generalized unit commitment problem that includes not only the operating characteristics of conventional generators but also the technical constraints of industrial users' electrical equipment.
In that case, it would be necessary to establish start-up and shut-down variable models for industrial equipment such as steel plants, similar to modeling thermal generator commitment variables, and incorporate them into the dispatch system\cite{ref16}.

However, real-world systems contain numerous industrial users, each comprising many equipment units, and each unit may have multiple operating states; adopting conventional integer variable modeling would lead to a rapid increase in the number of variables.
For example, modeling the start-up and shut-down constraints of equipment such as electric arc furnaces within a single steel plant may require thousands of integer variables, and as the number of industrial users increases, the computational complexity grows exponentially, rendering the model intractable.
Even from the perspective of a VPP operator, constraints of this scale remain difficult to embed in its operational optimization model.

\begin{figure}[htbp]
  \centering
  \includegraphics[width=\textwidth]{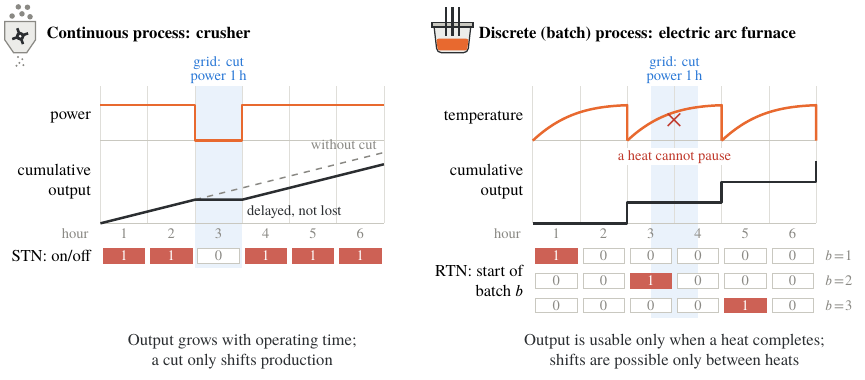}
  \caption{Examples of continuous (left) and discrete, batch-based (right) production processes}
  \label{fig_discrete_ipp}
\end{figure}

\begin{figure}[htbp]
  \centering
  \includegraphics[width=\textwidth]{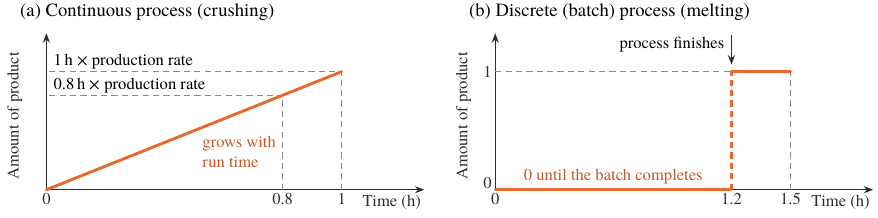}
  \caption[Product available to the next stage in continuous and batch processes]{Product available to the next stage in (a) a continuous process and (b) a discrete, batch-based process}
  \label{fig_discrete_ipp2}
\end{figure}

Although different types of industrial processes have different technical constraints, most industrial equipment shares certain commonalities in their operational characteristics, which can be described mathematically using similar models (i.e., general-purpose models).
From a mathematical perspective, industrial processes can be broadly categorized into continuous processes and discrete processes (as shown in Fig.~\ref{fig_discrete_ipp} and Fig.~\ref{fig_discrete_ipp2}, where the typical continuous process is the crushing process and the typical discrete process is the melting process).
The typical representative of continuous processes is the crushing process, characterized by a linear relationship between product quantity and operating time; for example, crushing for 8 minutes produces 8 units of product, and crushing for 10 minutes produces 10 units.
Such processes are relatively straightforward to model using input-output relationships or matrix formulations.
Discrete processes are more complex; for instance, in secondary steelmaking, the electric arc furnace melting process requires heating for a certain duration (e.g., 1.2 hours) before the scrap steel can be fully melted and made available for subsequent stages.
Before that point, the product is completely unusable, exhibiting a step characteristic that necessitates the use of discrete variables for modeling.

\subsection{Problem Description}
Existing general-purpose modeling approaches for industrial processes primarily fall into two categories: the State Task Network (STN) model and the Resource Task Network (RTN) model\cite{ref11}.
The STN model has been widely applied in modeling process industries such as chemical and cement manufacturing, describing production processes through tasks and states and using linear relationships for material conversion, making it suitable for continuous process modeling.
The RTN model is primarily used to describe discrete industrial processes, characterizing the technical constraints of batch processing, start-up/shut-down, and waiting times through the concept of batches, with modeling accuracy and expressiveness superior to those of the STN model.
Both models offer good modeling accuracy and expressiveness and remain the most mainstream industrial process modeling methods.
However, both models extensively use integer variables to describe the discrete operating states of equipment, resulting in high computational complexity that is difficult to scale to large-scale grid-interactive scenarios.

\begin{table}[!t]
  \centering
  \caption{Scope and typical use cases of industrial load modeling}
  \label{tab_industrial_load_scope}
  \setlength{\tabcolsep}{2.5mm}
  \begin{tabular}{p{1.8cm}p{3.8cm}p{3.2cm}p{4.2cm}}
    \toprule
    \textbf{Model}           & \begin{tabular}[c]{@{}l@{}}\textbf{Technical features}\\\textbf{modeled}\end{tabular}
                            & \begin{tabular}[c]{@{}l@{}}\textbf{Typical equipment/}\\\textbf{processes}\end{tabular}
                            & \begin{tabular}[c]{@{}l@{}}\textbf{Typical industrial}\\\textbf{load types}\end{tabular} \\
    \midrule
    STN & Continuous processes with linear input-output relationships & Crushers, compressors, assembly processes & Cement, textiles, steel powder, automotive, ceramics, air separation, lithium-ion battery assembly \\
    \midrule
    RTN & Batch-based discrete industrial processes & Electric arc furnaces, rolling mills & Secondary steelmaking (mini-mill), pulp and paper \\
    \midrule
    Others & Internal electrochemical and temperature dynamics & Electrolysis processes & Non-ferrous metals (aluminum electrolysis, etc.), chlor-alkali industry \\
    \bottomrule
  \end{tabular}
\end{table}

This dissertation focuses on general industrial processes represented by STN and RTN, i.e., processes whose technical constraints share mathematical forms that can characterize a class of equipment.
Industrial processes with complex temperature-efficiency coupling relationships, such as aluminum electrolysis, are not discussed in this dissertation; such processes often possess unique mechanisms that require dedicated models.
Table~\ref{tab_industrial_load_scope} compares the types of industrial loads covered and not covered by the models discussed in this dissertation.
Although the numerical cases use specific secondary-steelmaking, steel-powder, and cement-production processes, the modeling techniques can be transferred to other process industries by retaining the model structure and replacing the equipment and material parameters.

\subsection{Contributions of This Chapter}

\begin{figure}[!t]
  \centering
  \includegraphics[width=0.92\textwidth]{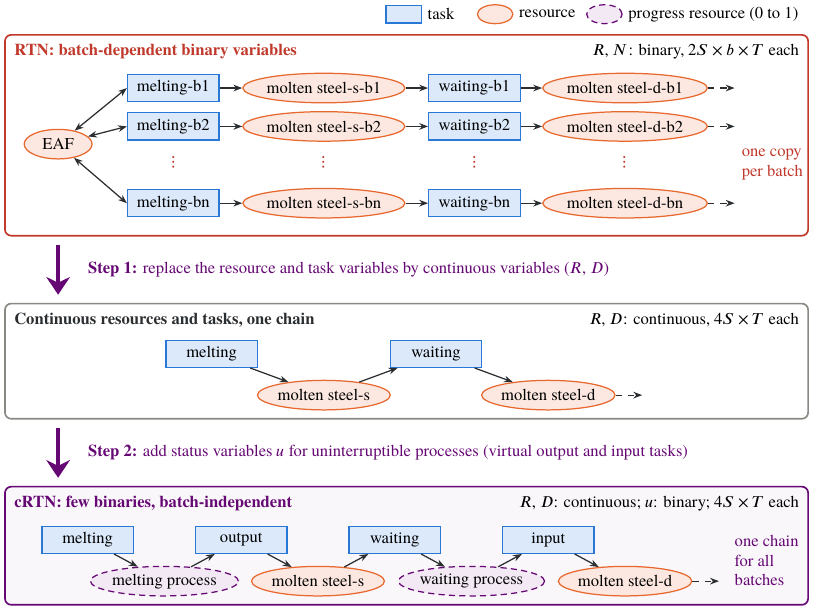}
  \caption{Reformulation of process-based industrial load models}
  \label{fig_framework_chap02}
\end{figure}

To make general-purpose industrial process models usable in grid-interactive optimization, this chapter reformulates their variables and constraints (Fig.~\ref{fig_framework_chap02}). The first objective is to replace discrete variables with continuous variables where the market time scale permits, reducing model complexity while retaining the represented process behavior. The second is to identify the common mathematical structure of STN and RTN rather than treating them as unrelated model families.
The main content of this chapter is based on the published paper\cite{lyu_efficient_2025}.

Specifically, for the STN model, this chapter proposes the Linearized State Task Network (LSTN) model, which uses the operating time of equipment at each state within each time period as decision variables, converting the original integer variable modeling into continuous variable modeling.
In the reported self-scheduling tests, LSTN solves the coordinated optimization of 2,000 industrial users within minutes, whereas the conventional STN model fails to converge within two hours for 20 users.
Furthermore, this chapter extends the linearization approach to the RTN model, proposing the continuous Resource Task Network (cRTN) model.
The cRTN model introduces status variables to describe uninterruptible processes and virtual tasks to handle inter-batch dependencies. In the secondary-steelmaking test, it reduces the mean solution time from 156.7 to 17.1 minutes (89.0\%) while retaining the represented batch constraints. Its continuous-time reformulation also avoids the rounding error created when conventional RTN forces task transitions onto discrete time-slot boundaries.

Theoretical analysis reveals that after mathematical transformation, the core constraints of the STN model are consistent with those of the RTN model, with the RTN model only adding additional constraints to describe discrete industrial processes on top of the STN model.
Therefore, the STN model can be viewed as a degenerate form of the RTN model, and the RTN model is the more general industrial process modeling framework. In other words, the two approaches can be expressed within a common continuous-time reformulation rather than treated as independent modeling systems.
This insight unifies the two modeling methodologies under a common mathematical framework.

\section{Modeling Challenges and Solution Approach for Industrial Production Processes}\label{sec_problem_description}

To incorporate the flexibility of industrial loads into the optimization decisions of VPPs, it is necessary to establish mathematical models that can accurately describe the technical constraints of industrial production processes.
This section first analyzes the modeling challenges of discrete industrial production processes and then presents the overall solution approach of this dissertation.

\subsection{Characteristics of Discrete Industrial Production Processes}

In typical industrial production processes, raw materials sequentially undergo multiple stages of processing on a production line to become the final product.
Compared to continuous industrial processes, discrete industrial production processes have the following main characteristics (summarized in Table~\ref{tab:ch02_complexity_source}):

\begin{table}[htbp]
  \centering
  \caption{Summary of modeling challenges and complexity sources for discrete industrial production processes}
  \label{tab:ch02_complexity_source}
  \setlength{\tabcolsep}{1.2mm}
  \begin{tabular}{p{3.3cm}p{3.2cm}p{4.3cm}p{2.7cm}}
    \toprule
    \textbf{Challenge} 
    & \textbf{Complexity source} 
    & \textbf{Mechanism} 
    & \begin{tabular}[c]{@{}l@{}}\textbf{Solution}\\ \textbf{approach}\end{tabular} \\
    \midrule
    Production process  
    & Batch-based, minimum unit requirements  
    & Integer variables needed for batch description 
    & \begin{tabular}[c]{@{}l@{}}Constraint\\reformulation\end{tabular} \\
    Equipment operating states  
    & Strict switching conditions, coupled with production progress 
    & State switching coupled with batches
    & \begin{tabular}[c]{@{}l@{}}Variable\\reformulation\end{tabular} \\
    Inter-stage coupling  
    & Sequential processing, batch completion required
    & Time-sensitive inter-stage coordination 
    & \begin{tabular}[c]{@{}l@{}}Constraint\\reformulation\end{tabular} \\
    \bottomrule
  \end{tabular}
\end{table}

(1) \textbf{Batch-based production}:
In continuous industrial processes, the production, storage, consumption, and transportation of materials is a continuous process without minimum processing unit restrictions.
For example, a steel-powder grinder can process 60 kg of steel powder in one hour or 1 kg in one minute, naturally lending the product to continuous variables in modeling.
However, in discrete industrial processes, each task has a minimum processing unit (batch).
For instance, an EAF in secondary steelmaking processes several tons of scrap steel at once, heating the entire batch until melting.
While the EAF processes 6 tons per hour, this cannot be modeled as 0.1 tons per minute because the scrap steel is processed as a whole batch---no steel is available for the next process until the full 60-minute EAF operation completes.
This characteristic necessitates the introduction of integer variables to describe the production process.

(2) \textbf{Discrete operating states and state switching constraints}:
This is a common characteristic in industrial production processes that typically requires integer variables to describe equipment operating states, similar to the state variables in thermal unit commitment.
In continuous industrial processes, although equipment can only operate in discrete states (e.g., on and off), the operating time, energy consumption, and material production in each state are continuously adjustable (e.g., operating for 47 minutes within an hour).
This continuous nature forms the basis for linearizing continuous industrial process models at demand response time scales.
However, in discrete industrial processes, equipment states are not only discrete but also cannot switch freely, being coupled with production progress.
For example, once an EAF starts processing a batch of scrap steel, it must complete the batch processing before returning to an idle state and cannot be interrupted midway, similar to the minimum start-up and shut-down time constraints in thermal unit commitment.

(3) \textbf{Coupling between preceding and subsequent production stages}:
For continuous industrial processes, the coupling between stages is modeled through material transformation matrices, and the storage limits for intermediate materials can be represented by linear constraints due to the continuous nature of the materials.
However, in discrete industrial processes, due to the batch processing mode, subsequent stages can only begin after the complete completion of previous stages, necessitating additional integer variables.
Furthermore, after processing in the prior stage, the product needs to be transferred and may have to wait before being further processed in the subsequent stage.
The transfer and waiting times may be subject to limitations; for example, molten steel from an EAF needs to be processed before it cools, which is time-sensitive and typically requires the introduction of status variables.

\subsection{Modeling Objectives}

Given the characteristics of discrete industrial production processes described above, the modeling approach proposed in this dissertation needs to achieve the following two objectives:

(1) \textbf{Effective representation of technical constraints for discrete industrial processes}:
The model variables satisfy the mathematical constraints proposed by the model if, and only if, the corresponding production schedule can be executed in reality without violating the technical constraints of the industrial process.
Additionally, the variables within the model should be easily comprehensible so that factory managers can conveniently implement production schedules based on the variable values provided by the model.

(2) \textbf{Computability and scalability}:
The computational complexity introduced by integrating the constraints of the proposed model into optimization problems should not be excessively high.
In other words, it should be solvable within an acceptable time frame through commercial solvers to obtain an optimal solution, and the model complexity should not increase with the number of production batches.

\subsection{Overall Solution Approach}

To achieve the above modeling objectives, this dissertation proposes improving existing industrial process models through two levels of reformulation: variable reformulation and constraint reformulation (Fig.~\ref{fig_framework_chap02}).

(1) \textbf{Variable reformulation: from discrete to continuous}:
The core idea is to convert the discrete operating state variables (integer variables) used in traditional industrial process modeling into continuous variables representing the operating time of equipment at each state within each time period.
This transformation exploits the relatively large time scale of demand response, enabling the start and stop times of equipment within a time period to be continuously adjusted at the hourly scheduling level, making the operating time at each state effectively continuous.
Through this variable substitution, the original mixed-integer programming problem can be transformed into a linear programming problem (for continuous industrial processes), or its number of integer variables can be significantly reduced (for discrete industrial processes).

(2) \textbf{Constraint reformulation: preserving discrete characteristics}:
For discrete industrial processes, variable substitution alone is insufficient, as batch-based production, uninterruptibility, and other discrete characteristics must be strictly guaranteed.
Therefore, this dissertation introduces the concepts of virtual tasks and status variables, designing solver-friendly mathematical formulations to describe the technical constraints of discrete industrial processes while using continuous variables to describe resources and tasks.
Specifically, output tasks and input tasks are introduced to express inter-batch dependencies, and status variables are introduced to express the operating states of uninterruptible processes.

Based on the above approach, this dissertation will improve the models for continuous industrial processes (STN model) and discrete industrial processes (RTN model) respectively, proposing the Linearized State Task Network (LSTN) model and the continuous Resource Task Network (cRTN) model.
Subsequent sections will detail the specific implementation methods and numerical validation results of these two improved models.

\section{Variable Reformulation: Linearized State Task Network Model}\label{sec_LSTN}

As analyzed above, conventional industrial process models extensively use integer variables, leading to computational difficulties.
This section demonstrates how to achieve model linearization through variable reformulation using continuous industrial processes as an example.
Specifically, for the State Task Network (STN) model, this dissertation proposes the Linearized State Task Network (LSTN) model, which replaces the conventional integer start-up variables with continuous operating time variables, thereby completely eliminating integer variables.
The LSTN model primarily focuses on the energy consumption characteristics of the industrial user's production process rather than the specific operational details of the equipment.

\subsection{Basic Formulation of the STN Model}

\begin{figure}[htbp]
  \centering
  \includegraphics[width=0.8\textwidth]{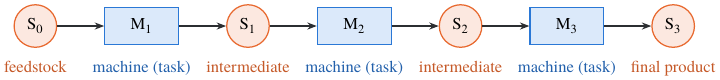}\\[0.6em]
\caption{Example of a state-task network (STN), with production equipment and material storage}
  \label{fig_LSTN}
\end{figure}

The general-purpose state task network model can describe a wide range of production processes arising in multiproduct/multipurpose industrial facilities (as shown in Fig.~\ref{fig_LSTN}, where M represents production equipment and S represents material storage).
Consider an arbitrary factory $f \in \mathcal{F}$ (for brevity, the subscript $f$ is omitted in this section, e.g., $E_{t}$ for $E_{f,t}$).
Let $i/i^{\rm end}$ denote the index/total number of production tasks, and $\mathcal{I}^{\rm P}/\mathcal{I}^{\rm S}$ denote the set of production tasks/states ($\mathcal{I}^{\rm P} = \{1, 2, ..., i^{\rm end}\}$, $\mathcal{I}^{\rm S} = \{0\} \cup \mathcal{I}^{\rm P}$, $i=0$ for the feedstock).
Let $t$ and $\mathcal{T}$ denote the index of time intervals and the whole time horizon, respectively ($\mathcal{T} = \{0, 1, 2, ..., t^{\rm end}\}$, $t=0$ for the initial or current time interval).
Let $k/\mathcal{K}_i$ denote the index/set of operating points of task $i$.

\subsection{Variable Reformulation in the LSTN Model}

\begin{figure}[htbp]
  \centering
  \includegraphics[width=\textwidth]{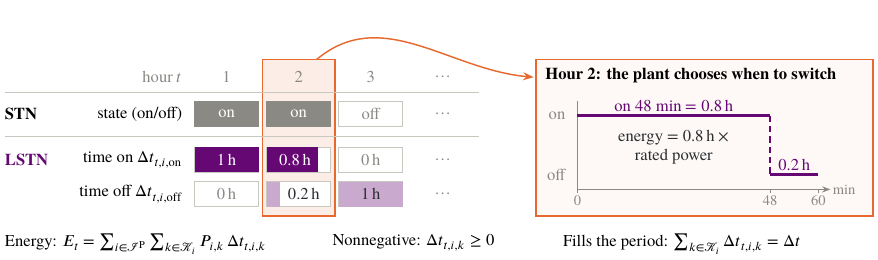}
  \caption[Operating time at each operating point as a continuous variable in the LSTN]{Operating time at each equipment operating point as a continuous variable in the linearized state-task network (LSTN)}
  \label{fig_continous_time}
\end{figure}

The conventional STN model assumes that industrial devices can only operate at one operating point within a time period\cite{ref21}.
In the context of demand response, although industrial tasks (equipment) usually operate at discrete operating points, the time to switch their operating points (e.g., seconds to minutes) is negligible compared to the time period during which the electricity price is implemented (e.g., an hour).
Considering this feature, this dissertation treats the operating time of the tasks at each operating point as a continuous variable (Fig.~\ref{fig_continous_time}), thereby obtaining a linear-formed model as follows.

Factory $f$ aims to minimize its total energy cost $Cost$ across $\mathcal{T}$:
\begin{equation}\label{primal_cost_chap02}
  Cost = \underset{t \in \mathcal{T}}{\Sigma}{Pr_{t}} E_{t},
\end{equation}
where $Pr_{t}$ (\$/kWh) is the electricity price and $E_{t}$ (kWh) is the energy consumption of factory $f$ at time $t$.
Actual production scheduling often considers output, quality, delivery, and staffing in addition to electricity expenditure. These objectives can be incorporated through weighted terms, hierarchical objectives, or hard and soft constraints. This dissertation uses electricity-cost minimization because it is aligned with the market-interaction problem studied here. If a firm's decision criterion differs materially from cost minimization, the optimal schedule and the interpretation of identified parameters will change, although the feasible-region and computational arguments remain applicable.
$E_{t}$ is decomposed into the energy consumption at each operating point of each task:
\begin{equation}\label{primal_constraint_E_chap02}
  E_{t} = \underset{i \in \mathcal{I}^{\rm P}}{\Sigma} \underset{k \in K_{i}}{\Sigma} P_{i,k} \Delta t_{t,i,k}, \ t \in \mathcal{T},
\end{equation}
where $P_{i,k}$ (kW) is the power of task $i$ at operating point $k$, and $\Delta t_{t,i,k}$ (h) is the continuous time duration that task $i$ operates at point $k$ within time period $t$.
Naturally, the operating time of a task at all its operating points (including the off state) within a time period is nonnegative (\ref{primal_constraint_timeNonNegative}) and sums up to the length of the time period $\Delta t$ (\ref{primal_constraint_timeSum}):
\begin{equation}\label{primal_constraint_timeNonNegative}
  \Delta t_{t,i,k} \ge 0, \  t \in \mathcal{T}, i \in \mathcal{I}^{\rm P}, k \in \mathcal{K}_i.
\end{equation}
\begin{equation}\label{primal_constraint_timeSum}
  \underset{k \in K_{i}}{\Sigma} \Delta t_{t,i,k} = \Delta t, \  t \in \mathcal{T}, i \in \mathcal{I}^{\rm P}.
\end{equation}
Note that we only focus on the duration of time that the devices operate at their different operating points (e.g., ``on''/``off'') within the time periods, rather than the specific time to switch operating points.
After determining the operating duration, the feasibility of the specific switching time is guaranteed by (\ref{primal_constraint_timeNonNegative})-(\ref{primal_constraint_timeSum}), and operations are left for the factory to implement.

\subsection{Material Balance Constraints of the LSTN Model}

Let $S_{t,i}$ (kg) denote the amount of material $i$ at the end of time period $t$, and $S^{\rm 0/max/tar}_i$ denote the initial/upper limit/target amount of material $i$.
The factory must meet its production target (\ref{primal_constraint_tar_chap02}) and buffer limit (\ref{primal_constraint_storageLimit_chap02}), with initial buffer states given by (\ref{primal_constraint_S0_chap02}) ($S^{\rm 0}_i$ is the material state at the initial time $t^{\rm 0}$, analogous to the initial state of charge of energy storage):
\begin{equation}\label{primal_constraint_tar_chap02}
  S_{t,i} \ge S^{0}_{i} + S^{\rm tar}_i, \  t = t^{\rm end}, i \in \mathcal{I}^{\rm S}.
\end{equation}
\begin{equation}\label{primal_constraint_storageLimit_chap02}
  0 \le S_{t,i} \le S^{\rm max}_{i}, \  t \in \mathcal{T}, i \in \mathcal{I}^{\rm S}.
\end{equation}
\begin{equation}\label{primal_constraint_S0_chap02}
  S_{t,i} = S^{\rm 0}_i, \ i \in \mathcal{I}^{\rm S}, t = t^{\rm 0}
\end{equation}

Let $G_{i,k}/C_{i,k}$ (kg/h) denote the material production/consumption rate of task $i$ at operating point $k$.
The change in buffer states across time for the feedstock, intermediate products, and final products is given by (\ref{primal_constraint_changeofS1_chap02}), (\ref{primal_constraint_changeofS2_chap02}), and (\ref{primal_constraint_changeofS3_chap02}), respectively:
\begin{equation}\label{primal_constraint_changeofS1_chap02}
  S_{t,i} = S_{t-1,i}
  - \underset{k \in K_{i+1}}{\Sigma} C_{i+1,k}  \Delta t_{t,i+1,k}, \ t \in \mathcal{T}, i = 0.
\end{equation}
\begin{equation}\label{primal_constraint_changeofS2_chap02}
  \begin{aligned}
    & S_{t,i} = S_{t-1,i} + \underset{k \in K_{i}}{\Sigma} G_{i,k}  \Delta t_{t,i,k}
  - \underset{k \in K_{i+1}}{\Sigma} C_{i+1,k}  \Delta t_{t,i+1,k}, \\
  & t \in \mathcal{T}, i \in \mathcal{I}^{\rm P} \setminus \{i^{\rm end}\}.
\end{aligned}
\end{equation}
\begin{equation}\label{primal_constraint_changeofS3_chap02}
  S_{t,i} = S_{t-1,i} + \underset{k \in K_{i}}{\Sigma} G_{i,k}  \Delta t_{t,i,k}, \ t \in \mathcal{T}, i = i^{\rm end}.
\end{equation}

Through the above modeling, we obtain the linear programming formulation of the LSTN model, in which the decision variables are $\{\Delta t_{t,i,k} | \forall i, \forall t, \forall k\}$ and the dependent variables are $\{E_{t,i}, S_{t,i} | \forall i, \forall t\}$.
The problem parameters include electricity prices $\{Pr_t|\forall t\}$ and equipment parameters $\{G_{i,k}, C_{i,k}, P_{i,k}, S^{\rm 0/max/tar}_i | \forall i, \forall k\}$, which can be obtained directly through the device nameplate or the factory inventory.

Although LSTN is derived for devices with discrete operating points, it can also model nonadjustable devices and devices with inherently continuous operating points.
The former can be modeled with only one operating point, and the latter can be modeled with two operating points, corresponding to ``on'' (i.e., operating at rated power) and ``off'' states.
For devices whose switching time between operating points is not negligible, integer variables can be introduced in conventional ways.

\subsection{Use of the LSTN Model in Demand Response}

The LSTN model formulates the factory's energy cost and production process constraints in a linear form, so the most straightforward approach is to embed LSTN directly into existing demand response models.
For example, the LSTN primal problem can be embedded into an economic dispatch model, or the optimality conditions of LSTN (such as KKT conditions, dual constraints, and strong duality conditions) can be embedded into demand response pricing problems.
Since the optimality conditions of linear programming problems are easy to express and computationally tractable, this direct embedding approach offers good scalability.

However, direct embedding implies a centralized control architecture that requires access to the factories' internal parameters, which may be impractical due to privacy concerns in real-world deployment.
To further reduce the computational burden and protect industrial users' privacy, this dissertation proposes the following hierarchical control architecture:

(1) \textbf{Adjustable capacity aggregation}:
The VPP operates between the factories and the grid operator (or electricity market), and is responsible for computing the aggregated demand response capacity using the parameters of the factory set $\mathcal{F}$, reporting this capacity to the grid operator, receiving dispatch signals from the grid, and distributing control signals to individual factories.

Different models can be used to describe the demand response capacity of factories; this dissertation uses the Virtual Battery (VB) model to approximate the external characteristics of the aggregated resources.
Let $\overline{P}^{\rm B}_t/\underline{P}^{\rm B}_t$ and $\overline{E}^{\rm B}_t/\underline{E}^{\rm B}_t$ denote the charging/discharging power capacity and charging/discharging energy capacity at time period $t \in \mathcal{T}^{\rm DR}$, respectively.
Without loss of generality, these parameters may have different values at different time periods, determined by the following optimization problem:
\begin{equation}\label{VB_objective}
  {\rm max.} \ \eta^{\top}[\overline{P}^{\rm B}_t;\underline{P}^{\rm B}_t;
  \overline{E}^{\rm B}_t;\underline{E}^{\rm B}_t]
\end{equation}
\begin{equation}\label{VB_cons_individual}
  {\rm s.t.} \ (\ref{primal_constraint_E_chap02})-(\ref{primal_constraint_changeofS3_chap02}), \ \forall f \in \mathcal{F},
\end{equation}
\begin{equation}\label{VB_cons_P}
    \overline{P}^{\rm B}_{t} \le \underset{f \in \mathcal{F}}{\Sigma}(E_{f,t} - \tilde{E}_{f,t})/\Delta t \le - \underline{P}^{\rm B}_t, 
\end{equation}
\begin{equation}\label{VB_cons_E}
    \overline{E}^{\rm B}_t \le \underset{(\tau \in \mathcal{T}^{\rm DR}, \tau \le t)}{\Sigma} \underset{f \in \mathcal{F}}{\Sigma}(E_{f,\tau} - \tilde{E}_{f\tau}) \le -\underline{E}^{\rm B}_t,
\end{equation}
where $\eta$ in (\ref{VB_objective}) is a coefficient vector characterizing parameter preference, e.g., $\eta = [1;0;0;0]$ is used for computing $\overline{P}^{\rm B}_{t}$; $\tilde{E}_{f,t}$ is the baseline load of factory $f$ at time $t$, determined, for example, by the production schedule the factory makes based on day-ahead electricity prices; and $\mathcal{T}^{\rm DR}$ is the time horizon of the demand response event, e.g., $\mathcal{T}^{\rm DR}=\{1, 2, 3, 4\}$, corresponding to a typical 4-hour demand response duration.
Constraint (\ref{VB_cons_individual}) represents the energy consumption constraints of each factory described by LSTN.
Constraints (\ref{VB_cons_P}) and (\ref{VB_cons_E}) define the virtual battery parameters.

(2) \textbf{Optimal dispatch}:
The grid operator sends control signals to the VPP based on the aggregated demand response capacity reported through the virtual battery parameters, using predefined strategies.
For the process by which the grid determines control signals, various existing strategies can be directly applied, such as participation of conventional batteries in unit commitment, wind power smoothing, and economic optimization in electricity markets.

Upon receiving the control signal $\delta_t, t \in \mathcal{T}^{\rm DR}$ (kW), the VPP solves the following problem to optimize the factories' energy consumption and respond to the control signal:
\begin{equation}\label{OD_objective}
  {\rm min.} \ \underset{t \in \mathcal{T}}{\Sigma} \underset{f \in \mathcal{F}}{\Sigma} Cost_{f} + M\underset{t \in \mathcal{T}^{\rm DR}}{\Sigma} \Delta E_{t}^{2}
\end{equation}
\begin{equation}\label{OD_cons_individual}
  {\rm s.t.} \ (\ref{primal_cost_chap02})-(\ref{primal_constraint_changeofS3_chap02}), \ \forall f \in \mathcal{F}
\end{equation}
\begin{equation}\label{OD_cons_mismatch}
  \underset{f \in \mathcal{F}}{\Sigma} (E_{f,t} - \tilde{E}_{f,t}) - \delta_{t}\Delta t = \Delta E_t, \ t \in \mathcal{T}^{\rm DR}
\end{equation}
where $M$ is a large positive number, $\Delta E_{t}$ is the energy mismatch defined by (\ref{OD_cons_mismatch}), and the second term in objective (\ref{OD_objective}) is a penalty term that enforces the factories to follow the load reduction control signal.
Without loss of generality, the control signal within $\mathcal{T}^{\rm DR}$ must be followed, while the factory costs and operational constraints over the entire time horizon $\mathcal{T}$ must be considered.
The above aggregation and dispatch models are both linear programming problems that can be efficiently solved.

\section{Numerical Tests of the LSTN Model}\label{sec_numerical}

Optimization problems were solved using Gurobi (V10.0.0) and MATLAB (R2021a) with YALMIP~\cite{Lofberg2004}.
Computation was executed on a workstation with an Intel Core i9-10900X CPU (3.7 GHz) and 128 GB RAM.
For consistency with electricity meter specifications, unless otherwise stated, figures and tables in this dissertation uniformly label the electricity consumption (kWh) of industrial loads in each time period rather than power (kW).

\subsection{Accuracy Verification of the LSTN Model}

\begin{table}[!t]
  \caption{Parameters of the original steel-powder manufacturing plant}
  \label{tab_parameter_lstn}
  \centering 
  \begin{tabular}{lllll}
  \toprule
  Name &
  \begin{tabular}[c]{@{}l@{}}Operating\\point\end{tabular} &
  \begin{tabular}[c]{@{}l@{}}Production rate\\(ton/h)\end{tabular} &
  \begin{tabular}[c]{@{}l@{}}Energy demand\\(kWh)\end{tabular} &
  \begin{tabular}[c]{@{}l@{}}Buffer capacity\\(ton)\end{tabular} \\ \midrule
                             & off  & 0  & 0  &                          \\ \cline{2-4}
\multirow{-2}{*}{Reduction}  & on & 15 & 75 & \multirow{-2}{*}{100}    \\ \hline
                             & off  & 0  & 0  &                          \\ \cline{2-4}
\multirow{-2}{*}{Atomizer}   & on & 30 & 60 & \multirow{-2}{*}{180}    \\ \hline
                             & off  & 0  & 0  &                          \\ \cline{2-4}
\multirow{-2}{*}{Dehydrator} & on & 15 & 10 & \multirow{-2}{*}{100}    \\ \hline
                             & off  & 0  & 0  &                          \\ \cline{2-4}
\multirow{-2}{*}{Dryer} & on & 15 & 30 & \multirow{-2}{*}{150} \\ \hline
                             & off  & 0  & 0  &                          \\ \cline{2-4}
\multirow{-2}{*}{Separator}  & on & 15 & 10 & \multirow{-2}{*}{100}    \\ \hline
                             & gear 1   & 0  & 0  &                          \\ \cline{2-4}
                             & gear 2   & 10 & 15 &                          \\ \cline{2-4}
\multirow{-3}{*}{Crusher}    & gear 3   & 15 & 20 & \multirow{-3}{*}{100}    \\ \hline
                             & gear 1   & 0  & 0  &                          \\ \cline{2-4}
                             & gear 2   & 10 & 15 &                          \\ \cline{2-4}
\multirow{-3}{*}{Classifier} & gear 3   & 20 & 25 & \multirow{-3}{*}{150}    \\ \hline
                             & gear 1   & 0  & 0  &                          \\ \cline{2-4}
                             & gear 2   & 10 & 6  &                          \\ \cline{2-4}
\multirow{-3}{*}{Blender}    & gear 3   & 15 & 10 & \multirow{-3}{*}{200} \\  \bottomrule
\end{tabular}
\end{table}

\begin{figure}[!t]
  \centering
  \includegraphics[width=\textwidth]{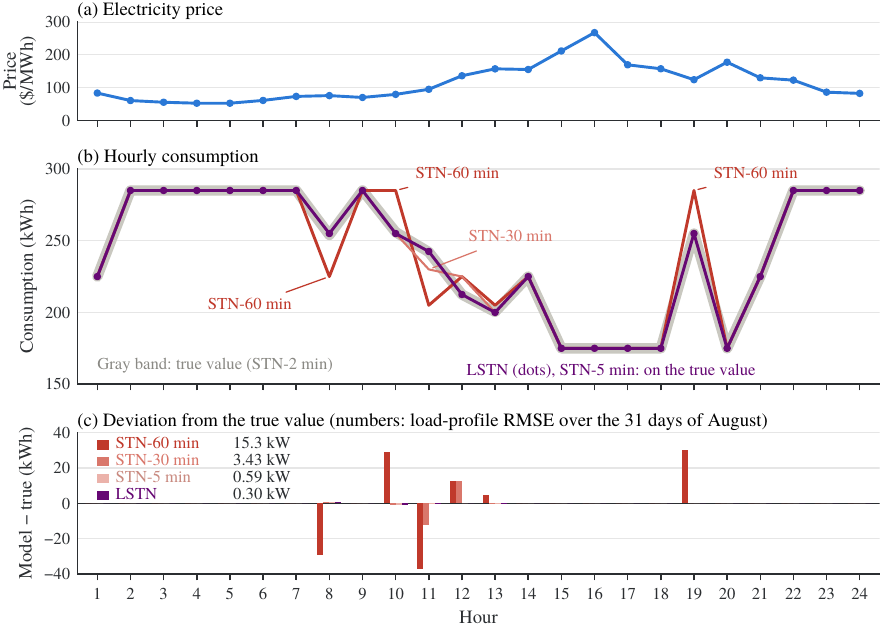}
\caption[Optimal hourly electricity consumption of the steel-powder plant on August 5]{Optimal hourly electricity consumption of the steel-powder plant on August 5: (a) electricity price; (b) model load profiles; (c) deviations from the two-minute state-task network (STN) benchmark}
  \label{fig_baseline_load}
\end{figure}

We first tested the modeling accuracy for the scenario of a single industrial user using hourly electricity price data for August 2022 in the PJM network and the site parameters of a steel-powder manufacturing plant~\cite{ref22} (Table~\ref{tab_parameter_lstn}).
The production target of the factory is set to 24 times the lowest production rate across the production tasks (i.e., 15 ton/h), and the initial material is half the buffer limit.
We assume that the time required for switching the operating state of the plant equipment is 2 minutes.
Therefore, the STN model with a scheduling time interval of 2 minutes (STN-2 min) is treated as an accurate industrial user model, with load profiles and daily energy costs generated using the MILP-based production process model~\cite{ref22} serving as the true values for calculating the error of other models.
The errors measured by RMSE across the 31 days of the month are listed in Table~\ref{tab_rmse_models_lstn}.
As an example, Fig.~\ref{fig_baseline_load} presents the load profiles given by each model on August 5.
$${\rm RMSE}=\sqrt{\frac{\sum^N_{i=1} (y^*_i - y_i)^2}{N}}$$

\begin{table}[!t]
  \caption{Performance comparison between LSTN and conventional STN models}
  \label{tab_rmse_models_lstn}
  \centering
  \begin{tabular}{ccc}
    \toprule
    Model &
    \begin{tabular}[c]{@{}c@{}}Load profile RMSE (kW)\\ /Ratio to max.\ value \end{tabular} &
    \begin{tabular}[c]{@{}c@{}}Energy cost RMSE (\$)\\ /Ratio to max.\ value \end{tabular} \\ \midrule
      STN-60min  & 15.3(5.37\%) & 1.979(0.175\%) \\
  STN-30min & 3.43(1.21\%) & 0.398(0.035\%) \\
  STN-5min  & 0.59(0.21\%) & 0.034(0.003\%) \\
  \textbf{LSTN}  & \textbf{0.30(0.17\%)}   & \textbf{0.033(0.003\%)}  \\ \bottomrule
  \end{tabular}
  \end{table}

Table~\ref{tab_rmse_models_lstn} shows that the proposed LSTN model achieves an RMSE of 0.30 compared to the high-granularity ground truth, substantially lower than the standard STN-60 min model (RMSE 15.3). This supports the assumption that the operating time of industrial devices can be treated as continuous, and that the derived LSTN model is more accurate than the MILP-based STN model.

\subsection{Computational Efficiency of the LSTN Model}

\begin{figure}[!t]
  \centering
  \includegraphics[width=\textwidth]{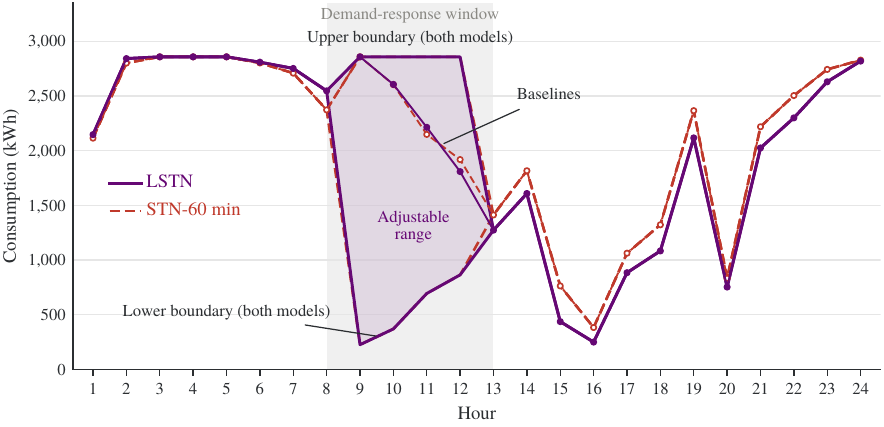}
\caption[Aggregate load-adjustment limits on August 5 from the STN and LSTN models]{Aggregate load-adjustment limits on August 5 from the state-task network (STN) and linearized state-task network (LSTN) models}
  \label{fig_boundary}
\end{figure}

To evaluate the computational gains of LSTN, we conducted tests under two scenarios: aggregating the adjustable capacity of factories and jointly optimizing multiple industrial users.
To test the model's performance under different parameters, the device parameters of different factories are uniformly distributed between [0.8, 1.2] times the values given in Table~\ref{tab_parameter_lstn}, and the demand response duration is 4 hours.
STN-60 min was also evaluated for comparison because this is the commonly used model and is the most computationally efficient among the MILP-based STN models.
We used the load profile on August 5 as the load baseline and then calculated the adjustable capacity of the aggregated factories given by STN-60 min and LSTN (Fig.~\ref{fig_boundary}).
Although the load baselines given by the two models are different, their results for the adjustable boundary are very similar.

\begin{figure}[!t]
  \centering
    \includegraphics[width=\textwidth]{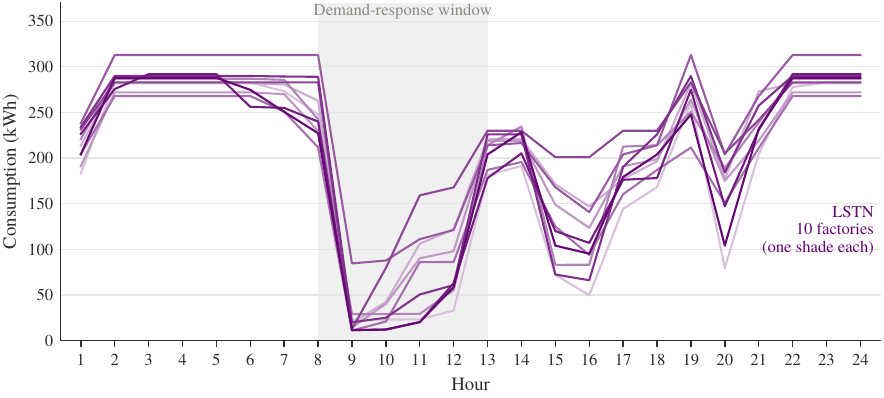}
\caption[Allocation of grid load-reduction requests among factories using the LSTN model]{Allocation of grid load-reduction requests among factories using the linearized state-task network (LSTN) model}
  \label{fig_dispatch_lstn}
\end{figure}

\begin{figure}[!t]
  \centering
    \includegraphics[width=\textwidth]{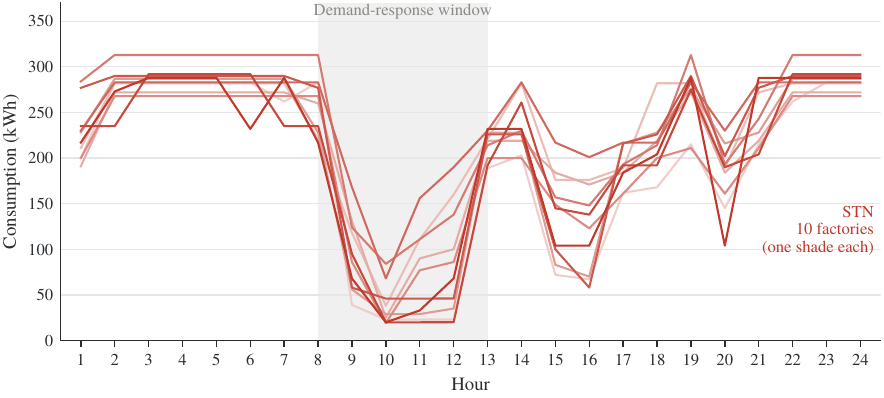}
\caption{Allocation of grid load-reduction requests among factories using the state-task network (STN) model}
  \label{fig_dispatch_stn}
\end{figure}

\begin{figure}[!t]
  \centering
  \includegraphics[width=\textwidth]{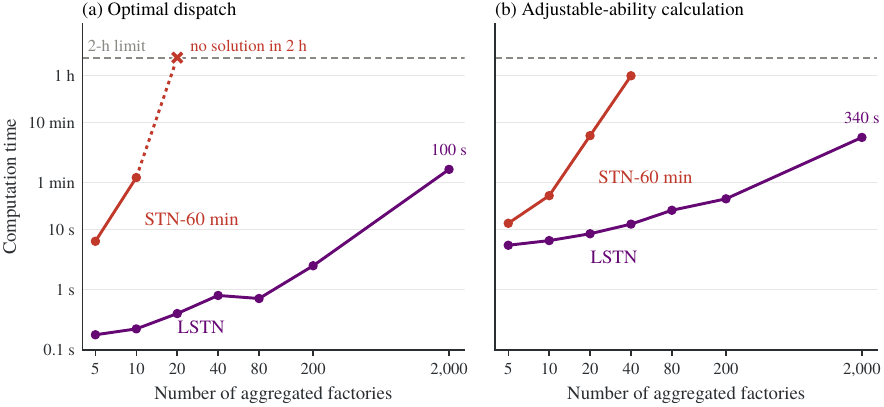}
\caption[Computation time versus number of factories for the STN and LSTN models]{Computation time versus number of factories for the hourly state-task network (STN) and linearized state-task network (LSTN) models: (a) optimal dispatch; (b) calculation of load-adjustment limits, with logarithmic axes}
  \label{fig_calculation_time_lstn}
\end{figure}

Based on the above method, the grid sends instructions for reducing energy consumption according to the adjustable boundary of the aggregated factories.
The dispatch instructions for the factories are shown in Fig.~\ref{fig_dispatch_lstn} and~\ref{fig_dispatch_stn}, where each line is one factory, drawn in shades of the model's color.
Fig.~\ref{fig_calculation_time_lstn} shows the difference in computation time: while the traditional STN-60 min failed to converge within 2 hours when scaling to 20 factories (over 10,000 integer variables), the LSTN model solved the optimization for 2,000 factories in minutes, satisfying the speed requirements of real-world economic dispatch.

\section{Constraint Reformulation: Continuous Resource Task Network Model}\label{sec_rtn}

The previous section demonstrated how to fully linearize continuous industrial process models through variable reformulation.
However, for discrete industrial processes, variable reformulation alone is insufficient, as batch-based production, uninterruptibility, and other discrete characteristics must be strictly guaranteed.
This section describes how to further perform constraint reformulation on top of variable reformulation to address the modeling requirements of discrete industrial processes.
Specifically, this dissertation systematically reformulates the Resource Task Network (RTN) model, proposing the continuous Resource Task Network (cRTN) model.
For ease of comparison, this section first introduces the basic formulation of the conventional RTN model and then demonstrates how the cRTN model achieves higher computational efficiency while maintaining the same modeling capabilities.

The conventional RTN model used here follows Zhang et al.\cite{ref10}, who adapted and extended Castro et al.'s formulation\cite{ref23} to incorporate flexible adjustable tasks.
While we maintain the mathematical essence of their formulation, some notation adjustments have been made for consistency with the cRTN model proposed later.

\subsection{Basic Concepts of the Conventional RTN Model}

\begin{figure}[!t]
  \centering
  \includegraphics[width=0.94\textwidth]{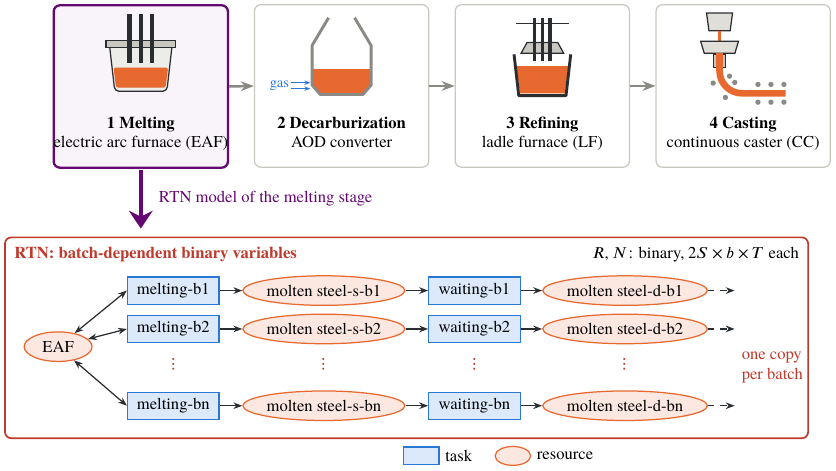}
  \caption{Production stages of the electric arc furnace steelmaking plant}
  \label{fig_rtn_process}
\end{figure}

We use the production process of short-process steelmaking as an example to illustrate the modeling method of the RTN.
The production process of a short-process steelmaking plant is shown in Fig.~\ref{fig_rtn_process}, which can be divided into four stages: 1.\ melting, 2.\ decarburization, 3.\ refining, and 4.\ casting.
In the RTN modeling framework, the four production stages and the transfer tasks between them are modeled as tasks indexed by $i$.
The scheduling of tasks is based on discrete time slots indexed by $t$, with a slot length of $\Delta t$ (e.g., $\Delta t=5$ minutes).
The integer variable $N_{i, t}$ models the start of task $i$, with $N_{i, t} = 1$ representing that task $i$ starts at time slot $t$.

The production equipment and products (intermediate and final products) are modeled as resources indexed by $r$.
$R_{r, t}$ represents the value of resource $r$ at time $t$.
For instance, if $r$ represents the EAF, $R_{r, t} = 1$ means that the EAF is idle at time slot $t$ and can be used for melting.
Since the product needs to be transported between different production stages, to distinguish the product before and after transportation, it is modeled as different resources, represented by superscripts s and d, respectively; i.e., $r^{\rm s}$ ($r^{\rm d}$) is the index of the product located at the transfer start point (destination).

The interaction between tasks and resources is modeled by the interaction matrix $\mathbf{M}$, in which an entry $\gamma_{r, i, \theta}$ represents the amount of resource $r$ consumed/generated by task $i$, $\theta$ time slots after the start of task $i$.
For example, the melting task (task $i$) generates molten steel (resource $r$) at 80 minutes (16 time slots with a slot length of $\Delta t=5$ minutes) after the starting time, so entry $\gamma_{r, i, 16}$ is set to 1.

Note that in the conventional RTN model, tasks are bound to batches; that is, the processing and transfer of different batches are modeled as different tasks.
This modeling method means that the scale of the RTN model of the same production line will increase with the production target (the number of batches), limiting the model's scalability.

\subsection{Mathematical Formulation of the Conventional RTN Model}

The modeling approaches herein were selected from Ref.~\inlinecite{ref10} but using the notation of this dissertation with some simplifications.

(1) \textbf{Resource balance}:
The amount of resources at the end of each time slot $R_{r, t}$ is determined by the initial amount (i.e., the amount at the end of the previous time slot) $R_{r, t-1}$ and the changes to the resources caused by the tasks within the time slot.
This principle of resource balance can be represented as a constraint for both equipment and products in the following form:
\begin{equation}\label{rtn_balance}
  R_{r, t}=R_{r, t-1} + \sum_i \sum_{\theta=0}^{\tau_i} \gamma_{r, i, \theta} N_{i, t-\theta} \quad \forall r, t
\end{equation}
Imposing this constraint on the equipment limits the use of a device to a single task within a given time period.
However, this leads to rounding errors, which will be further analyzed in the case study.
Similarly, the energy consumption of the production process $E^{\rm EL}_{t}$ can be represented with a comparable constraint, although there is no direct coupling between the energy consumption across different time slots:
\begin{equation}\label{rtn_electricity_balance}
  E^{\rm EL}_{t}=\sum_i \sum_{\theta=0}^{\tau_i} \gamma_{{\rm EL}, i, \theta} N_{i, t-\theta} \quad \forall t
\end{equation}
where $\gamma_{{\rm EL}, i, \theta}$ is the energy consumption of task $i$, $\theta$ time slots after the task starts.

(2) \textbf{Task execution}:
The RTN models each batch of products individually, and task modeling is also based on batches.
Therefore, each task for each batch needs to be executed only once:
\begin{equation}\label{rtn_task_execution}
  \sum_{t} N_{i, t} = 1 \quad \forall i
\end{equation}
For transfer tasks, the following constraint is added to ensure immediate execution:
\begin{equation}\label{rtn_transfer_exc}
  R_{r^{\rm s}, t} = 0 \quad \forall r^{\rm s}, t
\end{equation}
Constraint (\ref{rtn_transfer_exc}) requires no waiting time for intermediate products at the transfer start point.
The rationale is twofold: first, this is a common requirement in industrial processes such as steelmaking; second, it avoids more complex modeling approaches (e.g., completing manufacturing in one stage, followed by waiting, then transfer, then waiting again), thereby maintaining the model's simplicity.

(3) \textbf{Waiting time limit}:
Intermediate products can wait for some time between two production stages rather than immediately proceeding to the next stage.
This time flexibility is one source of the energy flexibility of industrial processes.
However, this waiting time usually has an upper limit because materials such as molten steel must be further processed before their temperature drops to a certain threshold.
The duration limit of the waiting process (including the transfer time) for resource $r$ is expressed as:
\begin{equation}\label{rtn_transfer_time}
  \Delta t \sum_t R_{r^{\rm d}, t} + w_{r^{\rm d}} \le W_{r^{\rm d}} \quad \forall i
\end{equation}
where $R_{r^{\rm d}, t} = 1$ indicates that the product has arrived at the transfer destination at time slot $t$ and is waiting; therefore, $\sum_t R_{r^{\rm d}, t}$ is the waiting time, $w_{r^{\rm d}}$ is the transfer time, and $W_{r^{\rm d}}$ is the maximum waiting time of resource $r^{\rm d}$.

(4) \textbf{Product delivery}:
By the end of the final time slot, each batch needs to reach the final stage, i.e., complete all the manufacturing processes:
\begin{equation}\label{rtn_product_delivery}
  R_{r^{\rm f}, \mathcal{T}} = 1, \forall r^{\rm f}
\end{equation}
where $r^{\rm f}$ is the index of the product located at the final stage.

(5) \textbf{Objective function}:
The goal of factory production scheduling does not belong to the scope of RTN modeling, but for the completeness of the model, we use an objective function that minimizes the production energy cost as an example:
\begin{equation}\label{rtn_objective}
  {\rm min.} \sum_{t} Pr_{t}  E^{\rm EL}_{t}
\end{equation}
where $Pr_{t}$ is the electricity price at time slot $t$.

(6) \textbf{Modeling of flexible tasks in the conventional RTN}:
In actual operation, a device may operate in one of several states, not just the ``on'' and ``off'' states.
The basic RTN model mentioned above can model this feature by treating the different operating states of the device as different tasks.
For example, if a device can operate at 50\% of its rated power, the interaction matrix needs to model its processing time as twice the rated operating state (i.e., $\gamma_{r, i, 2\tau} = 1$, where $\tau$ is the nominal processing time), and the energy consumption per time slot is half the nominal value.

If a device can adjust its operating point in each period or even continuously adjust within the operating boundaries, then its processing time can be adjusted within a feasible range.
This constraint can be expressed in terms of the start time of the subsequent task (i.e., the end time of the task, as we assume the transfer tasks are executed immediately) as follows:
\begin{equation}\label{rtn_flexible_mode_1}
  \sum_{t' = t+\tau^{\rm L}_i}^{t+\tau^{\rm H}_i} N_{i+, t'} \ge N_{i, t}
\end{equation}
where $\tau^{\rm L}_i$ and $\tau^{\rm H}_i$ are the minimum and maximum processing times of task $i$, respectively, and $i$ and $i+$ are the indices of the flexible task and its subsequent task, respectively.

Under the flexible operating mode, the power variable of the device operation (a continuous variable) must also be introduced, subject to the following constraints:
\begin{equation}\label{rtn_flexible_mode_2}
  P^{\rm L}_{i} \cdot S_{i, t} \le P_{i, t} \le P^{\rm H}_{i} \cdot S_{i, t} \quad \forall i, t
\end{equation}
where $P^{\rm L}_{i}$ and $P^{\rm H}_{i}$ are the minimum and maximum processing powers of task $i$, respectively;
$P_{i, t}$ is the power of task $i$ at time slot $t$;
$S_{i, t}$ is the processing status of task $i$ at time slot $t$;
and $S_{i, t} = 1$ indicates that task $i$ is being processed at time slot $t$, which means the device is on and usually cannot be interrupted until the end of the task.
Naturally, $S_{i, t}$ is determined by the start times of the task and the subsequent task:
\begin{equation}\label{rtn_flexible_mode_3}
  S_{i, t} - S_{i, t-1} = N_{i, t} - N_{i+, t} \quad \forall t
\end{equation}

Finally, different control methods need to comply with the energy-material conversion conditions of processing itself, requiring a certain amount of energy to complete processing:
\begin{equation}\label{rtn_energy_requirement}
  \sum_t \Delta t P_{i, t} \ge E^{\rm req}_i \quad \forall i
\end{equation}
where $E^{\rm req}_i$ is the energy required to complete task $i$, which can be determined by the nominal power multiplied by the nominal processing time of the task if the same amount of energy is required for different processing modes.

\subsection{Overall Framework of the cRTN Model}

Having introduced the conventional RTN model, this section proposes the continuous Resource Task Network (cRTN) model.
While maintaining the same representational capabilities as the conventional RTN, the model introduces two fundamental improvements:

(1) The modeling of tasks and resources is decoupled from batches.
In other words, the same product or process in the same stage is modeled with only one resource or task.
Thus, unlike the conventional RTN, the scale of the model will not increase with the production target (number of batches).

(2) The decision variables of the model are changed from the starting time slot, operating states, and power at each time slot to the operating time for each state within each time slot.
This idea is consistent with the LSTN model proposed in this dissertation for continuous industrial processes, which utilizes the relatively large time scale of demand response to make the decision variables continuous, thereby achieving better computational performance.
The specific mathematical form of the continuous Resource Task Network (cRTN) model is given below.

\subsection{Mathematical Formulation of the cRTN Model}

In the following model, $i$ and $r$ represent the task and resource indices, respectively.
We still use $R_{r, t}$ to represent the quantity of resources, but relax it to a continuous variable.
$R_{r, t}$ can be understood as the progress of the task generating resource $r$.
$R_{r, t}=0(1)$ represents no progress (task completion), and $R_{r, t} \in (0, 1)$ represents that processing is ongoing.
This design enables the cRTN model to represent ordinary processing procedures and flexible (power-adjustable) production processes in a unified form, avoiding the separate modeling required in conventional RTNs.

The continuous variable $D_{i, k, t}$ represents the time that task $i$ runs in state $k$ within time slot $t$ (similar to LSTN).
The task-resource association matrix $\mathbf{G}$ is introduced to represent the resource change for each task.
The matrix entry $g_{r,i,k}$ represents the amount of resource $r$ generated (or consumed, if negative) per unit time when task $i$ operates in state $k$.

For ease of description, we assume by default that the state $k=0$ corresponds to the idle state of the task ($g_{r, i, 0}=0$), and $k=1$ and $k=2$ represent the states corresponding to the minimum and maximum processing speeds after the task starts running, respectively.
For example, for the melting process, under the flexible adjustment mode, the processing time can be controlled to between 75 and 125 minutes.
Hence, the generation rate, i.e., the amount of product (in batches) generated per unit time, is $g_{r, i, 1}=1/125$ to $g_{r, i, 2}=1/75$ (per minute).
For tasks with only one processing speed, it is only necessary to model two states: $k=0$ (off) and $k=1$ (on).
This design allows the model to uniformly represent regular and flexible production processes, without requiring separate modeling as in conventional RTNs.

(1) \textbf{Resource balance}:
Using continuous decision variables of tasks, the resource changes by task are also continuous.
The resource balance constraint can be expressed by (\ref{crtn_balance_1}):
\begin{equation}\label{crtn_balance_1}
  R_{r, t} \le R_{r, t-1} + \sum_{i} \sum_{k} g_{r, i, k} D_{i, k, t} \quad \forall r, t.
\end{equation}
Note that we use ``$\le$'' rather than ``$=$'' for two reasons.
On the one hand, this approach avoids the problem of conflict with other constraints for modeling based on discrete time slots (e.g., Equation (\ref{crtn_r_limit})).
Using ``$\le$'' does not change the effect of the constraint since equality must be enforced when minimizing the energy cost.
On the other hand, ``$<$'' can be satisfied by idling the equipment for a short time.
To validate this relaxation, we analyzed the gap between its two sides under different optimality gap settings (see Fig.~\ref{fig_resource_balance_gap} in the case study), demonstrating that our relaxation does not compromise the model's accuracy.

A task can only process products in batches, and usually multiple batches will not be allowed to accumulate at the same stage; therefore, we have:
\begin{equation}\label{crtn_r_limit}
  0 \le R_{i, t} \le 1 \quad \forall i, t.
\end{equation}

(2) \textbf{Task execution}:
By definition, $D_{i, k, t}$ is nonnegative and does not exceed the length of the time slot (\ref{crtn_task_execution_1}), and the total operating time of task $i$ in time slot $t$ equals the length of the time slot (\ref{crtn_task_execution_2}):
\begin{equation}\label{crtn_task_execution_1}
  0 \le D_{i, k, t} \le \Delta t \quad \forall i, t
\end{equation}
\begin{equation}\label{crtn_task_execution_2}
  \sum_{k} D_{i, k, t} = \Delta t \quad \forall i, t
\end{equation}
This formulation is quite different from the conventional RTN model, where task execution is modeled as an integer variable.
The formulation is inspired by the LSTN model for continuous industrial processes, where task execution is modeled as a continuous variable.
The equivalence of ``continuous adjustment within a time slot'' is stated from the perspective of a VPP participating in the electricity market. The VPP is primarily concerned with total energy consumption and cost at the hourly or 15-minute level, while task start and stop times can be arranged at the minute level (e.g., a device operates for 48 minutes and idles for 12 minutes within a time slot). At the VPP level, this schedule has the same interval energy as uniform consumption within that time slot.
However, from the grid operation perspective, ``12 minutes idle followed by 48 minutes active'' versus ``60 minutes of uniform consumption'' within the same time slot, while consistent in total energy, would produce a significant power step within the time slot for the former, and the specific timing of adjustment carries uncertainty that may affect real-time balancing and ancillary service requirements; such issues are better addressed at the VPP-grid interface (e.g., day-ahead/intraday bidding and real-time decomposition) or through grid-side flexibility characterization, whereas the cRTN model focuses on the VPP-side energy space and bidding decisions.
By using this formulation, we build a continuous RTN model on the basis of the linearized version of the state task network (STN) model, demonstrating that continuous production processes are a degenerate case of discrete production processes.
In other words, discrete industrial process models can be constructed by adding constraints that describe discrete characteristics to continuous industrial process models.

(3) \textbf{Product delivery}:
The amount of the final product $i^{\rm end}$ needs to reach the target by the end of the ending time slot, which can be expressed as:
\begin{equation}\label{crtn_product_delivery}
  R_{i^{\rm end}, \mathcal{T}} \ge R^{\rm tg}_{i^{\rm end}}
\end{equation}
where $R^{\rm tg}_{i^{\rm end}}$ is the target amount of product at the final stage.

(4) \textbf{Objective function}:
The objective function is the same as that of the conventional RTN model:
\begin{equation}\label{crtn_objective}
  {\rm min.} \sum_{hr} Pr_{hr} \sum_{t \in T_{hr}} E^{\rm EL}_{t}
\end{equation}
Let $P_{i, k}$ represent the power consumption of task $i$ operating in state $k$; then, the energy consumption of the whole production process in time slot $t$ is given by:
\begin{equation}\label{crtn_electricity_balance}
  E^{\rm EL}_{t} = \sum_i \sum_{k} P_{i, k} D_{i, k, t} \quad \forall t
\end{equation}

\subsection{Modeling of Discrete Industrial Processes in cRTN}

\begin{figure}[!t]
  \centering
  \includegraphics[width=\textwidth]{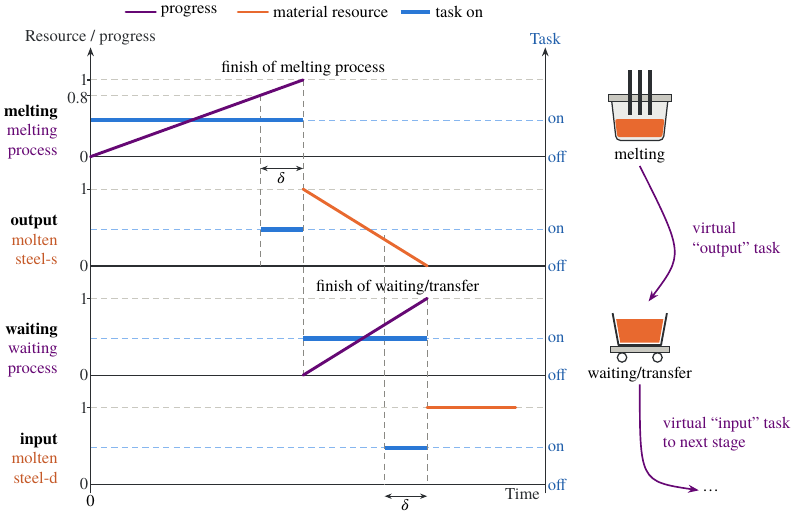}
  \caption{Virtual tasks for transferring completed batches between production stages}
  \label{fig_process_chap02}
\end{figure}

Compared to conventional RTN models, the task characteristics of discrete industrial processes are modeled differently in the cRTN model.
The proposed approach is a reformulation rather than a relaxation of RTN. Although continuous variables represent resources and task progress, the added constraints preserve the discrete nature of the industrial processes considered here.
The key modeling aspects are as follows:

(1) \textbf{Uninterruptible tasks}:
Once some tasks start running, they cannot stop until the batch is processed.
This type of task (such as the melting task) needs to satisfy $R_{r^{i+}, t}D_{i, 0, t} = 0$, where $r^{i+}$ is the resource produced by task $i$.
It is easy to verify that when $R_{r^{i+}, t}>0$, $D_{i, 0, t}=0$; that is, the task producing $R_{r^{i+}, t}$ has started, and it cannot be in the idle state ($k=0$), so it can only operate between the minimum ($k=1$) and maximum ($k=2$) processing speeds.
To avoid bilinear terms, we introduce integer variables $u_{i, t}$ to characterize the processing state of the task, where $u_{i, t}=1$ represents that the current process is ongoing.
For uninterruptible tasks, the above constraint can be rewritten as:
\begin{equation}\label{crtn_task_execution_3}
  D_{i, 0, t}  \le 1 - u_{i, t} \quad \forall t
\end{equation}
\begin{equation}\label{crtn_task_execution_4}
  R_{r^{i+}, t} \le u_{i, t} \quad \forall t
\end{equation}
These constraints mathematically ensure that once $R>0$, the task cannot be interrupted, preserving the discrete nature of the original constraints.

(2) \textbf{Batch-based processing}:
The subsequent task that consumes resource $r$ can only be carried out when the previous task is completely processed (i.e., $R_{r, t}=1$).
To reflect this characteristic of discrete industrial processes, we add virtual ``output'' tasks and ``input'' tasks between waiting tasks and processing tasks (Fig.~\ref{fig_process_chap02}).
Virtual tasks do not consume or generate actual resources, but mathematically guarantee the discreteness requirement of batch processing.
The virtual tasks must satisfy: (1) the output (input) task remains idle until the resource it consumes reaches 1; (2) in the same time slot when the preceding task is completed, it sets the consumed resource to 0 and the generated resource to 1, thus meeting the discreteness requirement of batch processing.
After introducing $u_{i, t}$, this characteristic of output (input) tasks can be expressed by:
\begin{equation}\label{crtn_task_execution_5}
  D_{i, 1, t} = u_{i, t} \Delta t \quad \forall t
\end{equation}
For resource $r^{i-}$ that the output (input) tasks consume, set $g_{r^{i-},i,1} = - 1/\Delta t$; for resource $r^{i+}$ that the output (input) tasks generate, set $g_{r^{i+},i,1} = 1/\Delta t$.
It is easy to verify that under constraints (\ref{crtn_balance_1}-\ref{crtn_task_execution_4}), (\ref{crtn_task_execution_5}) naturally meets the above requirements.

(3) \textbf{Waiting time limit}:
The waiting of intermediate products between production stages is also modeled as a task.
The waiting task consumes resources passed on by the preceding output task and generates resources for the input task.
Physically, waiting itself does not generate resources, so the resources here only represent the progress of waiting, and the parameters of the consumption (generation) rate are set to ensure that the total time for transferring and waiting for the intermediate products meets the requirements.
Therefore, the minimum and maximum generation rates of the waiting tasks are set to $1/(w_i+W_i)$ and $1/w_i$, respectively, where $w_i$ and $W_i$ are the transfer time and maximum waiting time of task $i$, respectively.
Naturally, the waiting tasks are also uninterruptible (satisfying Equations (\ref{crtn_task_execution_3}-\ref{crtn_task_execution_4})) because once the product reaches this task, waiting cannot be ``interrupted.''

Finally, since the waiting tasks already account for the waiting time, other tasks must start running immediately when their input materials arrive and cannot remain idle.
This can be represented by $R_{r^{i-}, t}D_{i, 0, t}=0$, where $r^{i-}$ represents the resources consumed by task $i$.
Similarly, this constraint can be replaced by (\ref{crtn_task_execution_3}) and (\ref{crtn_task_execution_6}):
\begin{equation}\label{crtn_task_execution_6}
  R_{r^{i-}, t} \le u_{i, t} \quad \forall t
\end{equation}

This section presents the principles and illustrative formulations of the modeling approach.
In practice, production scheduling problems with finite time horizons require additional constraints for the initial and final time slots.
For example, constraints must be added to prevent tasks with durations exceeding two time slots from starting in the final time slot.
For detailed implementation of these specific cases, readers can refer to the open-source code implementation of the cRTN model~\cite{rick10119_crtn_2024}.

\subsection{Unification of the LSTN and cRTN Models}

Having introduced both the LSTN and cRTN models, this section describes their underlying mathematical connection and the unification of continuous and discrete industrial process modeling.

In the LSTN model, resource balance constraints are used to track the evolution of material states over time.
Let $S_{t,i}$ (kg) denote the amount of material $i$ at the end of time period $t$.
The model uses $\Delta t_{t,i,k}$ (h) to represent the continuous time that task $i$ operates at operating point $k$ within time period $t$.
For each task $i$, let $G_{i,k}$ and $C_{i,k}$ (kg/h) denote the material production and consumption rates at operating point $k$, respectively.
The evolution of material states is governed by three different balance equations for feedstock, intermediate products, and final products, as given by Equations (\ref{primal_constraint_changeofS1_chap02})--(\ref{primal_constraint_changeofS3_chap02}).

These equations can be generalized using more compact notation.
By introducing $g_{r, i, k}$ to denote the consumption/production rate of task $i$ at operating point $k$ for resource $r$ (where $g_{r, i, k} > 0$ indicates production and $g_{r, i, k} < 0$ indicates consumption), the resource balance can be expressed in a unified form:
\begin{equation}\label{stn_balance}
  R_{r, t} = R_{r, t-1} + \sum_{i} \sum_{k} g_{r, i, k} \Delta t_{t,i,k} \quad \forall r, t
\end{equation}
where $R_{r, t}$ represents the quantity of resource $r$ at time $t$.

Comparing Equation (\ref{stn_balance}) with the resource balance constraint (\ref{crtn_balance_1}) of the cRTN model, one can observe that the two are mathematically identical in form.
This reveals an important theoretical insight: the LSTN and cRTN models share the same core constraint structure, and their essential difference is that the cRTN model adds additional constraints (Equations (\ref{crtn_task_execution_3})--(\ref{crtn_task_execution_6})) on top of this structure to describe the characteristics of discrete industrial processes.
These additional constraints ensure that batch-based production, uninterruptibility, and other discrete characteristics are strictly satisfied.

Therefore, from a mathematical perspective, the cRTN model is the more general industrial process modeling framework, while the LSTN model can be viewed as a special case of the cRTN model---when all discrete constraints are removed, the cRTN model degenerates into the LSTN model.
In other words, continuous industrial processes are a degenerate case of discrete industrial processes, and both can be described within a unified modeling framework.
This theoretical unification provides a new perspective for understanding the essence of industrial process modeling and establishes a theoretical foundation for subsequent research.

\section{Illustrative Comparison of cRTN and RTN}\label{sec_illustrative}

\begin{figure}[!t]
  \centering
  \includegraphics[width=0.5\textwidth]{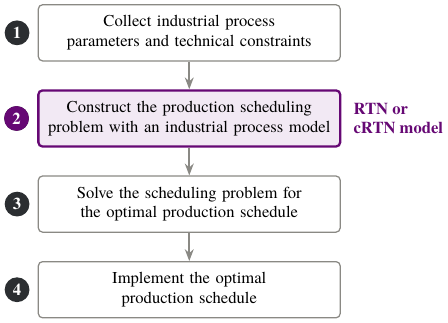}
  \caption{Industrial production scheduling procedure}
  \label{fig_flowchart}
\end{figure}

To illustrate the conceptual and mathematical differences between RTN and cRTN modeling approaches, we present a simple example of modeling the same discrete industrial process using both methods.
Specifically, we model an electric arc furnace (EAF) in a secondary steelmaking plant to minimize electricity costs while satisfying technical constraints and production targets.
The EAF processes scrap steel in batches through heating and melting.
Its original parameters are shown in Table~\ref{tab_parameter_illustrative}, where time parameters are expressed in terms of time slots with a slot length of 1 hour.
We consider a 6-hour scheduling horizon, resulting in a single production stage ($S=1$) and time horizon $\mathcal{T}=6$.
The production target is to provide 2 batches of molten steel ($b=2$) for subsequent production stages by the end of the time horizon.
Time-of-use electricity prices are presented in Table~\ref{tab_price_scenarios} (unit: \$/MWh).
The flowchart of the industrial process scheduling is shown in Fig.~\ref{fig_flowchart}, where Step 2---construction of the production scheduling problem---is where the RTN or cRTN model is applied.

\begin{table}[!t]
  \caption{Parameters of the electric arc furnace}
  \label{tab_parameter_illustrative}
  \centering
  \begin{tabular}{ll}
    \toprule
    Parameter                   & Value \\ \midrule
    Nominal power (MW)           & 1  \\ 
    Processing time (minute/time slot)      & 120/2  \\
    Transfer time (minute/time slot)      & 60/1  \\
    Maximum waiting time (minute/time slot)   & 120/2 \\ \bottomrule
  \end{tabular}
\end{table}

\begin{table}[!t]
  \caption{Time-of-use electricity prices for two scenarios}
  \label{tab_price_scenarios}
  \centering
  \begin{tabular}{lcccccc}
    \toprule
    Hour & 1 & 2 & 3 & 4 & 5 & 6 \\ \midrule
    Scenario 1 & 100 & 100 & 100 & 100 & 200 & 200 \\
    Scenario 2 & 100 & 100 & 100 & 200 & 200 & 100 \\ \bottomrule
  \end{tabular}
\end{table}

\subsection{Conventional RTN Modeling}

\begin{figure}[!t]
  \centering
  \includegraphics[width=0.93\textwidth]{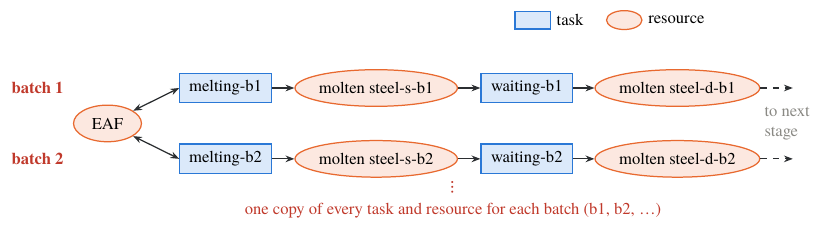}
  \caption{Resource-task network (RTN) representation of the electric arc furnace}
  \label{fig_rtn_model}
\end{figure}

In the RTN framework (Fig.~\ref{fig_rtn_model}), EAF-related tasks include: 1.\ processing (melting), 2.\ transferring, and 3.\ waiting.
To simplify the model, transferring and waiting tasks can be combined into a single waiting task.
This results in the RTN requiring modeling of 5 types of resources and 4 types of tasks, as listed in Tables~\ref{tab_rtn_resources} and~\ref{tab_rtn_tasks}, where b1 and b2 represent the two batches.
The number of resources, tasks, and corresponding variables scales with the number of batches (Table~\ref{tab_rtn_variables}).
Table~\ref{tab_rtn_tasks} also provides examples of vectors in the interaction matrix modeling task-resource relationships.
For instance, $\gamma_{1,1,[\theta]}= [-1, 0, 1]$ indicates that the melting task for batch 1 occupies the EAF at its start time slot and releases it after two time slots.
Similarly, $\gamma_{2,1,[\theta]}= [0, 0, 1]$ shows that the melting task for batch 1 produces one unit of molten steel-s-1 two time slots after initiation, where s(d) denotes the product located at the transfer start (destination).

\begin{table}[!t]
  \caption{Tasks and their resource interactions in the RTN model}
  \label{tab_rtn_tasks}
  \centering
  \begin{tabular}{llll}
    \toprule
    Index (i) & Task & \begin{tabular}[c]{@{}l@{}}Related\\ resources (r)\end{tabular} & \begin{tabular}[c]{@{}l@{}}Interaction matrix\\ $\gamma_{r,i,[\theta]}$\end{tabular} \\ 
    \midrule
    1 & melting-b1 & 1, 2 & \begin{tabular}[c]{@{}l@{}}$[-1, 0, 1]$, $[0, 0, 1]$\end{tabular} \\
    2 & melting-b2 & 1, 3 & \begin{tabular}[c]{@{}l@{}}$[-1, 0, 1]$, $[0, 0, 1]$\end{tabular} \\
    3 & waiting-b1 & 2, 4 & \begin{tabular}[c]{@{}l@{}}$[-1, 0]$, $[0, 1]$\end{tabular} \\
    4 & waiting-b2 & 3, 5 & \begin{tabular}[c]{@{}l@{}}$[-1, 0]$, $[0, 1]$\end{tabular} \\
    \bottomrule
  \end{tabular}
\end{table}

\begin{table}[!t]
  \caption{Resources and their task associations in the RTN model}
  \label{tab_rtn_resources}
  \centering
  \begin{tabular}{lll}
    \toprule
    Index (r) & Resource & Related tasks (i) \\ 
    \midrule
    1 & EAF & 1, 2 \\
    2 & molten steel-s-b1 & 3, 4 \\
    3 & molten steel-s-b2 & 3, 4 \\
    4 & molten steel-d-b1 & 3, 4 \\
    5 & molten steel-d-b2 & 3, 4 \\
    \bottomrule
  \end{tabular}
\end{table}

\begin{table}[!t]
  \caption{Variables in the RTN model}
  \label{tab_rtn_variables}
  \centering
  \begin{tabular}{lll}
    \toprule
    Variable & Type & Number \\ 
    \midrule
    $R$: value of the resources & integer & $S(1+2b)\mathcal{T}=30$ \\
    $N$: start time of the tasks & integer & $S(2b)\mathcal{T}=24$ \\
    \bottomrule
  \end{tabular}
\end{table}

Under this model, the RTN constraints are given by Equations (\ref{rtn_balance})--(\ref{rtn_product_delivery}), while the plant's electricity cost is represented by Equation (\ref{rtn_objective}), which calculates the total electricity procurement cost under the given time-of-use prices (Table~\ref{tab_price_scenarios}).
The production scheduling problem based on this RTN formulation is an integer programming problem.
We denote the scheduling results from this model as the Basic RTN solution.

For flexible tasks, additional power variables $P$ must be introduced to model the adjustable range (we assume power can be adjusted between 1/2 and 4/3 of the nominal value).
This introduces new continuous variables and additional constraints (Equations (\ref{rtn_flexible_mode_1})--(\ref{rtn_energy_requirement})), transforming the original integer programming problem into a mixed-integer linear programming (MILP) problem.

\subsection{cRTN Modeling}

\begin{table}[!t]
  \caption{Parameters of the cRTN model for the electric arc furnace}
  \label{tab_model_parameter_illustrative}
  \centering
  \begin{tabular}{lllll}
  \toprule
  Task (index)                    & \begin{tabular}[c]{@{}l@{}}Operating\\ state (k)\end{tabular} & \begin{tabular}[c]{@{}l@{}}Processing\\ rate (/$\Delta t$)\end{tabular} & \begin{tabular}[c]{@{}l@{}}Operating\\ power (MW)\end{tabular} \\
  \midrule
  melting (1) & 0                          & 0                          & 0                          \\
                 & 1                          & $1/4$              & $1/4$                  \\
                 & 2                          & $2/3$             & $2/3$                 \\ \hline
  output (2)     & 0                          & 0                          & 0                          \\
                 & 1                          & 1                          & 0                          \\ \hline
  waiting (3)    & 0                          & 0                          & 0                          \\
                 & 1                          & 1/3                      & 0                          \\
                 & 2                          & 1/2                        & 0                          \\ \hline
  input (4)      & 0                          & 0                          & 0                          \\
                 & 1                          & 1                          & 0                          \\  \bottomrule
\end{tabular}
\end{table}

\begin{figure}[!t]
  \centering
  \includegraphics[width=0.93\textwidth]{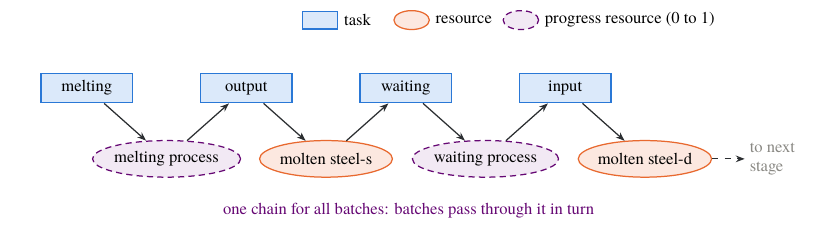}
  \caption{Continuous resource-task network (cRTN) representation of the electric arc furnace}
  \label{fig_cRTN_model}
\end{figure}

In the cRTN model (Fig.~\ref{fig_cRTN_model}), we similarly combine the transferring and waiting tasks to simplify the model.
Thus, the maximum and minimum processing rates of the combined task (referred to as the waiting task) are set to 1/3 and 1/2, respectively (Table~\ref{tab_model_parameter_illustrative}).
The cRTN requires modeling of 4 types of resources and 4 types of tasks, as shown in Tables~\ref{tab_crtn_tasks} and~\ref{tab_crtn_resources}.
Notably, unlike the conventional RTN, the number of resources, tasks, and corresponding variables is independent of the number of batches (Table~\ref{tab_crtn_variables}).
The association matrix entries $g_{r,i,k}$ represent the rate of resource generation (negative values indicate consumption).
For the EAF with flexible power adjustment between 1/4 and 2/3 of nominal power, the modeling approach remains consistent with non-flexible tasks: $g_{1,1,[k]}= [0, 1/4, 2/3]$ represents the progress rate of the melting task per time slot in states 0 (idle), 1 (minimum power), and 2 (maximum power), respectively.

\begin{table}[!t]
  \caption{Tasks and their resource associations in the cRTN model}
  \label{tab_crtn_tasks}
  \centering
  \begin{tabular}{llll}
    \toprule
    Index (i) & Task & \begin{tabular}[c]{@{}l@{}}Related\\ resources (r)\end{tabular} & \begin{tabular}[c]{@{}l@{}}Association matrix\\ $g_{r,i,[k]}$\end{tabular} \\ 
    \midrule
    1 & melting & 1 & $[0, 1/4, 2/3]$ \\
    2 & output & 1, 2 & \begin{tabular}[c]{@{}l@{}}$-[0, 1]$, $[0, 1]$\end{tabular} \\
    3 & waiting & 2, 3 & \begin{tabular}[c]{@{}l@{}}$-[0, 1/3, 1/2]$, $[0, 1/3, 1/2]$\end{tabular} \\
    4 & input & 3, 4 & \begin{tabular}[c]{@{}l@{}}$-[0, 1]$, $[0, 1]$\end{tabular} \\
    \bottomrule
  \end{tabular}
\end{table}

\begin{table}[!t]
  \caption{Resources and their task associations in the cRTN model}
  \label{tab_crtn_resources}
  \centering
  \begin{tabular}{llll}
    \toprule
    Index (r) & Resource & Generated by task & Consumed by task \\ 
    \midrule
    1 & melting process & 1 & 2 \\
    2 & molten steel-s & 2 & 3 \\
    3 & waiting process & 3 & 4 \\
    4 & molten steel-d-1 & 4 & next stage \\
    \bottomrule
  \end{tabular}
\end{table}

\begin{table}[!t]
  \caption{Variables in the cRTN model}
  \label{tab_crtn_variables}
  \centering
  \begin{tabular}{lll}
    \toprule
    Variable & Type & Number \\ 
    \midrule
    $R$: value of the resources  & continuous & $S(4)\mathcal{T}=12$  \\
    $D$: operating duration of the resources & continuous & $S(3\times4)\mathcal{T}=36$ \\
    $u$: operating status of the tasks & integer & $S(4)\mathcal{T}=24$ \\
    \bottomrule
  \end{tabular}
\end{table}

Under this formulation, the production scheduling model based on cRTN (Equations (\ref{crtn_balance_1})--(\ref{crtn_task_execution_6})) is an MILP problem.
Since not all processes on the production line are necessarily uninterruptible, the cRTN model requires at most $4S|\mathcal{T}|$ integer variables, independent of the number of batches.
In contrast, the RTN model requires $4bS|\mathcal{T}|$ integer variables, which scale linearly with the number of batches $b$.
For larger-scale problems, such as when $b=10$, the number of integer variables in the RTN model would be ten times that of the cRTN model.

\subsection{Comparison of Scheduling Results}

\begin{figure}[!t]
  \centering
  \includegraphics[width=0.8\textwidth]{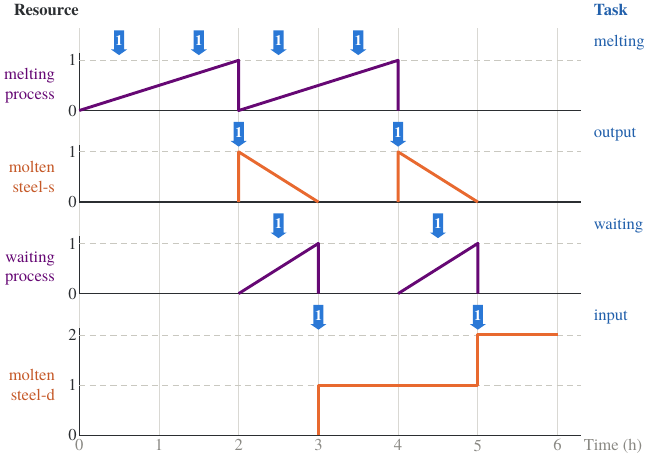}
  \caption{Electric arc furnace schedule from the continuous resource-task network (cRTN) model in Scenario 1}
  \label{fig_cRTN_result_scenario1}
\end{figure}

\begin{figure}[!t]
  \centering
  \includegraphics[width=0.8\textwidth]{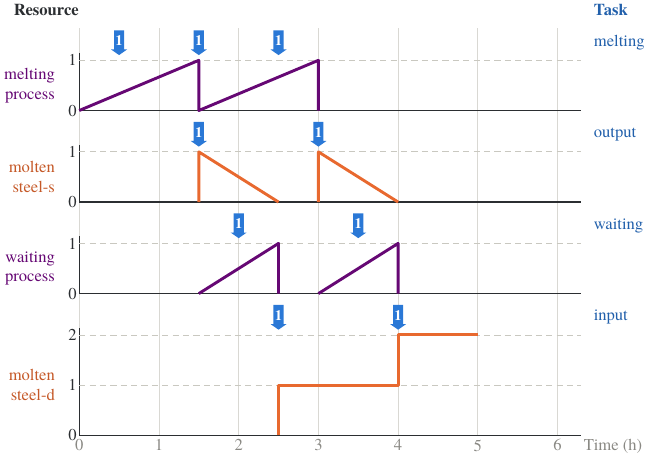}
  \caption{Electric arc furnace schedule from the continuous resource-task network (cRTN) model in Scenario 2}
  \label{fig_cRTN_result_scenario2}
\end{figure}

We compared the production scheduling results based on RTN and cRTN under two electricity price scenarios, with the main decision variables and production status listed in Tables~\ref{tab:scenario1} and~\ref{tab:scenario2} (values at the end of each time slot).
For RTN, the earliest start time N can be at the end of time slot 0 (beginning of time slot 1) to complete the production of 1 batch by the end of time slot 2.
The cRTN scheduling results are also visualized in Figs.~\ref{fig_cRTN_result_scenario1} and~\ref{fig_cRTN_result_scenario2}.

\begin{table}[!t]
  \caption{Comparison of RTN and cRTN scheduling results in Scenario 1}
  \label{tab:scenario1}
  \centering
  \begin{tabular}{llcccccc}
    \toprule
    Model & Time slot & 1 & 2 & 3 & 4 & 5 & 6 \\
    \midrule
    \multirow{4}{*}{RTN} 
    & EAF power/MW & 1 & 1 & 1 & 1 & 0 & 0 \\
    & molten steel-s & 0 & 1 & 0 & 1 & 0 & 0 \\
    & waiting & 0 & 1 & 0 & 1 & 0 & 0 \\
    & molten steel-d & 0 & 0 & 1 & 1 & 2 & 2 \\
    \midrule
    \multirow{4}{*}{cRTN}
    & EAF power/MW & 1 & 1 & 1 & 1 & 0 & 0 \\
    & molten steel-s & 0 & 1 & 0 & 1 & 0 & 0 \\
    & waiting & 0 & 1 & 0 & 1 & 0 & 0 \\
    & molten steel-d & 0 & 0 & 1 & 1 & 2 & 2 \\
    \bottomrule
  \end{tabular}
\end{table}

\begin{table}[!t]
  \caption{Comparison of RTN and cRTN scheduling results in Scenario 2}
  \label{tab:scenario2}
  \centering
  \begin{tabular}{llcccccc}
    \toprule
    Model & Time slot & 1 & 2 & 3 & 4 & 5 & 6 \\
    \midrule
    \multirow{4}{*}{RTN} 
    & EAF power/MW & 4/3 & 2/3 & 4/3 & 2/3 & 0 & 0 \\
    & molten steel-s & 0 & 1 & 0 & 1 & 0 & 0 \\
    & waiting & 0 & 1 & 0 & 1 & 0 & 0 \\
    & molten steel-d & 0 & 0 & 1 & 1 & 2 & 2 \\
    \midrule
    \multirow{4}{*}{cRTN}
    & EAF power/MW & 4/3 & 4/3 & 4/3 & 0 & 0 & 0 \\
    & molten steel-s & 0 & 1 & 1 & 0 & 0 & 0 \\
    & waiting & 0 & 1 & 1 & 0 & 0 & 0 \\
    & molten steel-d & 0 & 0 & 1 & 2 & 2 & 2 \\
    \bottomrule
  \end{tabular}
\end{table}

\begin{figure}[!t]
  \centering
  \includegraphics[width=0.8\textwidth]{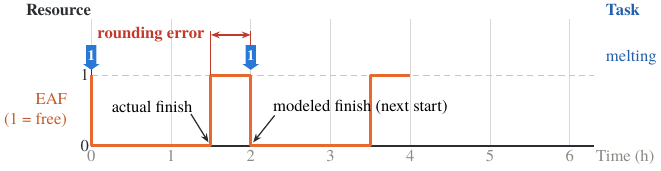}
  \caption{Illustration of time-grid rounding error in the resource-task network (RTN) model}
  \label{fig_rounding_error}
\end{figure}

In Scenario 1, the first four time slots have lower electricity prices, which exactly matches the time needed to produce 2 batches at nominal power.
Both RTN and cRTN yield identical scheduling results, validating that cRTN maintains RTN's expressiveness, particularly in how the designed output and input tasks preserve the discrete nature of batch-based processing even after resource continualization.

In Scenario 2, while the first three time slots have lower electricity prices, RTN still requires four time slots to complete production due to rounding errors (Fig.~\ref{fig_rounding_error}).
This occurs because in RTN, a resource (EAF) can only be occupied by one task within a time slot.
Therefore, in time slot 2, although the first batch's production could be completed at the midpoint, the second batch's production can only begin in the next time slot.
In contrast, cRTN can complete the production of 2 batches within three time slots by operating at 4/3 of nominal power, thereby reducing the electricity cost for the second batch by 25\%.

\section{Numerical Tests of the cRTN Model}

We present a more realistic use case for the cRTN model and validate through numerical results that cRTN can significantly reduce computation time while maintaining model accuracy.
Optimization problems were solved using Gurobi (V12.0.0)~\cite{gurobi} and MATLAB (R2024b)~\cite{matlab} with YALMIP~\cite{Lofberg2004}.
Computation was executed on a workstation with an Intel Core i9-10900X CPU (3.7 GHz) and 128 GB RAM.
The default optimality gap setting for Gurobi is 1e-4.
Code and data are available at \inlinecite{rick10119_crtn_2024}.
For consistency with electricity meter specifications, unless otherwise stated, figures and tables in this dissertation uniformly label the electricity consumption (kWh) of industrial loads in each time period rather than power (kW).

\subsection{Test Settings for the cRTN Model}

We borrow the scenario of the intraday production schedule of a steel plant under spot electricity prices from Ref.~\inlinecite{ref10} and use the same parameters for comparison with the conventional RTN model.
The steel plant production process to be modeled includes four stages (Fig.~\ref{fig_rtn_process}), namely, the electric arc furnace (EAF), argon oxygen decarburization (AOD), ladle furnace (LF), and continuous casters (CC), the original technical parameters of which are listed in Table~\ref{tab_parameter_crtn}.
All four stages are typical discrete processes carried out in batches, and each batch, once started, cannot be interrupted.
The power of the EAF can be adjusted between 75\% and 125\% of the nominal power, providing additional flexibility.
Without loss of generality, we do not consider equipment maintenance time or coupling between parallel production lines, as our model does not make significant improvements in these areas and instead uses existing modeling techniques.

We set the time slot length to 5 minutes, which is consistent with the pricing intervals of some real-time markets.
According to the proposed modeling method, the model parameters converted from the original technical parameters are listed in Table~\ref{tab_model_parameter}, where the processing rate represents the progress of tasks within each time slot and VT represents the virtual tasks.
For example, the processing rate of the waiting task (3) in operating state 2 is 1/3, which means the process needs to be in this state for 3 time slots to complete the transportation between the EAF and AOD (the shortest waiting time is the transportation time, (3-1)*5=10 minutes).
The parameters corresponding to the idle state ($k = 0$) are 0.

\begin{table}[!t]
  \caption{Parameters of the original steel plant}
  \label{tab_parameter_crtn}
  \centering
  \begin{tabular}{lllll}
    \toprule
    Process                   & EAF & AOD & LF  & CC \\ \midrule
    Nominal power (MW)        & 85  & 2   & 2   & 7  \\
    Processing time (minute)  & 80  & 75  & 35  & 50 \\
    Transfer time (minute)    & 10  & 5   & 10  & /  \\
    Max waiting time (minute) & 120 & 120 & 120 & /  \\ \bottomrule
  \end{tabular}
\end{table}

\begin{table}[!t]
  \caption{Parameters of the continuous RTN model}
  \label{tab_model_parameter}
  \centering
  \begin{tabular}{lllll}
    \toprule
  Process              & Task (index)                    & \begin{tabular}[c]{@{}l@{}}Operating\\ state (k)\end{tabular} & \begin{tabular}[c]{@{}l@{}}Processing\\ rate (/$\Delta t$)\end{tabular} & \begin{tabular}[c]{@{}l@{}}Operating\\ power (MW)\end{tabular} \\
  \midrule
  \multirow{6}{*}{EAF} & \multirow{3}{*}{processing (1)} & 0                          & 0                          & 0                          \\
                       &                                 & 1                          & $75\% \times (5/75)$              & 75\%$\times$85                  \\
                       &                                 & 2                          & $125\% \times (5/75)$             & 125\%$\times$85                 \\ \cline{2-5}
                       & \multirow{3}{*}{waiting (3)}    & 0                          & 0                          & 0                          \\
                       &                                 & 1                          & 1/49                       & 0                          \\
                       &                                 & 2                          & 1/3                        & 0                          \\ \hline
  \multirow{4}{*}{AOD} & \multirow{2}{*}{processing (5)} & 0                          & 0                          & 0                          \\
                       &                                 & 1                          & 1/15                       & 2                          \\ \cline{2-5}
                       & \multirow{3}{*}{waiting (7)}    & 0                          & 0                          & 0                          \\
                       &                                 & 1                          & 1/49                       & 0                          \\
                       &                                 & 2                          & 1/2                        & 0                          \\ \hline
  \multirow{4}{*}{LF}  & \multirow{2}{*}{processing (9)} & 0                          & 0                          & 0                          \\
                       &                                 & 1                          & 1/7                        & 2                          \\ \cline{2-5}
                       & \multirow{3}{*}{waiting (11)}   & 0                          & 0                          & 0                          \\
                       &                                 & 1                          & 1/25                       & 0                          \\
                       &                                 & 2                          & 1/3                        & 0                          \\ \hline
  \multirow{2}{*}{CC}  & \multirow{2}{*}{processing (13)} & 0                          & 0                          & 0                          \\
                       &                                 & 1                          & 1/10                       & 7                          \\ \hline
                       \multirow{2}{*}{VT}                    & \begin{tabular}[c]{@{}l@{}}output \\ (2, 6, 10, 14)\end{tabular}      & 0                          & 0                          & 0                          \\
                       &                                                              & 1                          & 1                          & 0                          \\ \hline
                       \multirow{2}{*}{VT}                    & \begin{tabular}[c]{@{}l@{}}input \\ (4, 8, 12)\end{tabular}      & 0                          & 0                          & 0                          \\
                       &                                                              & 1                          & 1                          & 0                          \\  \bottomrule
                      \end{tabular}
                    \end{table}

\begin{figure}[!t]
  \centering
  \includegraphics[width=\textwidth]{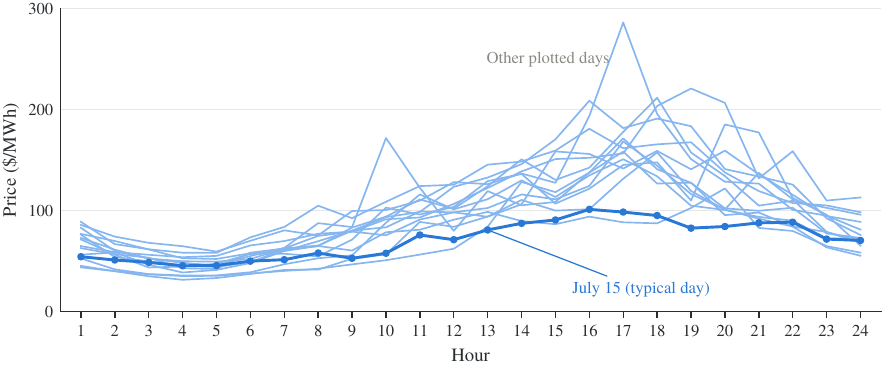}
  \caption{PJM real-time electricity prices used in the July 2022 scheduling tests}
  \label{fig_price_chap02}
\end{figure}

In the numerical tests, model accuracy is measured by the difference in the daily load curve and energy cost given by the cRTN model and the optimal scheduling results given by the conventional RTN model, calculated as the RMSE:
$${\rm RMSE}=\sqrt{\frac{\sum^N_{i=1} (y^*_i - y_i)^2}{N}}$$
To account for the differences in electricity prices on different days, we used the real-time system electricity prices of PJM on 15 days in July 2022 (Fig.~\ref{fig_price_chap02}) and tested the average error.
The original hourly prices were linearly interpolated to obtain prices every 5 minutes.
We avoided dates with several hours of very close electricity prices because, in these cases, the optimal production scheduling may not be unique (e.g., production can be arbitrarily shifted between two hours with the same electricity price without affecting the optimality of the result), which would lead to an overestimation of the model error.

\subsection{Scheduling Results of the cRTN Model}

\begin{table}[!t]
  \caption{Performance comparison of the cRTN model with a production target of 8 heats}
  \label{tab_rmse_models_crtn}
  \centering
  \begin{tabular}{llll}
    \toprule
    Model                      &
    \begin{tabular}[c]{@{}c@{}}Rounding \\ error (RMSE)  \end{tabular} &
    \begin{tabular}[c]{@{}c@{}}Average daily \\ energy cost (\$) \end{tabular} &
    \begin{tabular}[c]{@{}c@{}}Mean solution\\ time (minutes) \end{tabular}                                                                      \\ \midrule
    RTN (gap=0.1)                        & -     & 54571                    & -                    \\
    RTN (gap=0.0001)                        & 2.13\%     & 50763(-6.9\%)                    & 156.7                    \\
    cRTN (gap=0.0001)                      & 0 & \textbf{50421 (-7.4\%)} & \textbf{17.1 (-89.0\%)} \\\bottomrule
  \end{tabular}
\end{table}

\begin{figure}[!t]
  \centering
  \includegraphics[width=\textwidth]{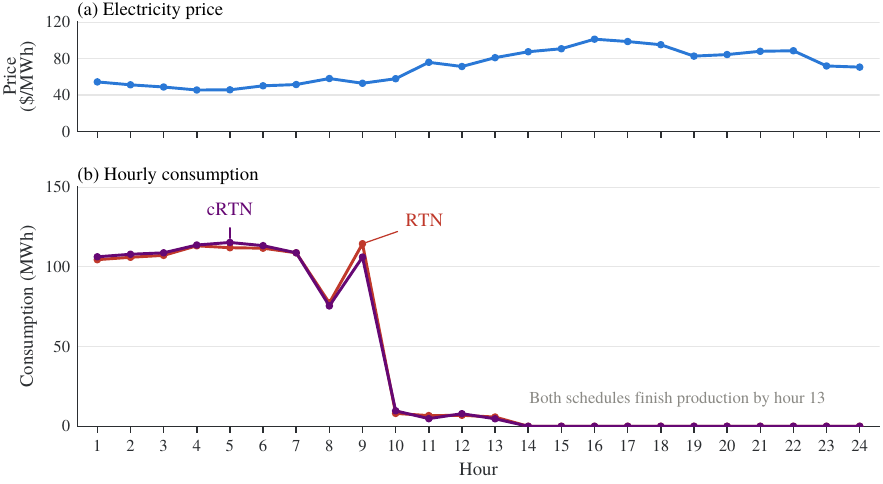}
  \caption[Optimal hourly electricity consumption from the RTN and cRTN models on July 15]{Optimal hourly electricity consumption from the resource-task network (RTN) and continuous resource-task network (cRTN) models on July 15: (a) electricity price; (b) load profiles}
  \label{fig_typical_load}
\end{figure}

(1) \textbf{Overall performance}:
The overall performance of cRTN is described in Table~\ref{tab_rmse_models_crtn}.
When comparing against the unoptimized case with an optimality gap of 0.1 (which serves as the baseline), both the RTN model and the cRTN model with a tighter gap (0.0001) demonstrate significant improvements in energy cost reduction (-6.9\% and -7.4\%, respectively).
Moreover, cRTN significantly reduces the solution time for the production scheduling problem compared with the conventional RTN modeling method.
In particular, using the cRTN method, the solution time can be reduced from 156.7 minutes for the conventional RTN model to 17.1 minutes (-89.0\%).

\begin{figure}[!t]
  \centering
  \includegraphics[width=\textwidth]{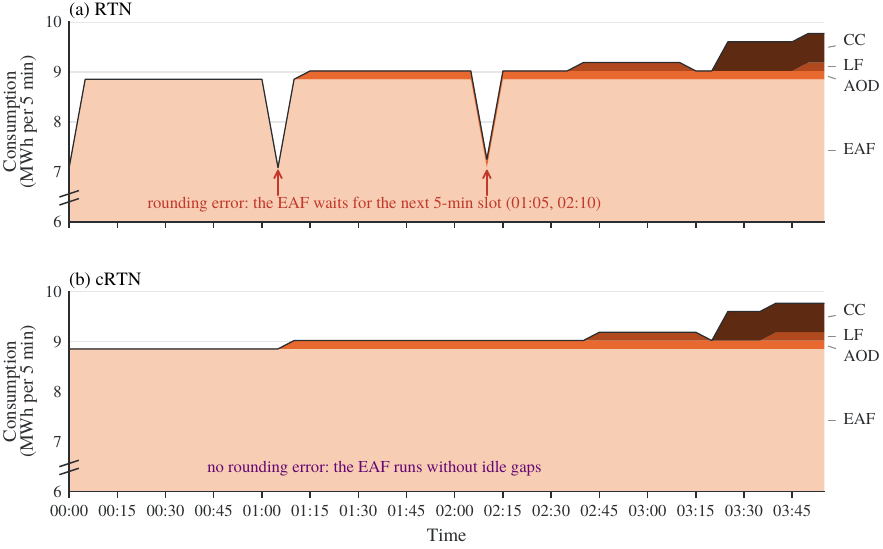}
  \caption[Equipment-level electricity consumption from 00:00 to 04:00 on July 15]{Equipment-level electricity consumption from 00:00 to 04:00 on July 15: (a) resource-task network (RTN) model; (b) continuous resource-task network (cRTN) model}
  \label{fig_typical_load_decp_compare}
\end{figure}

(2) \textbf{Rounding error}:
The relative difference between RTN and cRTN's optimal production schedules under 15 different electricity price scenarios is approximately 2.13\%, which can be attributed to the RTN's rounding errors.
For example, due to the constraints of Equation (\ref{rtn_balance}) under the discrete time horizon framework, the equipment (e.g., EAF) in the RTN model must wait until the next time slot (at 01:05) to start processing the next batch after completing the previous batch (e.g., at 01:01).
This means the device is ``idle'' from 01:01 to 01:05 due to insufficient model representation, which in turn means the energy use of the RTN model in the corresponding period is not fully optimized.
This source of error was also reported by Zhang et al.~\cite{ref10} and was referred to as the rounding error.
In contrast, the modeling method of cRTN eliminates the rounding error (see Figs.~\ref{fig_typical_load_decp_compare}), allowing the discrete industrial process to fully utilize its energy flexibility, resulting in a further reduction of approximately 0.5\% in energy costs.
The impact of the rounding error on energy cost reduction would be more significant if the length of the time slot were longer.
To visualize these differences, we provide the energy consumption curves of the steel plant under the optimal production schedules given by various methods on a typical day (July 15) in Fig.~\ref{fig_typical_load}.

\begin{figure}[!t]
  \centering
  \includegraphics[width=\textwidth]{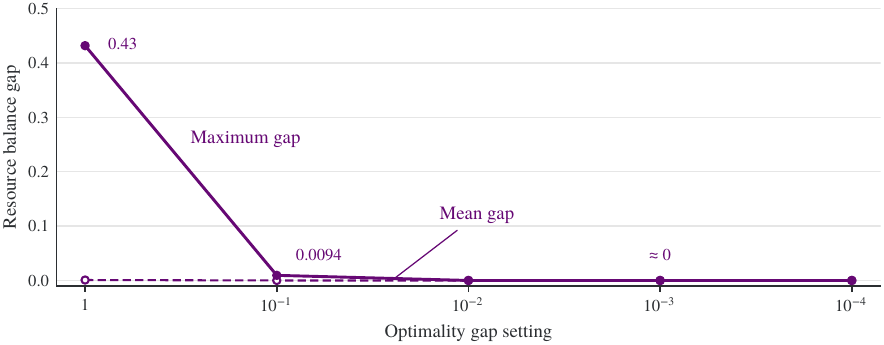}
  \caption[Resource-balance residuals in the cRTN model under different optimality gaps]{Resource-balance residuals in the continuous resource-task network (cRTN) model under different solver optimality-gap settings}
  \label{fig_resource_balance_gap}
\end{figure}

To validate our treatment of the resource balance constraint (Equation (\ref{crtn_balance_1})), we analyzed the gap between its two sides under different optimality gap settings, demonstrating that our relaxation does not compromise the model's accuracy.
As shown in Fig.~\ref{fig_resource_balance_gap}, the maximum resource-balance residual decreases from approximately 0.43 at an optimality-gap setting of 1 to 0.0094 at 0.1, and is indistinguishable from zero in the plot at settings of 0.01 or smaller.

\begin{figure}[!t]
  \centering
  \includegraphics[width=\textwidth]{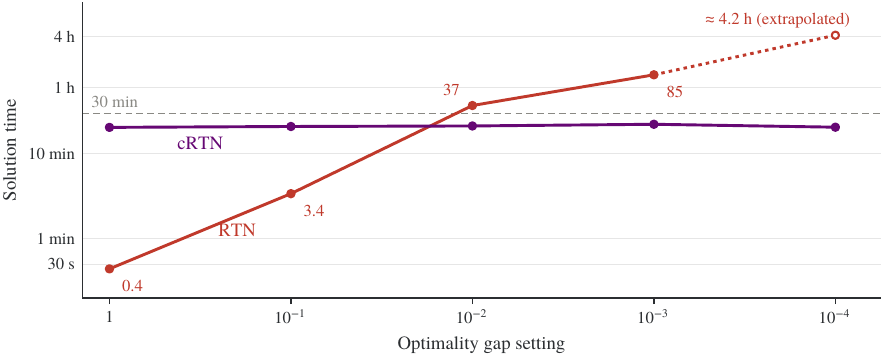}
  \caption[Solution times of the RTN and cRTN models under different optimality gaps]{Solution times of the resource-task network (RTN) and continuous resource-task network (cRTN) models under different solver optimality-gap settings}
  \label{fig_solution_time}
\end{figure}

(3) \textbf{Impact of optimality gap}:
Fig.~\ref{fig_solution_time} illustrates this performance comparison, showing solution times for both models across different gap settings.
For the RTN model, the solution time increases exponentially as the optimality gap requirement becomes stricter.
With gaps below 1e-2, computation time exceeds one hour.
The cRTN model demonstrates stable performance across different gap settings.
Even with very strict gap requirements (e.g., 1e-4), the increase in solution time remains marginal.
From the perspective of solution quality, under a fixed time limit of 30 minutes, cRTN consistently achieves gaps smaller than 1e-4, ensuring high solution quality.
In contrast, RTN can only reach gaps around 1e-2 within the same time constraint.
The superior convergence properties of cRTN make it particularly suitable for industrial production scheduling problems where both computational efficiency and solution precision are critical requirements.

\subsection{Scalability of the cRTN Model}

\begin{figure}[!t]
  \centering
  \includegraphics[width=\textwidth]{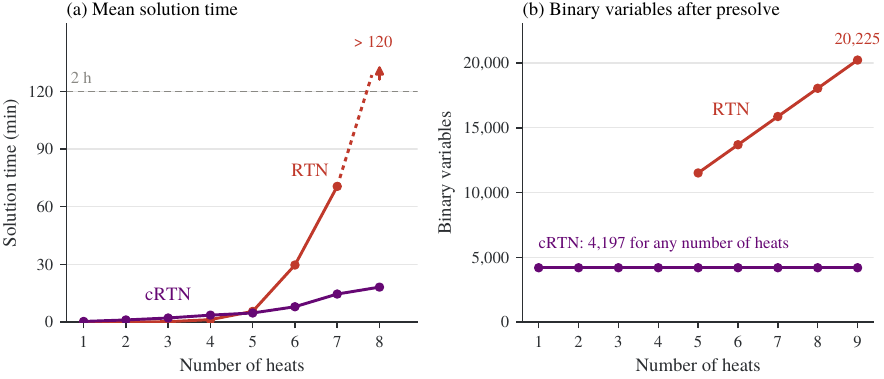}
  \caption[RTN and cRTN models under different production targets]{Resource-task network (RTN) and continuous resource-task network (cRTN) models under different production targets: (a) mean solution time over 15 price scenarios; (b) number of binary variables after presolve}
  \label{fig_calculation_time_crtn}
\end{figure}

\begin{table}[!t]
  \caption{Comparison of the number of variables in RTN and cRTN models}
  \label{tab_nofv_models}
  \centering
  \begin{tabular}{lll}
    \toprule
    Model      & \begin{tabular}[c]{@{}l@{}}Binary\\      variables\end{tabular} & \begin{tabular}[c]{@{}l@{}}Continuous\\      variables\end{tabular} \\ \midrule
    cRTN       & 4197                       & 8233                       \\
    RTN-5heats & 11513                      & 1262                       \\
    RTN-6heats & 13690                      & 1516                       \\
    RTN-7heats & 15870                      & 1772                       \\
    RTN-8heats & 18046                      & 2025                       \\
    RTN-9heats & 20225                      & 2280                       \\ \bottomrule
  \end{tabular}
\end{table}

Fig.~\ref{fig_calculation_time_crtn} shows the solution time (averaged over the 15 price scenarios) of the optimal production scheduling problem based on the cRTN model and the conventional RTN model.
The solution time of RTN shows close to an exponential increase.
When the number of heats reaches 10, the cRTN-based model can still be effectively solved within an hour, while the solution time of the RTN-based model exceeds 2 hours when the number of heats is more than 7, which is impractical for demand response participation.
The reason is, as mentioned before, under the conventional RTN modeling method, the scale of the model grows linearly with increasing production targets (the number of heats).
When heat $=9$, the number of (presolved) binary variables in the RTN model increases to over 20,000 (as shown in Table~\ref{tab_nofv_models}, where RTN-5heats denotes the RTN model with a production target of 5 heats, and so forth).
For MILP problems, an increase in the number of integer variables typically means exponential growth in solution time.
In contrast, the number of variables in the proposed cRTN model is 4,197, far less than that in the conventional RTN method.
When heat $=9$, the proposed cRTN model still has 4,197 binary variables, the same as when heat $=1$, and this number does not increase with the production target.
Nevertheless, the solution time of the cRTN does increase with the addition of heats (Fig.~\ref{fig_calculation_time_crtn}) because the solution time is also related to the parameter settings.
However, the increase in solution time for the cRTN model is much slower than that for the RTN model.

\section{Summary}\label{sec_chap02_conclusion}

This chapter makes mechanism-based industrial process models tractable for VPP and electricity-market optimization by reformulating the two dominant model families, STN and RTN.

For continuous industrial processes, the proposed Linearized State Task Network (LSTN) improves the widely used State Task Network (STN) model.
LSTN replaces conventional integer start-up variables with the time that equipment spends in each operating state during a market interval, thereby eliminating integer variables from the represented continuous-process model.
Numerical tests show that LSTN can optimize the energy-use decisions of 2,000 industrial users under electricity-price signals within a few minutes, whereas the conventional STN model fails to converge within two hours for only 20 users.

For discrete industrial processes, the chapter reformulates the widely used Resource Task Network (RTN) as cRTN.
Building upon the variable reformulation of the LSTN model, the cRTN model further performs constraint reformulation to address the modeling requirements of batch-based production, uninterruptibility, and other characteristics of discrete industrial processes.
In the secondary-steelmaking test, cRTN reduces mean solution time from 156.7 to 17.1 minutes, an 89.0\% reduction, while preserving the represented batch constraints. The number of cRTN binary variables also remains fixed as the production target increases, whereas the conventional RTN variable count grows with the number of heats.
By allowing task transitions within a market interval, cRTN also removes the rounding error caused by forcing transitions onto discrete time-slot boundaries. In the 15-day price test, this improves average daily energy cost by a further 0.5 percentage points relative to the tightly solved conventional RTN; the illustrative two-batch example in Section~\ref{sec_illustrative} shows a 25\% reduction in the second batch's electricity cost.

The analysis also links the two model families: adding constraints for discrete operating characteristics to LSTN yields cRTN. This common formulation clarifies which constraints are needed for continuous and batch processes and allows both to enter power-system optimization through a consistent modeling structure.


\chapter{Industrial Load Parameter Identification Adapted to Coarse-Granularity Measurements}

\section{Overview}

\subsection{Background}

\begin{figure}[htbp]
  \centering
  \includegraphics[width=\textwidth]{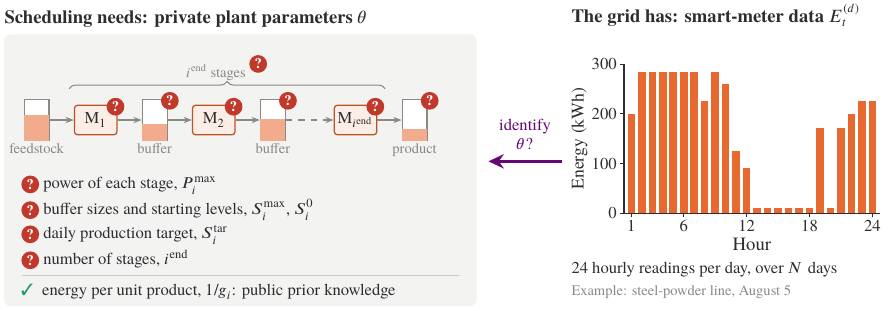}
\caption{Identification of private industrial process parameters from electricity prices and meter data}
  \label{fig_problem}
\end{figure}

Chapter 2 proposed computationally efficient, mechanism-based modeling methods for industrial loads, which alleviate the computational burden while ensuring model accuracy. However, Chapter 2 assumed that the model parameters were known. In practice, due to privacy protection and commercial confidentiality, internal model parameters are rarely directly accessible; they must be identified using externally observable data. This chapter relaxes the assumption of known model parameters and explores parameter identification methods for industrial loads operating under incomplete information (Fig.~\ref{fig_problem}).

\subsection{Problem Description}

Even when a suitable industrial-process model is available, its facility-specific parameters may remain unknown. In a steel plant with ten production stages, for example, one stage alone may require at least seven parameters, including production rates, operating power limits, and storage capacities.

When large industrial users directly participate in the market, they can theoretically access their own parameters to formulate bids. However, considering capacity thresholds and the technical complexity of market participation, most industrial users prefer to engage the electricity market through VPPs. In such arrangements, even though users and VPPs share the common goal of reducing electricity costs, factory parameters remain proprietary commercial secrets. Factories are typically reluctant to share this internal data with the grid or VPPs, requiring VPPs to perform parameter identification externally. 

Typically, the only data VPPs can access are coarse-granularity smart meter readings. In China, these are often hourly (or 15-minute) measurements, yielding merely 24 to 96 data points per day. Existing literature generally asserts that identifying load characteristics or equipment parameters requires minute-level or higher-frequency sampling to capture the dynamic transients of equipment start-up and shutdown. However, scaling such high-frequency data collection is cost-prohibitive and is typically restricted to isolated pilot projects. Consequently, achieving satisfactory parameter identification using widely available coarse-granularity smart meter data is generally considered a challenging task.

\subsection{Contributions of This Chapter}
To address these challenges, this chapter proposes Production Scheduling Identification (PSI), an industrial-load identification framework tailored to hourly smart meter data. The core content of this chapter is based on our published paper~\cite{lyu_production_2025}.

The core concept of PSI is to integrate available prior knowledge, rational behavioral assumptions, and external meter data to estimate model parameters via inverse reasoning. ``Prior knowledge'' refers to industry-general information accessible from public channels. For example, the energy intensity required to produce one kilogram of aluminum via electrolysis is known to be approximately 13 to 14 kWh. Such information remains stable for mature production lines and can be embedded into the optimization problem to narrow the parameter search space. 

Specifically, PSI adopts an inverse optimization framework. It treats a factory's smart meter data as the optimal solution that minimizes electricity costs, given the boundary conditions (electricity prices and production targets) and physical constraints. Inverse optimization then works backward from these observed optimal solutions to infer the underlying model parameters, including rated power, production rates, storage limits, and specific production targets.

The numerical tests evaluate whether the identified models reproduce a facility's external load response under previously unseen electricity prices. Using 21 training days, PSI achieves nRMSE values of 5.2\% for the cement case and 8.5\% for the steel-powder case, compared with 13.4\%--19.2\% for the machine-learning baselines. Thus, the process constraints and scheduling objective reduce the reported load-model error by more than half in both cases. This result concerns the external behavior of the identified model; individual non-binding parameters can remain weakly identified because they do not affect the observed optimal load curve.

\section{Concept and Framework of PSI}\label{sec_framework_chap03}

This section introduces the overall framework of the Production Scheduling Identification (PSI) method, detailing its core concept and methodological foundation. PSI leverages inverse optimization, treating smart meter data as the result of a factory's optimal production scheduling under historical electricity prices. By inferring the parameters that drive this optimization, we establish a load model. Note that while this chapter primarily builds upon the Linearized State-Task Network (LSTN) model proposed in Chapter 2, the core methodology of the PSI framework is general. Because Chapter 2 demonstrated the mathematical unification of LSTN and cRTN, the PSI method can be applied to the cRTN and other diverse model structures.

\subsection{Problem Description}

Our objective is to establish a load model when the internal parameters of industrial users are inaccessible. We investigate how to extract these required modeling parameters solely from external smart meter data. Specifically, we focus on modeling the industrial production process, dividing the necessary information into private facility parameters and public prior knowledge:

(1)~\textbf{Specific facility parameters (Private)}: These include internal equipment configurations, such as the rated power of machinery at each stage, material storage capacities, and daily production goals. Because these parameters constitute commercial secrets, we assume they are inaccessible to the VPP.

(2)~\textbf{Prior knowledge of the production process (Public)}: This includes macroscopic information obtainable via the Internet, expert knowledge, and industry standards, which is independent of a specific factory's private equipment. In this work, we assume the VPP knows the electricity and material consumption required to produce one unit of product at each specific stage. For mature, energy-intensive industries (e.g., steelmaking, cement, aluminum electrolysis), standardized information regarding the number of stages, inter-stage connections, and theoretical unit energy consumption is available in national standards or industry reports. For instance, a VPP can classify a facility as an aluminum electrolysis plant and ascertain the standard alumina and electricity required per kilogram of output without requiring onsite audits.

(3)~\textbf{Smart meter data}: This data consists of average power readings recorded every 15 minutes or hourly at the facility level. We assume that the entity conducting parameter identification (e.g., a VPP operator) possesses a historical dataset of this meter data (e.g., from the past month). While not publicly published, this data is naturally available to the VPP managing the user's market participation.

Furthermore, we assume the VPP knows the hourly electricity prices the user faced. Formally, we use a historical price-consumption dataset $HD = \{(Pr^{(1)}, E^{(1)}), ..., (Pr ^ {(n)}, E ^ {(n)})\}$ to identify the load model parameters $\theta$. Here, $Pr^{(n)}$ and $E ^ {(n)}$ denote the hourly electricity price and the user's hourly electricity consumption on day $n$, respectively. For conciseness, $\theta$ encompasses both the specific equipment parameters and the prior knowledge of the production process.

\subsection{Overall Framework}

\begin{figure}[!t]
  \centering
  \includegraphics[width=\textwidth]{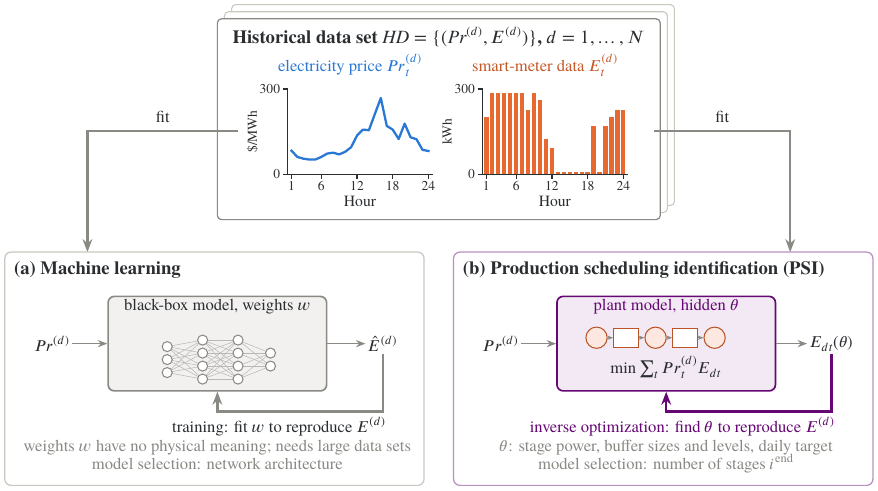}
\caption{Concept of production scheduling identification (PSI)}
  \label{fig_framework_chap03}
\end{figure}

(1)~\textbf{Philosophy}: PSI leverages historical electricity consumption and price data to fit (train) a load model capable of predicting user behavior under novel boundary conditions. This concept of modeling user behavior from historical data is inspired by machine learning—for example, training a black-box model using prices as inputs and meter data as outputs (Fig.~\ref{fig_framework_chap03}-a). However, purely data-driven methods require large volumes of high-quality data to perform adequately. The distinction of the PSI framework is its use of a structured load model governed by industrial production constraints, contrasting with the physics-agnostic nature of black-box models (Fig.~\ref{fig_framework_chap03}-b). Conceptually, PSI is a domain-knowledge-embedded data-driven framework. Because the structural model (e.g., mSTN) reflects the physics of the factory's energy consumption, the physical constraints reduce the model's reliance on large datasets. This mathematical and logical distinction is key to achieving accurate modeling using coarse-granularity smart meter data.

(2)~\textbf{Assumptions}: We build the PSI method primarily on two assumptions. First, we assume that industrial users facing fluctuating electricity prices optimize their production schedules to minimize electricity costs—a widely accepted assumption in grid-interactive research. Second, while the methodology does not limit itself to a specific industry, we assume we can model the user's load via an STN. While production processes vary significantly, generalized models like STN strike a balance between expressiveness and complexity. STN has been successfully applied to cement plants, steel-powder manufacturing, and battery assembly~\cite{ref22, golmohamadi_robust_2020, li_real-time_2017}. Therefore, we assume the user employs an STN-like structure to optimize energy consumption. Consistent with Chapter 2, electricity-cost minimization is adopted because it directly links production scheduling to market price signals. If the actual schedule also places material weight on output, quality, delivery, or staffing, those objectives or constraints must be represented before formulating the inverse problem. Otherwise, the inferred parameters describe only an equivalent preference under the selected objective, rather than the factory's actual decision process.

(3)~\textbf{Methodology}: Under these assumptions, the objective becomes: if the smart meter data $E ^ {(n)}$ under varying daily prices $Pr ^ {(n)}$ represent the optimal energy usage dictated by an STN model parameterized by $\theta$, how do we determine $\theta$? We propose a three-step inverse optimization approach: 
First, we enhance the LSTN model from Chapter 2 to derive the modified State-Task Network (mSTN) model (Section~\ref{sec_mSTN}). The mSTN uses the same parameters but adopts a mathematical form that allows the algorithm to determine the optimal number of production stages via model selection. 
Second, we construct an inverse optimization problem based on the mSTN, designing a loss function that measures how closely the $\theta$-parameterized model outputs match the actual smart meter data (Section~\ref{sec_problem_formulation}). 
Third, we design an iterative algorithm to minimize this loss function with respect to $\theta$ (Section~\ref{sec_iterative_algorithm}).

\begin{figure}[!t]
  \centering
  \includegraphics[width=\textwidth]{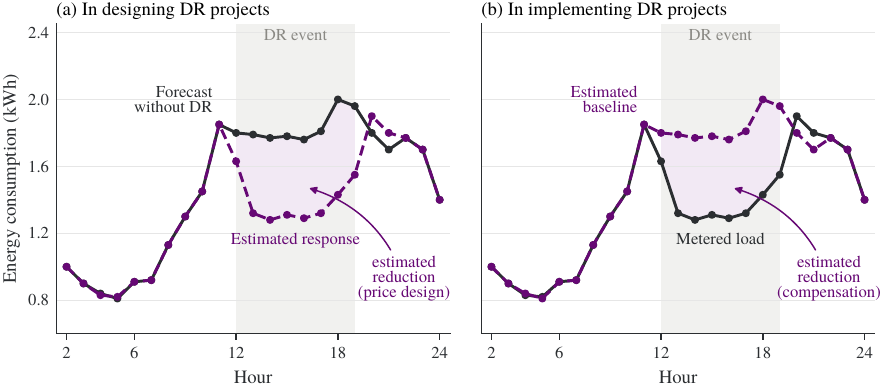}
  \par\smallskip\footnotesize Note: ``Hour'' on the horizontal axes denotes the hourly time intervals within a day; electricity consumption is reported in kWh.
\caption{Applications of industrial load models in demand response}
  \label{fig_illustration}
\end{figure}

(4)~\textbf{Use of identified load models}: Once PSI extracts the load parameters $\theta$, we establish the explicit mSTN (or STN) load model. The VPP can deploy this identified model similarly to a known-parameter STN model. Typical uses include estimating demand response capacity, embedding the constraints into market clearing algorithms, or integrating them into VPP bidding strategies. For instance, before initiating demand response, a VPP can query the model to estimate hourly load shifts if specific incentives are applied between 12:00 and 17:00 (Fig.~\ref{fig_illustration}(a)). Alternatively, the VPP can use the model to establish a baseline load profile~\cite{li_precision_2022}, calculating what the user's load would have been without incentives, thereby enabling financial settlement based on actual load reduction (Fig.~\ref{fig_illustration}(b)).

\section{Modified State-Task Network Model}\label{sec_mSTN}

Having established the overall framework of the PSI method, we now detail the mathematical formulation of the modified State-Task Network (mSTN) model. Building upon the LSTN model from Chapter 2, we modify it to accommodate the requirements of inverse optimization. While mathematically consistent with the LSTN, the mSTN introduces flexibility that allows the algorithm to determine the optimal number of production stages via model selection, whereas the LSTN assumes a fixed stage count. We first explain these modifications and then present the mathematical formulation.

Because the objective is to address incomplete information in VPP operation rather than develop another general-purpose process model, we use STN~\cite{ref20}, an established model for industrial loads~\cite{ref31}. Linearization, per-unitization, and task aggregation limit the size of the resulting inverse problem. The PSI structure is not tied to STN: cRTN or a custom industrial model can be used by replacing the optimality conditions in the identification problem (Section~\ref{sec_training}).

\subsection{Modifications of the mSTN Relative to the LSTN}

First, similar to LSTN, we adopt a linearized form. For machines with rapid start-up sequences, operating time within an hourly interval is effectively continuous~\cite{zhang_demand_2018}, aligning continuous modeling with physical operations. Chapter 2 demonstrated that this linearization improves computational speed with a minor impact on model error. The optimality conditions (KKT) of linear programming are computationally tractable, which is necessary for solving the inverse optimization problem efficiently.

Second, to address the multivalue problem (where different parameter sets yield identical load curves) during model fitting, we unitize the task parameters. Observing the LSTN model, we note that state-related constraints remain equivalent if we multiply $g_i$ and $S^{\rm 0/max/tar}_i$ by a constant. To anchor the solution, the mSTN enforces per-unitized parameters scaled to the final product (see Section~\ref{app_per_unitization}). For example, $S^{\rm 0}_0$ represents the amount of final product the factory can produce using the initial feedstock.

Third, the mSTN supports the aggregation of adjacent production tasks. In practice, the VPP may not know the exact number of physical tasks inside a facility. By treating the number of tasks as a hyperparameter, we allow the PSI method to determine the aggregation level dynamically. Furthermore, theoretical analysis shows that while task aggregation technically reduces model expressiveness, the resulting error is acceptable for hourly dispatch (see Section~\ref{app_aggregating}).

\subsection{Mathematical Formulation of the mSTN}

\begin{figure}[!t]
  \centering
  \includegraphics[width=0.8\textwidth]{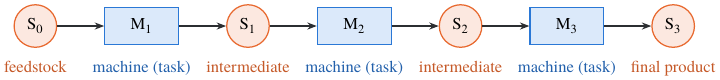}
\caption{Example of a state-task network (STN), with production equipment and material storage}
  \label{fig_mSTN}
\end{figure}

Fig.~\ref{fig_mSTN} maps a common industrial process, where ${\rm M}_i$ and ${\rm S_i}$ represent the $i$th machine (task) and its product (state). States capture feedstock, intermediate, or final products. To differentiate observed data from internal model variables, we use parenthetical superscripts for collected data. For instance, $E^{(d)}$ denotes the meter reading on day $d$, and $E^{(d)}_t$ is that reading at period $t$. For internal model variables, we use subscripts; thus, $E_{dt}$ represents the energy consumption calculated by our mSTN model.

On day $d$, the facility minimizes its total daily energy cost $Cost_d$, formulated as:
\begin{equation}\label{primal_cost_chap03}
  {\rm min.} Cost_d = \underset{t \in \mathcal{T}}{\Sigma}{Pr^{(d)}_{t}} E_{dt},
\end{equation}
where time intervals align with market pricing intervals (e.g., hourly).

Simultaneously, the facility must satisfy its daily production targets (\ref{primal_constraint_tar_chap03}) and assembly line technical constraints (\ref{primal_constraint_E_chap03})-(\ref{primal_constraint_storageLimit_chap03}). Equation (\ref{primal_constraint_E_chap03}) maps the facility's total electricity consumption to individual tasks. Equation (\ref{primal_constraint_Pmax}) bounds operating power by the equipment's rated capacity. Equation (\ref{primal_constraint_S0_chap03}) initializes the buffers. Equations (\ref{primal_constraint_changeofS1_chap03}), (\ref{primal_constraint_changeofS2_chap03}), and (\ref{primal_constraint_changeofS3_chap03}) govern the buffer state changes for feedstock, intermediate, and final products, respectively. Equation (\ref{primal_constraint_storageLimit_chap03}) enforces physical buffer limits. We append the respective Lagrange multipliers ($\mu^{\rm Star}_{di}$, $\mu^{\rm Smin/Smax}_{dti}$, $\mu^{\rm Pmin/Pmax}_{dti}$, and $\lambda^{S}_{ti}$) to each constraint. By convention, we assume multiplier inequalities are written as $\le 0$:
\begin{align}
  & S_{dti} \ge S^{0}_{i} + S^{\rm tar}_i \ : \mu^{\rm Star}_{di}, t = t^{\rm end}, i \in \mathcal{I}^{\rm S}. \label{primal_constraint_tar_chap03} \\
  & E_{dt} = {(\underset{i \in \mathcal{I}^{\rm P}}{\Sigma}P_{dti})}  \Delta t, t \in \mathcal{T}. \label{primal_constraint_E_chap03} \\
  & 0 \le P_{dti} \le P^{\rm max}_i \ : \mu^{\rm Pmin}_{dti}, \mu^{\rm Pmax}_{dti}, t \in \mathcal{T}, i \in \mathcal{I}^{\rm P}. \label{primal_constraint_Pmax} \\
  & S_{dti} = S^{\rm 0}_i : \lambda^{S}_{dti}, \ i \in \mathcal{I}^{\rm S}, t = 0. \label{primal_constraint_S0_chap03} \\
  & S_{dti} = S_{(t - 1)i} - P_{dt(i + 1)} g_{i + 1} \Delta t : \lambda^{S}_{dti}, t \in \mathcal{T}, i = 0. \label{primal_constraint_changeofS1_chap03} \\
  & S_{dti} = S_{d(t - 1)i} + P_{dti} g_{i}  \Delta t \notag \\
  & \quad - P_{dt(i + 1)} g_{i + 1}  \Delta t \ : \lambda^{S}_{dti}, t \in \mathcal{T}, i \in \mathcal{I}^{\rm P} \setminus \{i^{\rm end}\}. \label{primal_constraint_changeofS2_chap03} \\
  & S_{dti} = S_{d(t - 1)i} + P_{dti} g_{i} \Delta t : \lambda^{S}_{dti}, t \in \mathcal{T}, i = i^{\rm end}. \label{primal_constraint_changeofS3_chap03} \\
  & 0 \le S_{dti} \le S^{\rm max}_{i} \ : \mu^{\rm Smin}_{dti}, \mu^{\rm Smax}_{dti}, t \in \mathcal{T}, i \in \mathcal{I}^{\rm S}. \label{primal_constraint_storageLimit_chap03}
\end{align}

In the mSTN, the production scheduling decision variables are $\{P_{dti}, S_{dti}, E_{dt} | \forall i, \forall t\}$. Problem parameters consist of the external pricing signals $\{Pr^{(d)}_t|\forall t\}$ and the hidden facility data $\theta = \{g_i, P^{\rm max}_i, S^{\rm 0/max/tar}_i\}$. Under our PSI framework, we assume the user solves this mSTN optimization problem to minimize costs.

\subsection{Dual Problem of the mSTN}
We utilize the primal and dual forms of the mSTN to enforce optimality conditions during the PSI inverse identification phase. Because the mSTN is formulated as a linear program, formulating the dual is straightforward. The objective function is the Lagrangian dual function:
\begin{equation}\label{dual_g_function}
  \begin{aligned}
    {\rm max.} \ 
  & \underset{t \in \mathcal{T}}{\Sigma}{Pr^{(d)}_{t}}
  {P_{t, 0}}\Delta t
  - \underset{t \in \mathcal{T}}{\Sigma} \underset{i \in \mathcal{I}^{\rm P}}{\Sigma} P^{\rm max}_i \mu^{\rm Pmax}_{dti} 
  - \underset{i \in \mathcal{I}^{\rm S}}{\Sigma} S^{0}_{i} \lambda^{\rm S}_{d0i}\\
  &
  + \underset{t \in \mathcal{T}}{\Sigma} \underset{i \in \mathcal{I}^{\rm S}}{\Sigma} S^{\rm max}_i \mu^{\rm Smax}_{dti}
  + \underset{i \in \mathcal{I}^{\rm S}}{\Sigma} (S^{0}_{i} + S^{\rm tar}_i)\mu^{\rm Star}_{di}
\end{aligned}
\end{equation}
The dual constraints enforce that the coefficients of the primal variables in the Lagrangian equal 0:
\begin{align}
  & Pr^{(d)}_{t} \Delta t - \mu^{\rm Pmax}_{dti} + \mu^{\rm Pmin}_{dti}
  + g_{i-1}\lambda^{S}_{dti-1} - g_{i}\lambda^{S}_{ti} = 0, \label{dual_constraint_coefficient} \\
  & \notag \quad t \in \mathcal{T}, i \in \mathcal{I}^{\rm P} \\
  & \lambda^{S}_{ti} - \lambda^{S}_{d(t+1)i} - \mu^{\rm Smin}_{dti} + \mu^{\rm Smax}_{dti} = 0, \\
  & \notag \quad t \in \mathcal{T} \setminus \{t^{\rm end}\}, i \in \mathcal{I}^{\rm S} \\
  & \lambda^{S}_{ti} - \mu^{\rm Star}_{di} - \mu^{\rm Smin}_{dti} + \mu^{\rm Smax}_{dti} = 0, \\
  & \notag \quad  t = t^{\rm end}, i \in \mathcal{I}^{\rm S}
\end{align}
We enforce nonnegativity on the dual variables:
\begin{equation}\label{dual_constraint_non_negative}
  (\mu^{\rm Star}_{di}, \mu^{\rm Smin/Smax}_{dti}, \mu^{\rm Pmin/Pmax}_{dti}) \ge 0, \forall i, t \in \mathcal{T}
\end{equation}

\subsection{Per-Unit Values in the mSTN}\label{app_per_unitization}

Without loss of generality, we express the state change $\Delta S_i$ of intermediate $i$ (Equation \ref{primal_constraint_changeofS2_chap03}) as:
\begin{equation}
\Delta S_i = \Delta E_i  g_i - \Delta E_{i + 1}  c_{i + 1}
\end{equation}
where $\Delta E_i = P_i  \Delta t$ is the energy consumed by task $i$, and $g_i(c_{i +1})$ is the production (consumption) of intermediate $i$ per unit of energy consumed by task $i(i+1)$. 

Consider the conversion from intermediate $i$ to $i+1$: if only task $i+1$ operates and consumes $\Delta E_{i +1}$ energy, it consumes $\Delta E_{i + 1}  c_{i + 1}$ of intermediate $i$, producing $\Delta E_{i + 1}  g_{i + 1}$ of intermediate $i + 1$. Thus, the state conversion rate of task $i+1$ is $r_{i, i+1} = \frac{\Delta E_{i + 1}  g_{i + 1}}{\Delta E_{i + 1}  c_{i + 1}} = \frac{g_{i + 1}}{c_{i + 1}}$. Recursively, the total conversion rate from state $i$ to the final product $i^{\rm end}$ becomes:
\begin{equation}
  r_{i, i^{\rm end}} = \frac{\Delta S_{i^{\rm end}}}{\Delta S_{i}} =\prod_{j=i+1}^{i^{\rm end}} \frac{g_j}{c_j}
\end{equation}
We base the per-unit values of $g_i, c_{i+1}, S^{\rm 0/max/tar}_i$ on these values multiplied by $r_{i, i^{\rm end}}$. This per-unitization enforces $g_i = c_i, \forall i$. Therefore, we use $g_i$ to denote this unified parameter in the mSTN, effectively reducing parameter redundancy.

\subsection{Task Aggregation}\label{app_aggregating}
We designed the aggregation strategy to reduce the parameter footprint of the mSTN model, enhancing the convergence of the inverse identification algorithm. We prioritize aggregating production links with similar throughputs. Because these links do not present significant throughput bottlenecks, they are typically scheduled simultaneously during low-price periods. Aggregating such links introduces minimal mathematical error to the external load curve.

Taking tasks $i$ and $i + 1$ as an example, assume task $i$ processes $\Delta S_{i - 1}$ into intermediate $i$, which task $i+1$ then processes into $\Delta S_{i + 1}$. Using our per-unit values, $\Delta S_{i - 1} = \Delta S_{i + 1} = \Delta S$. The net electricity consumed by both tasks is $\Delta E = \sum_{j = i}^{i + 1} \frac{\Delta S}{g_j}$. We define the aggregated parameters as:
\begin{equation}
  g_{i'} = \left(\sum_{j = i}^{i + 1} \frac{1}{g_j}\right)^{-1}, P^{\rm max}_{i'} = \sum_{j = i}^{i + 1} P^{\rm max}_{j}
\end{equation}
A single new task $i'$, governed by parameters $g_{i'}$ and $P^{\rm max}_{i'}$, approximates the tandem operation of tasks $i$ and $i+1$, neglecting the internal buffer. We generalize this logic to aggregate multiple sequential tasks.

When engineering a balanced production line, adjacent tasks typically share similar throughputs ($g_{i}  P^{\rm max}_{i} \approx g_{i+1}  P^{\rm max}_{i+1}$). This balance implies that both tasks can run at rated power without overfilling the intermediate buffer. Therefore, aggregating adjacent tasks yields a sufficiently accurate approximation.

In our numerical tests, to test models with varying task numbers, we assume $g_i$ in the baseline 10-task model (Table~\ref{tab_parameter}) is known. We then aggregate tasks with their predecessors in the order (10, 3, 5, 9, 6, 2, 4, 8). Because machines with variable operating points exhibit varying $g_i$ values, we base our prior knowledge on the $g_i$ corresponding to rated power (e.g., for the crusher, $g=15/20=0.75$).

\section{Model Identification}\label{sec_training}

Having defined the mSTN model and its dual, we detail the inverse optimization-based model identification method. We frame parameter identification as an inverse optimization problem. We embed the primal and dual optimality conditions of the mSTN as constraints, treating the observed smart meter data as constants and the hidden model parameters as decision variables. Because this formulates a large-scale, non-convex problem—which is difficult for direct solvers to handle—we introduce an iterative algorithm. Inspired by Stochastic Gradient Descent (SGD) from machine learning, our algorithm approaches the parameters through iterative, batch-based updates.

\subsection{Problem Formulation}\label{sec_problem_formulation}
We utilize the historical price-consumption dataset to identify a load model that represents the facility's optimization strategy. We formulate the inverse optimization problem via equations (\ref{inverse_cost})-(\ref{inverse_dual_constraints}). Equation (\ref{inverse_cost}) is our loss function, measuring the average squared residual between the model output and the actual load profile. Equation (\ref{inverse_strong_duality}) enforces the strong duality theorem. Equations (\ref{inverse_primal_constraints}) enforce primal feasibility (treating facility parameters as variables), and (\ref{inverse_dual_constraints}) enforce dual feasibility. By combining these three conditions, we enforce optimality without relying on complementary slackness conditions (KKT), an approach used to facilitate inverse optimization solving~\cite{ruiz_2013_revealing}.
\begin{align}
  & {\rm min.} \ \frac{1}{n} \sum_{d = 1}^{n} \sum_{t \in \mathcal{T}} (E_{dt} - E^{(d)}_t)^2 \label{inverse_cost} \\
  & {\rm s.t.} \ {\rm strong \ duality:} (\ref{primal_cost_chap03})=(\ref{dual_g_function}), \forall d. \label{inverse_strong_duality} \\
  & \quad \ {\rm primal \ feasibility:}(\ref{primal_constraint_tar_chap03})-(\ref{primal_constraint_storageLimit_chap03}), \forall d. \label{inverse_primal_constraints} \\
  & \quad \ {\rm dual \ feasibility:} (\ref{dual_constraint_coefficient})-(\ref{dual_constraint_non_negative}), \forall d. \label{inverse_dual_constraints}
\end{align}

In this formulation, the facility parameters $\theta = \{g_i, P^{\rm max}_i, S^{\rm 0/max/tar}_i, \forall i\}$ act as the primary decision variables. The total variable set includes $\theta$, the primal operational variables $\{P_{dti}, S_{dti}, E_{dt} | \forall i, \forall t\}$, and the dual variables $\{\mu^{\rm Star}_{di}, \mu^{\rm Smin/Smax}_{dti}, \mu^{\rm Pmin/Pmax}_{dti}, \lambda^{S}_{ti}, \forall i, t \in \mathcal{T}\}$. The static problem parameters are the electricity prices $\{Pr^{(d)}_t|\forall t\}$ and historical smart meter data $\{E^{(d)}_t| \forall t, \forall d\}$. 

Because variable products exist in both constraints (\ref{inverse_strong_duality}) and (\ref{dual_constraint_coefficient}), the problem is nonlinear. We assume $\theta$ remains static over the observed dataset. Given that industrial production targets generally remain stable over short-to-medium horizons (e.g., a month), this assumption is reasonable.

\subsection{Assumptions for Model Identification}\label{sec_assumption}

To manage the scale and nonlinearity of the inverse problem, we introduce physical assumptions to constrain the parameter search space, improving computational feasibility.

(1)~\textbf{Upper limits}: We bound all variables below a sufficiently large real number (e.g., 1e3) to prevent numerical instability:
\begin{equation}\label{assumption_a}
  (g_i, P^{\rm max}_i, S^{\rm 0/max}_i, \mu^{\rm Star}_{di}, \mu^{\rm Smin/Smax}_{dti}, \mu^{\rm Pmin/Pmax}_{dti}, \lambda^{S}_{ti}) \le M, \ \forall i
\end{equation}

(2)~\textbf{Physical state transitions}: After a 24-hour cycle, raw feedstock must decrease (\ref{asp_star0}), final products must increase (\ref{asp_star2}), and intermediate buffers must maintain their baseline or increase (\ref{asp_star1}):
\begin{align}
  & S^{\rm tar}_i \le 0, \ i = 0 \label{asp_star0} \\
  & S^{\rm tar}_i = 0, \ i \in \mathcal{I}^{\rm P} \setminus \{i^{\rm end}\}. \label{asp_star1} \\
  & S^{\rm tar}_i \ge 0, \ i = i^{\rm end} \label{asp_star2}
\end{align}

(3)~\textbf{Maximum capacity utilization}: We assume that over an extended historical period, there is at least one instance where the factory operated all machinery at maximum rated power:
\begin{equation}\label{assumption_c}
  \underset{i \in \mathcal{I}^{\rm P}}{\Sigma} P^{\rm max}_{i}  \Delta t = \underset{t}{\rm max} (E_t).
\end{equation}

(4)~\textbf{Prior knowledge integration}: We assume the vector $\{g_i, \forall i\}$ is known. Because $1/g_i$ represents the theoretical electricity consumption required per unit of product output for a specific technology class, it constitutes an industry-standard quantity rather than a private parameter.

Note that we do not assume the true number of physical tasks is known. The algorithm determines the optimal stage count via model selection.

\subsection{Iterative Model Identification Algorithm}\label{sec_iterative_algorithm}

\begin{algorithm}[!t]
\caption{Model identification of PSI}
  \label{alg_framework}
  \begin{algorithmic}[1]
    \renewcommand{\algorithmicrequire}{\textbf{Input:}}
    \REQUIRE
    historical electricity consumption and electricity price pairs $\{(Pr^{(1)},E^{(1)}),...(Pr^{(n)},E^{(n)}) \}$, model library $\mathcal{L}$, maximum number of iterations $K$.
    \renewcommand{\algorithmicrequire}{\textbf{Output:}}
    \REQUIRE a load model parameterized by $\theta$.
    \STATE Partition the dataset into a training set $TRAIN$ and a cross-validation set $CV$.
    \FOR{model $m \in \mathcal{L}$ (number of tasks $i^{\rm end} = m$)}
      \STATE Calculate $\theta^{(k)}$ via (\ref{alg_initialize}), $k=1,...,n$.
      \STATE Set $k = n + 1$, initialize $\theta = \frac{1}{n} \sum_{k=1}^{n} \theta^{(k)}$.
      \WHILE{$k \le K$}
        \STATE Calculate $\theta^{(k)}$ via (\ref{alg_update})
        \STATE Update $\theta = \theta^{(k-1)} + \frac{1}{n} (\theta^{(k)} - \theta^{(k-1)})$.
        \STATE \textbf{if} $\theta$ converges, \textbf{break}.
        \STATE Set $k = k + 1$.
      \ENDWHILE
    \ENDFOR
    \STATE Select model $m^*$ that performs best on $CV$.
  \end{algorithmic}
\end{algorithm}

Solving the nonlinear inverse problem directly across the full dataset is computationally intractable. Therefore, we designed an algorithm inspired by Zero-Order Stochastic Gradient Descent (ZO-SGD)~\cite{liu_primer_2020}. We optimize over small batches (a single day's data) rather than the entire dataset, reducing the scale of each iteration. We estimate the gradient via programmatic search instead of analytical calculation.

Algorithm~\ref{alg_framework} details the process.
Steps 1--2: We partition the data into $TRAIN$ and $CV$ sets. We train every model $m$ in library $\mathcal{L}$ on the $TRAIN$ set, selecting the model $m^*$ using the $CV$ set (Step 12).
Steps 3--4: For initialization, we randomly select a day $d$ and solve the inverse problem to isolate $\theta^{(k)}$:
\begin{align}
  & {\rm min.} \sum_{t \in \mathcal{T}} (E_{dt} - E^{(d)}_t)^2, \label{alg_initialize} \\
  & \notag {\rm s.t.} (\ref{inverse_strong_duality}), (\ref{inverse_primal_constraints}), (\ref{inverse_dual_constraints}).
\end{align}
We average these to establish our starting reference $\theta^{\rm ref} = \frac{1}{n} \sum_{k=1}^{n} \theta^{(k)}$.

Steps 5--9: In iteration $k$, we randomly select day $d$ and solve an augmented penalty problem:
\begin{align}
  & {\rm min.}  (\theta^{(d)} - \theta^{\rm ref})^2 + \alpha \sum_{t \in \mathcal{T}} (E_{dt} - E^{(d)}_t)^2, \label{alg_update} \\
  & \notag {\rm s.t.} (\ref{inverse_strong_duality}), (\ref{inverse_primal_constraints}), (\ref{inverse_dual_constraints}).
\end{align}
We update $\theta^{(k)} = \theta^{(d)}$ and adjust the parameters via $\theta = \theta^{(k-1)} + \frac{1}{n} (\theta^{(k)} - \theta^{(k-1)})$. The objective incorporates a Euclidean distance penalty $(\theta^{(k)} - \theta)^2$ to restrict large parameter fluctuations. The weight $\alpha$ adaptively decays from 1 to 0, helping the algorithm converge smoothly. We step forward at a fixed rate of $\frac{1}{n}$.

To address non-convexity, we impose an execution time limit (e.g., 120 seconds) per iteration. The solver returns its feasible solution at the time limit to drive the parameter update forward. To accelerate execution, we recommend processing daily batches in parallel.

\section{Numerical Results}\label{sec_case_study_chap03}

We evaluate PSI using test datasets constructed from established cement and steel-powder process models. The comparison with standard machine-learning models examines load-model accuracy under unseen price scenarios and limited training data. The code and datasets are publicly available~\cite{rick10119_psi_2024}. All figures report discrete hourly energy consumption (kWh).

\subsection{Dataset and Partitioning}

\begin{figure}[!t]
  \centering
  \includegraphics[width=\textwidth]{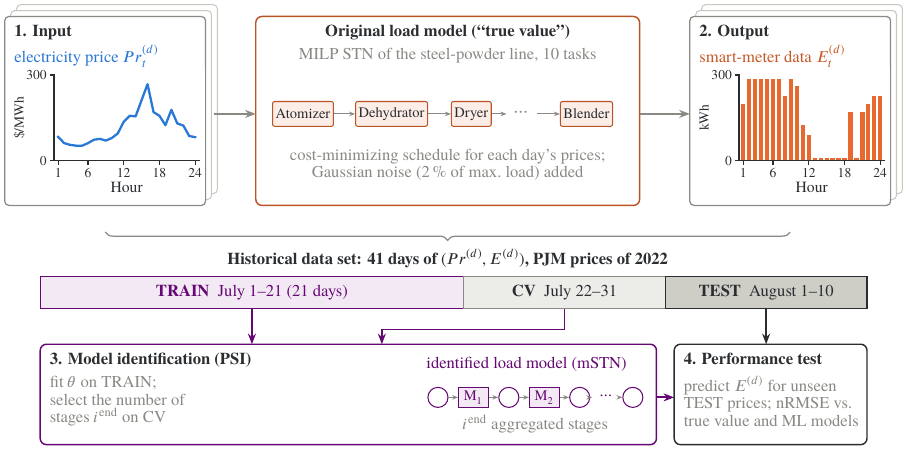}
\caption{Numerical test design for production scheduling identification (PSI)}
  \label{fig_case_study_design}
\end{figure}

\begin{figure}[!t]
  \centering
  \includegraphics[width=\textwidth]{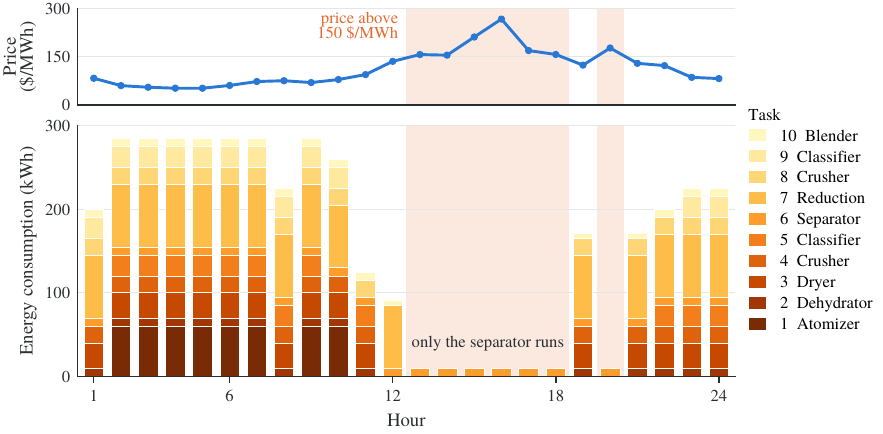}
\caption{Hourly electricity consumption by equipment in the steel-powder production line on August 5}
  \label{fig_price_chap03}
\end{figure}

\begin{table}[!t]
  \caption{Parameters of the original steel-powder manufacturing system}
  \label{tab_parameter}
  \centering 
  \begin{tabular}{lllll}
  \toprule
  \begin{tabular}[c]{@{}l@{}}Task name\\(number)\end{tabular} &
  \begin{tabular}[c]{@{}l@{}}Operating\\point\end{tabular} &
  \begin{tabular}[c]{@{}l@{}}Production\\rate\\(tons/h)\end{tabular} &
  \begin{tabular}[c]{@{}l@{}}Energy\\demand\\(kWh)\end{tabular} &
  \begin{tabular}[c]{@{}l@{}}Buffer\\capacity\\(tons)\end{tabular} \\ \midrule
                             & off  & 0  & 0  &                          \\ \cline{2-4}
\multirow{-2}{*}{Atomizer(1)}   & on & 30 & 60 & \multirow{-2}{*}{180}    \\ \hline
                             & off  & 0  & 0  &                          \\ \cline{2-4}
\multirow{-2}{*}{Dehydrator(2)} & on & 15 & 10 & \multirow{-2}{*}{100}    \\ \hline
                             & off  & 0  & 0  &                          \\ \cline{2-4}
\multirow{-2}{*}{Dryer(3)} & on & 15 & 30 & \multirow{-2}{*}{150} \\ \hline
                               & 1   & 0  & 0  &                          \\ \cline{2-4}
                             & 2   & 10 & 15 &                          \\ \cline{2-4}
\multirow{-3}{*}{Crusher(4,8)}    & 3   & 15 & 20 & \multirow{-3}{*}{100}    \\ \hline
                             & 1   & 0  & 0  &                          \\ \cline{2-4}
                             & 2   & 10 & 15 &                          \\ \cline{2-4}
\multirow{-3}{*}{Classifier(5,9)} & 3   & 20 & 25 & \multirow{-3}{*}{150}    \\ \hline
                             & off  & 0  & 0  &                          \\ \cline{2-4}
\multirow{-2}{*}{Separator(6)}  & on & 10 & 10 & \multirow{-2}{*}{100}    \\ \hline
 & off  & 0  & 0  &                          \\ \cline{2-4}
\multirow{-2}{*}{Reduction(7)}  & on & 15 & 75 & \multirow{-2}{*}{100}    \\ \hline
                             & 1   & 0  & 0  &                          \\ \cline{2-4}
                             & 2   & 10 & 6  &                          \\ \cline{2-4}
\multirow{-3}{*}{Blender(10)}    & 3   & 15 & 10 & \multirow{-3}{*}{200} \\  \bottomrule
\end{tabular}
\end{table}

\begin{table}[!t]
  \caption{Parameters of the original cement plant}
  \label{tab_parameter_cement}
  \centering 
  \begin{tabular}{lllll}
\toprule
\begin{tabular}[c]{@{}l@{}}Task\\ name\end{tabular}                               & \begin{tabular}[c]{@{}l@{}}Operating\\ point\end{tabular} & \begin{tabular}[c]{@{}l@{}}Production\\ rate (tons/h)\end{tabular} & \begin{tabular}[c]{@{}l@{}}Rated\\ power (kW)\end{tabular} & \begin{tabular}[c]{@{}l@{}}Buffer\\ capacity (tons)\end{tabular} \\ \midrule
\multirow{2}{*}{Crushing}                                                         & off                                                      & 0                                                                & 0                                                        & \multirow{2}{*}{2000}                                          \\
                                                                                  & on                                                       & 1000                                                             & 2200                                                     &                                                                \\ \hline
\multirow{2}{*}{\begin{tabular}[c]{@{}l@{}}Kiln feed\\ preparation\end{tabular}} & off                                                      & 0                                                                & 0                                                        & \multirow{2}{*}{2500}                                          \\
                                                                                  & on                                                       & 250                                                              & 11000                                                    &                                                                \\ \hline
\multirow{2}{*}{\begin{tabular}[c]{@{}l@{}}Clinker\\ production\end{tabular}}     & off                                                      & 0                                                                & 0                                                        & \multirow{2}{*}{1750}                                          \\
                                                                                  & on                                                       & 300                                                              & 11550                                                    &                                                                \\ \hline
\multirow{2}{*}{\begin{tabular}[c]{@{}l@{}}Finish\\ grinding\end{tabular}}        & off                                                      & 0                                                                & 0                                                        & \multirow{2}{*}{5000}                                          \\
                                                                                  & on                                                       & 350                                                              & 11220                                                    &                                                                \\ \bottomrule
\end{tabular}
\end{table}

We evaluated the model using 41 days of price-consumption pairs (Fig.~\ref{fig_case_study_design}), utilizing actual PJM real-time energy prices from July 1 to August 10, 2022. We ran the MILP manufacturing model to optimize production scheduling and generate hourly consumption traces. We injected Gaussian noise (zero mean, 2\% max load variance) to simulate meter data variations (Fig.~\ref{fig_price_chap03}). The foundational equipment parameters are sourced from references~\inlinecite{ref22} and \inlinecite{golmohamadi_robust_2020} (Tables~\ref{tab_parameter} and \ref{tab_parameter_cement}). We partitioned the data: the first 21 days formed the $TRAIN$ set, the next 10 days functioned as the $CV$ set, and the final 10 days served as the blind $TEST$ set to evaluate accuracy.

We generated the reference consumption data using the standard MILP STN model rather than our mSTN, ensuring that our PSI framework is evaluated against external data generation methods. ``True value'' refers to the output generated by the original STN running with the full parameter set.

\subsection{Hardware, Settings, and Task Descriptions}

We deployed Gurobi (V10.0.0) via YALMIP~\cite{Lofberg2004} in MATLAB (R2021a) on an Intel Core i9-10900X workstation. We enforce a 120-second timeout per ZO-SGD iteration, running a maximum of 210 iterations. The decay parameter $\alpha$ holds at 1 for the first $3n$ iterations, smoothly decaying to 0 thereafter.

\subsection{Configuration of Machine Learning Methods}\label{app_ml_methods}
We benchmark against three data-driven models:
\textbf{MLP}: A 24-48-48-24 deep network mapping 24-hour prices to 24-hour loads. Trained via Adam ($lr=2\times 10^{-5}$).
\textbf{LSTM}: Single LSTM layer exploiting time-series continuity. Trained via Adam ($lr=5\times 10^{-4}$).
\textbf{SVR}: 24 independent Support Vector Regression models, each predicting a single hour's consumption based on the full 24-hour price vector.

Because the dataset is limited to 21 training days, complex models are prone to overfitting. We selected these standard architectures to provide an appropriate baseline comparison for data-driven approaches.

\subsection{Results and Comparison}

We evaluate performance on the test set using the normalized root mean square error (nRMSE), obtained by normalizing the absolute RMSE by the maximum load.

\begin{figure}[!t]
  \centering
  \includegraphics[width=\textwidth]{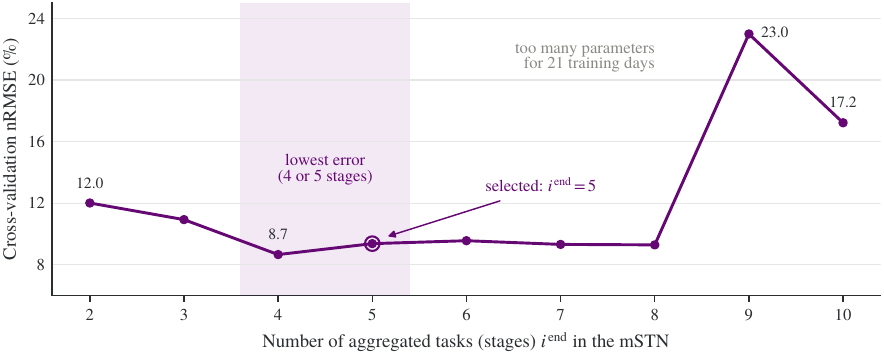}
\caption[Cross-validation error versus number of aggregated production stages]{Cross-validation error versus number of aggregated production stages in the steel-powder load model, measured by normalized root mean square error (nRMSE)}
  \label{fig_rmse_wrt_m}
\end{figure}

\textbf{Model selection and identification results.} Fig.~\ref{fig_rmse_wrt_m} demonstrates how PSI determines the appropriate aggregation level. For the 10-stage steel manufacturer, PSI achieves the lowest validation error with 4 or 5 aggregated stages. If we require PSI to model all 10 stages, performance decreases due to excessive decision variables relative to the 21-day dataset. By selecting $i^{\rm end} = 5$, PSI balances model expressiveness with mathematical tractability. Table~\ref{tab_identified_model} shows that PSI reconstructs the binding parameters reasonably well. While it shows larger errors with non-binding variables (e.g., max buffer of stage $3^*$), this is expected as non-binding constraints do not influence the optimal load curve. The cement plant required no aggregation due to its 4-stage structure.

\begin{table}[!t]
  \caption{Identified load model parameters for the steel-powder manufacturer}
  \label{tab_identified_model}
  \centering 
\begin{tabular}{llllll}
\toprule
\begin{tabular}[c]{@{}l@{}}Aggregated\\ task number\end{tabular} &
  \begin{tabular}[c]{@{}l@{}}Load model\\ parameter\end{tabular} &
  \begin{tabular}[c]{@{}l@{}}True\\ value\end{tabular} &
  \begin{tabular}[c]{@{}l@{}}Default\\ PSI\end{tabular} &
  \begin{tabular}[c]{@{}l@{}}Without as-\\ sumption (2)\end{tabular} &
  \begin{tabular}[c]{@{}l@{}}Without as-\\ sumption (3)\end{tabular} \\ \midrule
\multirow{2}{*}{$1^*$(1)}      & max. Power  & 60  & 56  & 44  & 39  \\ \cline{2-6} 
                        & max. Buffer & 90  & 80  & 106 & 36  \\ \hline
\multirow{2}{*}{$2^*$(2+3)}    & max. Power  & 40  & 37  & 55  & 272 \\ \cline{2-6} 
                        & max. Buffer & 75  & 67  & 320 & 56  \\ \hline
\multirow{2}{*}{$3^*$(4+5+6)}  & max. Power  & 55  & 54  & 55  & 53  \\ \cline{2-6} 
                        & max. Buffer & 50  & 8   & 301 & 8   \\ \hline
\multirow{2}{*}{$4^*$(7)}      & max. Power  & 75  & 80  & 69  & 85  \\ \cline{2-6} 
                        & max. Buffer & 50  & 72  & 199 & 264 \\ \hline
\multirow{2}{*}{$5^*$(8+9+10)} & max. Power  & 36  & 58  & 63  & 40  \\ \cline{2-6} 
                        & Target      & 240 & 222 & 85  & 218 \\ \bottomrule
\end{tabular}
\end{table}

\begin{table}[!t]
  \caption{Load model accuracy comparison (nRMSE)}
  \label{tab_rmse_methods}
  \centering
  \begin{tabular}{ccc}
    \toprule
    Approaches &
    \begin{tabular}[c]{@{}c@{}}Cement\\ plant\end{tabular} &
    \begin{tabular}[c]{@{}c@{}}Steel-powder\\ manufacturer\end{tabular} \\ \midrule
      MLP  & 13.4\% & 18.5\% \\
  LSTM & 13.9\% & 19.2\% \\
  SVR  & 13.7\% & 19.1\% \\
  \textbf{PSI}  & \textbf{5.2\%}   & \textbf{8.5\%}  \\ \bottomrule
  \end{tabular}
  \end{table}

\begin{figure}[!t]
  \centering
  \includegraphics[width=\textwidth]{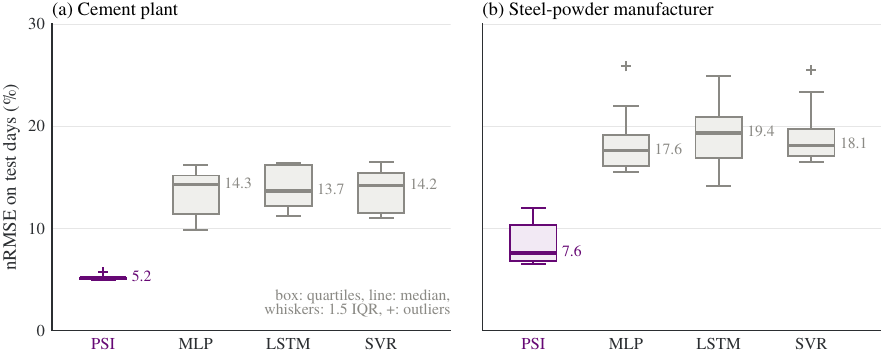}
\caption[Test-day nRMSE of load predictions from PSI and machine-learning models]{Test-day normalized root mean square error (nRMSE) of load predictions from production scheduling identification (PSI) and machine-learning models: (a) cement plant; (b) steel-powder plant}
  \label{fig_accuracy}
\end{figure}

\textbf{Performance of the identified load model.} Table~\ref{tab_rmse_methods} and Fig.~\ref{fig_accuracy} show that PSI achieves an nRMSE of 5.2\% for the cement plant and 8.5\% for the steel-powder manufacturer, whereas the machine-learning models yield 13.4\%--19.2\%. Relative to the best machine-learning result for each facility, PSI reduces error by 61.2\% in the cement case and 54.1\% in the steel-powder case. The lower cement-plant error is consistent with its smaller four-stage parameter search space.

\textbf{Comparison with machine learning models.} The ML baselines generally suffer from severe underfitting due to the limited training data (21 days), which is insufficient for these models to infer the underlying consumption patterns effectively. Conversely, PSI leverages the physical constraints of industrial production and the economic optimization objective to formulate its estimates, resulting in better performance even with small datasets.

\begin{figure}[!t]
  \centering
  \includegraphics[width=\textwidth]{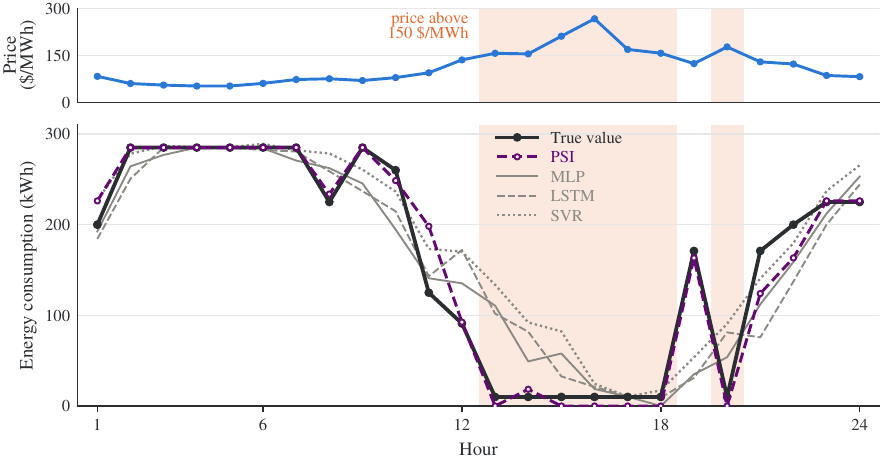}
\caption[Steel-powder plant load profiles on August 5 from PSI and machine-learning models]{Steel-powder plant load profiles on August 5 from production scheduling identification (PSI) and machine-learning models}
  \label{fig_typical_load_steel}
\end{figure}

\begin{figure}[!t]
  \centering
  \includegraphics[width=\textwidth]{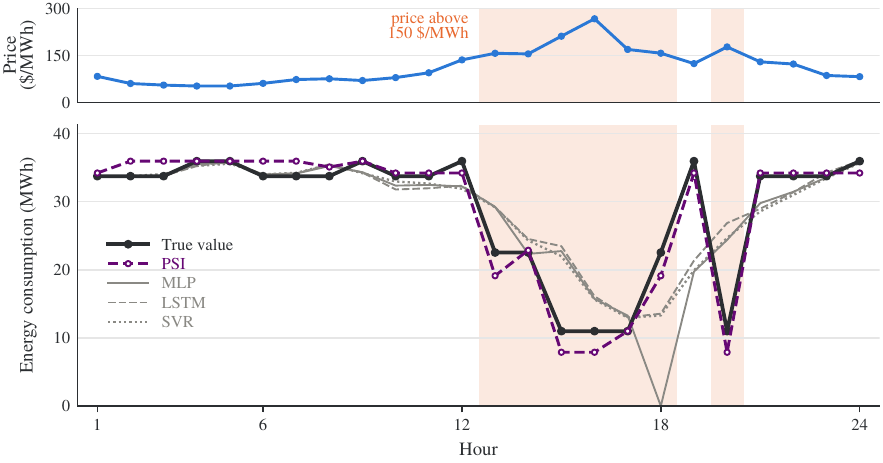}
\caption[Cement plant load profiles on August 5 from PSI and machine-learning models]{Cement plant load profiles on August 5 from production scheduling identification (PSI) and machine-learning models}
  \label{fig_typical_load_cement}
\end{figure}

Visualizing the outputs (Figs.~\ref{fig_typical_load_steel} and \ref{fig_typical_load_cement}) shows that PSI captures the consumption patterns well, particularly during high-price hours (e.g., 3:00 PM - 5:00 PM). Because PSI assumes cost-minimizing behavior, it tends to avoid scheduling loads during price peaks. The ML models, lacking an explicit economic objective, fail to accurately capture these load reductions.

\begin{figure}[!t]
  \centering
  \includegraphics[width=\textwidth]{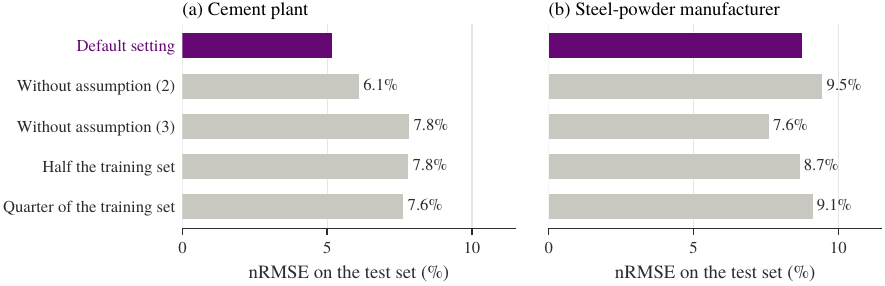}
\caption[Load-prediction errors of PSI under different assumptions and training-set sizes]{Load-prediction errors of production scheduling identification (PSI) under different assumptions and training-set sizes, measured by normalized root mean square error (nRMSE)}
  \label{fig_variation}
\end{figure}

\textbf{Performance under different settings.} Fig.~\ref{fig_variation} illustrates the impact of the physical assumptions. Removing the physical state constraints (Assumption 2) increases the reported error. Reducing the training dataset from 21 days to 5 days causes only a modest increase in this test, indicating data efficiency under the evaluated conditions.

\section{Discussion}\label{sec_discussion}

While the numerical tests validate the feasibility of PSI, scaling this method to field deployment involves addressing several issues that arise in practice:

First, the numerical tests calculate electricity costs as hourly consumption multiplied by an hourly price. The price may be a day-ahead or real-time wholesale price, or a time-of-use tariff set by a retailer or grid operator. In practice, many industrial facilities settle electricity purchases under long-term contracts or tariffs that change only monthly. If historical data contain no distinguishable response to price changes, or only a weak correlation, the price--consumption coupling assumed by PSI becomes poorly identifiable. Possible extensions include small controlled trials with designed incentives, change-point detection to estimate piecewise price sensitivity, robust or distributionally robust inverse optimization that jointly represents uncertain prices and production boundaries, and—where available—the use of observable proxies such as output, shifts, or operating rates to constrain the inverse problem.

Second, we assumed that users minimize electricity costs while meeting planned production targets. Factory managers may instead be risk averse or seek to limit deviations from an established production plan. Such objectives must be incorporated before constructing the inverse problem; otherwise, the apparently interpretable optimum and the inferred parameters reflect only an equivalent preference under the selected cost metric.

At the same time, we assume that the user's energy-consumption mechanism can be represented by an STN. Although STN is general, it omits details such as parameter uncertainty and nonlinear production efficiency. Chapter 2 shows that RTN is a more general model form, so embedding RTN in the PSI framework is an important direction for representing discrete batch interactions more comprehensively. More broadly, PSI can retain its inverse-optimization structure while replacing the default mSTN and behavioral assumptions. With sufficient computing resources, a library of candidate industrial-load and user-objective models could be screened through training and cross-validation.

The reported accuracy should therefore be interpreted within the data-generation mechanism and assumptions stated in Chapters 2 and 3. Extrapolation requires particular care when metering noise is large, contract prices or multi-objective decisions dominate, or the historical record contains little price-response information. For new or heterogeneous industrial parks, the identified parameters should be validated on independent test days and recalibrated periodically rather than transferred directly from another site.

\section{Summary}\label{sec_conclusion_chap03}

This chapter addresses the challenge of modeling industrial flexibility under incomplete parameter information. The Production Scheduling Identification (PSI) framework uses inverse optimization to estimate a facility-specific load model from hourly smart meter data.

PSI shows that a useful external load model can be identified without high-frequency sensor data. By treating meter data as the outcome of production scheduling and constraining the search space with process knowledge, the method estimates a model from hourly data without requiring the facility to disclose its full internal parameter set.

The process constraints also improve load-model accuracy when training data are limited. On test datasets generated from established cement and steel-powder process models using PJM prices, PSI achieves errors of 5.2\% and 8.5\% using 21 training days. These values are 61.2\% and 54.1\% lower, respectively, than the best result from the compared machine-learning models for each facility.

The identified model predicts both aggregate electricity consumption and its distribution across represented production stages. Subject to the behavioral and model assumptions discussed above, this information can support flexibility assessment, baseline estimation, and changes in production targets when VPPs cannot directly obtain private process parameters.


\chapter{Data-Driven Dimension Reduction for Industrial Load Modeling}

\section{Overview}

\subsection{Background}
A core challenge in incorporating massive industrial loads into power system dispatch---or, equivalently, into the operation of Virtual Power Plants (VPPs)---is how to approximate high-dimensional nonconvex feasible regions with low-dimensional linear constraints.
Taking the Gansu provincial electricity market as an example, the province has only 46 thermal generators, yet the number of high-voltage users reaches 100,000 and low-voltage users reaches 980,000~\cite{GansuPowerMarket2024}, a difference of several orders of magnitude.
Moreover, each industrial user contains numerous pieces of equipment, each with multiple operating states. If modeled directly and incorporated into the dispatch system as done for conventional thermal generators, the problem size would grow explosively, rendering the optimization problem unsolvable within a reasonable time frame.
Even in scenarios where VPPs participate in the electricity market, if the number of industrial users is too large, it is difficult to directly incorporate all their constraints into VPP-level decision-making.
Therefore, dimension reduction of industrial load model constraints, approximating complex high-dimensional feasible regions with simple low-dimensional linear constraints, is essential for real-world application in system-level optimization and VPP-level resource aggregation.

\begin{figure}[htbp]
  \centering
  \includegraphics[width=0.9\textwidth]{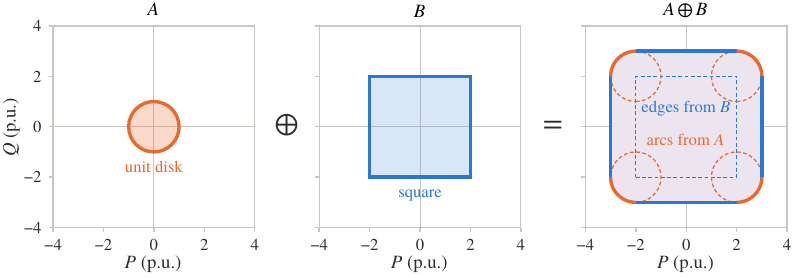}
  \caption{Illustration of feasible-region aggregation through the Minkowski sum}
  \label{fig_minkowski_sum}
\end{figure}

Among existing research, the concept closest to dimension reduction is aggregation.
Aggregation typically refers to summing multiple feasible regions, i.e., computing the Minkowski sum of sets (as illustrated in Fig.~\ref{fig_minkowski_sum}); dimension reduction, in addition to encompassing the aggregation process, can also refer to the approximation of high-dimensional constraints within a single device.
In other words, even when considering a single device, if its feasible region is high-dimensional, the process of approximating it with low-dimensional constraints also constitutes dimension reduction.
When multiple devices are involved, dimension reduction encompasses both the approximation of internal constraints within each device and the further simplification after aggregation across devices.
Therefore, dimension reduction is a more general concept than aggregation, covering constraint simplification problems from the device level to the user level and ultimately to the system level.

However, the dimension reduction and aggregation of feasible regions are highly complex.
Taking the simplest case of aggregation as an example (Fig.~\ref{fig_minkowski_sum}), even for the most basic geometric shapes, such as the Minkowski sum of a circle and a square, a concise formula is difficult to express, and multiple piecewise constraints are often needed to fully describe its feasible boundary.
It can be theoretically proven that as the number of sets increases, the representation complexity or solution complexity of computing their Minkowski sum grows exponentially. This exponential growth makes exact analytical solutions impractical, motivating the development of approximation algorithms.

\subsection{Motivation}

Existing dimension reduction methods for flexibility resource control constraints can be broadly classified into two categories: analytical methods and data-driven methods.
Analytical methods are based on mathematical derivation from the original constraint formulations, achieving dimension reduction through mathematical transformations and a certain degree of approximation.
Although such methods perform well for certain specific constraint forms, they have notable limitations: they exhibit poor adaptability (requiring new mathematical derivations for different constraints); a change in constraints necessitates redesigning the method; and high implementation difficulty, making them hard to generalize across the diversity of industrial users.

Data-driven methods adopt machine learning concepts, obtaining sample points inside and outside the feasible region through sampling, then training classifiers to distinguish feasible from infeasible points.
However, traditional data-driven methods are susceptible to the curse of dimensionality: while dense sampling is feasible in low dimensions, in high dimensions---for instance, for 96-dimensional industrial user electricity data (24 hours with one point per hour, or 96 points at 15-minute intervals)---the required number of samples is prohibitively large.

\begin{figure}[!t]
  \centering
  \includegraphics[width=0.9\textwidth]{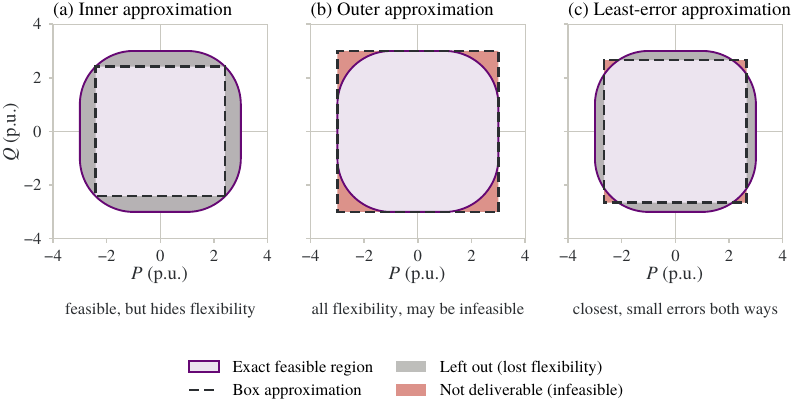}
  \caption[Approximations of the feasible region of demand-side resources]{Approximations of the feasible region of demand-side resources: (a) inner approximation; (b) outer approximation; (c) least-error approximation. Gray shows feasible operation excluded by the approximation; red shows operation allowed by the approximation but infeasible for the resource}
  \label{fig_demo_chap04}
\end{figure}

The approximation requirements for industrial users differ from the strict feasibility requirements typically imposed on generator models.
For generators, strict feasibility guarantees are typically required, i.e., inner approximation (where the approximated feasible region is entirely contained within the original) or outer approximation (where the approximated feasible region entirely contains the original).
However, for demand-side resources, what matters more is having a reasonable incentive mechanism to encourage load participation, rather than requiring the approximation results to be perfectly accurate (as illustrated in Fig.~\ref{fig_demo_chap04}).
According to China's demand-side response and virtual power plant policies~\cite{JiangsuDR2023, GuangdongDR2024}, demand-side response allows a 20\% error tolerance and virtual power plants allow a 10\% tolerance, indicating that approximation of demand-side resources can accommodate certain errors as long as the overall error probability is controlled within a reasonable range.
This relaxation opens new possibilities for designing approximation algorithms.

\subsection{Contributions of This Chapter}

\begin{figure}[!t]
  \centering
  \includegraphics[width=\textwidth]{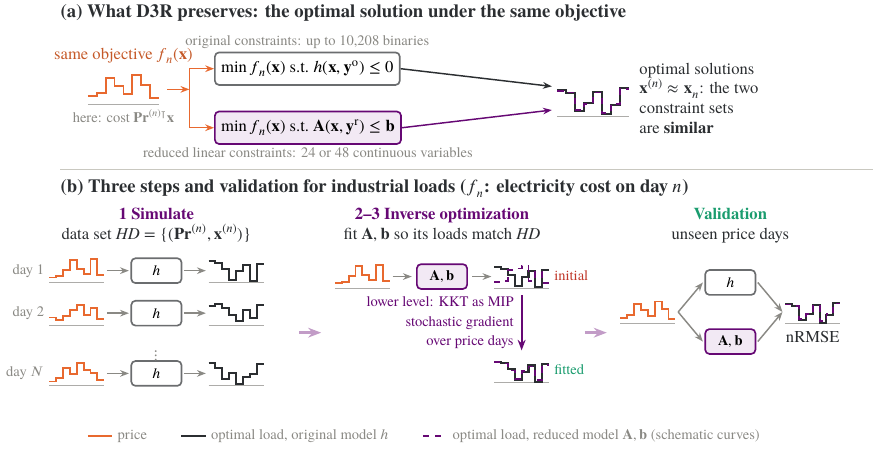}
\caption[Data-driven constraint dimension reduction (D3R) for industrial load modeling]{Data-driven constraint dimension reduction (D3R) for industrial load modeling: (a) similarity measured by the distance between the optimal solutions under the original and reduced constraints with the same objective $f_n$; (b) for industrial loads, with $f_n$ the electricity cost under the prices of day $n$: simulation of the original model to construct the dataset $HD$, formulation and solution of the inverse optimization problem for $\mathbf{A},\mathbf{b}$, and validation on unseen price days. Curves are schematic}
  \label{fig_framework_industrial}
\end{figure}

To address the above challenges, this chapter proposes Data-Driven Dimension Reduction (D3R) for industrial loads. The main content of this chapter is based on the published paper~\cite{lyu_data-driven_2025}.
Rather than derive a new approximation for every process model or densely sample a high-dimensional feasible region, D3R uses inverse optimization to learn least-error linear constraints from operating samples generated by the original model (Fig.~\ref{fig_framework_industrial}, where $h$ denotes the high-dimensional original constraints, $HD$ denotes the dataset, and $A$ denotes the reduced constraints).

Specifically, the method adopts a three-step framework: the first step embeds the original complex constraints into an optimization problem and solves it to obtain optimal solutions.
For industrial users, the primary concern is typically the net electricity consumption profile, i.e., 24 or 96 data points of electricity usage; regardless of how complex the original constraints are or how many variables there are, the resulting data are of these dimensions.
The second step assumes that these optimal solutions across multiple scenarios are generated by some simple linear constraints, and then derives the constraint parameters through inverse optimization.
The third step employs optimization algorithms to solve the inverse optimization problem, obtaining the linear constraint parameters that best approximate the original feasible region.
In particular, this chapter selects the adjustable load fleet (ALF) as the form of the linear constraints, which can capture the aggregate coupling characteristics (daily total electricity consumption constraint), load shifting capability (flexible adjustment of electricity consumption across time periods), and equipment composition diversity (combinations of multiple types of loads), while maintaining constraint simplicity for embedding in system-level optimization problems.
In solving the inverse optimization problem, the lower-level optimality conditions (KKT conditions) introduce nonconvexity and nonlinearity, resulting in high computational complexity.
To improve solution efficiency, the KKT conditions are transformed into a mixed-integer programming (MIP) problem, and a stochastic gradient descent approach is adopted, selecting only a subset of scenarios for optimization in each iteration, progressively approaching the optimal parameters through the iterative process.

Numerical tests cover a cement plant, a steel-powder manufacturer, and a steelmaking plant with different original model structures. D3R represents their optimal electricity responses with nRMSE values of 3.6\%--10.3\%, within or close to the 10\%--20\% tolerances reported in current demand-side participation rules. For the steelmaking case, the grid-facing representation uses 24 or 48 continuous variables instead of the original 10,208 integer variables. The same algorithmic structure is used for all three factories because its input is a 24- or 96-dimensional electricity profile rather than the original process-constraint form.

\section{Problem Description}\label{sec_problem_description_industrial}

The preceding chapters address the precise modeling of industrial loads (Chapter 2) and parameter identification under incomplete information (Chapter 3).
However, even with accurate models and parameters obtained, these high-dimensional complex constraints remain difficult to directly apply in system-level optimization.
Therefore, this section formally describes the constraint dimension reduction problem, elucidating how to approximate high-dimensional complex constraints with low-dimensional linear constraints and how to evaluate the quality of the approximation.

\subsection{Mathematical Formulation of Constraint Dimension Reduction}

Without loss of generality, our objective is to replace the original constraints (OCs) of a given industrial load, denoted as $h(\mathbf{x}, \mathbf{y}^{\rm o}) \leq 0$, with a set of dimension-reduced linear constraints (reduced constraints, RCs) $\mathbf{A}(\mathbf{x}, \mathbf{y}^{\rm r}) \leq \mathbf{b}$.
Here, $h(\cdot)$ is the original constraint function which typically involves complex mixed-integer constraints from resource-task network (RTN)~\cite{ref10} or state-task network (STN)~\cite{golmohamadi_robust_2020, ref22} models of industrial processes;
$x \in \mathbb{R}^T$ represents the hourly energy consumption (continuous), $T$ is the number of time intervals;
$\mathbf{y}^{\rm o}$ represents other variables of the OCs (including both continuous variables for states/resources and integer variables for operating modes/task scheduling);
$\mathbf{y}^{\rm r}$ represents other variables of the RCs (continuous);
superscripts o (original) and r (reduced) are used to distinguish between the original constraints and reduced constraints, respectively;
$A$ and $b$ are the parameter matrix and the right-hand side of the RCs, respectively.

If we require strict equivalence between the RCs and OCs, i.e., projecting the variable space associated with the OCs onto the RCs, then the following condition must hold:
\begin{equation}\label{model_strict_projection}
  \forall \mathbf{x}: \exists \mathbf{y}^{\rm r}, \mathbf{A}(\mathbf{x}, \mathbf{y}^{\rm r}) \leq \mathbf{b} \Leftrightarrow \exists \mathbf{y}^{\rm o}, h(\mathbf{x}, \mathbf{y}^{\rm o}) \leq 0.
\end{equation}
However, strict projection may be computationally infeasible in cases where $h(\cdot)$ has high dimensionality, $\mathbf{y}^{\rm o}$ contains integer variables, or $T$ is large.
Meanwhile, considering the current market mechanisms for demand-side resources, strict projection may not be necessary.
In electricity markets, demand-side resources typically provide services such as demand response, reserves, and frequency regulation, where performance requirements are more flexible compared to generation resources.
For instance, a 10\% to 20\% tolerance is generally acceptable, which motivates us to pursue an approximation-based approach rather than strict equivalence or guaranteeing feasibility.

\subsection{Quantification of Approximation Error}

Considering these market and modeling factors, we relax Equ.~(\ref{model_strict_projection}) and introduce a loss function $J(h(\cdot), \mathbf{A}, \mathbf{b})$ to quantify the error of the RCs relative to the OCs.
Therefore, our objective transforms into determining appropriate $\mathbf{A}$ and $\mathbf{b}$ values to minimize the approximation error $J(h(\cdot), \mathbf{A}, \mathbf{b})$.

The key question is how to define a ``good approximation''.
Drawing on the concept of loss functions in machine learning, we use the Euclidean distance between the optimization results based on the simplified model and those based on the original model as the objective function.
The logic is that if the approximation closely matches the original optimal solutions under various scenarios, then the constraint is considered good.
Formally, let $f_n(\mathbf{x})$, $n = 1, \dots, N$, be a set of objective functions. The RCs approximate the OCs well if, for each $f_n$, the solution that minimizes $f_n$ subject to the RCs is close to the solution that minimizes $f_n$ subject to the OCs.
This criterion depends only on the optimal solutions, not on the form of $f_n$ or of $h(\cdot)$, so it applies to dimension reduction more generally.
For industrial loads, $f_n$ is the electricity cost $\mathbf{Pr}^{(n)\intercal}\mathbf{x}$ under the prices of day $n$; Section~\ref{sec_problem_description_aggregation} applies the same criterion to distributed resources in coupled energy and regulation markets.
Specifically, we do not derive $A$ and $b$ analytically on the basis of the form and parameters of the constraint $h(\cdot)$, but instead observe the historical energy usage of industrial users and parameterize (train) $A$ and $b$ in a data-driven manner.
In doing so, we assume access to historical data of industrial users $HD = \{(\mathbf{Pr}^{(n)}, \mathbf{x}^{(n)})\}$, where $\mathbf{Pr}^{(n)}$ and $\mathbf{x}^{(n)}$ represent the hourly electricity price and the user's hourly electricity consumption on day $n$, respectively, $n = 1, ..., N$.
We also assume that $\mathbf{x}^{(n)}$ are optimized on the basis of the OCs to minimize the energy cost $\mathbf{Pr}^{(n)\intercal} \mathbf{x}^{(n)}$.
Note that $HD$ can be realistic or simulated historical data.
In practice, historical energy consumption data from industrial users may not originate from optimized decisions. For instance, factories might employ heuristic strategies rather than complex optimization models to schedule their production and energy usage.
In such cases, the D3R framework remains applicable but requires an additional step: first, simulating optimal energy usage data under varying electricity prices using the original constraints, then using these simulated data to train the reduced constraints.

\subsection{Inverse Optimization Problem}

Under the above assumptions, the approximation error can be expressed as the distance between the optimal energy consumption under historical electricity prices given by the OCs and RCs, and the training problem is formulated as follows:
\begin{subequations}\label{model_inverse_fit_d3r}
  \begin{equation}\label{model_inverse_fit_lossFunction_d3r}
    \underset{\mathbf{A}, \mathbf{b}}{\rm min.} J = \frac{1}{N} \sum_{n = 1}^{N} ||\mathbf{x}^{(n)} - \mathbf{x}_n||^2
  \end{equation}
  \begin{equation}\label{model_optimality}
    {\rm s.t.} \quad \mathbf{x}_n = {\rm argmin.}\{\mathbf{Pr}^{(n)\intercal} \mathbf{x}_n \ : \ \mathbf{A}(\mathbf{x}_n, \mathbf{y}^{\rm r}) \leq \mathbf{b}\}, \ \forall n
  \end{equation}
\end{subequations}
where $\mathbf{x}_n$ is the variable for the optimization problem regarding day $n$ based on the RCs to distinguish it from $\mathbf{x}^{(n)}$ in the dataset.
Equ.~(\ref{model_inverse_fit_d3r}) is a data-driven inverse optimization problem, which will be further elaborated in Section~\ref{sec_method}.

\section{Methodology}\label{sec_method}

Having established the mathematical formulation of the constraint dimension reduction problem, this section presents the methodology for solving it in detail.
The primary computational challenge is that although D3R in principle allows for the exploration of suitable parameter matrix sizes and element values within the space $A/b$, the high-dimensional search space and the nonlinear lower-level constraints of the optimality conditions in (\ref{model_optimality}) may lead to unaffordable computational costs.
To address this challenge, we pursue two strategies: first, we selected an appropriate form of the parameter matrix on the basis of the energy consumption characteristics of industrial users (Section~\ref{sec_alf}); second, we devised a solution algorithm based on zeroth-order stochastic gradient descent (ZOSGD)~\cite{liu_primer_2020} to iteratively solve the inverse optimization problem, with a transformation of the optimality conditions to reduce the computational complexity at each iteration (Section~\ref{sec_training_industrial}).

\subsection{Adjustable Load Fleet-Based Reduced Constraints}\label{sec_alf}

\subsubsection{Selection of the Adjustable Load Fleet Model}

A key question is: what constraint form can adequately approximate complex industrial loads?
If a completely general $\mathbf{A}\mathbf{x} \leq \mathbf{b}$ form is adopted, the search space for the parameter matrix $\mathbf{A}$ becomes unbounded, since the horizontal dimension is 24 (multiplied by $\mathbf{x}$) and the vertical dimension is undetermined, making it impossible to solve for all possible $\mathbf{A}$.
While theoretically feasible, this could incur prohibitive computational costs.

Since the lower-level of the inverse optimization involves optimality conditions (KKT conditions) that include complementary slackness terms and form nonlinear constraints, the problem remains complex even with acceleration methods.

We therefore restrict the structure of $A$ and $b$; searching over the entire space of $A$ and $b$ is not viable.
The adopted approach is to design a specific form for $A$ and $b$ based on prior knowledge, thereby reducing the search space.
Two main considerations guide this design:
first, when using $\mathbf{A}\mathbf{x} \leq \mathbf{b}$ to approximate the original industrial user constraints, the flexibility characteristics of the original constraints determine the design form of $\mathbf{A}\mathbf{x} \leq \mathbf{b}$; second, the form of $\mathbf{A}\mathbf{x} \leq \mathbf{b}$ must also meet the needs of the power system, i.e., it cannot be overly complex, lest the grid system be unable to dispatch or embed it in optimization.

Based on these considerations, we chose an adjustable load fleet (ALF) composed of multiple adjustable loads as the form of the reduced constraints to be trained.
The motivation for choosing ALF is that, unlike virtual battery models that require explicit baseline definitions or bidirectional energy exchange assumptions~\cite{9979729}, it can directly model the actual composition and behavior of industrial loads and reflects industrial reality through the adjustable-load structure.
The ALF model captures three key physical characteristics of industrial loads through linear constraints:

1) Equipment composition diversity: Multiple pieces of equipment in industrial facilities are represented by parallel adjustable loads, where the total energy consumption is the sum of individual device consumption.

2) Equipment power rating: Each device's power consumption is bounded by its rated capacity.

3) Production goals: Daily energy consumption limits reflect production targets, as they are typically proportional to energy usage through product energy intensity factors.

\subsubsection{Mathematical Formulation of the Adjustable Load Fleet}

The reduced constraints in the form of an ALF essentially impose upper and lower bounds on the factory's hourly and cumulative energy consumption: the factory's total energy consumption at each time interval equals the sum of energy consumption of individual devices (adjustable loads); each adjustable load has power upper and lower limits at each time interval, and the total daily energy consumption of each adjustable load also has upper and lower limits:
\begin{subequations}\label{model_alf}
  \begin{equation}\label{model_alf_powerLimit}
    x_{t} = \sum_{i=1}^{I} p_{t, i} \Delta t, \quad \underline{P}_{i} \le p_{t, i} \le \overline{P}_{i} \ : \ \underline{\mu}^{\rm P}_{t, i}, \overline{\mu}^{\rm P}_{t, i}, \ \forall  t
  \end{equation}
  \begin{equation}\label{model_alf_energyLimit}
    \underline{E}_{i} \le \sum_{t=1}^{T} p_{t, i} \Delta t \le \overline{E}_{i} \ : \ \underline{\mu}^{\rm E}_{i}, \overline{\mu}^{\rm E}_{i}
  \end{equation}
\end{subequations}
where the subscript $n$ is omitted; $I$ is the number of adjustable loads as a hyperparameter that balances model complexity and approximation accuracy.
A larger $I$ allows the RCs to better approximate the OCs but increases computational costs, while limited training data may constrain the choice of $I$ to avoid overfitting.
$x_{t}$ is the net energy consumption at time $t$, which is the summation of $p_{t, i} \Delta t$, representing the average consumption power of adjustable load $i$ at time $t$ multiplied by the length of time interval $\Delta t$.
$\theta = \{\underline{P}_{i}, \overline{P}_{i}, \underline{E}_{i}, \overline{E}_{i}\}$ include the power and energy limits of adjustable load $i$, which are the parameters to be fitted.
The Lagrange multipliers $\underline{\mu}^{\rm P}_{t, i}, \overline{\mu}^{\rm P}_{t, i}, \underline{\mu}^{\rm E}_{i}, \overline{\mu}^{\rm E}_{i}$ are the dual variables associated with the power and energy limits, respectively.
The above linear constraints can be easily transformed into the form of $\mathbf{A}(\mathbf{x}, \mathbf{y}^{\rm r}) \leq \mathbf{b}$, where $\mathbf{x} = [x_t]$ and $\mathbf{y}^{\rm r} = [p_{t, i}]$.

Although this constraint form is simple, it captures several key characteristics:
first, it reflects the aggregate coupling property, since (\ref{model_alf_energyLimit}) constrains the total energy consumption across time intervals;
second, it reflects load shifting capability, allowing flexible adjustment of electricity consumption across different time intervals given a fixed total;
third, (\ref{model_alf_powerLimit}) also reflects the diversity of equipment composition, capable of representing multiple production lines, some adjustable and some not.

While this simplified model may overlook some temporal coupling between production processes and relax integer variables, it strikes a reasonable balance between modeling accuracy and computational tractability.
Moreover, the D3R framework is not limited to the ALF model --- other reduced constraint forms can be chosen based on specific needs while maintaining the framework's core methodology.

\subsection{Training of the Adjustable Load Fleet Model}\label{sec_training_industrial}

\subsubsection{Transformation of Optimality Conditions}

Substituting the form of the ALF, the optimality conditions (Karush-Kuhn-Tucker conditions) represented by (\ref{model_optimality}) for day $n$ include primal feasibility (\ref{model_alf}), stationarity condition derived from the Lagrangian function (\ref{model_stationarity}), dual feasibility (\ref{model_dual_feasibility}), and complementary slackness (\ref{model_complementary_slackness1}, \ref{model_complementary_slackness2}):
\begin{subequations}\label{model_optimality_conditions}
  \begin{equation}\label{model_stationarity}
    Pr^{\rm E}_{t} - \underline{\mu}^{\rm P}_{t, i} + \overline{\mu}^{\rm P}_{t, i} - \underline{\mu}^{\rm E}_{i} + \overline{\mu}^{\rm E}_{i} = 0, \ \forall t, \forall i
  \end{equation}
  \begin{equation}\label{model_dual_feasibility}
    (\underline{\mu}^{\rm P}_{t, i}, \overline{\mu}^{\rm P}_{t, i}, \underline{\mu}^{\rm E}_{i}, \overline{\mu}^{\rm E}_{i}) \ge 0, \ \forall t, \forall i
  \end{equation}
\begin{equation}\label{model_complementary_slackness1}
    \underline{\mu}^{\rm P}_{t, i} (\underline{P}_{i} - p_{t, i}) = 0, \quad \overline{\mu}^{\rm P}_{t, i} (p_{t, i} - \overline{P}_{i}) = 0, \quad \forall t, i
  \end{equation}
  \begin{equation}\label{model_complementary_slackness2}
    \underline{\mu}^{\rm E}_{i} (\underline{E}_{i} - \sum_{t=1}^{T} p_{t, i}) = 0, \quad \overline{\mu}^{\rm E}_{i} (\sum_{t=1}^{T} p_{t, i} - \overline{E}_{i}) = 0, \quad \forall i
  \end{equation}
\end{subequations}

The complementary slackness conditions (\ref{model_complementary_slackness1}, \ref{model_complementary_slackness2}) involve bilinear constraints, which can be transformed via the Fortuny-Amat transformation~\cite{fortuny-amat_representation_1981} into a more easily solvable form.
For example, the first term can be replaced by the following:
\begin{equation}\label{model_complementary_slackness3}
  \underline{\mu}^{\rm P}_{t, i} \le M(1 - z_{t, i}), \quad p_{t, i} - \underline{P}_{i} \le M z_{t, i}, \quad \forall t, i
\end{equation}
where $z_{t, i}$ is a binary variable and $M$ is a large positive number.
This transformation converts the non-convex KKT conditions into a Mixed-Integer Programming (MIP) format, which can be handled by commercial solvers.

\subsubsection{Iterative Solution Algorithm}

Since the number of integer variables scales linearly with the dataset size $N$, directly solving the problem across all scenarios would produce an excessive number of constraints and highly complex KKT conditions, with excessive memory and computational requirements.
To address this, we use ZOSGD~\cite{liu_primer_2020}, selecting only a subset of scenarios in each iteration and optimizing the objective function.
This is equivalent to exploration, using zeroth-order methods to estimate the gradient direction, followed by iterative convergence.

ZOSGD is well suited for the D3R framework because it avoids explicit gradient computation, controls memory usage via batch processing, and provides iterative descent towards the target parameter set through the optimization iterations.
In each iteration, certain days of data are randomly selected and parameters are updated using estimated gradients:

1: Initialize $j = 0$ and $\theta^{(0)} = \{\underline{P}_{i}, \overline{P}_{i}, \underline{E}_{i}, \overline{E}_{i}\} = [0]$.

2: Randomly select $B$ days from the dataset and solve the batch problem:
$\underset{\theta}{\rm min.} J = \frac{1}{B} \sum_{n = 1}^{B} ||\mathbf{x}^{(n)} - \mathbf{x}_n||^2 \ {\rm s.t.} (\ref{model_alf})-(\ref{model_optimality_conditions})$ to obtain $\theta^{*}$.

3: Update $\theta^{(j+1)} = (1 - \alpha)\theta^{(j)} + \alpha \theta^{*}$; if $\theta^{(j+1)}$ converges, break; otherwise, set $j = j + 1$ and go to step 2.

where $\alpha$ is the learning rate, which can be adaptively adjusted.
The ZOSGD algorithm can be easily implemented using commercial solvers, and the computational complexity is independent of the number of days in the dataset.

\section{Numerical Results}\label{sec_numerical_industrial}

This section evaluates D3R on a cement plant, a steel-powder manufacturer, and a steelmaking plant. Although the original models have different structures and parameters, D3R maps all three into the same adjustable-load-fleet form.

\subsection{Dataset and Test Settings}

We tested the D3R framework on three datasets from a cement plant~\cite{golmohamadi_robust_2020}, a steel-powder manufacturer~\cite{ref22}, and a steelmaking plant~\cite{ref10}.
The first two were originally modeled via the STN, whereas the steelmaking plant was modeled via the RTN. Both are high-dimensional models with integer variables.

We used historical hourly electricity market prices from PJM in July 2022 for the simulations.
The data from July 1st to 21st were contained in the training set, whereas the data from July 22nd to 31st were used to test the accuracy of the reduced constraints.
The detailed settings and codes can be found in~\cite{rick10119_data-driven_2024}.
We employed Gurobi (V11.0.0) with YALMIP in MATLAB to solve the optimization problems on a workstation equipped with an Intel Core i9-10900X CPU (3.7 GHz) and 128 GB of RAM.

For complex industrial load models, there are currently no established methods for dimension reduction.
Therefore, we compared the accuracy of D3R with the following methods:

1) A baseline approach using the simple adjustable load model (SAL), which assigns values to $\theta$ based on the maximum and minimum values of the corresponding quantities in the historical load curves.
For instance, the maximum power value from historical data is directly taken as $P^{\rm max} (\overline{P})$ for the SAL model.

2) An analytical approach using the optimal virtual battery model (OVB)~\cite{9979729}, an inner-approximation method.
For the STN model, we need to relax the decision variables to continuous variables as the OVB method only applies to linear constraints.
Moreover, the OVB method could not be applied to steelmaking plants modeled by RTN due to the absence of linear relaxation methods, and therefore no comparison is made for this case.

\subsection{Results and Comparison}

\begin{table}[!t]
  \caption{Performance comparison of different methods}
  \label{tab_combined}
  \centering
  \setlength{\tabcolsep}{4pt}
  \begin{tabular}{c|ccc|ccc}
    \toprule
    \multirow{2}{*}{Method} & \multicolumn{3}{c|}{Normalized RMSE (\%)} & \multicolumn{3}{c}{Continuous (Integer) Variables} \\
    \cmidrule{2-7}
    & Cement & \begin{tabular}[c]{@{}c@{}}Steel\\powder\end{tabular} & \begin{tabular}[c]{@{}c@{}}Steel-\\making\end{tabular} & Cement & \begin{tabular}[c]{@{}c@{}}Steel\\powder\end{tabular} & \begin{tabular}[c]{@{}c@{}}Steel-\\making\end{tabular} \\
    \midrule
    Original & - & - & - & 196(288) & 490(720) & 0(10208) \\
    OVB & 25.4 & 36.7 & N/A & 48(0) & 48(0) & N/A \\
    SAL & 10.2 & 23.0 & 8.3 & 24(0) & 24(0) & 24(0) \\
    D3R-1 & \textbf{8.9} & 17.2 & \textbf{3.6} & 24(0) & 24(0) & 24(0) \\
    D3R-2 & 9.5 & \textbf{10.3} & 4.2 & 48(0) & 48(0) & 48(0) \\
    \bottomrule
  \end{tabular}
\end{table}

\begin{figure}[!t]
  \centering
  \includegraphics[width=\textwidth]{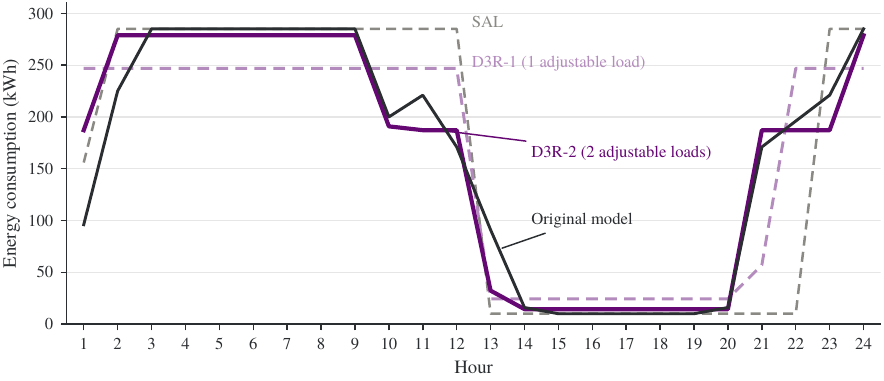}
  
  \caption{Optimal load profiles of the steel-powder plant from the detailed model and reduced models}
  \label{fig_dispatch_industrial}
\end{figure}

\begin{figure}[!t]
  \centering
  \includegraphics[width=\textwidth]{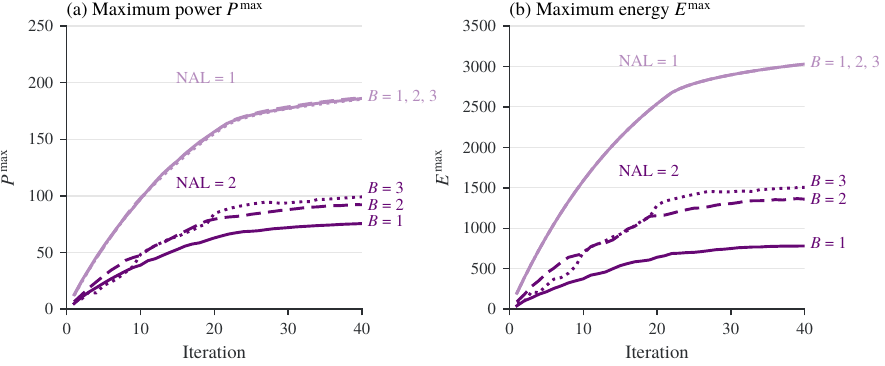}
  \caption[Training trajectories of the steel-powder reduced model]{Training trajectories of maximum power and daily-energy bounds for the steel-powder reduced model: (a) maximum power; (b) maximum daily energy, for one or two adjustable loads and batch sizes B = 1, 2, 3}
  \label{fig_convergence_training}
\end{figure}

\begin{figure}[!t]
  \centering
  \includegraphics[width=\textwidth]{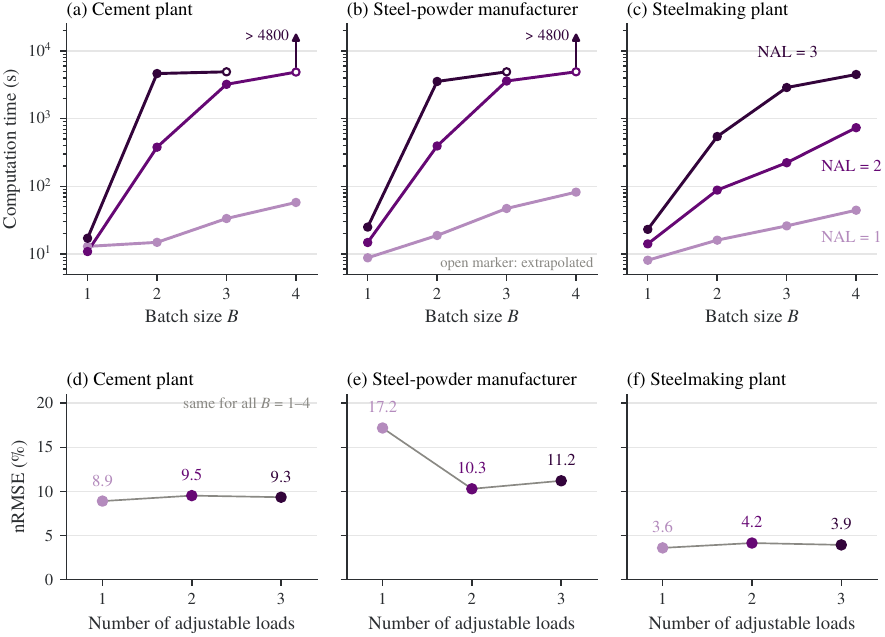}
  \caption[Training time and load-profile error of D3R for three plants]{Training time and load-profile error for the cement, steel-powder, and steelmaking plants (columns): (a)--(c) training time versus batch size B for different numbers of adjustable loads (NAL); (d)--(f) test-set normalized root mean square error (nRMSE) versus NAL, unchanged across the tested batch sizes. Open markers denote values reconstructed by extrapolating clipped source-figure curves; arrows denote values above 4800 s}
  \label{fig_convergence_results}
\end{figure}

Table~\ref{tab_combined} shows that D3R yields the lowest test-set nRMSE among the applicable methods for each industrial case.
Here, D3R-1(2) represents D3R via the ALF model with 1(2) adjustable loads.
The analytical OVB method yielded large errors ($>25\%$) because relaxing the integer variables degraded the accuracy.
Fig.~\ref{fig_dispatch_industrial} visually illustrates D3R's performance. The black curve represents the optimal response obtained from the original model, and the D3R-2 curve represents the approximate results obtained from a fleet of two adjustable loads, which tracks the main features closely.
Across the three cases, the best D3R configuration yields errors of 3.6\%--10.3\% (Table~\ref{tab_combined}), within or close to the cited 10\%--20\% tolerances for demand-side participation.

For the steelmaking plant, D3R replaces 10,208 binary variables with 24 or 48 continuous variables in the grid-facing representation. The D3R code requires no structural modification across the three factory models because the input is always a 24-dimensional electricity profile. This provides a common optimization interface while leaving each factory's precise process model unchanged for final execution.

Fig.~\ref{fig_convergence_training} and Fig.~\ref{fig_convergence_results} illustrate the training process and computational cost of D3R.
Fig.~\ref{fig_convergence_training} shows the convergence of several typical parameters during training, with training trajectories for one and two adjustable loads at batch sizes B = 1, 2, 3; Fig.~\ref{fig_convergence_results} compares the computation time and optimal energy consumption errors under different numbers of adjustable loads and batch sizes, where NAL denotes the number of adjustable loads and B denotes the batch size (number of scenarios selected per iteration).
In the cement plant case (4 production stages), the OVB method converged after 767 seconds.
However, in the steel-powder manufacturer case (10 production stages), the analytical OVB method failed to converge within 24 hours, whereas D3R remained tractable, indicating better scalability for complex industrial cases.

\section{Method Extension: Constraint Aggregation for Multiple Flexibility Resources}

The preceding sections established the D3R framework for industrial load constraint dimension reduction and validated its effectiveness on three factory datasets.
However, the previous focus was primarily on constraint dimension reduction for a single factory's energy consumption model.
In this section, we extend D3R to fleet-level aggregation across clusters of demand-side resources, i.e., the constraint aggregation of multiple flexibility resource models.
The penetration of emerging load resources such as electric vehicles and distributed energy storage continues to increase, and the operational model in which VPPs integrate heterogeneous resource clusters to participate in energy trading and ancillary service markets has become increasingly common.
The operational constraints of these resources also exhibit characteristics of high dimensionality, heterogeneity, and coupling, and the characterization of their aggregate feasible regions likewise faces the curse of dimensionality.
Therefore, extending the D3R framework from single industrial load model constraint dimension reduction to multiple flexibility resource aggregation holds significant theoretical and practical importance.

By extending D3R to multi-resource scenarios, this section addresses multiple demand-side flexibility resources (such as electric vehicles and energy storage) participating in energy and regulation markets.
Unlike industrial users whose market role is primarily energy trading, flexibility resources such as electric vehicles possess the dual capability of bidding for both energy and frequency regulation ancillary services, and their market bids need to characterize the feasible operating space under energy-regulation coupling constraints.
This extension further validates the framework's adaptability and provides a new technical pathway for VPPs to simplify the bidding process and reduce computational complexity.
It also highlights the usefulness of the D3R method for constraint dimension reduction across different resource types and market environments.

\subsection{Feasible Region Aggregation Problem for Distributed Resources}\label{sec_problem_description_aggregation}

Consider a VPP that aggregates distributed energy resources (DERs) to jointly provide energy and frequency regulation services to the power system.
According to the bidding model specified by the independent system operator (ISO), the VPP submits its feasible region (FR) for providing energy and regulation services.
Generally speaking, bidding models consist of two parts: the feasible region and the price.
Here we focus on the feasible region part, since distributed resources typically participate in the market as price takers.
Based on the submitted feasible region and the bids/offers of other units and loads, the ISO solves the dispatching problem (or the clearing problem in a market environment) to determine the VPP's hourly energy output and regulation capacity.
The VPP then schedules its internal DERs to meet the dispatch results as much as possible, as deviating from the dispatch results may incur additional penalties.

\begin{figure}[!t]
  \centering
  \includegraphics[width=\textwidth]{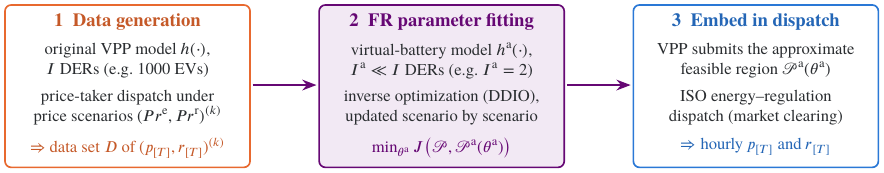}
  \caption{Aggregation of resource feasible regions for virtual power plant (VPP) operation}
  \label{fig_framework_aggregation}
\end{figure}

Without loss of generality, the feasible region $\mathcal{P}$ of the VPP is energy-regulation coupled, represented formally as:
\begin{align}\label{abstract_fr}
  \mathcal{P} := \{ & \ (p_{[T]}, r_{[T]}) : \exists y_{[I],[T]},                             \\
  \nonumber         & \ {\rm s.t.} \  h(p_{[T]}, r_{[T]}, y_{[I],[T]}; \theta_{[I]}) \le 0\}.
\end{align}
where $T$ and $I$ represent the number of scheduling time intervals and DERs, respectively.
The variables with subscript $[T]$ are a shorthand notation for a T-dimensional vector, for example, $p_{[T]} = (p_1, p_2, ..., p_T)$ and $r_{[T]} = (r_1, r_2, ..., r_T)$ represent the hourly (baseline energy) output and regulation capacity of the VPP, respectively.
$h(\cdot)$ represents the original operational constraints of the VPP for providing energy-frequency services, with $y_{[I][T]}$ and $\theta_{[I]}$ representing the variables and parameters for the operation of the DERs within the VPP, respectively.

In (\ref{abstract_fr}), $h(\cdot)$ includes the power and energy constraints of thousands of coupled DERs within the VPP.
The feasible region $\mathcal{P}$ can be expressed as a set of linear constraints on $(p, r)$, formally represented as $\mathbf{A}(p, r) \le \mathbf{b}$.
Computing $\mathcal{P}$ is a projection problem to eliminate the internal variables $y$ from the original constraints $h(\cdot)$, which is an NP-hard problem.
Computing an exact $\mathcal{P}$ is often infeasible when $I$ reaches thousands.
Our main idea is to first determine a suitable low-dimensional constraint form $h^{\rm a}(\cdot)$ to approximate the original constraints $h(\cdot)$, and then construct an approximate feasible region $\mathcal{P}^{\rm a}$ based on $h^{\rm a}(\cdot)$:
\begin{align}\label{abstract_fr_surrogate}
  \mathcal{P}^{\rm a} := \{ & \ (p_{[T]}, r_{[T]}) : \exists y^{\rm a}_{[I^{\rm a}],[T]},                                                     \\
  \nonumber                 & \ {\rm s.t.} \  h^{\rm a}(p_{[T]}, r_{[T]}, y^{\rm a}_{[I^{\rm a}],[T]}; \theta^{\rm a}_{[I^{\rm a}]}) \le 0\}.
\end{align}
where $I^{\rm a}$ is the number of DERs in the approximate model.
If $I^{\rm a} << I$, then using the constraints of the approximate model to derive $\mathcal{P}^{\rm a}$ or directly submitting $h^{\rm a}(\cdot)$ will be much less computationally intensive.
For example, for a VPP consisting of $I = 1000$ electric vehicles, we can approximate the original constraints using an approximate model with $I^{\rm a} = 2$.
It is necessary to carefully select the parameters $\theta^{\rm a}$ of the approximate model (which are the battery model parameters in this work) to ensure that $\mathcal{P}^{\rm a}$ is sufficiently close to $\mathcal{P}$ (Fig.~\ref{fig_framework_aggregation}), which is the main issue we focus on, represented as:
\begin{equation}\label{abstract_fr_approximation}
  \underset{\theta^{\rm a}}{\rm min.} J(\mathcal{P}, \mathcal{P}^{\rm a}(\theta^{\rm a}))
\end{equation}
where the loss function $J(\cdot, \cdot)$ measures the distance between $\mathcal{P}$ and $\mathcal{P}^{\rm a}$.

This problem is essentially consistent with the industrial load constraint dimension reduction problem described earlier, both approximating high-dimensional constraints with low-dimensional ones, except that the scenario setting extends from a single energy market to coupled energy-regulation markets.

\subsection{Feasible Region Aggregation Method for Distributed Resources}\label{sec_method_ev}

Having established the feasible region aggregation problem for distributed resources, this section describes how to apply the D3R framework to solve it.
Similar to the industrial load case, the key lies in selecting an appropriate approximate model form and designing an efficient parameter fitting algorithm.
The difference is that the distributed resource feasible region being aggregated here needs to simultaneously consider coupling constraints between the energy and regulation markets, and the approximate model needs to be able to characterize this coupling relationship.

\subsubsection{Feasible Region Approximate Model}

The virtual battery model describes the operation of DERs using time-coupled power and energy constraints, which can be applied to model common resources such as energy storage, electric vehicles, and thermostatically controlled loads.
We employ the virtual battery model to formulate the coupled energy-regulation constraints of the DERs.
For resource $i$, its hourly output needs to satisfy the operation constraints for power limits (\ref{model_ev_powerLimit}), energy limits (\ref{model_ev_energyLimit}), change of energy (\ref{model_ev_energyChange}), and initial energy (\ref{model_ev_energyInit}):
\begin{subequations}\label{model_energy}
  \begin{equation}\label{model_ev_powerLimit}
    0 \le p^{\rm dis(ch)}_{t, i} \le \overline{p}^{\rm dis(ch)}_{t, i} \ : \ \underline{\mu}^{\rm pd(c)}_{t, i}, \overline{\mu}^{\rm pd(c)}_{t, i}, \ \forall  t
  \end{equation}
  \begin{equation}\label{model_ev_energyLimit}
    \underline{e}_{t, i} \le e_{t, i} \le \overline{e}_{t, i} \ : \ \underline{\mu}^{\rm e}_{t, i}, \overline{\mu}^{\rm e}_{t, i},\ \forall t
  \end{equation}
  \begin{align}\label{model_ev_energyChange}
    e_{t+1, i} = & e_{t, i} +
    (\eta^{\rm ch} p^{\rm ch}_{t, i} -  \frac{1}{\eta^{\rm dis}} p^{\rm dis}_{t, i}) \Delta t \ : \ \lambda^{\rm e}_{t, i},\ \forall t
  \end{align}
  \begin{equation}\label{model_ev_energyInit}
    e_{t, i} = e_{0, i} \ : \ \lambda^{\rm e0}_{i},\ t = 1
  \end{equation}
\end{subequations}
where $p_{t, i}$ represents the baseline power of resource $i$ at time interval $t$;
$e_{t, i}$ represents the energy of resource $i$ at the end of time interval $t$;
the superscripts dis(d)/ch(c) represent discharging to/charging from the grid, respectively;
$e_{0, i}$ is the initial energy of resource $i$;
$\eta$ is the charging/discharging efficiency;
$\mu$ and $\lambda$ are the Lagrange multipliers of the corresponding constraints.

The constraints for resource $i$ providing regulation include non-negative capacity (\ref{model_reg_nonnegative}), power capacity limits (\ref{model_reg_powerLimit1}-\ref{model_reg_powerLimit2}), and maintenance time requirements (\ref{model_reg_req1}-\ref{model_reg_req2}):
\begin{subequations}\label{model_reg}
  \begin{equation}\label{model_reg_nonnegative}
    r_{t, i} \ge 0 \ : \ \underline{\mu}^{r}_{t, i}, \ \forall t
  \end{equation}
  \begin{equation}\label{model_reg_powerLimit1}
    p^{\rm dis}_{t, i} - p^{\rm ch}_{t, i} + r_{t, i} - \overline{p}^{\rm dis}_{t, i} \le 0
    \ : \ \mu^{rpd}_{t, i}, \ \forall t
  \end{equation}
  \begin{equation}\label{model_reg_powerLimit2}
    - p^{\rm dis}_{t, i} + p^{\rm ch}_{t, i} + r_{t, i} - \overline{p}^{\rm ch}_{t, i} \le 0
    \ : \ \mu^{rpc}_{t, i}, \ \forall t
  \end{equation}
  \begin{equation}\label{model_reg_req1}
    \eta^{\rm ch} (r_t - p^{\rm dis}_{t, i} + p^{\rm ch}_{t, i}) \Delta t^{\rm req} + e_{t, i} - \overline{e}_{t, i} \le 0
    \ : \ \mu^{rec}_{t, i}, \ \forall t
  \end{equation}
  \begin{equation}\label{model_reg_req2}
    \frac{1}{\eta^{\rm dis}} (r_t + p^{\rm dis}_{t, i} - p^{\rm ch}_{t, i}) \Delta t^{\rm req} - e_{t, i} + \underline{e}_{t, i} \le 0
    \ : \ \mu^{red}_{t, i}, \ \forall t
  \end{equation}
\end{subequations}
where $r_{t, i}$ represents the regulation capacity of resource $i$ at time interval $t$; $\Delta t^{\rm req}$ represents the duration required for the resource to maintain its maximum regulation output, for example, 15 minutes in the PJM market~\cite{he_optimal_2016}.
The bids of the VPP are the aggregated bids of the DERs, represented as:
\begin{equation}\label{model_vpp_aggBid}
  p_t = \sum_{i=1}^{I} (p^{\rm dis}_{t, i} - p^{\rm ch}_{t, i}), \ r_t = \sum_{i=1}^{I} r_{t, i}, \ \forall t
\end{equation}

In the above model, the parameters are $\theta = \{\overline{p}^{\rm dis(ch)}_{t, i}, \underline{e}_{t, i}, \overline{e}_{t, i}, e_{0, i}, \forall t \forall i\}$, and the internal variables are $y = \{p_{t, i}, r_{t, i}, e_{t, i}, \forall t \forall i\}$.
The constraints $h(\cdot)$ are the combination of (\ref{model_energy}) through (\ref{model_vpp_aggBid}).

Note that although we omitted the superscript a representing the approximate model above, which implies the assumption of formal consistency between the approximate model and the original model, our method does not depend on this assumption.
In the case study, we tested the generality of the proposed method in scenarios where the approximate model and the original model differ in formulations.

\subsubsection{Determining Parameters of the Approximate Model}

The parameters of the approximate model $\theta^{\rm a}$ need to be determined based on the original operational model of the VPP.
We still use data-driven inverse optimization to determine $\theta^{\rm a}$, which consists of two steps: data generation and parameter fitting.
The basic logic is that if the dispatching results given by the ISO based on $\mathcal{P}^{\rm a}(\theta^{\rm a})$ submitted by the VPP can approximate the dispatching results based on $\mathcal{P}$ under various dispatching scenarios, then $\mathcal{P}^{\rm a}$ can be considered sufficiently close to $\mathcal{P}$.

(1)~\textbf{Data generation}:
In theory, to obtain the dispatching results after submitting the original feasible region, it is necessary to model all units and loads within the scope of the ISO.
However, in practice, the capacity of a VPP is very small, so it can be modeled as a price taker.
Under the assumption of perfect competition, the dispatching scenarios can be represented by energy-regulation price scenarios~\cite{zhou_incentive-compatible_2023}.
The joint energy-regulation dispatching of the VPP is equivalent to the following optimization problem:
\begin{subequations}\label{model_bid_primal_ev}
  \begin{equation}\label{model_vpp_profit_ev}
    \underset{p_{[T]}, r_{[T]}}{\rm min.} - \sum_{t=1}^{T}(Pr^{\rm e}_{t} p_{t} + Pr^{\rm r}_{t} r_{t}
    - Pr^{\rm deg}(p_{t}))\Delta t
  \end{equation}
  \begin{equation}\label{model_vpp_constraint}
    {\rm s.t.} \ (p_{[T]}, r_{[T]}) \in \mathcal{P}
  \end{equation}
\end{subequations}
where $Pr^{\rm e}_{t}$ is the energy price;
$Pr^{\rm r}_{t} = s^{\rm perf} (Pr^{\rm cap}_{t} + Pr^{\rm mil}_{t} a^{\rm mil}_{t})$ is the equivalent regulation capacity price for the VPP;
$s^{\rm perf}$ is the performance score;
$Pr^{\rm cap(mil)}_{t}$ is the regulation capacity (mileage) price;
$a^{\rm mil}_{t}$ is the expected regulation mileage for time interval $t$.
$Pr^{\rm deg}$ is the cost function of the resources, which can be included in the bidding model along with $\mathcal{P}$, but its specific determination is not the focus of this work.
In the following, we assume that $Pr^{\rm deg}$ is directly proportional to the total discharge power of the VPP.

Given the energy-regulation price $(Pr^{\rm e}_{[T]}, Pr^{\rm r}_{[T]})$, solving (\ref{model_bid_primal_ev}) using the original VPP model can obtain the corresponding scheduling results $(p_{[T]}, r_{[T]})$, thus generating a price-dispatch dataset $D = \{(Pr^{\rm e(k)}_{[T]}, Pr^{\rm r(k)}_{[T]}, p^{(k)}_{[T]}, r^{(k)}_{[T]})\}$, where $k$ and $K$ represent the scenario index and total number of scenarios, respectively.

In practice, the original constraints $h(\cdot)$ can be directly substituted to replace (\ref{model_vpp_constraint}), as the optimal solution of the problem remains unchanged before and after projection~\cite{tan_non-iterative_2023}.

(2)~\textbf{Parameter fitting}:
Similar to the industrial load case, the generated scheduling results in dataset $D$ are treated as the optimal solutions (with some noise added) of the optimization problem (\ref{model_bid_primal_ev}) formulated by the approximate model, and $D$ is used to fit the problem parameters.
This can be represented by the following data-driven inverse optimization (DDIO) problem:
\begin{subequations}\label{model_inverse_fit_ev}
  \begin{equation}\label{model_inverse_fit_lossFunction_ev}
    \underset{\theta^{\rm a}}{\rm min.} J = \frac{1}{K} \sum_{k = 1}^{K} (
    ||p^{\rm a(k)}_{[T]} - p^{(k)}_{[T]}||^2 + ||r^{\rm a(k)}_{[T]} - r^{(k)}_{[T]}||^2
    )
  \end{equation}
  \begin{equation}\label{model_inverse_fit_constraint_1_ev}
    {\rm s.t.} \ {\rm strong \ duality:} (\ref{model_vpp_profit_ev}) = g(\mu^{(k)}, \lambda^{(k)}), \forall k.
  \end{equation}
  \begin{equation}\label{model_inverse_fit_constraint_2_ev}
    {\rm primal \ feasibility:} h^{\rm a}(p^{\rm a(k)}_{[T]}, r^{\rm a(k)}_{[T]}, y^{\rm a(k)}; \theta^{\rm a}) \le 0, \forall k.
  \end{equation}
  \begin{equation}\label{model_inverse_fit_constraint_3_ev}
    {\rm dual \ feasibility:} (\ref{model_dual_feasibility_ev}); {\rm stationarity:} (\ref{model_coefficients}), \forall k.
  \end{equation}
\end{subequations}
where $g(\mu^{(k)}, \lambda^{(k)})$ is the Lagrangian dual function of (\ref{model_bid_primal_ev});
(\ref{model_inverse_fit_constraint_1_ev}-\ref{model_inverse_fit_constraint_3_ev}) are the strong duality, primal feasibility, dual feasibility, and stationarity conditions, respectively, guaranteeing the optimality of $(p^{\rm a(k)}_{[T]}, r^{\rm a(k)}_{[T]})$.
The Lagrangian dual function and corresponding constraints are given in Section~\ref{app_dual}.

(3)~\textbf{Iterative algorithm for solving the DDIO problem}:

Determining the optimal parameters (\ref{model_inverse_fit_ev}) is a high-dimensional nonlinear optimization problem, especially challenging to solve when $K$ is large.
One approach is to update $\theta^{\rm a}$ step by step based on the structure of the problem using Newton's method~\cite{tan_data-driven_2023}, but this relies on specific forms of the optimization problem.
Instead, we adopt a commonly used method in machine learning, where we update the parameters iteratively using the data of one scenario at a time.
When updating the parameters with data of scenario $k$, the following problem is solved:
\begin{subequations}\label{model_inverse_fit_update}
  \begin{equation}\label{model_inverse_fit_update_lossFunction}
    \underset{\theta^{\rm a(k)}}{\rm min.} (
    ||p^{\rm a(k)}_{[T]} - p^{(k)}_{[T]}||^2 + ||r^{\rm a(k)}_{[T]} - r^{(k)}_{[T]}||^2
    ) + \alpha ||\theta^{\rm a(k)} - \theta^{\rm a}||^2
  \end{equation}
  \begin{equation}\label{model_inverse_fit_update_constraint}
    {\rm s.t.} (\ref{model_inverse_fit_constraint_1_ev}-\ref{model_inverse_fit_constraint_3_ev}), {\rm for} \ k.
  \end{equation}
\end{subequations}
where $\theta^{\rm a}$ is the parameter value from the previous iteration.
The last term in (\ref{model_inverse_fit_update_lossFunction}) is added to help convergence, where $\alpha$ adaptively increases with iterations.
(\ref{model_inverse_fit_update}) can be directly fed to commercial solvers.
In practice, a maximum computation time can be set for each iteration, and the obtained feasible solution when the maximum time is reached can be used for update.
This strategy can easily be implemented by commercial solvers.
The overall iterative procedure is given in Algorithm~\ref{alg_framework_aggregation}.
For initializing $\theta^{\rm a}$, we summed the original model parameters directly and distributed them equally to the approximate model in numerical tests.

\begin{algorithm}[!t]
  \caption{Iterative solution of the DDIO problem}
  \label{alg_framework_aggregation}
  \begin{algorithmic}[1]
    \renewcommand{\algorithmicrequire}{\textbf{Input:}}
    \REQUIRE
    maximum number of iterations $N$, price-dispatch dataset $\{(Pr^{\rm e(k)}_{[T]}, Pr^{\rm r(k)}_{[T]}, p^{(k)}_{[T]}, r^{(k)}_{[T]})\}$.
    \renewcommand{\algorithmicrequire}{\textbf{Output:}}
    \REQUIRE an approximate model parameterized by $\theta^{\rm a}$.
    \STATE Initialize the parameters $\theta^{\rm a}$, set $n = 0$.
    \WHILE{$n \le N$}
    \STATE Let $k = n \ {\rm mod} \ K + 1$, solve (\ref{model_inverse_fit_update}) to obtain $\theta^{\rm a(n)}$.
    \STATE Update $\theta^{\rm a}$ by the average of the last min$\{K, n\}$ $\theta^{\rm a(n)}$s
    \STATE \textbf{if} $\theta^{\rm a}$ converges, \textbf{break}.
    \STATE Set $n = n + 1$.
    \ENDWHILE
  \end{algorithmic}
\end{algorithm}

\subsection{Dual Function and Constraints of the Bidding Problem}\label{app_dual}

For completeness, this section presents the Lagrangian dual function and corresponding constraints of the bidding problem (\ref{model_bid_primal_ev}), which are needed for constructing the optimality conditions in the DDIO problem.
The superscript $k$ is omitted below.
$\forall k$, we have: Lagrangian dual function:
\begin{equation}\label{model_dual_function}
  \begin{aligned}
    g(\mu, \lambda) = & \sum_{i=1}^{I} (\sum_{t=1}^{T}(
    \underline{e}_{t, i} \underline{\mu}^{\rm e}_{t, i} - \overline{e}_{t, i} \overline{\mu}^{\rm e}_{t, i}
    - \overline{p}^{\rm dis}_{t, i} \mu^{\rm rpd}_{t, i}
    - \overline{p}^{\rm ch}_{t, i} \mu^{\rm rpc}_{t, i}            \\
                      & - \overline{e}_{t, i} \mu^{\rm rec}_{t, i}
    + \underline{e}_{t, i} \mu^{\rm red}_{t, i}) + e_{0, i} \lambda^{\rm e0}_{i})
  \end{aligned}
\end{equation}

Dual feasibility ($\forall i$):
\begin{equation}\label{model_dual_feasibility_ev}
  (\underline{\mu}_{t, i}^{\rm pd(c)}, \underline{\mu}_{t, i}^{\rm e}, \overline{\mu}_{t, i}^{\rm e}, \underline{\mu}_{t, i}^{\rm r}, \mu_{t, i}^{\rm r p d(c)}, \mu_{t, i}^{\rm r e d(c)}) \ge 0, \ \forall t, \forall i
\end{equation}

Stationarity ($\forall i$):
\begin{subequations}\label{model_coefficients}
  \begin{equation}\label{model_coefficients_pdis}
    \begin{aligned}
      Pr^{\rm deg} - Pr^{\rm e}_t - \underline{\mu}_{t, i}^{\rm p d}-\frac{\Delta t}{\eta^{\rm dis}} \lambda^{\rm e}_{t, i} + \mu_{t, i}^{\rm r p d} - \mu_{t, i}^{\rm r p c} \\
      + \frac{\Delta t^{\rm req}}{\eta^{\rm dis}} \mu_{t, i}^{\rm r e d} - \eta^{\rm ch} {\Delta t}^{\rm req} \mu_{t, i}^{\rm r e c}=0, \ \forall t
    \end{aligned}
  \end{equation}
  \begin{equation}\label{model_coefficients_pch}
    \begin{aligned}
      Pr^{\rm e}_{t} - \underline{\mu}_{t, i}^{\rm p c}+\Delta t \eta^{\rm ch} \lambda^{\rm e}_{t, i} - \mu_{t, i}^{\rm r p d} + \mu_{t, i}^{\rm r p c} \\
      - \frac{\Delta t^{\rm req}}{\eta^{\rm dis}} \mu_{t, i}^{\rm r e d} + \eta^{\rm ch} {\Delta t}^{\rm req} \mu_{t, i}^{\rm r e c} = 0, \ \forall t
    \end{aligned}
  \end{equation}
  \begin{equation}\label{model_coefficients_e0}
    \lambda_{t, i}^{\rm e} -\lambda^{\rm e0} + \mu_{t, i}^{\rm r e c} - \mu_{t, i}^{\rm r e d}=0, \ t = 1
  \end{equation}
  \begin{equation}\label{model_coefficients_e}
    -\underline{\mu}_{t, i}^{\rm e} + \overline{\mu}_{t, i}^{\rm e}+\lambda_{t, i}^{\rm e} - \lambda_{t - 1}^{\rm e}+\mu_{t, i}^{\rm r e c}-\mu_{t, i}^{\rm r e d}=0, \ 1 < t < T + 1
  \end{equation}
  \begin{equation}\label{model_coefficients_eEnd}
    -\underline{\mu}_{t, i}^{\rm e} + \overline{\mu}_{t, i}^{\rm e} - \lambda_{t - 1}^{\rm e}=0, \ t = T + 1
  \end{equation}
  \begin{equation}\label{model_coefficients_r}
    - Pr_{t}^{\rm r} - \underline{\mu}_{t, i}^{\rm r}+\mu_{t, i}^{\rm r p d}+\mu_{t, i}^{\rm r p c} + \eta^{\rm ch} {\Delta t}^{\rm req} \mu_{t, i}^{\rm r e c} + \frac{\Delta t^{\rm req}}{\eta^{\rm dis}} \mu_{t, i}^{\rm r e d}=0, \ \forall t
  \end{equation}
\end{subequations}

\subsection{Numerical Tests for Group Resource Model Aggregation}\label{sec_numerical_aggregation}

Having established the D3R framework for aggregating the joint feasible regions of distributed resources, this section uses electric vehicles (EVs) to verify that the method can be transferred to joint energy--regulation bidding. The test also examines whether D3R generalizes beyond industrial loads to other demand-side resources.

We employed Gurobi and YALMIP~\cite{Lofberg2004} to solve the optimization problems on a workstation equipped with an Intel Core i9-10900X CPU (3.7 GHz) and 128 GB of RAM.

We adopt the scenario in Reference~\inlinecite{lyu_co-optimizing_2023}, in which an aggregator coordinates 4000 EVs to provide energy and regulation services, and use historical PJM energy and regulation prices to construct the training and test datasets. Two resource-dispatch mechanisms are considered: decoupled operation~\cite{vagropoulos_optimal_2013} and coupled operation~\cite{lyu_co-optimizing_2023}. The classical outer-approximation method~\cite{xu_hierarchical_2016} is included as a benchmark.

\begin{table}[!t]
  \caption{Normalized error comparison of aggregate feasible regions}
  \label{tab_mae}
  \centering
  \begin{tabular}{cccc}
    \toprule
    \begin{tabular}[c]{@{}c@{}}DER operation\\model\end{tabular} & \begin{tabular}[c]{@{}c@{}}Outer\\Approx.~\cite{xu_hierarchical_2016}\end{tabular} & \begin{tabular}[c]{@{}c@{}}Proposed\\(I=1)\end{tabular} & \begin{tabular}[c]{@{}c@{}}Proposed\\(I=2)\end{tabular} \\ \hline
    \begin{tabular}[c]{@{}c@{}}Decoupled\\operation~\cite{vagropoulos_optimal_2013}\end{tabular} & 1.8\%                      & \textbf{1.6\%}             & 1.7\%                      \\ \hline
    \begin{tabular}[c]{@{}c@{}}Coupled\\operation~\cite{lyu_co-optimizing_2023}\end{tabular} & 13.6\%                     & 13.4\%                     & \textbf{7.5\%}             \\ \bottomrule
  \end{tabular}
\end{table}

The aggregate feasible regions are embedded in the dispatch problem and compared with the results of the original model. Table~\ref{tab_mae} reports normalized mean absolute error, defined as the mean absolute error divided by the maximum result from the original model. Under decoupled operation, all methods remain below 2\% error. When cross-resource coupling is introduced, the error of analytical outer approximation rises above 10\%, whereas D3R with $I=2$ reduces the error to 7.5\%, more than 40\% below that of outer approximation. The result shows that D3R can represent complex joint feasible regions more accurately and can provide a common modeling and bidding interface for multiple classes of distributed energy resources.

\section{Quantifying the Benefits of Industrial Load--Power System Interaction Using D3R}

D3R was originally developed to support interactions between aggregators and electricity markets, but it can also reduce model complexity in other large-scale optimization problems. This section applies D3R to industrial electricity-use constraints and embeds the reduced model in a power-system planning and operation model. A nationwide case analysis then quantifies the resulting system-level effects. Chapter 5 addresses the distinct challenges of using D3R specifically for aggregator--market interaction.

\subsection{Background and Problem}

Industrial loads are often regarded as high-quality flexibility resources because of their scale and advanced energy-management capabilities. Under conventional operating practices, however, most industrial facilities tend to operate at full capacity to maximize asset utilization and production value. They are therefore relatively insensitive to price signals, and changing their production schedules can be costly, leaving little practically available flexibility. Two structural changes are reshaping this situation. First, decarbonized power systems that rely on variable renewable energy are inherently prone to multi-timescale supply--demand mismatches, creating a growing need for both diurnal and seasonal flexibility. Second, industrial restructuring is leading to declining demand and overcapacity in energy-intensive industries such as steel, cement, and primary aluminum. By removing the need to operate at full capacity at all times, excess capacity allows these industries to align production more closely with electricity supply without reducing annual output.

The central challenge is the computational complexity of modeling and optimizing the intricate coupling between two already large and complex systems: the electricity system and industrial production. The analysis must simulate year-round power-system planning and operation at provincial nodes while retaining detailed plant-level production constraints. To address this challenge, we developed an integrated provincial-level power--smelting co-optimization model. Primary aluminum is used as the case study to quantify the value of flexible industrial electricity use under a 2050 power-sector carbon-neutrality constraint.

To solve this computationally heavy 8760-hour simulation, we adopted the D3R framework paired with an iterative decomposition algorithm. The solution process begins by embedding the linearized smelter model obtained through D3R in the electricity-system capacity-expansion model. Solving this integrated model yields an initial estimate of capacity expansion and dispatch as well as hourly locational marginal prices (LMPs) for each province. These LMPs are then fixed and passed to the detailed potline-level smelter models as exogenous price signals. Each smelter independently optimizes its 8760-hour production schedule, including discrete shutdown and restart decisions, and the resulting hourly load profile is returned to the capacity-expansion model. The power-system model is re-optimized to generate an updated set of LMPs, and the price--load feedback loop continues until the planning objective converges. This procedure bridges high-fidelity smelting physics with large-scale grid planning while making system-level scenario analysis computationally tractable. The detailed computational procedure and datasettings are reported in the corresponding published study~\inlinecite{lyu_industrial_accepted}.

\subsection{Scenario Design and Analysis Method}

The scenario analysis considers four dimensions that affect the system value of excess aluminum-smelting capacity:
\begin{enumerate}
  \item \textbf{Total demand for primary aluminum}, which determines the overall scale of primary aluminum production and hence the total amount of potential flexibility.
  \item \textbf{Potline operating flexibility}, which represents the physical minimum-load limit and the economic penalties associated with non-baseload operation, including shutdown and restart costs.
  \item \textbf{Power-system technology costs}, which determine the relative economic competitiveness of demand-side flexibility and supply-side alternatives such as battery storage and hydrogen.
  \item \textbf{Labor flexibility}, which determines whether smelters can reduce or reallocate labor costs during seasonal shutdowns.
\end{enumerate}

To ensure the robustness and traceability of the scenario settings, the bounds for each dimension were systematically derived from empirical literature and industry standards. The Mid technology-cost scenario adopts peer-reviewed trajectories from the PyPSA-China dataset, whereas the Low and High bounds are scaled by $-20\%$ and $+50\%$, respectively, following the AACE International Class 4 estimate guidelines.

The four dimensions yield 72 scenario combinations ($3\times4\times3\times2$). For each combination, we vary the retained overcapacity level in discrete steps from 0\% to 100\% of the initial excess capacity to identify the level that maximizes net system benefit. The 0\% retention level serves as the no-overcapacity baseline, in which excess capacity is fully decommissioned, leaving the sector with exactly enough capacity to meet projected annual demand under continuous, inflexible baseload operation. Parameterizing the overcapacity level across the 72 scenarios produces more than 1000 optimization results. Unless otherwise specified, the results below refer to the Mid--Mid--Mid--Inflexible core scenario in 2050.

\subsection{Numerical Results and Benefit Analysis}

\subsubsection{Projected Aluminum Demand and Excess Smelting Capacity in China}

\begin{figure}[!t]
  \centering
  \includegraphics[width=\textwidth]{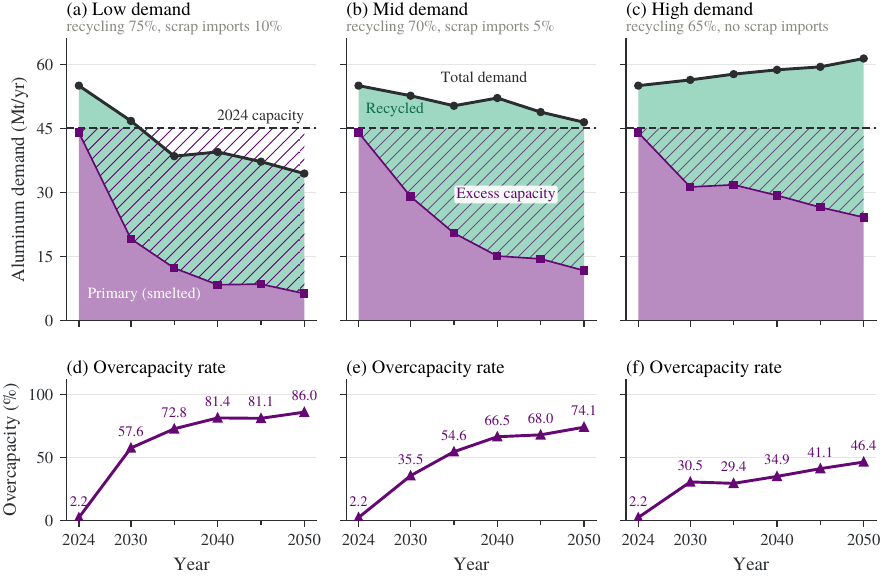}
  \caption[Projected aluminum demand and excess primary aluminum capacity in China]{Projected aluminum demand and excess primary aluminum capacity in China under low, mid, and high demand scenarios: (a)--(c) primary and recycled aluminum demand relative to 2024 smelting capacity; (d)--(f) the corresponding overcapacity rates}
  \label{fig_ne_demand}
\end{figure}

The projections indicate that China's primary aluminum demand will decline rapidly over the next one to two decades and remain low as recycled aluminum supplies a growing share of demand, creating unprecedented overcapacity in primary smelting. As shown in Fig.~\ref{fig_ne_demand}, recycled aluminum rises from approximately 25\% of total production in 2024 to 60\% by 2050. This growth reflects a delayed increase in available scrap after two decades of expanding aluminum use, because aluminum products typically remain in service for 10--20 years. Producing aluminum from scrap is much less energy intensive, so the transition reduces energy use and carbon emissions but leaves a large share of existing smelting capacity idle.

The resulting structural excess-capacity share generally exceeds 60\%. Because aluminum smelters typically operate for more than 50 years, the vast majority of China's existing smelting capacity will remain far from retirement age in 2050, locking in this structural baseline. Even under the High-demand scenario, the industry retains 46\% excess capacity, and the share reaches 86\% under the Low-demand scenario. Although such capacity is traditionally viewed as problematic and resource-wasteful, it provides the physical headroom required for seasonal demand response.

\subsubsection{Flexibility Value of Excess Capacity to the Power System}

\begin{figure}[htbp]
  \centering
  \includegraphics[width=\textwidth]{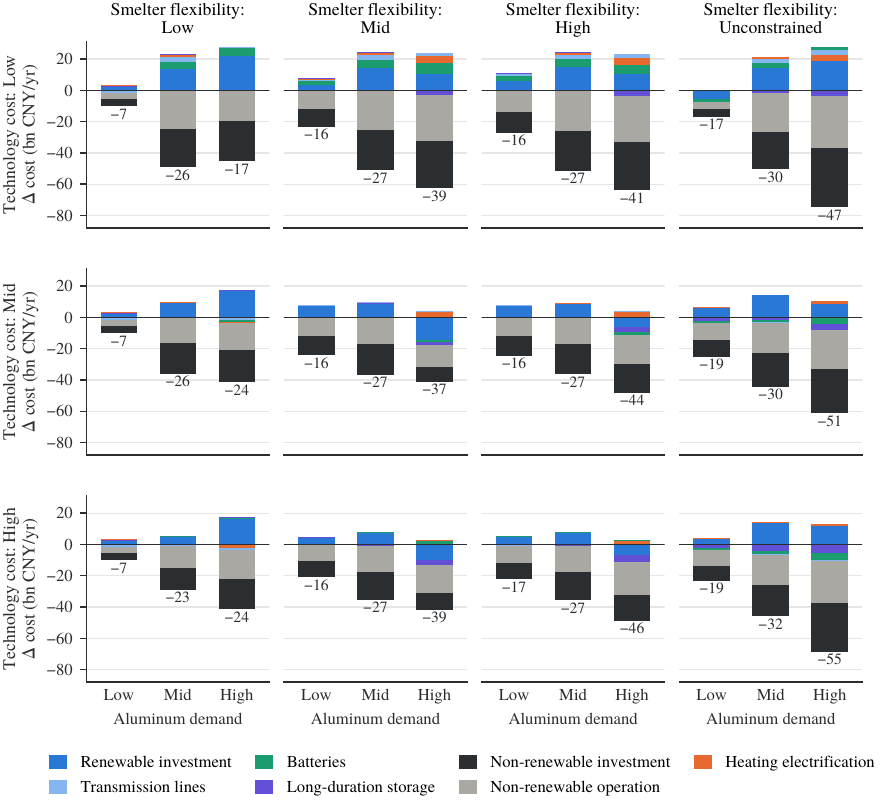}
  \caption{Components of the change in power-system cost enabled by excess capacity}
  \label{fig_ne_value}
\end{figure}

Fig.~\ref{fig_ne_value} shows the reduction in electricity-system costs in the 2050 decarbonized-power scenario when aluminum smelters with overcapacity are co-optimized with the power system, relative to the no-overcapacity case. This reduction represents the value of flexibility in smelting electricity use. Power-system cost includes operating costs and annualized capital investment in new generation, storage, and transmission capacity; smelter costs are considered separately below.

In the Mid-demand scenarios, smelter overcapacity provides a flexibility value of CNY 23--32 billion per year to the power system, equivalent to 11\%--15\% of the aluminum smelting industry's product value in 2050. Across the scenarios, the savings are driven by substantial decreases in the investment and operating costs of nonrenewable energy sources, principally coal, natural gas, and nuclear generation, reflecting a reduced reliance on firm generation resources.

Two interconnected mechanisms explain the shift toward a more renewable-dominated portfolio. First, without adequate long-duration demand-side flexibility, wind and solar expansion faces diminishing marginal benefits because variability and mismatch with electricity consumption lead to severe curtailment or the need for massive storage. Concentrating smelting electricity consumption during periods of abundant renewable generation in summer improves renewable utilization and makes greater renewable deployment economically attractive. Second, shifting smelting load to summer mitigates winter peaks, substantially reducing the need for costly clean-firm generation during periods of renewable scarcity.

\subsubsection{Cost Trade-off and the Optimal Range of Retained Excess Capacity}

\begin{figure}[!htbp]
  \centering
  \includegraphics[width=\textwidth]{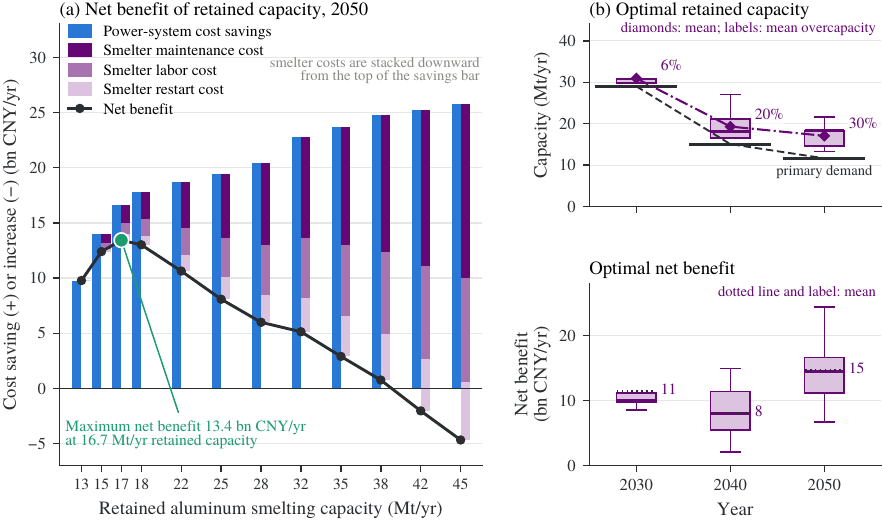}
  \caption[Trade-off analysis and cost-effective strategy for retaining overcapacity]{Trade-off analysis and cost-effective strategy for retaining overcapacity: (a) annual power-system cost savings, additional smelter costs, and net benefit versus retained smelting capacity in the 2050 core scenario; (b) distributions of optimal retained capacity and net benefit across scenarios in 2030, 2040, and 2050. Boxes span the interquartile range, center lines mark medians, and whiskers extend to the most extreme values within 1.5 times the interquartile range of the box}
  \label{fig_ne_tradeoff}
\end{figure}

While flexible aluminum smelting with overcapacity can reduce power-system costs, it also entails higher maintenance, labor, restart, and product-storage costs. As shown in Fig.~\ref{fig_ne_tradeoff}(a), power-system cost savings increase with retained capacity but plateau once smelting capacity exceeds approximately 35 million tonnes per year, corresponding to 60\% overcapacity, whereas smelter operating costs increase approximately in proportion to excess capacity. Retaining 30\% overcapacity maximizes net system benefit---the reduction in power-system cost minus the increase in smelter cost---at CNY 13 billion per year in the core scenario.

Fig.~\ref{fig_ne_tradeoff}(b) shows that the optimal overcapacity shares across the scenarios increase from 6\% in 2030 to 20\% in 2040 and 30\% in 2050. As the electricity system approaches net-zero emissions, the need for seasonal flexibility becomes more acute, increasing the value of excess industrial capacity.

\subsubsection{Seasonal Operating Pattern and Complementarity with Residual Load}

\begin{figure}[!htbp]
  \centering
  \includegraphics[width=\textwidth]{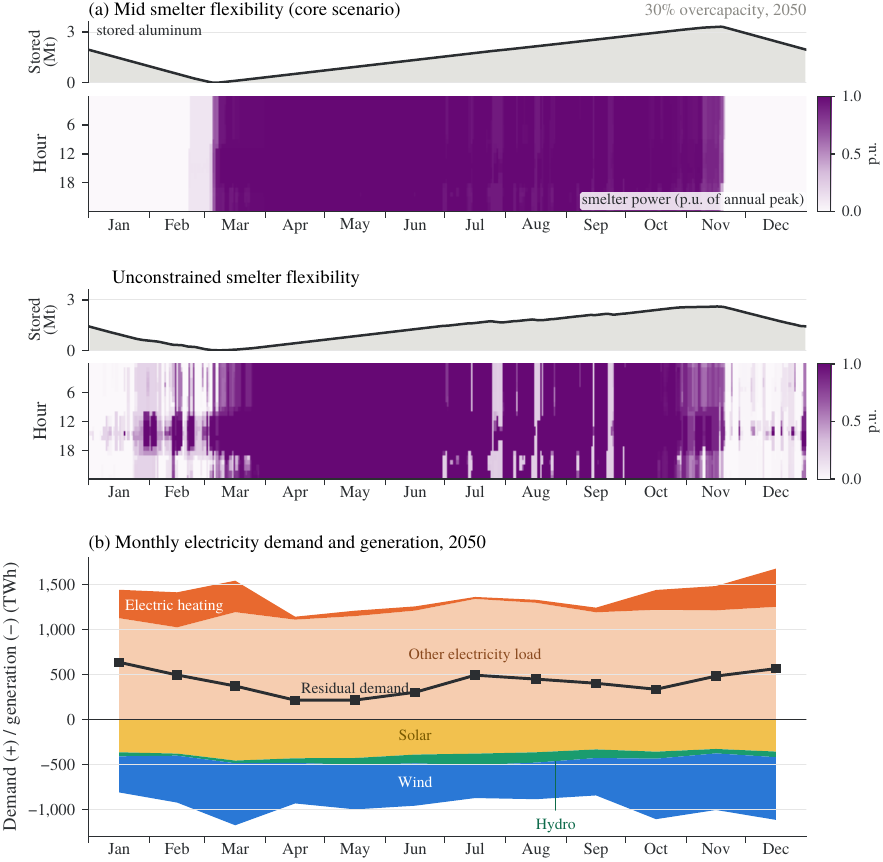}
  \caption[Seasonal operation of aluminum smelters is complementary to energy-system seasonality]{Seasonal operation of aluminum smelters and the power system in 2050: (a) hourly smelting power normalized by the annual peak (color) and aluminum inventory (line) with 30\% overcapacity under mid (core scenario) and unconstrained smelter flexibility; (b) monthly electricity demand (positive), wind, solar, and hydro generation (negative), and residual demand (demand minus these sources of generation)}
  \label{fig_ne_operation}
\end{figure}

Fig.~\ref{fig_ne_operation}(a) illustrates how renewable-energy seasonality and heating electrification alter smelter operation by encouraging production schedules that follow the seasonality of the energy system. Production largely ceases from mid-November to mid-March, with demand met by inventory, whereas smelters operate at or near rated capacity from late March to early November. Even with unconstrained smelter flexibility, this seasonal operating pattern remains dominant. Short-duration peaks can be balanced more economically by batteries or vehicle-to-grid resources, whereas high restart costs and strong thermal constraints make long-duration operation with infrequent shutdowns more suitable for potlines.

Fig.~\ref{fig_ne_operation}(b) reveals the underlying system driver: the emergence of a ``seasonal duck curve.'' Driven by high renewable penetration and heating electrification, China's residual load exhibits a distinct and prolonged winter peak. Renewable output, particularly hydropower, is lowest when heating demand peaks, forcing the system to rely more heavily on gas and coal generation to ensure supply. Aluminum smelting with overcapacity fills the need for dedicated seasonal flexibility, creating a strong economic incentive to avoid production during high-cost winter periods and rely on inventory to meet demand.

\subsubsection{Aluminum Production Cost under Different Excess-Capacity Shares}

\begin{figure}[!htbp]
  \centering
  \includegraphics[width=\textwidth]{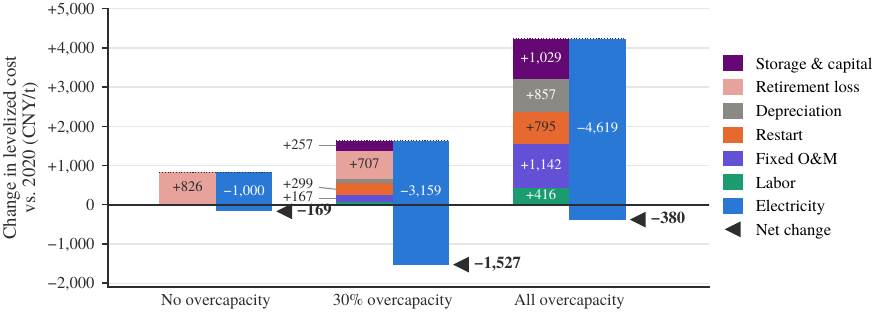}
  \caption{Aluminum production cost under different excess-capacity shares}
  \label{fig_ne_cost}
\end{figure}

This seasonal operating paradigm can also decrease the cost of producing aluminum. In the core scenario, maintaining 30\% overcapacity reduces the levelized production cost by more than CNY 1,500 per tonne, or approximately 9\%, relative to the no-overcapacity case (Fig.~\ref{fig_ne_cost}). By retaining 30\% overcapacity, smelters can use low-cost renewable electricity during high-generation periods. The resulting electricity-cost savings---approximately CNY 2,800 per tonne---are sufficient to offset the additional costs of overcapacity maintenance, depreciation, restarts, and product storage.

Maintaining a moderate level of overcapacity outperforms both complete decommissioning and full retention. If no excess capacity is retained, smelters cannot shift production to absorb low-cost clean electricity in summer. Conversely, retaining 100\% of the initial excess capacity introduces fixed maintenance and capital burdens that outweigh the benefits of seasonal flexibility. The reduced aluminum production cost, together with the power-system cost savings reported above, indicates a potential win--win outcome for the grid and aluminum producers.

\subsubsection{Cross-Industry Labor Coordination and Social Effects}

\begin{figure}[htbp]
  \centering
  \includegraphics[width=\textwidth]{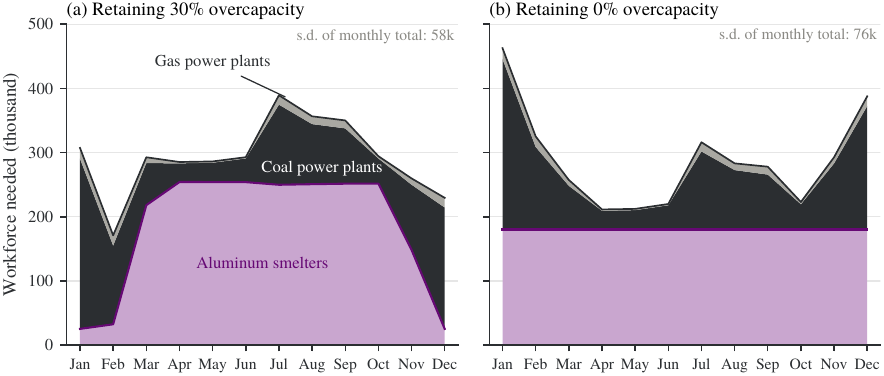}
  \caption[Complementary labor demand in aluminum smelting and thermal power generation]{Complementary labor demand in aluminum smelting and thermal power generation under seasonal operation: monthly workforce requirements in 2050 with (a) 30\% and (b) 0\% smelting overcapacity}
  \label{fig_ne_employment}
\end{figure}

As Fig.~\ref{fig_ne_employment} illustrates, seasonally operated aluminum smelters and the power sector have complementary workforce-demand patterns: coal and gas power plants require more workers in winter, whereas aluminum smelters require more workers in spring, summer, and autumn.

Because China's largest aluminum companies have historically operated captive power plants, some may be able to move workers between smelters and power plants seasonally. This workforce-management paradigm could reduce total seasonal employment fluctuation, calculated as the standard deviation of monthly employment across the two sectors, by 25\%. Such mobility would require worker training and flexible employment arrangements but could mitigate job-loss disruption associated with the energy transition and industrial restructuring.

\section{Summary}

This chapter addresses the constraint-dimension-reduction problem that arises when industrial loads participate in power-system dispatch. It proposes the data-driven dimension reduction (D3R) framework and develops a three-layer argument spanning individual industrial-load modeling, extension to multi-resource aggregation, and system-level application. Unlike analytical approaches based directly on constraint forms and parameters, D3R uses optimal solutions generated under the original complex constraints to train low-dimensional linear constraints that best approximate the original feasible behavior. This shift from analytical derivation to data-driven inverse optimization provides a new way to reduce and approximate complex constraints involving integer variables.

For individual industrial loads, the original process constraints are mapped into an adjustable-load-fleet model whose parameters are fitted through inverse optimization. Across the cement, steel-powder, and steelmaking cases, the best D3R configurations yield errors of 3.6\%--10.3\%, within or close to the cited 10\%--20\% tolerance range for demand-side participation. The original steelmaking model contains 10,208 binary variables, whereas its reduced grid-facing representation uses only 24 or 48 continuous variables.

D3R was then extended from the dimension reduction of individual industrial-load models to the aggregation of distributed-resource fleets, providing a unified interface for joint energy and regulation feasible regions. When the operation of distributed energy resources included coupling constraints, D3R reduced approximation error by more than 40\% relative to conventional outer approximation. The result shows that D3R applies not only to industrial electricity-use feasible regions but also to bidding approximations for other distributed resources with joint feasible regions.

At the system level, the analysis examines how industrial flexibility can address seasonal supply--demand mismatches in power systems with high renewable penetration. As low-cost but variable wind and solar generation becomes a primary energy source, industrial production can adapt its operating schedule to renewable-energy availability. In the aluminum-smelting case, retaining approximately 30\% overcapacity allows smelters to avoid winter peaks, reduces both power-system and aluminum-production costs, and may create complementary seasonal labor demand across industries. The case demonstrates how this advanced computational method can help evaluate the system value of industrial flexibility beyond short-term demand response.

A broader implication is that relatively simple mathematical structures can capture the principal features of complex demand-side flexibility. The adjustable-load-fleet model represents the aggregation, coupling, and time-shifting capabilities needed for system interaction. Standardized low-dimensional representations could therefore allow heterogeneous demand-side resources to report flexibility through a common market interface, much as generators with different physical characteristics submit bids using standardized forms.

In practice, the composition of aggregated resources changes dynamically as resources enter or leave. Further research should examine generalization under dynamic model updates and determine appropriate update cycles. Nonstationarity caused by extreme events, price spikes, or rule changes may shift the distribution represented by historical operating data. Under such conditions, shorter recalibration cycles, scenario-based or robust fitting objectives, and explicit checks of consistency between training and deployment conditions may be required before approximation accuracy can be extended to previously unobserved operating regimes.

Although D3R does not directly use the methods and models developed in Chapters 2 and 3, that research remains necessary.
The reason is that the dimension-reduced model obtained in this chapter is ultimately an approximation of the original feasible region: while it can satisfy the deviation assessment requirements of the electricity market, the actual execution of industrial processes must strictly satisfy their technical constraints (such as equipment start-up/shutdown sequences and state transitions), and decisions made under the reduced constraints cannot be directly used as final execution plans.
Therefore, the bidding or dispatch results obtained based on the reduced constraints from this chapter need to be restored and decomposed through the precise model, yielding executable power profiles that satisfy industrial constraints; the specific methods for this restoration and decomposition are detailed in Chapter 5.


\chapter{Optimization Decision-Making for Aggregators Coordinating Massive Resources}

\section{Overview}

\subsection{Background}
This chapter considers the optimization decision problem faced by industrial users interacting with electricity markets within a load-aggregator framework.

\begin{figure}[h]
  \centering
  \includegraphics[width=\textwidth]{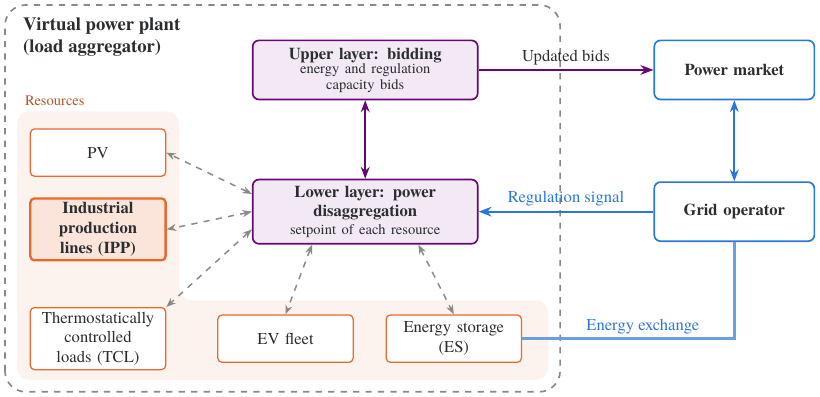}
  \caption{Operating framework of a load aggregator coordinating industrial users and other resources}
  \label{fig_problem_chap05}
\end{figure}

Specifically, this optimization involves two key decision problems: the bidding strategy and the power disaggregation strategy (Fig.~\ref{fig_problem_chap05}).
The bidding strategy refers to the regulation capacity and price that the load aggregator submits to the grid in the day-ahead or intraday market, which directly determines whether it can obtain ancillary-service opportunities and the resulting economic compensation.
The power disaggregation strategy refers to how a load aggregator including industrial users optimally disaggregates the overall regulation power requested by the grid among individual resources, so as to maximize overall revenue or minimize costs while satisfying the operational constraints of each resource.
A load aggregator may coordinate industrial production lines together with energy storage, electric vehicles, and thermostatically controlled loads. Their different operating constraints, state dynamics, and response costs make both bidding and power disaggregation difficult.

Existing research on VPP bidding and power disaggregation typically treats them as independent optimization problems solved separately, i.e., first optimizing the bidding strategy and then optimizing the power disaggregation strategy based on the obtained grid commands.
However, bidding and power disaggregation are inherently coupled and should be co-optimized.
The main reasons are as follows: first, as a result of the market operation process, bidding and power disaggregation are alternating and process-coupled, where the bids for a certain time period determine the power required by the grid to be disaggregated in that period, and the disaggregation results determine the feasible bidding range for subsequent periods by affecting resource state variables (such as energy storage state of charge, temperature, production progress, etc.). This coupling is more important in the real-time market environment, where the VPP can modify its bids based on the latest information, including price forecasts, grid command forecasts, and resource states, to increase its profits.
Second, bidding and power disaggregation optimize the VPP's profits in the long term (within the future hours) and in the short term (within the next few seconds), respectively, and profits at different time scales need to be balanced. Any approach that optimizes short-term profits without considering long-term impacts is inherently a greedy algorithm that naturally deviates from the optimal.

\subsection{Problem Description}
Specifically, there are three technical challenges in VPP power disaggregation: the uncertainty of grid commands, the temporal coupling of VPP operation, and the online control requirements for large-scale resources.
First, the grid commands for typical ancillary services such as peak shaving, frequency regulation, and reserves cannot be accurately known in advance and involve multiple possible scenarios, requiring disaggregation strategies to be formulated under uncertainty.
Second, the objective of the operational strategy of VPP should be to maximize the total profit of the industrial user over the entire time window (e.g., the next 24 hours), while state variables such as temperature, energy storage energy, and production status exhibit temporal coupling, i.e., control decisions in the current period affect the operating states and available regulation capacity in future periods, resulting in a joint optimization problem of enormous scale.
Finally, in practice, a VPP may aggregate tens of thousands of resources with different characteristics, and typical ancillary services such as secondary frequency regulation require second-level response, necessitating online control of large-scale resources with high communication and computation demands.

Due to the complexity of this problem, existing research typically adopts simplified online control strategies.
The most classic strategy is proportional allocation, which distributes the grid command to each resource in proportion to its available power adjustment capacity.
Although simple and easy to implement, this strategy clearly fails to exploit the complementary characteristics of different resources to reduce VPP operation costs and cannot achieve optimal economic performance.
If the randomness of grid commands and temporal coupling characteristics are further considered, solving the optimal control strategy exactly would face an enormous computational burden.
For instance, assuming 20 grid command scenarios, 24 future time periods, 10,000 resources, and 5 constraints per resource type, the optimization problem to be solved online (once every 2 seconds) would contain more than 20 million constraints.
Even with sufficient computational resources, the solution time for such a scale of online optimization problem would reach several minutes or longer, and the deployment cost of the algorithm would be very expensive, making it difficult to meet the real-time control requirements for ancillary service participation.

\subsection{Contributions of This Chapter}

This chapter develops a joint optimization framework for the bidding and power-disaggregation decisions of load aggregators that include industrial users. The framework extends our published work~\cite{lyu_co-optimizing_2023} to this setting.
During bidding, the dimension-reduced constraints from Chapter 4 are embedded in a stochastic multi-period, multi-scenario model. The model determines the bids and the corresponding resource allocations and records the shadow price of each resource state. During real-time disaggregation, these prices adjust the immediate dispatch cost to reflect how current decisions affect future profit. They can be reused throughout the time period, so the large stochastic problem does not need to be solved again for every regulation command. For any given regulation signal, the optimal disaggregation solution is proved to match the corresponding variables in the bidding solution.
The disaggregation stage uses the mechanism-based constraints from Chapter 2 and the parameter estimates from Chapter 3 when actual parameters are unavailable. A tailored fast algorithm recovers executable decisions from the bidding results.

The numerical tests evaluate economic performance and computation time using PJM real-time energy prices and RegD signals. Relative to proportional disaggregation, the proposed strategy increases average market income by 53.3\%, reduces operating costs by 11.4\%, and increases average VPP profit from \$254 to \$1,333 in the reported July 15--28 simulation.
In terms of computational efficiency, when the number of resources exceeds 10,000, the solution time of the original co-optimization problem (with more than 20 million constraints) exceeds several minutes, which can be solved offline in the day-ahead or intraday stage but is difficult to apply directly online.
The shadow-price-based power disaggregation model (with tens of thousands of constraints) has a solution time on the order of seconds but still requires online invocation of a solver.
The fast disaggregation algorithm (involving only algebraic operations) has a solution time on the order of milliseconds, meeting the requirements of real-time control.
The practical effect is to move the large stochastic optimization problem to the bidding stage. During regulation deployment, the stored shadow prices allow tens of thousands of resource setpoints to be updated through millisecond-level arithmetic calculations while reproducing the joint model's allocation under the stated assumptions.

\section{Overall Framework}\label{sec_framework_chap05}

To address uncertainty in grid commands, temporal coupling, and the online control of large resource portfolios, this chapter adopts ``simplified bidding, precise disaggregation.'' The dimension-reduced model is used to optimize long-term profit during bidding, while shadow prices connect that result to the precise model during real-time disaggregation. Under the assumptions stated in this chapter, the resulting allocation matches the corresponding decision in the joint optimization model.
This section first introduces the electricity market framework and the overall framework of the VPP operation strategy, clarifying the coupling relationship between bidding and power disaggregation and its impact on the economic performance of the VPP's participation in the electricity market.

\subsection{Market Framework}

This chapter focuses on secondary frequency regulation, also known as automatic generation control, which maintains the system frequency close to the nominal value by correcting frequency and power mismatches that persist after primary regulation.
Compared to primary regulation, secondary frequency regulation has a relatively slower response, typically taking place over a period of a few seconds to several minutes, making it suitable for the participation of demand-side resources.
The same formulation may be adapted to other ancillary services with uncertain deployment and fast response requirements, such as spinning reserves.
The challenges faced by VPPs in providing these ancillary services have commonalities, namely, the random nature of their calls and the need for a quick response.

\begin{figure}[!t]
  \centering
  \includegraphics[width=\textwidth]{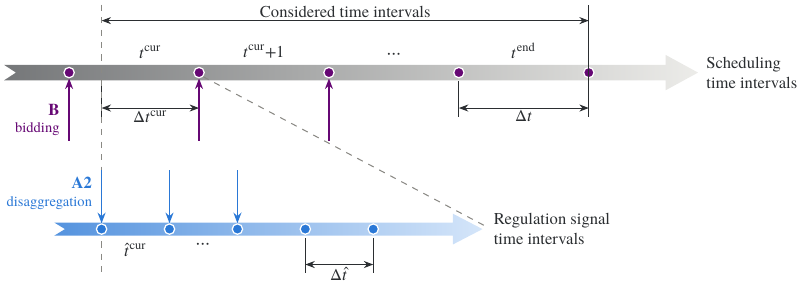}
  \caption{Timeline of virtual power plant (VPP) bidding and power disaggregation}
  \label{fig_framework_time}
\end{figure}

Existing research on VPP participation in electricity market ancillary services typically adopts simplified strategies such as proportional or uniform power allocation for bidding, failing to exploit the differentiated characteristics of heterogeneous resources to optimize regulation costs.
This chapter argues that bidding and power disaggregation strategies are inherently coupled and should be co-optimized, for the following main reasons:

(1)\textbf{Process coupling}:
As a result of the market operation process, bidding and power disaggregation are alternating and process-coupled (see Fig.~\ref{fig_framework_time}).
The bids for a certain time period determine the regulation power to be disaggregated in that period, and the disaggregation results determine the feasible bidding range for subsequent periods by affecting resource state variables (such as energy storage state of charge, temperature, production progress, etc.).
Considering this coupling is more important in the real-time market environment, where the VPP can modify its bids based on the latest information, including price forecasts, regulation signal forecasts, and resource states, to increase its profits~\cite{ortegavazquez_optimal_2014}.

(2)\textbf{Balancing profits across different time scales}:
Bidding and power disaggregation optimize the VPP's profits in the long term (within the future hours) and in the short term (within the next few seconds), respectively, and profits at different time scales need to be balanced (as shown in Fig.~\ref{fig_framework_time}).
Conceptually, any approach that optimizes short-term profits without considering long-term impacts is inherently a greedy algorithm that naturally deviates from the long-term optimal.

\subsubsection{Load Aggregator Operation and Bidding}

Because of the relatively small capacity of the resources, this chapter regards the VPP as a price taker that jointly optimizes its bids for energy (baseline power) and regulation capacity in the market, with the forecast market prices as the boundary condition.
Without loss of generality, this chapter uses a real-time market framework, in which the VPP can modify bids for future time periods, while the energy and frequency regulation capacity bids for the current period cannot be modified.
It is assumed that the VPP can conveniently obtain the cost characteristics, operating parameters, and states of its internal resources, and formulates the bidding strategy and controls their power consumption/production based on this information.
This chapter focuses on the optimal operation strategy of the VPP; the compensation or profit allocation among resource owners~\cite{chen_bargaining_2021} is beyond the scope of this chapter.

\subsubsection{Real-Time Regulation Service Process}
Before actually providing frequency regulation, the VPP receives its accepted regulation capacity $r$ after market clearing, which should be consistent with its bids since a price taker can declare a regulation capacity with a minimum price to guarantee acceptance.
When actually providing frequency regulation (known as regulation deployment), the grid operator sends regulation signals $\delta \in [-1, 1]$ to the VPP (e.g., every 2 seconds in PJM~\cite{bjm2022bpm11}).
The product of the regulation signal and the regulation capacity $\delta r$ is the to-be-adjusted output of the VPP.

\subsubsection{Uncertainty of the Regulation Signals}
Regulation signals are generated by an automatic closed-loop feedback control system.
From the perspective of the VPP, the parameters of the frequency control system are unknown, and load fluctuations are random.
Therefore, a common modeling approach is to divide the possible values of the regulation signal into discrete intervals and predict the probability of each regulation signal scenario based on historical data~\cite{vagropoulos_optimal_2013}.
This chapter uses $s \in \mathcal{S}$ to denote an arbitrary regulation signal scenario, $\mathcal{S}$ to denote the set of scenarios, $\delta_s$ to denote the representative signal value, and $\pi_s$ to denote the probability of scenario $s$.
For example, if scenario $s$ represents the case where the regulation signal $\delta$ lies in $[0.9, 1)$, then $\delta_s$ could be set to 0.95 and $\pi_s$ can be estimated based on historical data.

\subsection{Framework of the Load Aggregator Operation Strategy}

The VPP operation strategy proposed in this chapter is divided into an upper layer for bidding strategy and a lower layer for power disaggregation.
At the upper layer, the VPP maximizes its bidding profit in the market based on the operational constraints of its internal resources and the boundary conditions of the market.
In this process, the VPP must consider the uncertainty of regulation signals and what kind of disaggregation strategy to adopt when deploying frequency regulation, and estimate the corresponding operational costs (Section~\ref{sec_model}).

At the lower layer, the VPP considers how to control the output of each resource in a fast and optimized manner to respond to the regulation signals sent by the grid operator.
In this process, due to the heterogeneous characteristics of resources and the existence of state variables (such as the state of charge of energy storage), different control (power disaggregation) strategies not only affect the immediate operating costs produced but may also affect the feasible operating domain of the VPP in future periods, thereby impacting its overall revenue (Section~\ref{sec_disaggregation}).

\begin{figure}[!t]
          \centering
          \includegraphics[width=\textwidth]{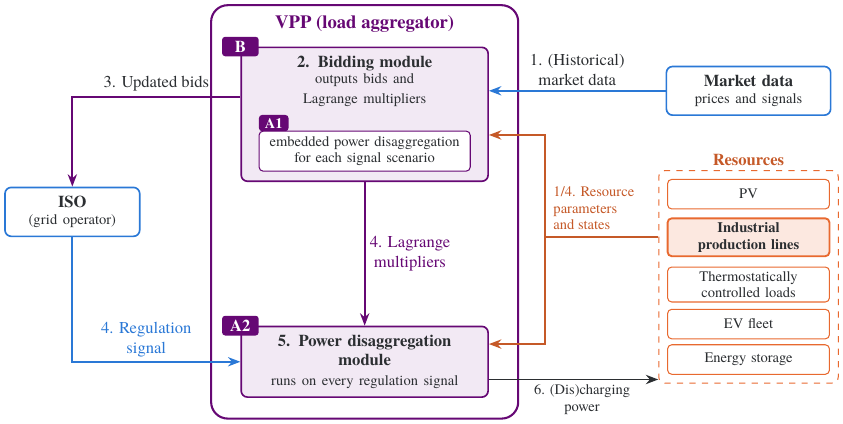}
  \caption{Joint bidding and power disaggregation framework for a virtual power plant (VPP)}
  \label{fig_overview}
\end{figure}

Specifically, the framework proposed in this chapter (Fig.~\ref{fig_overview}) includes the optimal bidding module and its embedded power disaggregation submodule.

The bidding module (B) runs periodically (e.g., every 30 minutes).
The VPP collects the operating parameters and current states of each resource (such as arrival and departure times, current state of charge, charging demand, and battery parameters of electric vehicles, or production progress of industrial production lines, etc.), and generates forecasts for prices and regulation signals based on historical market data.
Based on these parameters, the bidding module outputs the latest optimal bids, and the Lagrange multipliers can be obtained without additional computational overhead.
The power disaggregation submodule (A1) is embedded in the bidding module (B).
This chapter divides the value region of regulation signals into discretized scenarios (see Fig.~\ref{fig_signal_scenario} in the case study as a demonstration, where the granularity of the discretized regulation signal scenarios, i.e., the length of the value intervals, is 0.1), and the power disaggregation in each scenario is simultaneously optimized.

Nevertheless, the power disaggregation results in the bidding model only correspond to the discretized regulation signal scenarios (e.g., $\delta_{\hat{s}} \in$ [0, 0.1), ..., [0.9, 1),...), while the actual regulation signal may take any value in its nearly continuous range (e.g., $\delta_{\hat{s}} = 0.9998$).
Therefore, the bidding model cannot be directly used to respond to regulation signals.
To address this problem, an intuitive way is to adopt a sufficiently small granularity of the discretized regulation signal scenarios, making them close to continuous (e.g., [0, 0.0001), ..., [0.9998, 0.9999), ...).
However, the resulting large-scale problem cannot be solved online or is even computationally intractable.
Instead, this chapter designs a power disaggregation module (A2) to quickly decide the power of each resource in regulation deployment.

The power disaggregation module (A2) runs upon receiving the latest regulation signal issued by the grid operator (e.g., every 2 seconds).
Using the Lagrange multipliers from the bidding model as problem parameters, the power disaggregation problem outputs the power of each resource to follow the regulation signal.
The mathematical models of the bidding module and the power disaggregation module are introduced below.

\section{Upper Layer: Load Aggregator Market Bidding Optimization}\label{sec_model}

To achieve co-optimization of bidding and power disaggregation, this section first establishes the mathematical model for VPP bidding.
The key aspect of this model is how to embed the consideration of power disaggregation strategies within the bidding stage.
This section first introduces the standardized operation model of heterogeneous resources for providing regulation services, and then formulates the bidding optimization problem based on this model.
The bidding model accounts for the uncertainty of regulation signals, adopts a stochastic programming formulation, and embeds a simplified power disaggregation subproblem, thereby providing key parameters such as shadow prices for the precise power disaggregation at the lower layer.

\subsection{Standardized Operation Model}\label{subsec_standardizedModel}

In the following content, this chapter uses subscripts $t, s, i$ for the relevant variables regarding the time period $t$, scenario $s$, and resource $i$, respectively.
For charging, discharging, and initial states, the subscripts ${\rm ch(dis)}$ and ${\rm init}$ are used, respectively.
For example, $p^{\rm ch}_{t, s, i}$ represents the charging power of resource $i$ in time period $t$ in scenario $s$, where ``charging'' refers to drawing power from the grid.
For the denotation of parameters, underlines/overlines are used for lower/upper limits.
This chapter uses $e$ to denote the state of the resources, which can be the battery state of charge, indoor temperature, or products of industrial production lines.
$e_{t, i}$ represents the state of resource $i$ at the beginning of time period $t$; for formal consistency, let $e_{T+1, i}$ denote the state at the end of the time horizon, where $T$ is the number of time periods on the horizon.

\subsubsection{Individual Resource Operation Constraints}
The output $p_{t, s, i}$ of a single resource is the algebraic sum of the power in both directions of discharging to the grid and charging from the grid (\ref{model_std_balance2}).
This design can consider different characteristics of the two directions while keeping the model linear.
Naturally, we have $\underline{p}^{\rm dis(ch)}_{t, i} \ge 0$, and after optimization, the output in both directions will not be positive at the same time.
In general, the power limit (\ref{model_std_powerLimit}) and energy limit (\ref{model_std_energyLimit}) of the resources can be time-varying.
The expected change in state across time is given by (\ref{model_std_energyChange}), which is affected by the operation of all the associated units.
This feature can be observed in industrial production processes, where operating one process can consume the material of other processes.
Also in (\ref{model_std_energyChange}), the dissipation of heat and the effect of ambient temperature on the room temperature are expressed by introducing dissipation rate $\rho$ and environmental influence coefficient $w$, respectively.
For details of the method to obtain the parameter values, refer to~\inlinecite{liu_optimal_2021, ref_patent_lyu}.
The initial states of the resources are given by (\ref{model_std_energyInit}).
        \begin{subequations}\label{model_std}
          \begin{equation}\label{model_std_balance2}
            p_{t, s, i} =  p^{\rm dis}_{t, s, i} - p^{\rm ch}_{t, s, i}, \forall t, \forall s, \forall i
          \end{equation}
          \begin{equation}\label{model_std_powerLimit}
            \underline{p}^{\rm dis(ch)}_{t, i} \le p^{\rm dis(ch)}_{t, s, i} \le \overline{p}^{\rm dis(ch)}_{t, i}, \ \forall  t, \forall s, \forall i
          \end{equation}
          \begin{equation}\label{model_std_energyLimit}
            \underline{e}_{t, i} \le e_{t, i} \le \overline{e}_{t, i},\ \forall t, \forall i
          \end{equation}
          \begin{align}\label{model_std_energyChange}
            e_{t+1, i} = & \rho_i e_{t, i} +
            \underset{s}{\sum}\pi_{t, s} \underset{j}{\sum} (\eta^{\rm ch}_{ij} p^{\rm ch}_{t, s, j} -  \frac{1}{\eta^{\rm dis}_{ij}} p^{\rm dis}_{t, s, j}) \Delta t \\
            \nonumber    & + w_{t, i} \Delta t \ : \ \lambda^{e}_{t, i}, \ \forall t, \forall i
          \end{align}
          \begin{equation}\label{model_std_energyInit}
            e^{\rm init}_{i} - e_{t, i} = 0,\ t = 1, \forall i
          \end{equation}
        \end{subequations}

The standardized model encompasses the operating characteristics of all typical resources, and for a specific resource, its operating model can be seen as a degeneration of the standardized model:

The output of renewable energy sources is constrained by the maximum available generation, which can be expressed by (\ref{model_std_powerLimit}).
For the operation of energy storage, the power boundary parameters $\underline{p}^{\rm dis(ch)}_{t, i}$ are typically constants.
The electric vehicle operation model is similar to that of energy storage except that the plug-in state indicating whether the EV is connected to the charging pile changes over time; therefore, $\underline{p}^{\rm dis(ch)}_{t, i}$ becomes time-varying.
The characteristics of deferrable loads are similar to those of EVs, with the primary feature being that the total electricity consumption over a period must reach a minimum acceptable value.
Due to thermal inertia, the temperature of thermostatically controlled loads can be analogous to the state of charge in energy storage, thereby building a battery-like model of TCLs~\cite{liu_optimal_2021}.
The room temperature requirements may be time-varying, leading to varying $\underline{e}_{t, i}/\overline{e}_{t, i}$; for example, an office building may have a looser temperature restriction during off-hours.

Industrial production lines are modelled based on the linearized state-task network model proposed in Chapter 2~\cite{lyu_lstn_2023}, including the power limits, material buffer capacity, initial values, and production targets of the production processes.
The key difference from other resources is that the material buffer state of a production process is affected not only by its own operating power but also by the operation of related processes, because the product of one stage often serves as the raw material for the subsequent stage.
This inter-process coupling can be described by (\ref{model_std_energyChange}), where $\eta^{\rm ch}_{ij}$ represents the change in material $i$ caused by the energy consumed by production link $j$.
This relationship can be expressed in the form of an incidence matrix (see Appendix~\ref{app_cptDOFR}), where for other types of resources such as energy storage, only the diagonal elements are non-zero (i.e., $\eta^{\rm ch}_{ii}$), making it a degenerate case of the general model.

For resources that cannot meet the frequency regulation service requirements, it is assumed that their power is constant within a specific time period and does not participate in tracking regulation signals.
This only requires adding (\ref{model_nc_power}) for the corresponding resource to the bidding model and fixing its output at the reference power when deploying regulation:
        \begin{equation}\label{model_nc_power}
          p^{\rm dis(ch)}_{t, s, i} = p^{\rm dis(ch)}_{t, 0, i}, \ \forall t, \forall s
        \end{equation}

        \subsubsection{Aggregate Power Balance}
In each dispatching scenario, the aggregate power of all resources must follow the command sent by the grid (\ref{model_std_balance}).
        \begin{equation}\label{model_std_balance}
          \underset{i}{\sum} p_{t, s, i} = \tilde{p}_t + \delta_s r_t, \forall t, \forall s
        \end{equation}
where $\tilde{p}_t$ is the energy bid (baseline power) of the VPP in time period $t$ and $r_t$ is the regulation capacity of the VPP in time period $t$.
The product of the regulation signal $\delta_s$ and $r_t$ is the to-be-adjusted output of the VPP in scenario $s$.

        \subsubsection{Required Maintenance Time for Regulation}
To ensure the reliability of the regulation capacity, i.e., that the resources can continuously provide the deployed regulation under extreme conditions, the grid operator usually requires that the regulation resources can maintain maximum output for a duration of $\Delta t^{\rm req}$ (e.g., 15 minutes in PJM~\cite{masiello_business_2014}), which can be modelled by (\ref{model_std_req1}--\ref{model_std_req2}) $(\forall t, \forall i)$, where the scenarios $\delta_s = 1$ (\ref{model_std_req1}) and $\delta_s = -1$ (\ref{model_std_req2}) represent the extreme up-regulation and down-regulation cases, respectively.
For resources with energy state constraints, the stricter the maintenance time requirement, the smaller the available regulation capacity usually is.
Essentially, this constraint increases the conservatism of the capacity declaration to cope with the uncertainty of the regulation signal:
        \begin{subequations}\label{model_std_req}
          \begin{equation}\label{model_std_req1}
            (1 - \frac{(1 - \rho_i)\Delta t^{\rm req}}{\Delta t}) e_{t, i} - \underset{j}{\sum}  \frac{1}{\eta^{\rm dis}_{ij}} p^{\rm dis}_{t, s, j} \Delta t^{\rm req} + w_{t, i} \Delta t^{\rm req} \ge \underline{e}_{t, i}, \forall i, \delta_s = 1
          \end{equation}
          \begin{equation}\label{model_std_req2}
            (1 - \frac{(1 - \rho_i)\Delta t^{\rm req}}{\Delta t}) e_{t, i} + \underset{j}{\sum} \eta^{\rm ch}_{ij} p^{\rm ch}_{t, s, j} \Delta t^{\rm req} + w_{t, i} \Delta t^{\rm req} \le \overline{e}_{t, i}, \forall i, \delta_s = - 1
          \end{equation}
        \end{subequations}

        \subsubsection{Operation Cost of the Load Aggregator}
The cost function of the VPP is given in the piecewise linear form (\ref{model_std_cost}), where the cost coefficient $Pr$ can also be understood as the compensation price negotiated between the VPP and the resource owners, such as rewarding at 120\% of the actual cost, rather than the actual cost itself.
If needed, a more accurate piecewise linear approximation can also be adopted~\cite{han_practical_2014}, without interfering with the linearity of the model.
        \begin{equation}\label{model_std_cost}
          Cost_{t} = \underset{i}{\sum} \underset{s \in \mathcal{S}}{\sum}\pi_{t, s} (Pr^{\rm dis}_{i} p^{\rm dis}_{t, s, i} + Pr^{\rm ch}_{i} p^{\rm ch}_{t, s, i}) \Delta t, \ \forall t
        \end{equation}

        \subsubsection{Operation Profit of the Load Aggregator}
Over the time horizon, the expected total profit of the VPP in the market is formulated as follows:
        \begin{subequations}\label{model_vpp_profit}
          \begin{equation}\label{model_vpp_profit_total}
            Profit = \sum_{t=1}^{T}(Income^{\rm e}_{t}
            + Income^{\rm r}_{t} - Cost_{t})
          \end{equation}
          \begin{equation}\label{model_vpp_profit_e}
            Income^{\rm e}_{t} =  Pr^{\rm e}_{t} (\tilde{p}_t + \sum_{s} \pi_s  \delta_s r_t) \Delta t
          \end{equation}
          \begin{equation}\label{model_vpp_profit_r}
            Income^{\rm r}_{t} =  s^{\rm perf} r_{t} (Pr^{\rm cap}_{t} + Pr^{\rm mil}_{t} a^{\rm mil}_{t})   \Delta t
          \end{equation}
        \end{subequations}
where $Pr$ denotes the market price, with superscripts e, r, cap, and mil representing energy, regulation, regulation capacity, and regulation mileage, respectively.
In the income from the energy market (\ref{model_vpp_profit_e}), $\sum_{s} \pi_s  \delta_s r_t$ is the energy exchanged with the grid following regulation signals, which is settled at the energy price.
In the income from the regulation market (\ref{model_vpp_profit_r}), $s^{\rm perf}$ is the performance score of the VPP, which is a measure of the regulation tracking accuracy, and $a^{\rm mil}_{t}$ is the expected regulation mileage at time period $t$.
Both are treated as given parameters from the grid side.

\subsection{Compact Formulation of the Standardized Operation Model}\label{app_cptDOFR}
To facilitate readers' comprehension, Section~\ref{sec_model} presented the standardized operational model for the VPP in an element-wise form.
This section provides a compact matrix-form model (\ref{model_cpt}).
The meaning of the model is the same as before and will not be reiterated.
In the following model, $\mathbf{P}^{\rm dis(ch)}_{t} \in R^{S \times I}$, i.e., $[p^{\rm dis(ch)}_{t,s,i}]$ denotes the power matrix for the regulation signal scenarios and the resources at time $t$.
Similarly, $\mathbf{E}_{t} \in R^{I}$ denotes the state vector of the resources at time $t$.
$\mathbf{H}^{\rm dis} \in R^{I \times I}$($[\frac{1}{\eta^{\rm dis}_{ij}}]$) and $\mathbf{H}^{\rm ch} \in R^{I \times I}$($[\eta^{\rm ch}_{ij}]$) are the incidence matrices reflecting the power-state transition relationship between the $I$ resources, while $\mathbf{F}$ is the state-state transition matrix.
Underlined/overlined are the corresponding parameters for lower/upper limits.
$\mathbf{1}_{\rm I} \in R^{I}$ denotes a vector of all ones.
$\boldsymbol{\delta} = [\delta_s]$ is the vector of regulation signal values, and $\boldsymbol{\Pi}_t$ is the vector whose elements are the probability of the corresponding regulation signal scenario at time $t$.
$\mathbf{F}' = (\frac{\Delta t - \Delta t^{\rm req}}{\Delta t}\mathbf{1}_{\rm I} - \frac{\Delta t^{\rm req}}{\Delta t}\mathbf{F})$.
$\mathbf{w}_t = [w_{t, i}]$.
$\mathbf{P}^{\rm dis(ch)}_{t,(1/-1)}$ denotes the vector from the power matrix for the regulation signal scenarios of $\delta_s = 1/-1$.
\begin{subequations}\label{model_cpt}
  \begin{equation}\label{model_cpt_powerLimit}
    \underline{\mathbf{P}}^{\rm dis(ch)}_{t} \le \mathbf{P}^{\rm dis(ch)}_{t} \le \overline{\mathbf{P}}^{\rm dis(ch)}_{t}, \ \forall  t
  \end{equation}
  \begin{equation}\label{model_cpt_energyLimit}
    \underline{\mathbf{E}}_{t} \le \mathbf{E}_{t} \le \overline{\mathbf{E}}_{t},\ \forall t
  \end{equation}
  \begin{equation}\label{model_cpt_energyChange}
    \mathbf{E}_{t} = \mathbf{F} \mathbf{E}_{t - 1} +
    (\mathbf{H}^{\rm ch} \mathbf{P}^{\rm ch}_{t} -  \mathbf{H}^{\rm dis} \mathbf{P}^{\rm dis}_{t}) \boldsymbol{\Pi}_{t} \Delta t, \ \forall t
  \end{equation}
  \begin{equation}\label{model_cpt_balance}
    (\mathbf{P}^{\rm ch}_{t} - \mathbf{P}^{\rm dis}_{t})^{\top} \mathbf{1}_{\rm I} = \tilde{p}_t \mathbf{1}_{\rm S} + r_t\boldsymbol{\delta}, \forall t
  \end{equation}
  \begin{equation}\label{model_cpt_req1}
    \mathbf{F}' \mathbf{E}_t - \mathbf{H}^{\rm dis}\mathbf{P}^{\rm dis}_{t,(1)} \Delta t^{\rm req} + \mathbf{w}_t \Delta t^{\rm req} \ge \underline{\mathbf{E}}_{t}, \forall t
  \end{equation}
  \begin{equation}\label{model_cpt_req2}
    \mathbf{F}' \mathbf{E}_t + \mathbf{H}^{\rm ch}\mathbf{P}^{\rm ch}_{t,(-1)} \Delta t^{\rm req} + \mathbf{w}_t \Delta t^{\rm req} \le \overline{\mathbf{E}}_{t}, \forall t
  \end{equation}
  \begin{equation}\label{model_cpt_cost}
    cost_{t} = \mathbf{1}_{\rm I}^{\top} (Pr^{\rm dis}_{i} \mathbf{P}^{\rm dis}_{t} + Pr^{\rm ch}_{i} \mathbf{P}^{\rm ch}_{t}) \boldsymbol{\Pi}_t \Delta t, \ \forall t
  \end{equation}
\end{subequations}

\subsection{Consideration of Disaggregation Strategy in Bidding}
The VPP must model its disaggregation strategy appropriately during bidding, as different disaggregation strategies may result in different operating costs and available regulation capacities, thereby affecting the optimality of the bidding decision.
Proportional disaggregation is a commonly used strategy in the existing literature: it pre-allocates regulation capacity at the resource level for each resource and assumes that each resource's response to the regulation signal (i.e., bias relative to its baseline output $p_{t, i, 0}$) is proportional to the allocated regulation capacity $r_{t, i}$ (\ref{model_ppdis_allocation}).
The total regulation capacity of the VPP is the sum of each resource's regulation capacity (\ref{model_ppdis_rch}).
        \begin{subequations}\label{model_ppdis}
          \begin{equation}\label{model_ppdis_allocation}
            p_{t, s, i} = p_{t, 0, i} + r_{t, i} \delta_s, \forall t, \forall s, \forall i
          \end{equation}
          \begin{equation}\label{model_ppdis_rch}
            r_{t} = \underset{i}{\sum} r_{t, i}, \forall t
          \end{equation}
        \end{subequations}

From an economic perspective, proportional disaggregation fails to prioritize the use of low-cost resources to reduce response costs, which can decrease the profits of the VPP.
Moreover, in terms of implementation difficulty, the VPP must adjust the output of all resources each time it receives a regulation signal, which may bring a greater communication and control burden, and frequent changes to energy use may greatly inconvenience electricity users.
For secondary frequency regulation services, the response time of up to minutes is sufficiently long for the operation of demand-side flexible resources, such as energy storage and electric vehicles.
In this case, the regulation disaggregation strategy of the VPP can be optimized.
If the output of each resource in different regulation signal scenarios (i.e., $p_{t, s, i}$) in the operation model is directly regarded as an independent variable, then the optimal solution of the following bidding problem is equivalent to adopting the optimal disaggregation strategy:
        \begin{subequations}\label{model_bidding_opdis}
          \begin{equation}\label{model_bidding_opdis_objective}
            {\rm max.} \ Profit
          \end{equation}
          \begin{equation}\label{model_bidding_opdis_constraints}
            {\rm s.t.} \ (\ref{model_std} - \ref{model_vpp_profit})
          \end{equation}
        \end{subequations}
In the bidding model, the decision variables for bids are $\{\tilde{p}_{t}, r_{t} | t \in \mathcal{T}\}$.
The power disaggregation strategy is embodied in the decision variables $\{p_{t, s, i} | t \in \mathcal{T}, \forall i, \forall s\}$.
In practice, the bids for time periods that have passed the gate closure time (e.g., $\tilde{p}_{1}, r_{1}$) should be set to the most recently declared value, considered as problem parameters rather than decision variables.

This chapter focuses on disaggregating the total power requested from the VPP into setpoints for individual resources. Uncertainty in resource parameters, such as EV availability or owner behavior, is outside its scope~\cite{kaur_coordinated_2019}.
If specific requirements necessitate the consideration of uncertainties, the bidding model and power disaggregation model can be customized accordingly.
In practice, when the VPP comprises resources such as air conditioning in commercial buildings and industrial loads, the accompanying EV resources are typically owned by individuals working in these establishments.
The EVs considered in this chapter exhibit fleet-like characteristics, with predetermined arrival and departure times, and therefore the model neglects uncertainties related to resource parameters.

\section{Lower Layer: Power Disaggregation Optimization Among Resources}\label{sec_disaggregation}

The shadow prices determined during bidding summarize the effect of current resource use on future profit. They therefore reduce the real-time problem's temporal dimension while preserving the joint model's allocation under the stated assumptions.
This section first establishes the optimal power disaggregation problem based on shadow prices, then proves the optimality and time invariance of this disaggregation strategy, and finally exploits the problem structure to propose a fast disaggregation algorithm involving only algebraic operations, achieving millisecond-level precise control of large-scale resources.

\subsection{Optimal Disaggregation Problem}

When receiving the regulation signal $\delta_s$ at time $\hat{t}$ within time period $t$, the VPP solves an optimal power disaggregation problem to determine the output of each resource within the effective duration $\Delta \hat{t}$ of the regulation signal, until the next regulation signal is received.
This section focuses on the current time period ($t=1$) and a specific regulation signal, but the subscripts of $t$ and $s$ are retained for formal consistency.
The optimal disaggregation problem is formulated as follows:

          \begin{subequations}\label{model_opdis_objective}
            \begin{equation}\label{model_opdis_objective_total}
              {\rm min.} \ \underset{i}{\sum} (Cost_{i}^{\rm imd} + Cost_{i}^{\rm prf})
            \end{equation}
            \begin{equation}\label{model_opdis_objective_immediate}
              {\rm s.t.} \ Cost_{i}^{\rm imd} = (Pr^{\rm dis}_{i} p^{\rm dis}_{t, s, i} + Pr^{\rm ch}_{i} p^{\rm ch}_{t, s, i}) \Delta \hat{t}, \ \forall i
            \end{equation}
            \begin{equation}\label{model_opdis_objective_profit}
              Cost_{i}^{\rm prf} =   \underset{j}{\sum} \lambda^{e}_{t, j} (\eta^{\rm ch}_{ij} p^{\rm ch}_{t, s, i} -  \frac{1}{\eta^{\rm dis}_{ij}} p^{\rm dis}_{t, s, i})\Delta \hat{t}, \ \forall i
            \end{equation}
            \begin{equation}\label{model_opdis_balance}
              \underset{i}{\sum} p_{t, s, i} = \tilde{p}_t + r_t \delta_s
            \end{equation}
            \begin{equation}\label{model_opdis_disaggregation}
              \ p_{t, s, i} =  p^{\rm dis}_{t, s, i} - p^{\rm ch}_{t, s, i}, \ \forall i
            \end{equation}
            \begin{equation}\label{model_opdis_powerLimit}
              \underline{p}^{\rm dis(ch)}_{t, i} \le p^{\rm dis(ch)}_{t, s, i} \le \overline{p}^{\rm dis(ch)}_{t, i}, \ \forall i
            \end{equation}
          \end{subequations}

The objective of the VPP is to minimize the cost generated by responding to the regulation signal during $\Delta \hat{t}$ (\ref{model_opdis_objective_total}).
This cost includes the immediate operating costs $Cost_{i}^{\rm imd}$ (\ref{model_opdis_objective_immediate}) induced within $\Delta \hat{t}$, as well as the impact on the VPP's total profit over the time horizon $Cost_{i}^{\rm prf}$ (\ref{model_opdis_objective_profit}) after the states of the resources are changed (e.g., energy storage state of charge).
In (\ref{model_opdis_objective_profit}), $\lambda^{e}_{t, j}$ is the Lagrange multiplier corresponding to the state constraint (\ref{model_std_energyChange}) for resource $j$ and time $t$ at the optimum of the bidding problem.
If $Cost_{i}^{\rm prf}$ is ignored, the resulting control strategy will essentially be a myopic greedy algorithm.
This chapter will show that by incorporating $Cost_{i}^{\rm prf}$ into the objective function, the control strategy becomes globally optimal.

The constraints (\ref{model_opdis_balance}--\ref{model_opdis_powerLimit}) are basically consistent with the constraints in the bidding model regarding $t$, but the constraints related to the state variables (\ref{model_std_energyLimit}, \ref{model_std_energyChange}) are omitted because this chapter assumes the resources to be flexible (Assumption~\ref{assumption_feasible}), implying that their state constraints usually do not take effect in the middle of the time period.
On one hand, considering that the duration of a regulation signal is typically only a few seconds, it is reasonable to assume that the state of each resource will not suddenly violate the constraints after responding to a signal.
On the other hand, if the state constraint of a resource during the current period is indeed tight, or it must use electricity at a power not lower than a specific value to ensure the demand for the next period, then the parameter value of $\underline{p}^{\rm dis(ch)}_{t, i}/\overline{p}^{\rm dis(ch)}_{t, i}$ in (\ref{model_opdis_powerLimit}) can be modified while retaining the form.
For example, if the state of charge of an energy storage reaches the allowed lower limit in advance due to model accuracy issues, its discharge power upper limit $\underline{p}^{\rm dis}_{t, i}$ can be set to 0.

\subsection{Optimality of the Disaggregation Strategy}
This section analyzes the optimality of the above disaggregation strategy (\ref{model_opdis_objective}), meaning that the obtained power disaggregation result is consistent with the result given by the optimal disaggregation strategy in the bidding problem (\ref{model_bidding_opdis}).

Intuitively, at the optimum of the bidding problem, $\lambda^{e}_{t, i(j)}$ represents the shadow price of the state variable, e.g., the change in VPP operation profit caused by every 1 kWh change in energy storage state of charge.
$Cost_{i}^{\rm prf}$ given by (\ref{model_opdis_objective_profit}) as the product of $\lambda^{e}_{t, i}$ and the change in $e_{t, i}$ is exactly equal to the effect of the power disaggregation result on the VPP's profit over the time horizon.
Therefore, even though the decision variables of the disaggregation strategy (\ref{model_opdis_objective}) include only the output of each resource in the current period, it achieves global optimization of VPP profit.

The rigorous mathematical proof is based on an assumption on the flexibility of the resources and the following propositions (see Section~\ref{theorem_optimality_app} for the proofs):
        \begin{assumption}\label{assumption_feasible}
          The bidding problem (\ref{model_bidding_opdis}) and the power disaggregation problem (\ref{model_opdis_objective}) are strictly feasible.
        \end{assumption}
        \begin{proposition}\label{theorem_optimality}
          \textbf{Optimality.} $\forall s \in \mathcal{S}$, the optimal solution to the power disaggregation problem (\ref{model_opdis_objective}), i.e., $\{p_{t, s, i}, \forall i\}$ corresponds to the optimal solution to the bidding problem (\ref{model_bidding_opdis}).
        \end{proposition}
        \begin{proposition}\label{theorem_invariance}
          \textbf{Time invariance.} If the occurrence frequency of the regulation signals is consistent with the forecasted probability and the VPP controls the output of the resources according to the optimal solution of the bidding problem from $\hat{t}$ to $\hat{t}'$ $(\hat{t}' > \hat{t})$ within the current time period, then the optimal solution to the bidding problem remains unchanged.
        \end{proposition}
According to Propositions~\ref{theorem_optimality} and~\ref{theorem_invariance}, the proposed optimal disaggregation strategy can optimize the net profit of the VPP over the time horizon, and this optimality is time-invariant.
Theoretically, the bidding problem needs to be solved only once before entering the current period to obtain $\lambda^{e}_{t, i}$, and the optimal disaggregation problem can be invoked after receiving a regulation signal.
If the occurrence of the regulation signals deviates significantly from the forecast, the bidding problem can be solved again to update $\lambda^{e}_{t, i}$.

\subsection{Proof of Proposition~\ref{theorem_optimality}}\label{theorem_optimality_app}

The idea is to verify that the optimality conditions of the bidding problem can imply the optimality conditions of the optimal disaggregation problem.
For the sake of expression, this chapter uses (\ref{model_std_balance2}) to eliminate $p_{t,i,s}$.
Following convention, the objective function is written as min. and the inequalities as $\le 0$, despite their expression in the aforementioned format.
First, the parts where $\{p^{\rm ch(dis)}_{t,i,s} \ (t=1)\}$ appear in the KKT conditions of the bidding problem are written out:

a) Primal feasibility: (\ref{model_std_powerLimit}) and (\ref{model_std_balance}).
The dual variables of the two constraints are denoted as $\underline{\mu^{\rm dis(ch)}_{t, s, i}} / \overline{\mu^{\rm dis(ch)}_{t, s, i}}$ and $\lambda^{\rm bal}_{t,s}$, respectively.

b) Dual feasibility: $\underline{\mu^{\rm dis(ch)}_{t, s, i}} \ge 0, \overline{\mu^{\rm dis(ch)}_{t, s, i}} \ge 0, \ \forall i$.

c) Complementary slackness: $\underline{\mu^{\rm dis(ch)}_{t, s, i}} (\underline{p}^{\rm dis(ch)}_{t, i}- p^{\rm dis(ch)}_{t, s, i}) = 0$, $\overline{\mu^{\rm dis(ch)}_{t, s, i}} (p^{\rm dis(ch)}_{t, s, i} - \overline{p}^{\rm dis(ch)}_{t, i}) = 0, \ \forall i$

d) Stationarity: ${\partial L}/{\partial p^{\rm dis(ch)}_{t, s, i}} = 0, \ \forall i$, where $L$ is the Lagrangian function of the bidding problem. We have:
        $$\lambda^{\rm bal}_{t,s} - \underline{\mu^{\rm ch}_{t, s, i}} + \overline{\mu^{\rm ch}_{t, s, i}} +  \pi_{t, s}  (Pr^{\rm ch}_{i} + \sum_{j} \lambda^{\rm e}_{t, j} \eta^{\rm ch}_{ij} )\Delta t = 0, \forall i$$
        $$- \lambda^{\rm bal}_{t,s} - \underline{\mu^{\rm dis}_{t, s, i}} + \overline{\mu^{\rm dis}_{t, s, i}} +  \pi_{t, s}  (Pr^{\rm dis}_{i} - \sum_{j} \lambda^{\rm e}_{t, j} \frac{1}{\eta^{\rm dis}_{ij}} )\Delta t = 0, \forall i$$
        
The KKT conditions of the optimal disaggregation problem are as follows:
        
a) Primal feasibility: (\ref{model_opdis_powerLimit}) and (\ref{model_opdis_balance}).
The dual variables are denoted as $\underline{\mu^{\rm dis(ch)}_{t, i}} / \overline{\mu^{\rm dis(ch)}_{t, i}}$ and $\lambda^{\rm bal}_{t, s}$, respectively.
        
b) Dual feasibility and c) Complementary slackness are in the same form as in the bidding problem.
        
d) Stationarity: ${\partial L}/{\partial p^{\rm dis(ch)}_{t, i}} = 0, \ \forall i$. We have:
        $$\lambda^{\rm bal}_{t, s} - \underline{\mu^{\rm ch}_{t, i}} + \overline{\mu^{\rm ch}_{t, i}} + (Pr^{\rm ch}_{i} + \sum_{j} \lambda^{\rm e}_{t, j} \eta^{\rm ch}_{ij} )\Delta \hat{t} = 0, \forall i$$
        $$- \lambda^{\rm bal}_{t, s} - \underline{\mu^{\rm dis}_{t, i}} + \overline{\mu^{\rm dis}_{t, i}} + (Pr^{\rm dis}_{i} - \sum_{j} \lambda^{\rm e}_{t, j} \frac{1}{\eta^{\rm dis}_{ij}} )\Delta \hat{t} = 0, \forall i$$
        
At the optimum of the strictly feasible (Assumption~\ref{assumption_feasible}) bidding problem, $\exists$ $p^{\rm ch(dis)*}_{t,i,s}$, $\lambda^{\rm e*}_{t, i}$, $\underline{\mu^{\rm dis(ch)*}_{t, s, i}}$, $\overline{\mu^{\rm dis(ch)*}_{t, s, i}}$, and $\lambda^{\rm bal*}_{t, s}$, such that the above KKT conditions are satisfied.
Then, it can be verified that by setting $p^{\rm ch(dis)}_{t,i,s} = p^{\rm ch(dis)*}_{t,i,s}$, and $(\underline{\mu^{\rm dis(ch)}_{t, s, i}}, \overline{\mu^{\rm dis(ch)}_{t, s, i}}, \lambda^{\rm bal}_{t, s})=
\Delta \hat{t} / (\Delta t^{\rm cur} \pi_{t^{\rm cur}, s}) (\underline{\mu^{\rm dis(ch)*}_{t, s, i}}, \overline{\mu^{\rm dis(ch)*}_{t, s, i}}, \lambda^{\rm bal*}_{t, s})$, the KKT conditions of the optimal disaggregation problem are also satisfied.
Since both problems are convex, the optimal value is unique, and the proposition is proved.
        
        \subsection{Proof of Proposition~\ref{theorem_invariance}}\label{theorem_invariance_app}
        
The idea is to verify that the optimal solution to the bidding problem Bid($\hat{t}$) also satisfies the optimality conditions for the bidding problem Bid($\hat{t}'$), where only the differences in KKT conditions regarding (\ref{model_std_energyChange}), (\ref{model_std_energyInit}), and the objective function for the current time period need to be investigated.
        
Denote the optimal solution for Bid($\hat{t}$) as $\{p^{\rm dis(ch)*}_{t, s, i}\}$, $\{\tilde{p}^{*}_{t}, r^{*}_{t}\}$, and $\{e^{*}_{t, i}\}$.
From $\hat{t}$ to $\hat{t}'$, due to responding to regulation signals, the states of the resources change by:

          \begin{align}\label{app_energyChange_1}
            \Delta e_{i} =
            \underset{s}{\sum}\pi_{t, s} \underset{j}{\sum} (\eta^{\rm ch}_{ij} p^{\rm ch*}_{t, s, j} -  \frac{1}{\eta^{\rm dis}_{ij}} p^{\rm dis*}_{t, s, j}) \Delta t', \ \forall i
          \end{align}

where $\Delta t'$, the remaining time of the current period, is given by $\Delta t - (\hat{t}' - \hat{t})$.
After substituting $\{p^{\rm dis(ch)*}_{t, s, i}\}$, $\Delta t'$, and (\ref{model_std_energyInit}), constraint (\ref{model_std_energyChange}) at $\hat{t}'$ is replaced by:
          \begin{align}\label{app_energyChange_2}
            e_{t+1, i} = & \rho_i e^{init}_{i} +
            \underset{s}{\sum}\pi_{t, s} \underset{j}{\sum} (\eta^{\rm ch}_{ij} p^{\rm ch*}_{t, s, j} -  \frac{1}{\eta^{\rm dis}_{ij}} p^{\rm dis*}_{t, s, j}) \Delta t \\
            \nonumber    & + w_i \omega_t \ : \ \lambda^{e}_{t, i}, \ \forall t, \forall i
          \end{align}
which is identical to (\ref{model_std_energyChange}) at $\hat{t}$ and therefore satisfied.
The same transformation can be applied to the relevant terms in the objective function.
Therefore, the primal constraints of Bid($\hat{t}'$) are satisfied.

For the KKT conditions with dual variables, take those related to $\{p^{\rm ch(dis)}_{t,i,s}\}$ as an example.
As presented in the proof of Proposition~\ref{theorem_optimality}, in the KKT conditions (a--d), the differences in Bid($\hat{t}$) and Bid($\hat{t}'$) lie only in the parameter $\Delta \hat{t}'$.
The ratio $\Delta \hat{t}' / \Delta \hat{t}$ is multiplied by the dual variables of Bid($\hat{t}$), and these dual variables satisfy the KKT conditions of Bid($\hat{t}'$).
Similarly, other dual constraints of Bid($\hat{t}'$) can be satisfied.
Therefore, the optimal solution to Bid($\hat{t}$) also satisfies the optimality conditions for Bid($\hat{t}'$).
The proof is complete.

\subsection{Fast Disaggregation Algorithm}
The above optimal disaggregation problem involves constraints and variables of only the current time period and the specific regulation signal, so the problem size is much smaller than that of the bidding problem.
Nevertheless, its solution as a linear program generally requires the use of a commercial solver, and the solver needs to be invoked frequently and online in response to regulation signals, placing high demands on hardware.
This section further exploits the structural properties of this problem to propose a fast algorithm involving only algebraic operations, reducing the online computational complexity.

For the sake of clarity, variable substitution is first performed.
The original power segments are re-indexed with $k$, with $p_k$ denoting the output of an arbitrary power segment, taking injection into the grid as positive.
For a discharge segment, $p_k(i) := p^{\rm dis}_{t, s, i}$.
For a charging segment, $p_k(i) := - p^{\rm ch}_{t, s, i}$.
For resources with piecewise linear costs, each segment can be represented separately.
Correspondingly, the equivalent cost coefficients for the discharging and charging power segments are rewritten as:
          \begin{subequations}\label{model_substitution}
            \begin{equation}\label{model_substitution_dis}
              c_{k}(i) := Pr^{\rm dis}_{i} - \sum_j \lambda^{e}_{t, j} (\frac{1}{\eta^{\rm dis}_{ij}}),
            \end{equation}
            \begin{equation}\label{model_substitution_ch}
              c_{k}(i) := - Pr^{\rm ch}_{i} - \sum_j \lambda^{e}_{t, j} \eta^{\rm ch}_{ij},
            \end{equation}
          \end{subequations}
After substituting $p_k$ and $c_k$, the optimal disaggregation problem (\ref{model_opdis_objective}) can be rewritten as:
          \begin{subequations}\label{model_alg}
            \begin{equation}\label{model_alg_objective}
              {\rm min.} \ \sum_k c_{k} p_{k}
            \end{equation}
            \begin{equation}\label{model_alg_balance}
              {\rm s.t.} \ \sum_k p_{k} = p^{\rm req}
            \end{equation}
            \begin{equation}\label{model_alg_powerLimit}
              \underline{p}_{k} \le p_{k} \le \overline{p}_{k}, \ \forall k
            \end{equation}
          \end{subequations}
where $p^{\rm req} := \tilde{p} + \delta_s r_t$ is the required net output of the VPP, and $\underline{p}_{k} / \overline{p}_{k}$ is the lower/upper limit of power segment $k$, which can be obtained by substituting $\underline{p}^{\rm dis}_{t, i} / \overline{p}^{\rm dis}_{t, i}$ and $- \overline{p}^{\rm ch}_{t, i} / -\underline{p}^{\rm ch}_{t, i}$, respectively.

\begin{figure}[!t]
  \centering
  \includegraphics[width=\textwidth]{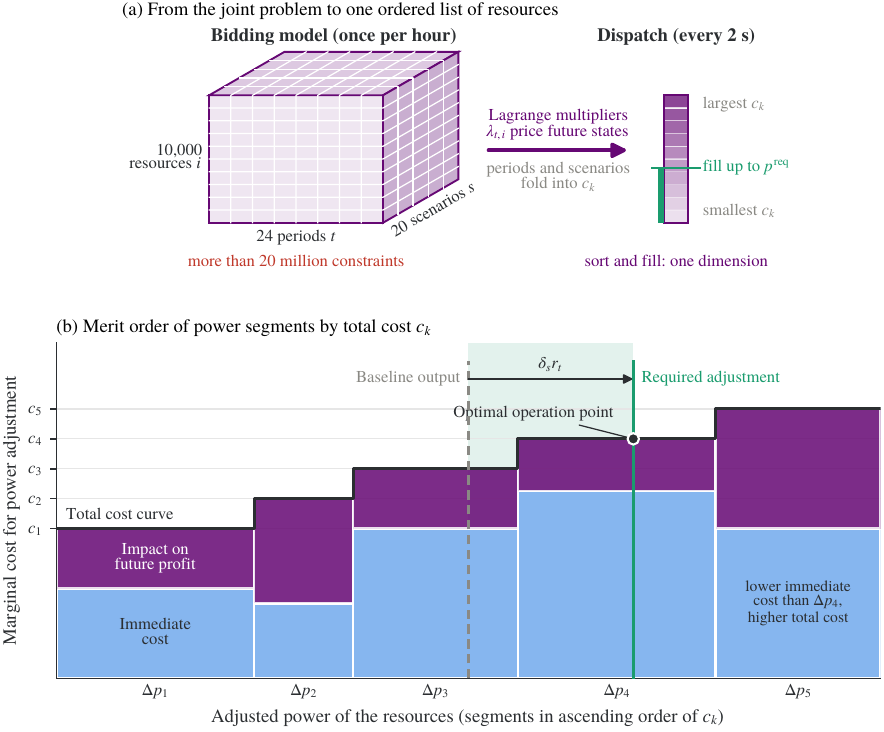}
  \caption[Schematic of fast allocation of regulation power among resources]{Schematic of fast allocation of regulation power among resources: (a) Lagrange multipliers from the bidding model incorporate the value of future resource states into the dispatch cost coefficients; (b) allocation among power segments in ascending order of total cost $c_k$ until the required output $p^{\rm req}$ is met}
  \label{fig_fast_disaggregation}
\end{figure}

The structure of the problem is now more apparent, namely, its inequality constraints (\ref{model_alg_powerLimit}) and cost function (\ref{model_alg_objective}) of the power segments are decoupled, with the variables only coupled through the equality constraint (\ref{model_alg_balance}).
Therefore, a straightforward solution approach is to start by adjusting the power from the segment with the smallest cost coefficient until the total power reaches $p^{\rm req}$ (Fig.~\ref{fig_fast_disaggregation}).
Based on the substituted variables, this chapter implements this process in Algorithm~\ref{alg_control}.

\begin{algorithm}[!t]
  \caption{Fast disaggregation algorithm}
  \label{alg_control}
  \begin{algorithmic}[1]
    \renewcommand{\algorithmicrequire}{\textbf{Input:}}
    \REQUIRE
    Required power $p^{\rm req}$, power segment parameters $\{c_k, \underline{p}_k, \overline{p}_k\}$ (indexed in ascending order of $c_k$), number of power segments $K$
    \renewcommand{\algorithmicrequire}{\textbf{Output:}}
    \REQUIRE Updated output of the resources $\{p_{t, s, i}\}$.
    \renewcommand{\algorithmicensure}{\textbf{Initialisation:}}
    \ENSURE $p_k \leftarrow \underline{p}_k, \forall k$, $\Delta p \leftarrow p^{\rm req} - \sum_k p_k$, $k \leftarrow 1$.
    \WHILE{$\Delta p > 0$ and $k \le K$}
    \IF{$\overline{p}_k - p_k < \Delta p$}
    \STATE $p_k \leftarrow \overline{p}_k$, $\Delta p \leftarrow \Delta p - \overline{p}_k + \underline{p}_k$, $k \leftarrow k + 1$.
    \ELSE
    \STATE $p_k \leftarrow \underline{p}_k + \Delta p$, $\Delta p \leftarrow 0$.
    \ENDIF
    \ENDWHILE
    \STATE Recover $\{p_{t, s, i}\}$ through $\{p_k\}$ and (\ref{model_opdis_disaggregation}).
  \end{algorithmic}
\end{algorithm}

For the input of Algorithm~\ref{alg_control}, the operations involved are parameter substitution based on the optimal solution of the bidding problem and sorting of each resource parameter (with computational complexity O(log(K))), which needs to be calculated only once before entering the current period.
After receiving a regulation signal, the complexity of the fast disaggregation algorithm is linear, with only O(K), where K is the number of power segments.
In actual operation, the output of each power segment from the last response can be used as the initial value, rather than re-initializing each time, further reducing the computational overhead.
When the rate of change of the regulation signal is small relative to the width of the power segments, only the output of the marginal resource needs to be updated while other resources operate at the upper/lower power boundaries, which considerably simplifies the implementation of real-time control.

In cases where the actual distribution of the regulation signal deviates significantly from the expected values, resulting in the VPP's inability to fully track the regulation signal, such as when an excessively high up-regulation is required, Algorithm~\ref{alg_control} adjusts the output of all resources (considering injection into the grid) to their maximum and terminates at $k=K$.
This approach aligns with the existing literature, which relaxes the constraint of tracking the regulation signal by introducing a deviation variable and minimizes the deviation using the Big-M method~\cite{lyu_co-optimizing_2023}.
In practice, due to the limitations imposed on maintenance time (\ref{model_std_req}), the reported regulation capacity by the VPP exhibits a certain level of conservatism, thereby minimizing the probability of such scenarios occurring.

\section{Case Study}\label{sec_case_study_chap05}

\begin{figure}[!t]
  \centering
  \includegraphics[width=\textwidth]{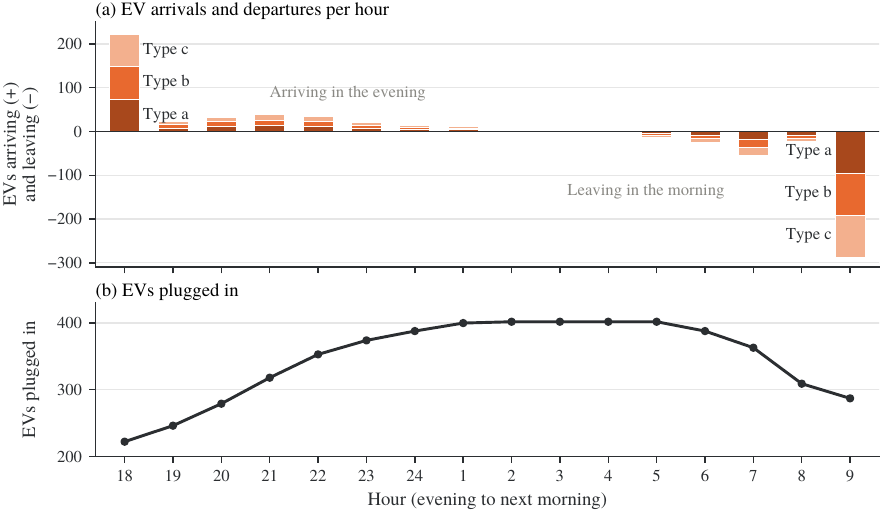}
  \caption[Electric vehicle (EV) arrivals, departures and plugged-in EVs on July 21]{Electric vehicle (EV) arrivals and departures on July 21: (a) arrival and departure times by type; (b) number of plugged-in EVs}
  \label{fig_ev_arrive_leave_chap05}
\end{figure}

\begin{figure}[!t]
  \centering
  \includegraphics[width=\textwidth]{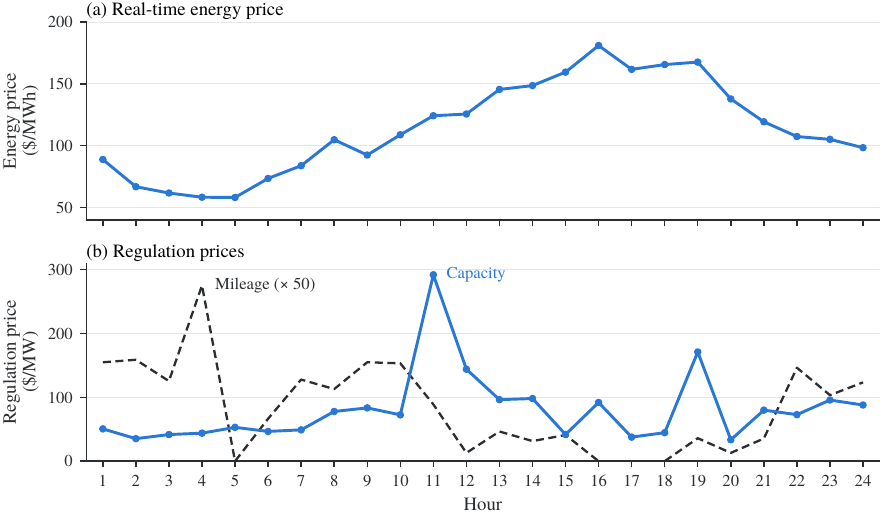}
  \caption{Real-time energy and regulation market prices on July 21}
  \label{fig_price}
\end{figure}

This section demonstrates the effectiveness of the proposed method in enhancing the profits of a VPP providing regulation services through case studies.
The VPP consists of photovoltaic (PV), energy storage (ES), a fleet of electric vehicles (EVs; arrival and departure patterns in Fig.~\ref{fig_ev_arrive_leave_chap05}), thermostatically controlled loads (TCLs), and industrial production line (IPP) resources, with a PV installation capacity of 2.5 MW and ES capacity of 1 MW/2 MWh.
The parameters for the other resources were taken from~\inlinecite{chen_scheduling_2021, ref22, li_emission-concerned_2013}, with the original data available in~\inlinecite{lyu_test_data_2023}.
This chapter uses the system real-time energy and frequency regulation prices of PJM in July 2022 as boundary conditions (Fig.~\ref{fig_price}).
Other market parameters include the length of (scheduling) time period $\Delta t = 1$ hour, the required maintenance time $\Delta t^{\rm req} = 0.5$ hours, and the performance score $s^{\rm perf} = 0.984$ given by example 1 provided by PJM in~\cite{example_pjm}.
The historical regulation signals (RegD) published by PJM are used for the simulation.

\begin{figure}[!t]
  \centering
  \includegraphics[width=\textwidth]{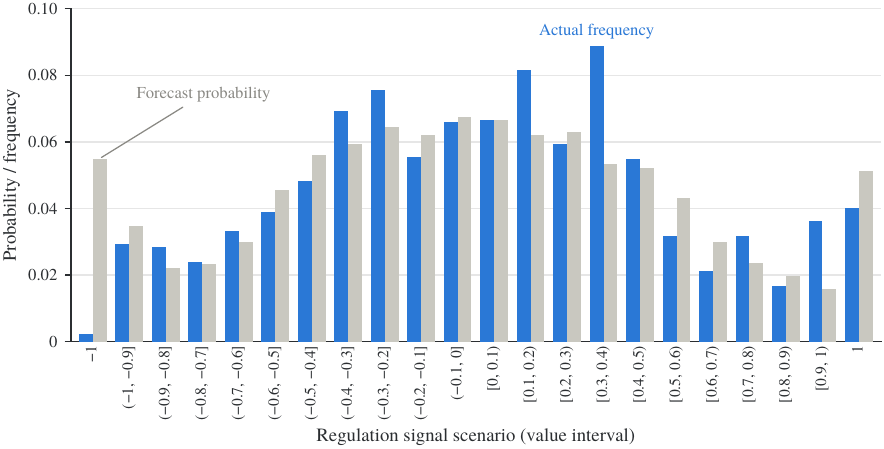}
  \caption{Observed and forecast regulation-signal distributions from 1:00 to 2:00 on July 21}
  \label{fig_signal_scenario}
\end{figure}

\begin{figure}[!t]
  \centering
  \includegraphics[width=\textwidth]{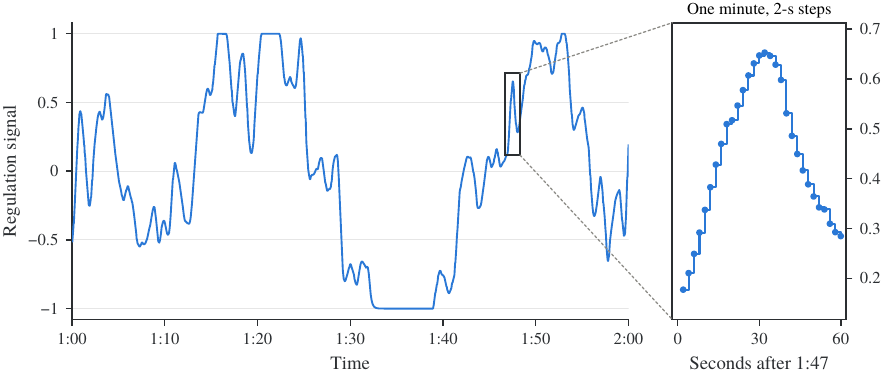}
  \caption[Regulation signal from 1:00 to 2:00 on July 21]{Regulation signal from 1:00 to 2:00 on July 21, with a one-minute detail at two-second resolution on the right}
  \label{fig_signal_time}
\end{figure}

This chapter does not assume that the regulation signals can be accurately predicted; it is necessary to estimate the probability of each signal scenario based on historical data.
For simplicity and easy reproduction, the nearly continuously distributed regulation values in the range $[-1, 1]$ are divided into 22 scenarios (signal value ranges): $\{-1, (-1, -0.9], …, 1\}$, and the frequency of the same time period over the past 14 days is used to estimate the probability of the corresponding scenario.
The regulation mileage is estimated in the same way.
In the bidding model, the median of the range is used to represent the signal value of the corresponding scenario, for example, $\delta_s= - 0.95$ for $s = (-1, -0.9]$.
Note that the actual regulation signal is used to simulate the VPP's response, which may deviate from the predicted distribution (Fig.~\ref{fig_signal_scenario} and Fig.~\ref{fig_signal_time}).
To cope with this, the VPP updates its bids for future periods every half hour based on the latest resource states.


\subsection{Results and Comparisons}

  \begin{table}[!t]
    \caption{Comparison of VPP profits using different methods}
    \label{tab_profit_method}
    \centering
    \begin{tabular}{cccc}
      \toprule
      \multicolumn{1}{c}{\multirow{2}{*}{\begin{tabular}[c]{@{}c@{}}Bidding-Disaggregation \\      Method\end{tabular}}} &
      \multicolumn{1}{c}{\begin{tabular}[c]{@{}c@{}}Income from\\      the Market\end{tabular}}                  &
      \multicolumn{1}{c}{\begin{tabular}[c]{@{}c@{}}Operational\\      Cost\end{tabular}}                  &
      \multicolumn{1}{c}{\begin{tabular}[c]{@{}c@{}}VPP \\      Profit\end{tabular}}                                                                                                   \\
      \multicolumn{1}{c}{}                                            & \multicolumn{1}{c}{(\$)} & \multicolumn{1}{c}{(\$)} & \multicolumn{1}{c}{(\$)} \\ \midrule
      \multirow{2}{*}{Proportional Disaggregation}                    & {1709}                   & {1455}                   & {254}                    \\
                                                                      & (0.0\%)                  & (0.0\%)                  & (0.0)                    \\ \cline{2-4}
      \multirow{2}{*}{Greedy Disaggregation}                          & 1316                     & 1845                     & -529                     \\
                                                                      & (-23.0\%)                & (+26.8\%)                & (-782)                   \\\cline{2-4}
      \multirow{2}{*}{\textbf{Optimal Disaggregation}}                & \textbf{2622}            & \textbf{1289}            & \textbf{1333}            \\
                                                                      & \textbf{(+53.3\%)}       & \textbf{(-11.4\%)}       & \textbf{(+1079)}         \\ \bottomrule
    \end{tabular}
  \end{table}

\begin{figure}[!t]
  \centering
  \includegraphics[width=\textwidth]{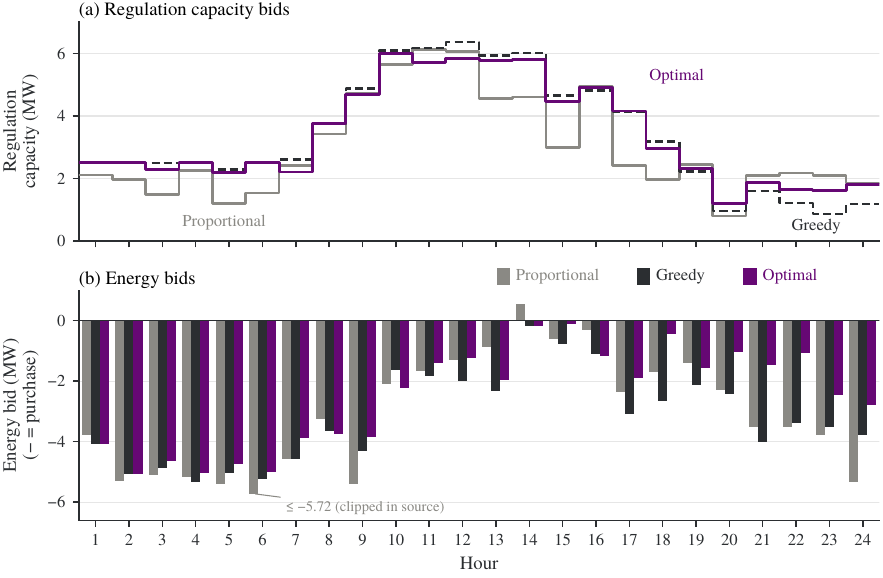}
  \caption{Virtual power plant (VPP) bids on July 21: (a) regulation capacity; (b) energy}
  \label{fig_bids}
\end{figure}

Under the above settings, three different bidding strategies for providing regulation services are run and the actual responses are simulated, calculating the costs and revenues of the VPP, with the average results over July 15--July 28 summarized in Table~\ref{tab_profit_method} and the bidding results presented in Fig.~\ref{fig_bids}.
Here, \textbf{Optimal Disaggregation} refers to the proposed method, i.e., using (\ref{model_bidding_opdis}) to decide the VPP's bids and running Algorithm~\ref{alg_control} to respond to regulation signals.
The difference between \textbf{Greedy Disaggregation} and Optimal Disaggregation is that $\lambda^e_{t, j}$ in (\ref{model_substitution}) is set to 0, so the strategy aims to minimize the immediate response cost.
\textbf{Proportional Disaggregation} is a commonly used strategy in the existing literature~\cite{ref34}.

\begin{figure}[!t]
  \centering
  \includegraphics[width=\textwidth]{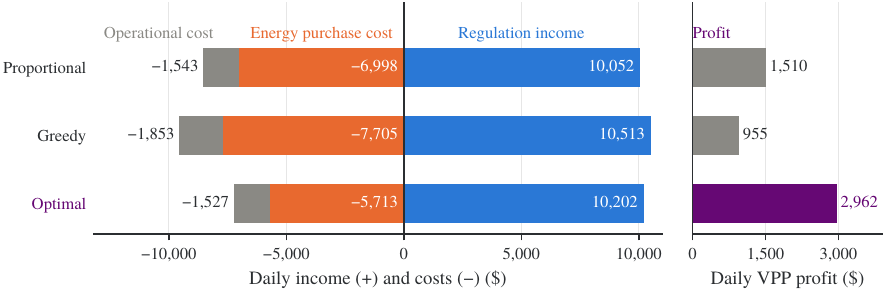}
  \caption[Market income, operating cost, and VPP profit on July 21 by disaggregation method]{Market income, operating cost, and virtual power plant (VPP) profit on July 21 under optimal, proportional, and greedy disaggregation}
  \label{fig_profit_com}
\end{figure}

\begin{figure}[!t]
  \centering
  \includegraphics[width=\textwidth]{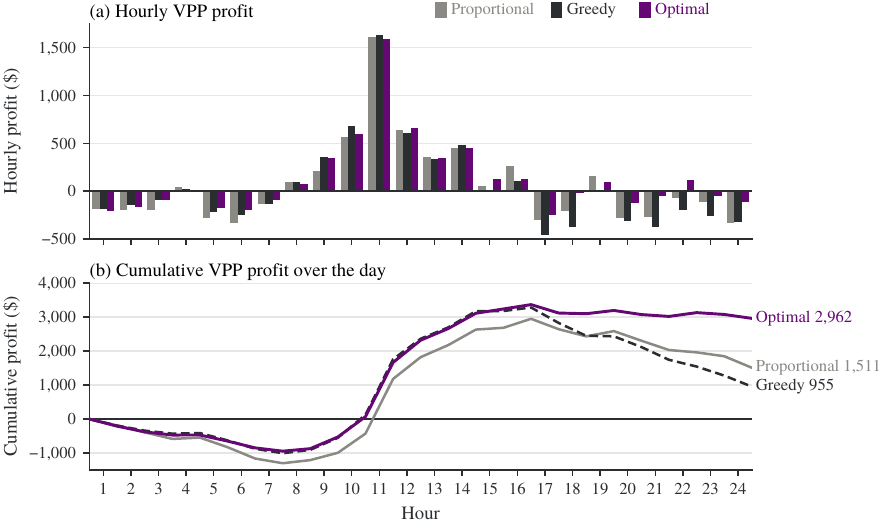}
  \caption{Virtual power plant (VPP) profit on July 21: (a) hourly profit; (b) cumulative profit}
  \label{fig_profits}
\end{figure}

Table~\ref{tab_profit_method} shows that the proposed strategy increases average market income from \$1,709 to \$2,622, a 53.3\% increase relative to proportional disaggregation. Average operating cost falls from \$1,455 to \$1,289, an 11.4\% reduction, and average VPP profit rises from \$254 to \$1,333.
Interestingly, using greedy disaggregation, i.e., merely minimizing the immediately induced operational costs, leads to the lowest VPP profits, which highlights the necessity of considering the temporal-coupling characteristics of VPP operation.
In contrast, optimal disaggregation enables maximization of the VPP profit throughout the day (Fig.~\ref{fig_profit_com}), although its profits may not be the highest in some individual time periods (Fig.~\ref{fig_profits}). 

\subsection{Regulation Deployment Results}

\begin{figure}[!t]
  \centering
  \includegraphics[width=\textwidth]{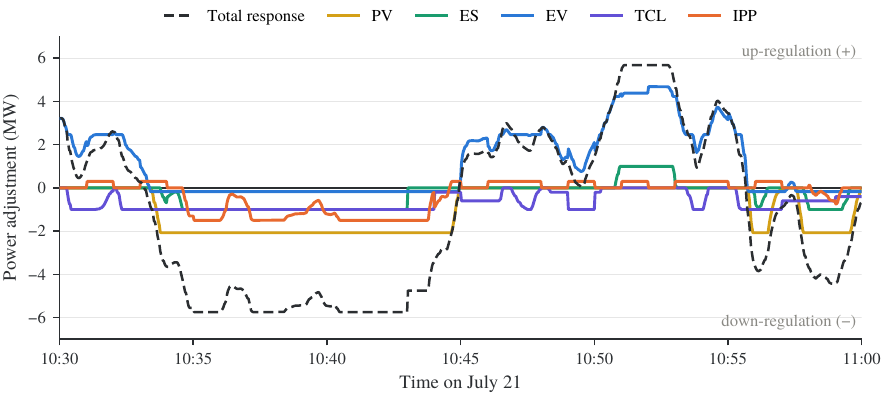}
  \caption[Regulation power by resource type under optimal disaggregation on July 21]{Regulation power allocated to each resource type under optimal disaggregation from 10:30 to 11:00 on July 21}
  \label{fig_typical_hour_optimal}
\end{figure}

\begin{figure}[!t]
  \centering
  \includegraphics[width=\textwidth]{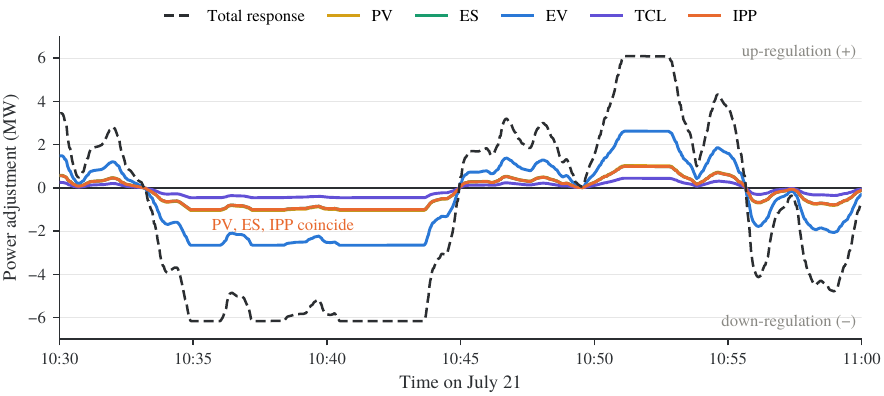}
  \caption[Regulation power by resource type under proportional disaggregation on July 21]{Regulation power allocated to each resource type under proportional disaggregation from 10:30 to 11:00 on July 21}
  \label{fig_typical_hour_prop}
\end{figure}

Figs.~\ref{fig_typical_hour_optimal} and~\ref{fig_typical_hour_prop} show the disaggregation results for 10:30--11:00 on July 21.
The optimal disaggregation strategy (Fig.~\ref{fig_typical_hour_optimal}) prioritizes the use of low-cost resources.
For example, after receiving an up-regulation (injecting energy into the grid) command, EVs respond first, followed by ES.
Note that the power adjustment here is relative to the baseline power; therefore, the \textit{increase} in EV power does not necessarily imply discharging: it may be a reduction in charging power, which does not incur additional costs.
On the other hand, under proportional disaggregation (Fig.~\ref{fig_typical_hour_prop}), all resources respond to regulation signals simultaneously, regardless of their cost characteristics.

\begin{figure}[!t]
  \centering
  \includegraphics[width=0.7\textwidth]{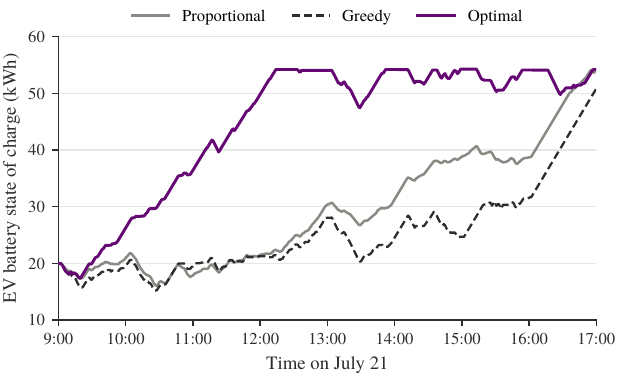}
  \caption{Electric vehicle (EV) state of charge during regulation provision on July 21}
  \label{fig_state_ev}
\end{figure}

\begin{figure}[!t]
  \centering
  \includegraphics[width=0.7\textwidth]{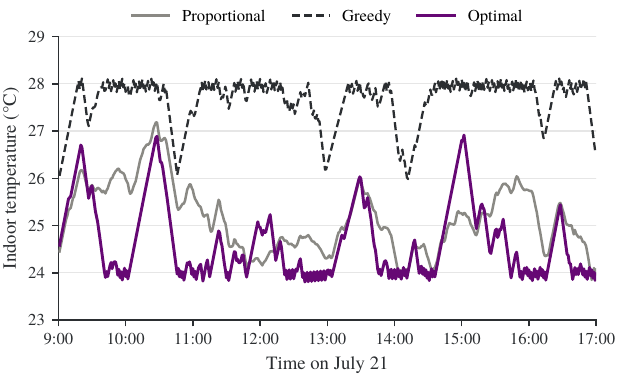}
  \caption{Indoor temperature of thermostatically controlled loads (TCLs) during regulation provision on July 21}
  \label{fig_state_tcl}
\end{figure}

\begin{figure}[!t]
  \centering
  \includegraphics[width=0.7\textwidth]{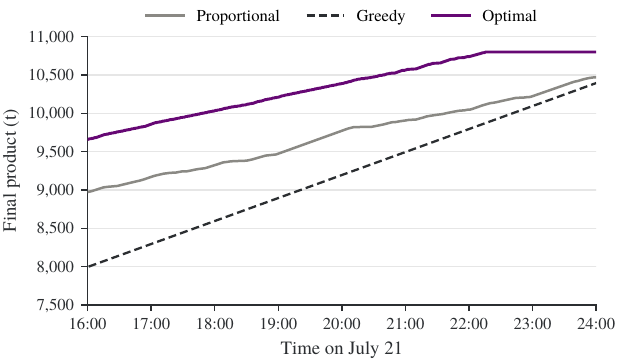}
  \caption{Final-product quantities of industrial production processes (IPPs) during regulation provision on July 21}
  \label{fig_state_ipp}
\end{figure}

Figs.~\ref{fig_state_ev},~\ref{fig_state_tcl}, and~\ref{fig_state_ipp} illustrate the influence of regulation deployment on the EV battery state of charge, the indoor temperature of TCLs, and the final product quantity of IPPs under different disaggregation strategies.
From these, the advantages of optimal disaggregation compared to greedy disaggregation can be intuitively understood.
From Fig.~\ref{fig_state_ev}, under greedy disaggregation, the EV battery state of charge remains at a low level at 16:00, which is caused by frequent reduction of EV charging power.
To achieve the target state of charge before departure, the EV must maintain maximum charging power until 17:00 and is thus unable to provide regulation capacity.
Similarly, Fig.~\ref{fig_state_tcl} shows that greedy disaggregation reduces TCL power at 10:00, causing the indoor temperature to reach the set threshold by 11:00, affecting its regulation capacity at that time.
The results for IPPs (Fig.~\ref{fig_state_ipp}) are similar, where the last production line under greedy disaggregation must operate at rated power in the last few hours to achieve production targets.
This can explain the higher profits observed in the last few hours of the day under optimal disaggregation, as shown in Fig.~\ref{fig_profits}.

\subsection{Sensitivity Analysis}
\begin{figure}[!t]
  \centering
  \includegraphics[width=\textwidth]{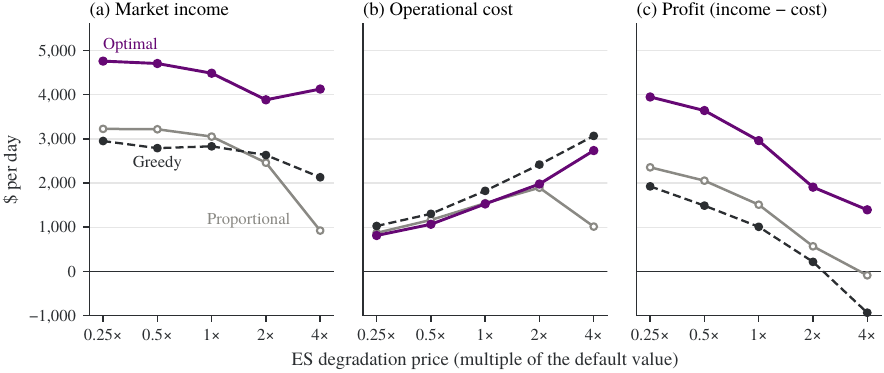}
  \caption[VPP income, cost, and profit versus the degradation price of energy storage]{Virtual power plant (VPP) market income, operating cost, and profit versus the battery degradation price of energy storage: (a) market income; (b) operating cost; (c) profit}
  \label{fig_wtr_deg}
\end{figure}

The fluctuation in battery degradation cost can impact the profitability of the VPP.
An investigation is conducted to assess the influence of these parameters on the performance of the proposed method.
Fig.~\ref{fig_wtr_deg} illustrates the VPP's market income and operational costs corresponding to different degradation prices of ES, ranging from 0.25 to 4 times the default value.
As the degradation price escalates, the market income diminishes across all methods.
Across the tested degradation-cost range, the proposed approach retains higher net performance than the alternatives.

\subsection{Computation Time}

\begin{figure}[!t]
  \centering
  \includegraphics[width=0.8\textwidth]{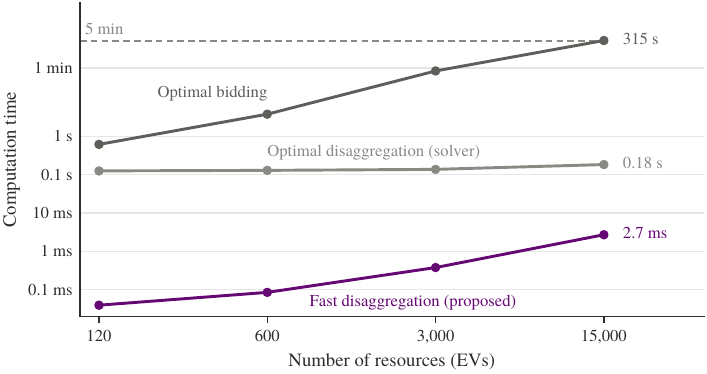}
  \caption[Computation time versus the number of electric vehicles (EVs)]{Computation time for bidding, optimization-based power disaggregation, and fast disaggregation versus the number of electric vehicles (EVs)}
  \label{fig_calculation_time}
\end{figure}

This chapter uses CPLEX (V12.4) and MATLAB (R2021a) with YALMIP to solve the optimization problems on a workstation with an Intel Core i9-10900X CPU (3.7 GHz) and 128 GB RAM.
Fig.~\ref{fig_calculation_time} shows the variation in computation time for solving the optimal bidding problem, optimal disaggregation problem, and fast disaggregation algorithm with the number of resources in the VPP.
Without loss of generality, the number of EVs is increased to scale up the number of resources, since EVs are typical resources with small individual capacities but large quantities.
In the test, the computation time for the original optimal bidding problem exceeded 5 minutes after reaching 15,000 EVs, which is unacceptable for online response to regulation signals but still satisfactory for determining the bids.
The optimal disaggregation problem reduced the computation time to less than 1 second but relies on powerful solvers and computing platforms.
The fast disaggregation algorithm, on the other hand, does not rely on these resources and further reduces the computation time by several orders of magnitude.
The results verify that the proposed method is in line with the response time requirements for providing regulation services.

\section{Discussion}\label{sec_conclusion_chap05}

Bidding determines the regulation capacity sold by an aggregator, while power disaggregation determines the resource costs and future flexibility associated with delivering that capacity. The PJM-based case shows that accounting for this coupling can improve VPP profit. The improvement arises through three mechanisms:

First, the method fully exploits the heterogeneity among different resources.
These differences include not only different arrival and departure times and charging demands of EVs, but also different parameters such as (dis)charging efficiency and battery degradation costs, as well as the thermal inertia of TCLs and the production coupling of industrial production lines.
Intuitively, the proposed method can prioritize the use of low-cost resources.
Existing methods, especially the commonly used proportional disaggregation, fail to take advantage of this heterogeneity.

Second, the method achieves correspondence between the bidding model and the power disaggregation model.
In other words, the consideration of power disaggregation in bidding is consistent with the actual power disaggregation method in regulation deployment.
In the greedy algorithm and heuristic weight methods, proportional disaggregation is used as an assumption in the bidding problem, while the actual power disaggregation is based on degradation cost or heuristic weights.
This mismatch leads to logical inconsistency and loss of VPP profits.

Third, the method balances short-term and long-term profits.
The greedy algorithm optimizes short-term profits without considering long-term impacts and therefore necessarily deviates from the optimal.
In the proposed method, the impact of power disaggregation on future profits is reflected through Lagrange multipliers (shadow prices), thereby achieving theoretical optimality.
The optimal operation strategy for the VPP constructed in this chapter, which accounts for the uncertainty of grid commands and the temporal coupling, can substantially improve the operational profit of the VPP compared to the traditional proportional allocation and greedy strategies.

\section{Summary}

This chapter addresses the coupling, computational complexity, and response-time requirements that arise when a load aggregator containing industrial users jointly determines its bidding and power-disaggregation strategies. It constructs a co-optimization framework that coordinates industrial users and related demand-side resources across market and real-time control timescales.

In the bidding stage, the constraint dimension reduction method proposed in Chapter 4 is employed to reduce the high-dimensional complex industrial load model to concise linear constraints and embed them in the market interaction model, thereby resolving the computational complexity issue arising from directly using the precise model.
In the disaggregation stage, the Lagrange multipliers from the bidding problem are used as parameters, and power-disaggregation decisions are made with the precise process model. A tailored fast algorithm uses these models to recover executable decisions in real time.
Shadow prices link the two stages. Each price records the marginal value of preserving a resource state for future periods. Once calculated during bidding, these values can be reused for subsequent real-time allocations instead of resolving the full multi-period problem.
It is theoretically proved that for any given regulation signal, the optimal solution to the power disaggregation problem corresponds to the optimal solution of the corresponding variables in the bidding problem.

Using reduced constraints for bidding and the precise model for execution keeps the market problem solvable without treating the approximate schedule as a final control command.
Using PJM real-time energy prices and RegD signals, the case study shows that the proposed strategy increases average market income by 53.3\% and reduces operating costs by 11.4\% relative to proportional disaggregation. The representative comparison also gives a 40\% reduction in interaction costs relative to the simplified strategy used for that assessment.
The full bidding problem takes more than five minutes when the portfolio reaches 15,000 EVs, which is acceptable for periodic bidding but not for online response. The solver-based disaggregation problem takes less than one second, and the algebraic algorithm further reduces the calculation to the millisecond level for portfolios with tens of thousands of resources while reproducing the joint model's allocation under the chapter's assumptions.

This separation between periodic optimization and fast execution allows aggregators to account for resource costs and temporal coupling while meeting real-time response requirements.
The same decomposition may also be useful for other electricity-market participants that must translate a large offline optimization problem into fast online decisions.
Future research should focus on incorporating other realistic operational constraints (such as resource ramping rate limitations) and exploring profit allocation strategies among aggregator participants within the proposed framework, which will further enhance the usefulness and efficacy of the method in real-world deployment.

At the system level, this chapter focuses on profit maximization by an individual aggregator and explains the associated solvable structure and shadow-price mechanism. When many industrial loads operate under similar information architectures and bidding rules, however, their strategies may converge and create synchronized power variations over short intervals. Such collective behavior can increase net-load ramping and reduce regional voltage- and frequency-security margins.

In practice, these risks can be mitigated through market competition, capacity-admission limits, performance assessment, network-constrained clearing, and decentralized decision-making. Future work could incorporate system sensitivities and chance constraints into aggregator decisions or coordinate differentiated response priorities with system operators.


\chapter{Conclusion and Future Work}

High shares of variable renewable energy increase the flexibility needed to balance power-system supply and demand. Industrial users can provide flexibility at large scale, but their production constraints differ from generator models and their internal parameters are often unavailable to aggregators. This dissertation uses smart meters as the external measurement interface and maps industrial process constraints into a concise flexibility space for electricity-market operation. The resulting chain of mechanism modeling, parameter identification, constraint dimension reduction, and optimal decision-making makes industrial-load participation computationally tractable while retaining precise process models for final execution.

Specifically, this dissertation makes the following four core contributions:

\section{Contributions}

(1) \textbf{Developing a unified, scalable modeling framework for industrial loads:}
This dissertation reformulates general-purpose models for complex industrial processes such as steel and cement production so that their nonconvex constraints can be handled more efficiently in power-system optimization. The reformulation reduces the number of binary variables and links the mathematical structures of STN and RTN. In a standard grid-interaction test case, it reduces solution time from 24 hours to 30 minutes while preserving the accuracy of the original constraints and supports coordinated optimization of 2,000 industrial users. This scale allows mechanism-based industrial constraints to enter aggregator and power-system scheduling problems that conventional formulations cannot solve within the available time.

(2) \textbf{Overcoming the difficulty of obtaining industrial equipment parameters:}
This dissertation proposes a Production Scheduling Identification (PSI) framework based on hourly smart meter data, which narrows the parameter search space by incorporating industrial process physics and cost-minimizing behavior. In the cement and steel-powder tests, the identified models reproduce the external load response of the true-parameter models with errors of 5.2\% and 8.5\%, compared with 13.4\%--19.2\% for the machine-learning baselines. The method therefore provides an external flexibility model from hourly data without requiring the facility to disclose its full private parameter set.

(3) \textbf{Describing massive numbers of user constraints through data-driven dimension reduction:}
This dissertation proposes a data-driven constraint-dimension-reduction framework that uses operating samples generated by the original industrial process model to learn least-error low-dimensional linear constraints. It reduces a steelmaking model with 10,208 integer variables to 24--48 continuous variables through data-driven inverse optimization. Across the three industrial cases, the best configurations yield errors of 3.6\%--10.3\%, within or close to the 10\%--20\% tolerances for demand-side participation. The reduced model provides a common grid-facing interface, while the original model remains available to enforce process feasibility during execution.

(4) \textbf{Co-optimizing aggregator bidding and power disaggregation:}
Building on Chapters 2--4, this dissertation develops a joint bidding and power-disaggregation model within a ``dimension reduction--recovery'' framework. The large stochastic optimization is solved during bidding, and its shadow prices summarize how current resource use affects future profit. Reusing this information allows an aggregator to allocate real-time commands among tens of thousands of resources through millisecond-level arithmetic calculations. In the representative comparison, the method reduces interaction costs by 40\% relative to a simplified strategy while retaining the solution quality of the joint optimization model.

\section{Future Research Directions}

This dissertation develops modeling and decision tools for shifting power-system operation from passive load following toward active grid--load interaction. Applying these tools in industrial settings, however, also requires attention to techno-economic feasibility and user acceptance, as well as integration with market mechanisms, system-security requirements, and the broader energy transition.

Regarding market mechanisms, future research could design demand-side participation rules suited to national and regional conditions, including capacity-cost compensation, coordination with spot markets, and market monitoring. Regarding system security, dynamic models are needed to assess how large-scale industrial responses affect frequency stability, voltage stability, and power flows, followed by corresponding operating safeguards and emergency procedures. Modeling and parameter identification should also represent actual industrial decision logic and objectives beyond electricity-cost minimization.

At the energy-transition level, a particularly important question is the role of industrial flexibility over seasonal and other long time scales. Most existing studies focus on hourly or intraday adjustment, whereas systems with very high shares of variable renewables may face seasonal energy imbalances that are more consequential than short capacity shortages. The system-level scenario analysis in Chapter 4 indicates the economic feasibility of a long-cycle operating pattern for aluminum smelting: firms use excess capacity to produce and build inventory during renewable-rich, low-price seasons, then reduce or suspend production during winter periods with tight supply and high prices. Such operation may better respect the engineering preference of heavy industrial processes for stable within-season operation while helping the power system balance supply and demand across seasons. Further work should test this transition with data from different industries and develop quantitative tools for evaluating it.

A longer-term question is how cheap but highly variable renewable electricity will change equipment design, production organization, and industrial ecosystems. Historical changes in energy sources have repeatedly reshaped modes of production. Research that connects these industrial changes to electricity-market design and system planning would help decision-makers evaluate the transition.

\bibliography{ref/refs}  

\appendix
\listoffigures           
\listoftables            
\backmatter




\begin{resume}

Ruike Lyu received his bachelor's degree in Electrical Engineering from Tsinghua University, Beijing, China, in 2021, and is currently pursuing a Ph.D. degree there. He was a one-year visiting scholar at Princeton University. His research interests include demand-side flexibility from electric vehicles, commercial buildings, and industrial loads. Ruike has been recognized for his work with several awards, including Best Paper at multiple conferences such as CEEPE 2024, PESGM 2025, and EECT 2025.

  \subsection*{Publications related to this dissertation}

  \begin{achievements}
    \item \textbf{R. Lyu}, A. Li, J. Wang, H. Luo, Y. Shen, H. Guo, E. Du, C. Kang, and J. Jenkins, ``Industrial overcapacity can enable seasonal flexibility in electricity use,'' \textit{Nature Energy}, vol. 11 (2026): 1203--1215. https://doi.org/10.1038/s41560-026-02073-y

    \item \textbf{R. Lyu}, X. Su, E. Du, H. Guo, Q. Chen, and C. Kang, ``Efficient scheduling of discrete industrial processes through continuous modeling,'' \textit{IEEE Transactions on Smart Grid}, vol. 16, no. 6 (2025): 4726-4740.

    \item \textbf{R. Lyu}, H. Guo, Q. Tang, Q. Chen, and C. Kang, ``Production scheduling identification: An inverse optimization approach for industrial load modeling using smart meter data,'' \textit{IEEE Transactions on Smart Grid}, vol. 16, no. 2 (2025): 1207-1220.
        
    \item \textbf{R. Lyu}, H. Guo, G. Strbac, and C. Kang, ``Data-driven dimension reduction for industrial load modeling using inverse optimization,'' \textit{IEEE Transactions on Smart Grid}, vol. 16, no. 3 (2025): 2695-2698.

    \item \textbf{R. Lyu}, H. Guo, K. Zheng, M. Sun, and Q. Chen, ``Co-optimizing bidding and power allocation of an EV aggregator providing real-time frequency regulation service,'' \textit{IEEE Transactions on Smart Grid}, vol. 14, no. 6 (2023): 4594-4606.
    
    \item \textbf{R. Lyu}, Y. Gu, and Q. Chen, ``Electric vehicle charging right trading: Concept, mechanism, and methodology,'' \textit{IEEE Transactions on Smart Grid}, vol. 13, no. 4 (2022): 3094-3105.
    
    \item \textbf{R. Lyu}, H. Guo, and Q. Chen, ``Approximating energy-regulation feasible regions of virtual power plants: A data-driven inverse optimization approach,'' \textit{2024 IEEE Power \& Energy Society General Meeting (PESGM)}, Seattle, WA, USA, 2024, pp. 1-5.

    \item \textbf{R. Lyu}, H. Guo, Y. Zheng, Y. Bai, and Q. Chen, ``LSTN: A linear model of industrial production process for demand response,'' \textit{2023 IEEE PES Innovative Smart Grid Technologies Europe (ISGT EUROPE)}, Grenoble, France, 2023, pp. 1-5.

  \end{achievements}

  \subsection*{Patents}

  \begin{achievements}
    \item Q. Chen, Y. Chen, K. Zheng, and \textbf{R. Lyu}, ``Method for estimating state of power based on electrochemical model of lithium-ion battery,'' US Patent App. 18/919,593, 2025.

    \item Q. Chen, Y. Chen, K. Zheng, and \textbf{R. Lyu}, ``Method for updating state of charge based on power characteristic of electrochemical model of lithium-ion battery,'' US Patent App. 18/919,638, 2025.
  \end{achievements}

\end{resume}




\end{document}